\documentclass[12pt]{article}
\usepackage{amsfonts}
\usepackage{amsmath, amssymb}
\usepackage{youngtab}
\usepackage{hyperref}
\usepackage{psfrag}
\usepackage{feynmp}
\usepackage[pdftex]{graphicx}
\usepackage{braket}
\usepackage{bm}
\usepackage{subfigure}
\usepackage[utf8]{inputenc}

\allowdisplaybreaks[1]

\Yboxdim{5pt}

\newcommand{\nn}{\nonumber}

\makeatletter
    
    \@addtoreset{equation}{section}
  \makeatother

\renewcommand{\title}[1]{\vbox{\center\LARGE{#1}}\vspace{5mm}}
\renewcommand{\author}[1]{\vbox{\center\large#1}\vspace{5mm}}

\begin{document}
\bibliographystyle{utphys}


\begin{titlepage}
\begin{center}
\vspace{5mm}
\hfill {\tt UT-Komaba-19-3}\\
\hfill {\tt IPMU19-0133}\\
\vspace{20mm}

\title{\LARGE  SUSY localization for Coulomb branch operators
\\
\vspace{2mm}
 in omega-deformed 3d $\mathcal{N} = 4$ gauge theories 
 }
 
\vspace{7mm}

Takuya Okuda$^a$ and Yutaka Yoshida$^b$

\vspace{6mm}

$^a$Graduate School of Arts and Sciences, University of Tokyo\\
Komaba, Meguro-ku, Tokyo 153-8902, Japan \\
\vskip 1mm
\href{mailto:takuya@hep1.c.u-tokyo.ac.jp}{\tt takuya@hep1.c.u-tokyo.ac.jp}

\vspace{3mm}
$^b$Kavli IPMU (WPI), UTIAS, University of Tokyo \\
 Kashiwa, Chiba 277-8583, Japan \\
\href{mailto:yutaka.yoshida@ipmu.jp}{\tt yutaka.yoshida@ipmu.jp}

\end{center}

\vspace{10mm}
\abstract{We perform SUSY localization for Coulomb branch operators of 3d $\mathcal{N}=4$ gauge theories in $\mathbb{R}^3$ with $\Omega$-deformation.
For the dressed monopole operators whose expectation values do not involve non-perturbative corrections, our computations reproduce the results of the so-called abelianization procedure.
For the expectation values of other operators and the correlation functions of multiple operators in $U(N)$ gauge theories, we compute the non-perturbative corrections due to monopole bubbling using matrix models obtained by brane construction.
We relate the results of localization to algebraic structures discussed in the mathematical literature, and also point out a similar relation for line operators in 4d $\mathcal{N}=2$ gauge theories.
For $U(N)$ (quiver) gauge theories in 3d we demonstrate a direct correspondence between wall-crossing and the ordering of operators.
}
\vfill

\end{titlepage}

\tableofcontents

\section{Introduction and summary}


A generic $\mathcal{N}=4$ supersymmetric gauge theory in the flat three dimensional (3d) spacetime has several branches of vacua.
The Higgs branch
$\mathcal{M}_{H}$ is parametrized by the vacuum expectation values (vevs) of the scalars in hypermultiplets modulo global gauge transformations. 
The topology and the metric of~$\mathcal{M}_{H}$ do not receive quantum corrections and are determined by minimizing the classical potential.
The Coulomb branch $\mathcal{M}_{ C}$,
on the other hand, 
is classically parametrized by the vevs of the scalars in vector multiplets and 
the dual photons.
The space~$\mathcal{M}_{ C}$ receives perturbative as well as non-perturbative quantum corrections.
In this paper we study the Coulomb branch~$\mathcal{M}_{ C}$ quantitatively using supersymmetric localization.

When an $\mathcal{N}=2$ subalgebra in the $\mathcal{N}=4$ supersymmetry (SUSY) algebra is chosen, an $\mathcal{N}=4$ vector multiplet decomposes into a vector multiplet and a chiral multiplet in the adjoint representation.
At the same time a BPS monopole operator, the disorder operator defined by requiring the gauge field to have a Dirac monopole singularity, becomes the bottom component of a chiral multiplet.
The quantum corrected Coulomb branch is parametrized by the vevs of the {\it Coulomb branch operators}, {\it i.e.}, certain BPS gauge invariant combinations of the vector multiplet scalars and {\it dressed monopole operators}~\cite{Aharony:1997bx,Kapustin:1999ha,Borokhov:2003yu}. 
The subalgebra picks a complex structure on $\mathcal{M}_{C}$ (and one on $\mathcal{M}_{H}$).
In a non-abelian gauge theory the vevs of monopole operators receive not just perturbative but also non-perturbative corrections due to 't Hooft-Polyakov monopoles (instantons in 3d). 
The latter make it  non-trivial to study the Coulomb branch of a non-abelian gauge theory.

Under 3d mirror symmetry \cite{Intriligator:1996ex}, which is an infrared duality between two 3d $\mathcal{N}=4$ theories, 
the quantum corrected Coulomb branch $\mathcal{M}_{C}$ is isomorphic to the Higgs branch 
of the  dual theory.
When a 3d $\mathcal{N}=4$ theory is embedded in type IIB string theory as a low energy world-volume theory on D3-branes, 
mirror symmetry is identified with the S-duality of the whole system~\cite{Hanany:1996ie}.
Although such a brane construction can sometimes be used to find the dual theory and analyze the Coulomb branch by mirror symmetry,
a first principles path integral computation of the monopole operator vevs including non-perturbative corrections in an non-abelian gauge theory has been missing.

A crucial progress in the subject was the identifications of the Coulomb branch chiral rings for various $\mathcal{N}=4$ theories in physics~\cite{Bullimore:2015lsa} and in mathematics~\cite{Nakajima:2015txa, Braverman:2016wma}.
In~\cite{Bullimore:2015lsa} the Coulomb branch chiral ring was constructed via a procedure called {\it abelianization}; the algebra $\mathbb{C} [\mathcal{M}_{C}]$ of holomorphic functions on the Coulomb branch (or equivalently the cohomology of Coulomb branch operators, or the algebra of their vevs) is embedded in a larger algebra~$\mathbb{C} [\mathcal{M}^{\text{abel}}_{C}]$ of functions that are only holomorphic in the complement of the locus where, classically, some non-abelian gauge symmetry would be restored.
One determines the ring relations of  $\mathbb{C} [\mathcal{M}^{\text{abel}}_{C}]$ using mirror symmetry and other consistency conditions.
The generators of $\mathbb{C} [\mathcal{M}_{C}]$ are 
certain Weyl invariant combinations in  $\mathbb{C} [\mathcal{M}^{\text{abel}}_{C}]$. 
Abelianization also allows one to quantize the Coulomb branch, {\it i.e.}, to promote $\mathbb{C} [\mathcal{M}_{C}]$ into a non-commutative algebra, by postulating the Heisenberg commutation relations among the generators of~$\mathbb{C} [\mathcal{M}^{\text{abel}}_{C}]$.

An arbitrary element of $\mathbb{C} [\mathcal{M}_{C}]$ is a polynomial in the generators.
Even when 
the generators of~$\mathbb{C} [\mathcal{M}_{C}]$ do not receive non-perturbative corrections, 
the vev of a general monopole operator contains non-perturbative corrections, which indirectly can be determined algebraically by the polynomial.
It is still desirable to directly compute the non-perturbative contributions.
When the vevs of the generators of $\mathbb{C} [\mathcal{M}_{C}]$ do receive non-perturbative corrections, we need an intrinsically non-perturbative method to compute them.

Closely related to 3d $\mathcal{N}=4$ theories are 4d $\mathcal{N}=2$ supersymmetric gauge theories on $S^1 \times \mathbb{R}^3$~\cite{Seiberg:1996nz}.
The Coulomb branch is parametrized by the vevs of line operators \cite{Gaiotto:2010be} wrapping the circle.
Wilson, 't Hooft, and dyonic line operators can be regarded as the 4d counterparts of the polynomials in scalars, bare monopole operators, and dressed monopole operators, respectively.
The vevs of the line operators were computed by localization in~\cite{Ito:2011ea} with an $\Omega$-deformation, {\it i.e.}, in the background where a plane $\mathbb{R}^2\subset \mathbb{R}^3$ is rotated by an angle $2\pi \lambda$ when going around the $S^1$.
When $\lambda$ is non-zero, the expectation value of a product of line operators is a non-commutative product (the so-called {\it Moyal product}) of the vevs of the operators; this realizes the deformation quantization of the Coulomb branch.%
\footnote{%
Computations in~\cite{Gaiotto:2010be} and~\cite{Ito:2011ea} were done in IR and UV formulations respectively, and the relation between them was recently studied in~\cite{Brennan:2019hzm}.
}
The localization formula for  an 't Hooft operator vev involves non-perturbative contributions, denoted by~$Z_\text{mono}$, due to monopole bubbling effects where smooth 't Hooft-Polyakov monopoles screen the magnetic charges of Dirac monopoles that define the 't Hooft operator.
The bubbling contribution $Z_\text{mono}$ was originally computed using Kronheimer's relation~\cite{Kronheimer:MTh} between monopole configurations with Dirac singularities and instantons on the Taub-NUT space.
The references~\cite{Brennan:2018yuj,Brennan:2018rcn} introduced a new method to compute $Z_\text{mono}$ in 4d theories using supersymmetric quantum mechanics.
This approach was extended in~\cite{Assel:2019iae,Hayashi:2019rpw} to study aspects of line operators in 4d $\mathcal{N}=2$ gauge theories.

The first aim of our work is to provide a first-principles derivation of the results obtained by the abelianization procedure~\cite{Bullimore:2015lsa}.
The second aim is to extend to 3d the calculation and the study of the bubbling contributions $Z_\text{mono}$ originally done in 4d.
In particular we study wall-crossing in the matrix models that compute the 3d version of~$Z_\text{mono}$.
We restrict ourselves to those $\mathcal{N}=4$ gauge theories which are built from vector and (full rather than half) hypermultiplets and which do not have a Chern-Simons term in the action.
Below we summarize the results we obtain in this paper.

We define the 3d $\Omega$-background $\mathbb{R}^2_\epsilon\times \mathbb{R}$ by a suitable scaling limit of the 4d $\Omega$-background; in particular we write down explicitly the physical Lagrangians that define the theories in the background.
In this physical set-up we compute by localization the vevs and correlators of Coulomb branch operators  inserted at a point $(0,0,s)\in \mathbb{R}^2_\epsilon \times\mathbb{R}$.
The vev of a bare monopole operator  is independent of $s$ and takes the form
\begin{align}
 \langle {V}_{{\bm B}} \rangle =\sum_{\mathfrak{m} \in \Lambda_{\mathrm{cr}}(G)+{\bm B} \atop ||\mathfrak{m}|| \le ||{\bm B}|| }
 e^{  {\bm b} \cdot \mathfrak{m} }  Z_{1 \mathchar `-\text{loop}}({\bm \varphi}, {\bm m}; \mathfrak{m} ;  \epsilon) Z_{\text{mono}} ({\bm \varphi}, {\bm m};  {\bm B} ,\mathfrak{m}; \epsilon).
\label{eq:exbaremono-intro}
\end{align}
Let us explain the new symbols that appear in~(\ref{eq:exbaremono-intro}).
We denote the one-loop determinant by~$Z_{1 \mathchar `-\text{loop}}$.
We write ${\bm \varphi}$ for the vev of a complex combination of scalars in the vector multiplet.
Similarly $\bm{b}$ is the vev of a complex combination of another scalar in the vector multiplet and a dual photon (compact scalar obtained by dualizing the low-energy abelian gauge field).
The symbol $ {\bm m}$ collectively denotes mass parameters.
The magnetic charge of the monopole operator is $\bm{B}$, which is an element of the cocharacter lattice~$\Lambda_\text{cochar}(G)$ of the gauge group.
We sum over smaller magnetic charges ${\mathfrak m}$; they are elements of the coroot lattice $\Lambda_\text{cr}(G)$ (charges of smooth monopoles) shifted by $\bm{B}$.

Upon dimensional reduction from 4d to 3d, the supersymmetric quantum mechanics of~\cite{Brennan:2018yuj,Brennan:2018rcn} reduce to matrix models.
For $U(N)$ (possibly quiver) gauge theories we use these matrix models to compute the bubbling contributions $Z_\text{mono}$ for monopole operators in 3d.
We identify the natural brane configurations, related to those in~\cite{Brennan:2018yuj,Brennan:2018rcn} by a circle compactification and a T-duality involving a Taub-NUT space, that yield the matrix models as the world-volume theories on Euclidean D1-branes.

As in~\cite{Ito:2011ea}, we  show by localization that the vev of a product of operators  in some ordering is given by the Moyal product,%
\footnote{%
The Moyal product and the Weyl transform are discussed in a very recent mathematics paper~\cite{etingof2019short}.
} 
denoted by $*$ and defined later,  of  the vevs of the individual operators in that ordering.   For example, we have for two operators
\begin{equation}
\langle \mathcal{O}_1(s_1) \mathcal{O}_2(s_2)\rangle
=
\left\{
\begin{array}{cl}
\langle \mathcal{O}_1\rangle * \langle \mathcal{O}_2\rangle  & \text{for }\quad  s_1 > s_2 , \\
\langle \mathcal{O}_2\rangle * \langle \mathcal{O}_1\rangle  & \text{for }\quad  s_1 < s_2 .
\end{array}
\right.
\end{equation}
Furthermore it is natural as in~\cite{Ito:2011ea} to consider the {\it Weyl transform} of a vev $\langle\mathcal{O}\rangle$ to obtain an operator $\widehat{\mathcal{O}}$ that acts on an appropriate Hilbert space; the Moyal products transform into operator products.
As an illuminating example, let us consider the $U(1)$ gauge theory with two hypermultiplets of charge $+1$.
This theory has monopole operators with magnetic charges given by arbitrary integers.
For such operators $ {V}_{\pm 1}$ with charges $\pm 1$, we obtain the vevs
\begin{equation}
v_\pm :=\langle {V}_{\pm 1} \rangle = e^{\pm b} \prod_{f=1}^2 (\varphi-m_f)^{1/2} \,,
\end{equation}
Here $b$
is a holomorphic combination of the dual photon $\gamma$ and  a scalar 
 in the 
 vector multiplet.
 $m_f$ ($f=1,2$) are the masses of the two hypermultiplets and $\varphi := \langle   \Phi_{ww}  \rangle$ is the vev of the relevant scalar in the vector multiplet.
The Moyal product,  and consequently their corresponding operator product, satisfy the relations
\begin{equation}
v_+ * v_- = \prod_{f=1}^2 (\varphi-m_f+\epsilon/2) \,,\qquad
 \widehat V_{+1}  \widehat V_{-1}   = \prod_{f=1}^2 (\hat \varphi-m_f+\epsilon/2) \,.
\end{equation}
 Changing the ordering has the effect $\epsilon \rightarrow -\epsilon$ in these expressions.
These relations represent the deformation quantization of the classical Coulomb branch characterized by the relation $v_+ v_- = \prod_{f=1}^2 (\varphi-m_f)$, which itself is a complex deformation of the relation $v_+ v_- =  \varphi^2$ for the orbifold $\mathbb{C}^2/\mathbb{Z}_2$.
This is isomorphic to the Higgs branch of the theory, as expected from the self-mirror property of the theory~\cite{Intriligator:1996ex}.

 As can be seen from the example above, two monopole operators $V_{\bm{B}_1}$ and $V_{\bm{B}_2}$ in the $\Omega$-background  in general do not commute.
 For a charge $\bm{B} \in \Lambda_\text{cochar}(G)$ for which an $\mathfrak{m}\in \Lambda_\text{cr}(G)$ satisfying $||\mathfrak{m}||\leq ||\bm{B}||$ necessarily satisfies $||\mathfrak{m}||= ||\bm{B}||$, the monopole operator $V_{\bm{B}}$ is unambiguously specified by $\bm{B}$.
Even if $\bm{B}_1$ and $\bm{B}_2$ are such charges, {\it i.e.}, so-called minuscule cocharacters, the sum $\bm{B}_1+ \bm{B}_2$ does not uniquely specify the monopole operator $V_{\bm{B}_1+ \bm{B}_2}$, whose vev has a non-perturbative contribution $Z_\text{mono}$ that depends on the ordering of $V_{\bm{B}_1}$ and $V_{\bm{B}_2}$ when $V_{\bm{B}_1+ \bm{B}_2}$ is resolved into their product.
In terms of the matrix models that compute the bubbling contributions, the non-commutativity of  the bare monopole operators can be interpreted as a wall-crossing phenomenon.
This is similar to the relation between wall-crossing and operator ordering found in~\cite{Assel:2019iae,Hayashi:2019rpw}.

We apply our localization results to $U(1)$ theories, $U(N)$ theories with hypermultiplets in the fundamental and the adjoint representations, a $PSU(2)$ theory, and $U(N)$ linear quiver theories.
We reproduce the relations among generators and their quantization found earlier by abelianization~\cite{Bullimore:2015lsa}.
An important difference from abelianization is that actual SUSY localization yields explicit functions as quantized chiral ring elements, and the non-commutative structure is governed by Moyal multiplication, as explained above.
The relation with abelianization involves the so-called Weyl transform.
We also revisit the localization results for 4d $\mathcal{N}=2^*$ theory with gauge group $U(N)$; we point out that the generators of the spherical double affine Hecke algebra in the functional representation~\cite{Francesco2017aa} can be identified with the Wilson-'t Hooft line operators with magnetic charges given by minuscule coweights.

This paper is organized as follows. In Section~\ref{sec:set-up} we define the physical set-up in which we compute the vevs and the correlators of Coulomb branch operators.
In particular we write down the physical Lagrangians that define the 3d gauge theories in the $\Omega$-background, and also specify the boundary conditions that define the monopole operators.
In Section~\ref{sec:localization} we perform SUSY localization to compute the vevs of Coulomb branch operators including monopole operators, in the $\Omega$-deformed $\mathcal{N}=4$ gauge theories.
In Section~\ref{sec:matrix-branes} we explain how to use brane configurations to obtain matrix models whose partition functions are identified with~$Z_\text{mono}$.
In Section~\ref{sec:quantization} we study the non-commutative structure of the theories in the $\Omega$-background, and discuss its relation to the deformation quantization of the Coulomb branch.
In Sections~\ref{sec:rank-one} through~\ref{sec:linear-quiver}, we apply our localization results to various examples, including abelian, non-abelian, and quiver gauge theories.
We conclude with discussion in Section~\ref{sec:discussion}.
Appendix~\ref{app:JK} explains the Jeffrey-Kirwan prescription that we apply in this paper.
Appendix~\ref{app:QACB} explains the derivation of an equation that appears in Section~\ref{sec:quantization}.

Note added: As we were completing the draft we learned of a related work~\cite{Assel:2019yzd} by B.~Assel, S.~Cremonesi, and M.~Renwick.  We thank them for agreeing to coordinate submissions.

\section{Set-up}\label{sec:set-up}


In this section we explain what we mean by $\mathbb{R}^3$ with $\Omega$-deformation precisely.
We mostly follow the conventions of \cite{Dedushenko:2016jxl,Dedushenko:2017avn}.
For Euclidean spacetime indices we write $\mu,\nu,\ldots = 1,2,3$.
There are three types of $SU(2)$ doublet indices: $\alpha,\beta,\ldots=1,2$ for the Lorentz group $SU(2)_\text{rot}$, $a,b,\ldots=1,2$ for R-symmetry $SU(2)_H$, and $\dot a,\dot b,\ldots=1,2$ for another R-symmetry $SU(2)_C$.
We use the antisymmetric tensors $\varepsilon^{\alpha \beta}$, $\varepsilon^{a b}$, $\varepsilon^{\dot a \dot b}$, $\ldots$, $\varepsilon_{\alpha \beta}$, $\varepsilon_{a b}$, and $\varepsilon_{\dot a \dot b}$ (with $\varepsilon^{12}=-\varepsilon_{12}=1$) to raise and lower $SU(2)$ doublet indices ($\psi^\alpha = \varepsilon^{\alpha\beta}\psi_\beta$, $\lambda_{\dot a} = \varepsilon_{\dot a \dot b} \lambda^{\dot b}$, etc.).
Sometimes contracted indices are omitted with the convention $\psi\chi= \psi^\alpha  \chi_\alpha$.
The gamma matrices are taken to be the usual Pauli matrices: $\gamma_\mu=\sigma_\mu$.
The gauge field enters the covariant derivative as in $D_\mu = \partial_\mu - iA_\mu$.

\subsection{Flat space without $\Omega$-deformation}

In flat space without $\Omega$-deformation, the SUSY transformations of the vector multiplet fields are
\begin{equation} \label{SUSY-flat-vector}
\begin{aligned}
\delta_{\xi} A_{\mu} &= \textstyle \frac{i}{2} \xi^{a\dot{b}}\gamma_{\mu}\lambda_{a\dot{b}} \,, \\
\delta_{\xi} \lambda_{a\dot{b}} &= \textstyle - \frac{i}{2}\varepsilon^{\mu\nu\rho}\gamma_{\rho}\xi_{a\dot{b}}F_{\mu\nu} - D_a{}^c\xi_{c\dot{b}} -i\gamma^{\mu}\xi_a{}^{\dot{c}} D_{\mu}\Phi_{\dot{c}\dot{b}}  + \frac{i}{2}\xi_{a\dot{d}} [ \Phi_{\dot{b}}{}^{\dot{c}}, \Phi_{\dot{c}}{}^{\dot{d}}] \,, \\
\delta_{\xi}\Phi_{\dot{a}\dot{b}} &= 
\frac{1}{2} (\xi^c{}_{\dot{a}}\lambda_{c\dot{b}} + \xi^c{}_{\dot{b}}\lambda_{c\dot{a}} )
 \,,\\
\delta_{\xi} D_{ab} &= -i D_{\mu}(\xi_{(a}{}^{\dot c}\gamma^{\mu}\lambda_{b)\dot c}) + i [\xi_{(a}{}^{\dot{c}}\lambda_{b)}{}^{\dot{d}}, \Phi_{\dot{c}\dot{d}}] \,. 
\end{aligned}
\end{equation}
 Here indices in round parentheses  are symmetrized.
Two SUSY transformations with bosonic parameters $\xi$ and $\widetilde \xi$ anti-commute to
\begin{equation}
\{\delta_{\xi}, \delta_{\widetilde \xi} \} =  \mathcal{L}_{v}   + \mathcal{G}_{\Lambda} \,,
\end{equation}
where $\mathcal{L}_v$ is the gauge covariant Lie derivative%
\footnote{\label{footnote-Lie-der}
We have $\mathcal{L}_{v} A_\mu = v^\nu F_{\nu \mu}$ for the gauge field and $\mathcal{L}_{v} \Phi = v^\mu D_\mu \Phi$ for a scalar $\Phi$.
We note that in \cite{Dedushenko:2016jxl} $\mathcal{L}_v$ denotes the gauge non-covariant Lie derivative: $\mathcal{L}_v^\text{there}\Phi= v^\mu \nabla_\mu \Phi$, etc.
}
with respect to the vector field
\begin{equation}
v^\mu:= i \widetilde \xi{}^{a \dot a} \gamma^\mu \xi_{a \dot a} 
= i  \xi{}^{a \dot a} \gamma^\mu \widetilde \xi_{a \dot a} 
\end{equation}
and $\mathcal{G}_{\Lambda}$ is the infinitesimal gauge transformation%
\footnote{%
\label{footnote:GLambda-def}%
The definition is such that $\mathcal{G}_{\Lambda} q^a = i \Lambda q^a$,  $\mathcal{G}_{\Lambda}A_\mu = D_\mu \Lambda $.
} 
with parameter
\begin{equation}
\Lambda =   \widetilde{\xi}^{a\dot a} \xi_a{}^{\dot b} \Phi_{\dot a \dot b}\in {\rm Lie}(G)  \,.
\end{equation}
Here ${\rm Lie}(G)$ denotes the Lie algebra of the gauge group $G$.

For our SUSY localization calculations, we need an off-shell completion of the SUSY transformations generated by a fixed choice of $\xi_{\alpha a \dot a}$.
Following~\cite{Hama:2012bg,Dedushenko:2016jxl} we introduce auxiliary parameters $\nu_{\alpha \check{a} \dot a}$ constrained by $\xi_{\alpha a \dot a}$ as
\begin{equation} \label{nu-constrains}
\xi^{\alpha c}{}_{\dot a} \xi_{\beta c \dot b}  = \nu^{\alpha \check c}{}_{\dot a} \nu_{\beta \check c \dot b} 
\,,\quad
\xi_a{}^{\dot c}\nu_{\check b \dot c} =0 \,.
\end{equation}
The checked indices ($\check{a},\check{b},\ldots$), distinguished from the unchecked ones, represent the doublet of an auxiliary $SU(2)_{\check R}$ symmetry \cite{Hama:2012bg}.
The SUSY transformations of the hypermultiplet fields are
\begin{equation} \label{SUSY-flat-hyper}
\begin{aligned}
\delta_{\xi} q^a &= \xi^{a\dot{b}} \psi_{\dot{b}}\,,
\\
 \delta_{\xi} \psi_{\dot{a}} &= i\gamma^{\mu}\xi_{a\dot{a}} D_{\mu}q^a  - i\xi_{a\dot{c}}\Phi^{\dot{c}}{}_{\dot{a}}q^a
+i \nu^{\check a}{}_{\dot a}  G_{\check a} \,, \\
\delta_{\xi} \widetilde q^{a}& = \xi^{a\dot{b}} \widetilde \psi_{\dot{b}} \,,
\\
\delta_{\xi} \widetilde \psi_{\dot{a}} &= i\gamma^{\mu}\xi_{a\dot{a}} D_{\mu}\widetilde q^a  + i\xi_{a\dot{c}}\widetilde q^a\Phi^{\dot{c}}{}_{\dot{a}} +i \nu^{\check a}{}_{\dot a} \widetilde G_{\check a}\,,
\\
\delta_{\xi} G_{\check a} & =  \nu^{\alpha}{}_{\check a}{}^{ \dot a}\left( \gamma^\mu{}_\alpha{}^\beta D_\mu \psi_{\beta\dot a}+ \Phi_{\dot a}{}^{\dot b} \psi_{\alpha\dot b}  + \lambda_{\alpha  b\dot a} q^b \right) \,,
\\
\delta_{\xi} \widetilde{G}_{\check a} & =  \nu^{\alpha}{}_{\check a}{}^{ \dot a}\left( \gamma^\mu{}_\alpha{}^\beta D_\mu \widetilde \psi_{\beta \dot a}- \widetilde\psi_{\alpha\dot b} \Phi^{\dot b}{}_{\dot a}  - \widetilde{q}^b  \lambda_{\alpha  b\dot a} \right) \,.
\end{aligned}
\end{equation}
 Here  $q_a$, $\psi_{\dot{a}}$, and $G_{\check a} $ take values in a representation $\mathcal{R}$ of  the Lie algebra of  the gauge group~$G$. 
 Fields~$\widetilde q^{a}$, $\widetilde \psi_{\dot{a}}$, and $\widetilde G_{\check a}$ take  values in the complex conjugate representation $\mathcal{R}^*$.

\subsubsection{Topological twist and Coulomb branch operators}
\label{sec:twist-Coulomb}

We are interested in SUSY localization with Coulomb branch operators.
Let $\Phi_{\dot a \dot b} $ ($\dot a, \dot b=1,2$) be the scalars in the vector multiplet. 
The most basic example of a Coulomb branch operator is a gauge invariant function\footnote{%
For example, if $G$ is represented  by matrices $\mathrm{Tr} (\Phi_{\dot a \dot b} v^{\dot a} v^{\dot b} )^k$  is a gauge invariant function of  $\Phi_{\dot a \dot b} v^{\dot a} v^{\dot b}$.
} 
of the combination 
\begin{equation}
\Phi_{\dot a \dot b} v^{\dot a} v^{\dot b} \,,
\end{equation}
where $v^{\dot a}$ is a non-zero vector in the doublet of $SU(2)_C$.
We see from (\ref{SUSY-flat-vector}) that the operator is invariant if the SUSY parameters satisfy the condition
\begin{equation} \label{C-op-cond-v}
\xi_{\alpha a \dot a} v^{\dot a} =0 \,.
\end{equation}
For our purposes {\it Coulomb branch operators} are the local gauge invariant operators that are invariant under the SUSY transformation whose parameter satisfies $\eqref{C-op-cond-v}$ for fixed $v^{\dot a}$.

A simple way to satisfy~(\ref{C-op-cond-v}) is to let $\xi_{\alpha a \dot a}$ take the form
\begin{equation}\label{xi-M-v}
\xi_{\alpha a \dot a}= M_{\alpha a} v_{\dot a} 
\end{equation}
for a $2\times 2$ matrix $M_{\alpha a}$, normalized so that $\det (M_{\alpha a}) =1$.
A choice of $M_{\alpha a}$ breaks the $SU(2)_\text{rot}\times SU(2)_H$ symmetry to the stabilizer subgroup.
We choose
\begin{equation} \label{M-delta}
M_\alpha{}^a =  \delta_\alpha^a \,, \qquad (M_{\alpha a} =- \varepsilon_{\alpha a} \,, \ M^{\alpha a} = \varepsilon^{\alpha a})
\end{equation}
so that the stabilizer is isomorphic to $SU(2)$.%
\footnote{%
Conversely, if the stabilizer is isomorphic to $SU(2)$ the matrix $M_\alpha{}^a$ equals $\delta_\alpha^a$ up to a rescaling and an $SU(2)_\text{rot}$ (or $SU(2)_H$) rotation.
}
This choice corresponds to the mirror of the Rozansky-Witten topological twist~\cite{Rozansky:1996bq}, or equivalently to the dimensional reduction of the Donaldson-Witten twist~\cite{Witten:1988ze}, and has been widely studied in recent years.
The Coulomb branch operators as defined by the condition~(\ref{C-op-cond-v}) are observables of this topological field theory.

When dealing with hypermultiplets, we need $\nu_{\alpha \check{a} \dot a}$ that satisfy the constraints~(\ref{nu-constrains}), which can be solved by the ansatz
\begin{equation}
\nu_{\alpha \check{a} \dot a} = N_{\alpha \check{a}} v_{\dot a}
\end{equation}
with the $2\times 2$ matrix $ N_{\alpha \check{a}} $ normalized as $\det(N_{\alpha \check a}) = \det (M_{\alpha a})=1$.
We choose
\begin{equation}
N_\alpha{}^{\check a} =  \delta_\alpha^{\check a} \,,
\end{equation}
so that we have in particular $N^\alpha{}_{\check a} M_\alpha{}^a =  \delta_{\check a}^a$.%
\footnote{%
We lose little generality if we restrict ourselves to the stabilizer subgroup of $\delta_{\check a}^a$ in $SU(2)_H\times SU(2)_{\check H}$; we can then replace checked indices $\check{a}, \check{b},\ldots $ with unchecked ones $a, b, \ldots$.
}

\subsection{Flat space with $\Omega$-deformation}\label{sec:flat-omega}

\subsubsection{SUSY transformations}

In this subsection we use the symbol $ \delta_\xi $ to denote the flat space SUSY transformations~(\ref{SUSY-flat-vector}).
It should be understood that the SUSY parameters $\xi_{\alpha a \dot a}$ are specialized as in~(\ref{xi-M-v}).
We wish to study the flat space theory deformed by the vector field%
\footnote{%
More generally, we may take $V^\mu = \epsilon \varepsilon^{\mu\nu\rho} n^\nu x^\rho$, where $n^\mu=- \frac{1}{2} (\gamma^\mu)^{\alpha\beta} M_\alpha{}^a h_{ab} M_\beta{}^b$.
For $n^\mu$ to be a unit vector we need that $h_a{}^b h_b{}^c =\delta_a^c$.
}
\begin{equation}
V = \epsilon \partial_\phi =\epsilon  ( x^1\partial_2 - x^2 \partial_1)=:\epsilon \, \varepsilon^{\mu\nu\rho} n^\nu x^\rho \partial_\mu \,,
\end{equation}
where $x^1+ix^2=r e^{i\phi}$, $n^\mu=\delta^\mu_3$.

Our $\Omega$-deformation is obtained from the four-dimensional $\mathcal{N}=2$ theory on $S^1\times \mathbb{R}^3$ by taking the radius of $S^1$ to be small.
The deformation breaks the R-symmetry $SU(2)_H\times SU(2)_C$ to the subgroup $U(1)\times U(1)$, which is the stabilizer of%
\footnote{%
Our choices of $h_a{}^b$ and $\overline{h}{}^{\dot a}{}_{\dot b}$ are different from those of~\cite{Dedushenko:2016jxl,Dedushenko:2017avn}, but are related to them by conjugation.
}
\begin{equation}\label{h-hbar-choice}
h_a{}^b \equiv \text{diag}(1,-1) \in \mathfrak{su}(2)_H
\,, \qquad
\overline{h}{}^{\dot a}{}_{\dot b} \equiv \text{diag}(1,-1) \in \mathfrak{su}(2)_C \,.
\end{equation}

Let us assume that the SUSY parameter $\xi_{\alpha a \dot a}$ is given as~(\ref{xi-M-v}).
The $\Omega$-deformed SUSY transformation $\delta_\Omega$ is given by
\begin{equation} \label{delta-Omega-vec}
\begin{aligned}
\delta_\Omega A_{\mu} &= \delta_\xi A_{\mu} \,, \\
\delta_\Omega \Phi_{\dot{a}\dot{b}} &= \delta_\xi \Phi_{\dot{a}\dot{b}}
+ \overline{h}_{\dot a\dot b} V^\mu \delta_\xi A_\mu
\,,     \\
\delta_\Omega  \lambda_{\alpha a \dot a} &= \delta_\xi \lambda_{\alpha a \dot a} + i  \gamma^\mu{}_\alpha{}^\beta  \overline{h}_{\dot a}{}^{\dot b} \xi_{\beta a \dot b}   V^\nu F_{\nu\mu }
\\
&\qquad -\frac12 \xi_{\alpha a \dot a } V^\mu D_\mu \Phi_{\dot b \dot c} \overline{h}{}^{\dot b\dot c} +\xi_{\alpha a \dot b} V^\mu D_\mu \Phi_{\dot c \dot a} \overline{h}{}^{\dot b\dot c} \,, \\
\delta_\Omega  D_{ab} & = \delta_\xi D_{ab}+
\left(
 \frac{1}{2}\xi_a{}^{\dot a} \overline{h}_{\dot a}{}^{\dot b}  \Big(   \delta_b^c \mathcal{L}_{V}
+ \frac{i}{2}\epsilon    h_b{}^c \Big ) \lambda_{ c \dot b} + ( a\leftrightarrow b)
\right)
\end{aligned}
\end{equation}
for the vector multiplet, and by
\begin{equation} \label{delta-Omega-hyp}
\begin{aligned}
\delta_\Omega q^a &= \xi^{a\dot{b}} \psi_{\dot{b}} \,,
\\
\quad \delta_\Omega  \psi_{\dot{a}} 
&=
i\gamma^{\mu}\xi_{a\dot{a}} D_{\mu}q^a  - i\xi_{a\dot{c}}
(\delta^a_b \Phi^{\dot{c}}{}_{\dot{a}} -i \delta^a_b \overline{h}{}^{\dot c}{}_{\dot a}  V^\mu D_\mu
-\frac{\epsilon}{2} h^a{}_b  \overline{h}{}^{\dot c}{}_{\dot a} ) q^b +i \nu^a{}_{\dot a} G_a  \,,
\\
\delta_\Omega  \widetilde q^{a} & = \xi^{a\dot{b}} \widetilde \psi_{\dot{b}}\,,
\\
\delta_\Omega  \widetilde \psi_{\dot{a}} & =  i\gamma^{\mu}\xi_{a\dot{a}} D_{\mu}\widetilde q^a  
+ i\xi_{a\dot{c}} (\widetilde q^a \Phi^{\dot{c}}{}_{\dot{a}} -i  \overline{h}{}^{\dot c}{}_{\dot a} V^\mu D_\mu\widetilde q^a
-\frac{\epsilon}{2} h^a{}_b  \overline{h}{}^{\dot c}{}_{\dot a}  \widetilde q^b ) +i \nu^a{}_{\dot a} \widetilde G_a \,,
\\
\delta G_a & =  \nu^{\alpha}{}_ a{}^{ \dot a}\left( \gamma^\mu{}_\alpha{}^\beta D_\mu \psi_{\beta\dot a}+ \Phi_{\dot a}{}^{\dot b} \psi_{\alpha\dot b}  -i \overline{h}_{\dot a}{}^{\dot b} \mathcal{L}_{V}  \psi_{\alpha\dot b}   + \lambda_{\alpha  b\dot a} q^b \right) \,,
\\
\delta \widetilde{G}_a & =  \nu^{\alpha}{}_a{}^{ \dot a}\left( \gamma^\mu{}_\alpha{}^\beta D_\mu \widetilde \psi_{\beta \dot a}- \widetilde\psi_{\alpha\dot b} \Phi^{\dot b}{}_{\dot a}  -i \overline{h}_{\dot a}{}^{\dot b} \mathcal{L}_{V} \widetilde \psi_{\alpha\dot b} - \widetilde{q}^b  \lambda_{\alpha  b\dot a} \right) \,
\end{aligned}
\end{equation}
for the hypermultiplet.
These transformations deformed by $\epsilon$ coincide with the flat space SUSY transformations on $S^1\times\mathbb{R}^3$ of~\cite{Ito:2011ea} in the limit $R\rightarrow 0$ with identification $\epsilon={ -} \lambda/R$.  Here $R$ is the radius of $S^1$ and $\lambda$ is the $\Omega$-deformation parameter on $S^1 \times \mathbb{R}^3$.
If desired it is possible to give more covariant expressions by field redefinitions corresponding to the topological twist.

Let us define%
\footnote{%
More generally, we define $\check{h}_{\check{a}}{}^{\check{b}} :=N^\alpha{}_{\check a} M_{\alpha}{}^a h_a{}^{b} M^\beta{}_b N_\beta{}^{\check{b}}\in\mathfrak{su}(2)_{\check H}$.
}
\begin{equation}
\check{h}_{\check{a}}{}^{\check{b}} :=\text{diag}(1,-1) \in\mathfrak{su}(2)_{\check H} \,.
\end{equation}
The SUSY transformations square to
\begin{equation} \label{delta-Omega-squared}
\delta_\Omega^2 = \overline{h}{}^{\dot a \dot b} v_{\dot a} v_{\dot b}  \mathcal{L}_{V} + \mathcal{G}(\Lambda)  +  R_H(\Lambda_H) + R_{\check{H}}(\Lambda_{\check H}) \,,
\end{equation}
where
$ \mathcal{G}(\Lambda)=\mathcal{G}_\Lambda$, $R_H(\Lambda_H)$, and $ R_{\check{H}}(\Lambda_{\check H})$ denote the gauge,  $SU(2)_H$, and  $SU(2)_{\check H}$ transformations with parameters
\begin{equation}
\Lambda = v^{\dot a}v^{\dot b} \Phi_{\dot a\dot b} \in {\rm Lie}(G)\,,
\qquad
\Lambda_H = \frac{\epsilon}{2}    \overline{h}{}^{\dot a \dot b} v_{\dot a} v_{\dot b}   (h_a{}^b) \in \mathfrak{su}(2)_H \,,
\end{equation}
and
\begin{equation}
\Lambda_{\check H}=
 \frac{\epsilon}{2}    \overline{h}{}^{\dot a \dot b} v_{\dot a} v_{\dot b} (\check{h}{}_{\check a}{}^{\check b}) \in\mathfrak{su}(2)_{\check H} \,,
\end{equation}
respectively.%
\footnote{%
The gauge transformations act as in footnote~\ref{footnote:GLambda-def}.
We have similarly $R_H(\Lambda_H) q_a =i {\Lambda_H}_a{}^b q_b $, etc.
}

Polynomials of $\Phi_{\dot a \dot b} v^{\dot a} v^{\dot b}$ as well as monopole operators defined in Section~\ref{sec:monopole-op-def} are invariant under $\delta_\Omega$ if they are placed at points such that $x^1=x^2=0$, but with arbitrary values of $x^3$.

\subsubsection{Physical Lagrangians}

We define the $\Omega$-deformation of 3d $\mathcal{N}=4$ theories by the physical Lagrangians obtained by a twisted dimensional reduction of the 4d $\mathcal{N}=2$ Lagrangians.
These physical Lagrangians that define the $\Omega$-deformed theories depend on the matrices $h_a{}^b$, $\overline{h}{}^{\dot a}{}_{\dot b}$, and $M_{\alpha a}$, but not on the parameters $v^{\dot a}$ that enter the deformed SUSY transformations.

For the vector multiplet, the $\Omega$-deformed Lagrangian is%
\footnote{%
We have 
 $ \mathcal{L}_V \lambda_{\alpha a \dot a} =  V^\mu D_\mu \lambda_{\alpha a \dot a} +\frac{1}{4} \partial_{[\mu} V_{\nu]} (\gamma^{\mu\nu})_\alpha{}^\beta \lambda_{\beta a \dot a} $, etc.
The notion of a spinor Lie derivative is explained for example in~\cite{MR1016603}.
See also footnote~\ref{footnote-Lie-der}.
}
\begin{equation}\label{Lag-YM-Omega}
\begin{aligned}
 \mathcal{L}_\text{YM}^\Omega 
 &= 
   \frac{1}{g^2} {\rm Tr} \Big(
 F^{\mu\nu} F_{\mu\nu}
 - (D^\mu \Phi^{\dot a \dot b} + \overline{h}{}^{\dot a \dot b} V^\nu F_\nu{}^\mu)
 (D_\mu \Phi_{\dot a \dot b} + \overline{h}_{\dot a \dot b} V^\rho F_{\rho\mu})
 \\
 &
  + i \lambda^{a\dot a} \gamma^\mu D_\mu \lambda_{a\dot a} 
-     D^{ab}D_{ab}
-i \lambda^{a\dot a} [ \Phi_{\dot a}{}^{\dot b} ,\lambda_{a\dot b}] 
- \lambda^{ a \dot a} \overline{h}{}_{\dot a}{}^{\dot b}
 \left(
\delta_a{}^b \mathcal{L}_{V} + \frac{i}{2} \epsilon h_a{}^b
\right)
\lambda_{ b \dot b}
  \\
  &\quad
   - \frac14 [\Phi^{\dot a}{}_{\dot b} -i \overline{h}{}^{\dot a}{}_{\dot b}  V^\mu D_\mu,
   \Phi^{\dot c}{}_{\dot d} -i \overline{h}{}^{\dot c}{}_{\dot d}  V^\nu D_\nu]
   [\Phi^{\dot b}{}_{\dot a} -i \overline{h}{}^{\dot b}{}_{\dot a}  V^\rho D_\rho,
   \Phi^{\dot d}{}_{\dot c} -i \overline{h}{}^{\dot d}{}_{\dot c}  V^\sigma D_\sigma]
  \Big)
    \,.
\end{aligned}
\end{equation}
Here we assumed that the gauge group is simple.  Generalization to other cases is straightforward and leads to multiple gauge couplings.
The notation ${\rm Tr}$ can be thought of literally as a trace for a matrix group, but in general it is to be understood as given by a Killing form: ${\rm Tr}(\bullet\,\bullet)
= \bullet \cdot \bullet$.
Note that differential operators drop out of the last term in~(\ref{Lag-YM-Omega}) because $[V^\mu D_\mu,V^\nu D_\nu]=0$.
For the hypermultiplet, the deformed Lagrangian is
\begin{equation}\label{Lag-hyper-Omega}
\begin{aligned}
\mathcal{L}_\text{hyper}^\Omega 
&= D^\mu\widetilde{q}\,^a D_\mu q_a -i \widetilde{\psi}^{\dot a} \gamma^\mu D_\mu \psi_{\dot a} + i \widetilde{q}\,^{a}D_a{}^b q_b
\\
&
 - \frac12 \widetilde{q}\,^a \left(  \delta_a^b \Phi^{\dot a \dot b}  - i  \delta_a^b \overline{h}{}^{\dot a \dot b}  V^\mu D_\mu  + \frac{\epsilon}{2} h_a{}^b  \overline{h}{}^{\dot a \dot b} \right) 
\left( \delta_b^c \Phi_{\dot a \dot b} -i \delta_b^c  \overline{h}{}_{\dot a \dot b}  V^\nu D_\nu + \frac{\epsilon}{2} h_b{}^c  \overline{h}_{\dot a \dot b}\right) q_c
\\
&\qquad
 -i \widetilde{\psi}^{\dot a} (\Phi_{\dot a}{}^{\dot b} -i \overline{h}{}_{\dot a}{}^{\dot b}  \mathcal{L}_{V} ) \psi_{\dot b} + i\left( \widetilde{q}\,^a \lambda_a{}^{\dot b}\psi_{\dot b} + \widetilde{\psi}^{\dot a} \lambda^b{}_{\dot a} q_b \right)  + \widetilde{G}{}^a G_a\,.
\end{aligned}
\end{equation}
These Lagrangians are invariant under the SUSY transformations~(\ref{delta-Omega-vec}) and (\ref{delta-Omega-hyp}).

\subsection{Bare and dressed monopole operators}\label{sec:monopole-op-def}

To perform SUSY localization we need a workable definition of bare and dressed monopole operators.
From now on we impose the normalization condition%
\footnote{%
This means that $v^{\dot a}$ is of the form $v^{\dot a} =2^{-1/2} (\zeta^{1/2}, \zeta^{-1/2})$,  where $\zeta$ is the inhomogeneous coordinate of~$\mathbb{P}^1$.  
A  $\zeta$ chooses 
 a decomposition of the $\mathcal{N}=4$ vector multiplet to  $\mathcal{N}=2$ multiplets and fix a complex structure of the Coulomb branch. 
In later sections the symbol $\zeta$ has a different meaning (an FI parameter).
}
\begin{equation}
\overline{h}_{\dot a \dot b} v^{\dot a } v^{\dot b} = 1 \,,
\end{equation}
so that the equations we obtain are simpler.
Let us introduce a new $SU(2)_C$ doublet
\begin{equation}
w^{\dot a}:= \overline{h}{}^{\dot a}{}_{\dot b} v^{\dot b} \,.
\end{equation}
Then%
\footnote{%
The matrices $ h_a{}^b$ and $\overline{h}{}^{\dot a}{}_{\dot b} $ in~(\ref{h-hbar-choice}) satisfy the relations
$
h_a{}^b h_b{}^c = \delta_a^c$,
$\overline{h}{}^{\dot a}{}_{\dot b} \overline{h}{}^{\dot b}{}_{\dot c}  =\delta^{\dot a}_{\dot c}$.
}
\begin{equation} \label{h-decomposition}
\overline{h}{}^{\dot a}{}_{\dot b} =  - v^{\dot a} v_{\dot b} + w^{\dot a} w_{\dot b} \,.
\end{equation}

A (bare) monopole operator is defined by the singular boundary condition on the gauge field and a scalar in the vector multiplet~\cite{Borokhov:2002ib,Borokhov:2002cg,Borokhov:2003yu}.
We assume that the gauge group is compact and connected, but not necessarily simply connected.
Let ${\bm B}$ be a GNO charge~\cite{Goddard:1976qe}, or in other words an element of the cocharacter lattice $\Lambda_\text{cochar}(G)={\rm Hom}(U(1),\mathbb{T})$, where~$\mathbb{T}$ is the maximal torus of the gauge group $G$.
We will simply call ${\bm B} \in \Lambda_\text{cochar}(G)$ a cocharacter.  
To insert a monopole operator of charge ${\bm B}$ at the position $x^\mu=0$ means that in the path integral the gauge field $A_\mu$
is required to behave asymptotically near $x^\mu=0$ as
\begin{equation} \label{boundary-condition-vector}
A_\mu dx^\mu \sim - \frac{ {\bm B}}{2} \cos\theta d \phi ,
\end{equation}
where $(r,\theta,\phi)$ are the polar coordinates on $\mathbb{R}^3$.
In order to preserve some supersymmetry, an appropriate linear combination of scalars should also obey a singular boundary condition specified by $B$.
For localization with the SUSY parameter~(\ref{xi-M-v}) to be possible, the boundary condition should be compatible with the equations $\delta_\Omega \lambda_{\alpha a \dot a}=0$.
This leads to the boundary condition%
\footnote{%
The explicit expressions for the equations $\delta_\Omega \lambda_{\alpha a \dot a}=0$ will be given in~(\ref{FDPhiD})-(\ref{PhiwwPhivv}).
}
\begin{equation} \label{boundary-condition-Phi}
\Phi_{\dot a \dot b}\sim (v_{\dot a} w_{\dot b} + w_{\dot a} v_{\dot b})\frac{ \bm B}{2r} 
\end{equation}
as  $ r \rightarrow 0$.
Note that $\Phi_{ww}$ is regular.
All other fields are also taken to be regular as $r\rightarrow 0$.

The definition of a dressed monopole operator is similar to that of a dyonic line operator~\cite{Kapustin:2005py}.
A dressed monopole operator is constructed by imposing, as in the case of a bare monopole operator, the singular boundary conditions~(\ref{boundary-condition-vector}) and~(\ref{boundary-condition-Phi}) corresponding to the magnetic charge ${\bm B} \in\Lambda_\text{cochar}(G)$, and then by inserting a polynomial $f(\Phi_{ww})$ ({\it dressing factor}), invariant under the subgroup $G_{\bm B}\subset G$ that preserves ${\bm B}$, of the local field $\Phi_{ww}$ at the location of the Dirac monopole.

For gauge group $G=U(N)$, we can write the cocharacter lattice as
\begin{equation}\label{eq:cochar-UN}
\Lambda_\text{cochar}(U(N))  = \bigoplus_{i=1}^N \mathbb{Z} {\bm e}_i  \simeq \mathbb{Z}^N.
\end{equation}
The generator $\bm{e}_i$ ($i=1,\ldots,N$) corresponds to the diagonal matrix ${\rm diag}(0,\ldots,\stackrel{\text{$i$-th}}{1},\ldots,0)$ when we identify Lie($U(N)$) with the set of hermitian matrices.
A monopole operator is specified its magnetic charge up to the Weyl group action, and we will often use as a representative a ``dominant'' cocharacter~$\bm{B}$.  For $G=U(N)$ we choose a convention, motivated by the brane construction~\cite{Brennan:2018yuj},  such that a representative $\bm{B}=\sum_{i=1}^N B_i \bm{e}_i$ satisfies the inequalities $B_1\leq B_2\leq\ldots\leq B_N$.

For the gauge group~$G=\prod_{l=1}^n U(N_l)$ that will appear in a quiver gauge theory, the cocharacter lattice is given by
\begin{align} \label{cochar-quiver}
\Lambda_{\text{cochar}}  \left( \prod_{l=1}^n U(N_l) \right) 
=\bigoplus_{l=1}^n \mathbb{Z} \bm{e}_i^{(l)} \simeq \mathbb{Z}^{\sum_{l=1}^n N_l} \,.
\end{align}
 Here  $\{ \bm{e}^{(l)}_i | i=1,\ldots,N_l\}$ is an orthonormal basis of $\mathrm{Lie}(U(N_l))  \subset \mathrm{Lie}(G)$  given by diagonal matrices.

\section{SUSY localization}\label{sec:localization}


\subsection{Localization locus}\label{sec:locus}

We first assume that the parameter $v^{\dot a}$ satisfies the reality condition%
\footnote{%
This means that $v^{\dot a}$ is of the form $v^{\dot a} =2^{-1/2} (\zeta^{1/2}, \zeta^{-1/2})$ with $|\zeta|=1$.
}
\begin{equation}\label{v-reality-condition}
(v^{\dot a})^* =  w_{\dot a}.
\end{equation}
In this case we impose the reality conditions
\begin{equation} \label{reality-conditions}
(A_\mu)^\dagger = A_\mu \,,
\quad
(\Phi_{\dot a \dot b})^\dagger = - \Phi^{\dot a \dot b} \,,
\quad
(D_{ab})^\dagger = - D^{ab} \,,
\end{equation}
\begin{equation} \label{reality-conditions-hyper}
(q_a)^\dagger = \widetilde{q}\,{}^a\,,
\quad
(G_a)^\dagger = \widetilde{G}\,{}^a\,,
\end{equation}
on the bosonic fields, where the dagger ($\dagger$) denotes hermitian conjugation.%
\footnote{The matrices $h{}_a{}^b$ and $\overline{h}{}^{\dot a}{}_{\dot b}$ in~(\ref{h-hbar-choice}) have the properties
\begin{equation}
(\overline{h}_{\dot a \dot b})^* = - \overline{h}{}^{\dot a \dot b}  \quad ( \Longleftrightarrow  (\overline{h}{}^{\dot a}{}_{\dot b})^* = \overline{h}{}^{\dot b}{}_{\dot a} ) \,,
\qquad
(h_{a b})^* = - h{}^{a b}  \quad ( \Longleftrightarrow  (h{}_a{}^b)^* = h{}_b{}^a ) \,.
\end{equation}
}
The reality conditions and the boundary condition
 \eqref{boundary-condition-Phi}  are compatible when \eqref{reality-conditions} is satisfied.
We take
\begin{equation} \label{localization-action}
S_\text{loc} = \, \delta_\Omega \int d^3x
\left(
 \sum_{\alpha, a, \dot a} (\delta_\Omega\lambda_{\alpha a \dot a})^\dagger \lambda_{\alpha a \dot a}
 +
 \sum_{\alpha, \dot a} 
 \left(
  (\delta_\Omega\psi_{\alpha \dot a})^\dagger \psi_{\alpha \dot a}
  +
    (\delta_\Omega\widetilde \psi_{\alpha \dot a})^\dagger \widetilde\psi_{\alpha \dot a}
  \right)
 \right).
\end{equation}
Expressions with a dagger in $S_\text{loc}$ should be converted into expressions without one by~(\ref{reality-conditions}) before $\delta_\Omega$ in front of the integration symbol is applied.
We consider the path integral
\begin{equation}
\int \mathcal{D}(\text{fields}) \exp\left( -\int d^3x(\mathcal{L}_\text{YM}^\Omega + \mathcal{L}_\text{hyper}^\Omega) -{\rm t} \,S_\text{loc}  \right) (\delta_\Omega\text{-invariant insertions}) \,.
\end{equation}
In the limit $ {\rm t}\rightarrow +\infty$ the path integral localizes to the localization locus, {\it i.e.}, the space of solutions to the equations $\delta_\Omega (\text{fermions})=0$.
The variations $\delta_{ \Omega} \lambda_{\alpha a \dot a}$ are given in (\ref{delta-Omega-vec}), while~$\delta_{ \Omega} \psi_{\dot a}$ and~$\delta_{ \Omega} \widetilde \psi_{\dot a}$ are given in  (\ref{delta-Omega-hyp}).

When $v^{\dot a}$ does not satisfy the condition \eqref{v-reality-condition}, we do not impose the reality conditions \eqref{reality-conditions} because they are not compatible with the boundary condition \eqref{boundary-condition-Phi}  on the scalars.
We take as $S_\text{loc}$ the quantity obtained from the right hand side of~(\ref{localization-action}) by eliminating daggers formally applying~(\ref{reality-conditions}).
We assume that an appropriate contour is taken to make the path integral converge.
In the limit $ {\rm t}\rightarrow +\infty$ the saddle points that contribute to the path integral are still given the solutions to the equations $\delta_{ \Omega} \lambda_{\alpha a \dot a}=\delta_{ \Omega} \psi_{\dot a}=\delta_{ \Omega} \widetilde \psi_{\dot a}=0$.

To study the localization locus for the vector multiplet, let us define
\begin{equation}\label{Dmu-def}
\boldsymbol{D}_\mu:=\frac{1}{2} \gamma_{\mu}{}_\alpha{}^\beta M^{\alpha a} M_{\beta b} D_a{}^b \,.
\end{equation}
We define the scalars $\Phi_{vv}$, $\Phi_{vw}$, and $\Phi_{ww}$ by the decomposition
\begin{equation} \label{Phi-decomposition}
\Phi_{\dot a \dot b} =: \Phi_{vv} v_{\dot a } v_{\dot b} + \Phi_{vw} ( v_{\dot a } w_{\dot b} + w_{\dot a } v_{\dot b} ) +  \Phi_{ww} w_{\dot a } w_{\dot b}  \,.
\end{equation}
It is convenient to decompose the eight equations $\delta \lambda_{\alpha a \dot a} =0$ into six $M^{\beta a} \gamma_\mu{}_\beta{}^\alpha \delta \lambda_{\alpha a \dot a} =0$ and two $M^{\alpha a}\delta \lambda_{\alpha a \dot a} =0$, which are further decomposed into parts proportional to $v_{\dot a}$ and $w_{\dot a}$ by~(\ref{h-decomposition}) and~(\ref{Phi-decomposition}).
We obtain%
\footnote{%
An equation $ i  [   \Phi_{ww}, \Phi_{vv}       ]  +V^\mu   D_\mu (\Phi_{vv} +  \Phi_{ww}   ) =0 $ combined with $V^\mu D_\mu \Phi_{ww}=0$, a consequence of $(\ref{DmuPhiww})$, leads to~(\ref{PhiwwPhivv}).
}
\begin{equation} \label{FDPhiD}
\frac{1}{2} \varepsilon_\mu{}^{\nu\rho} F_{\nu\rho}  + D_\mu  \Phi_{vw} - i \boldsymbol{D}_\mu=0 \,,
\end{equation}
\begin{equation} \label{DmuPhiww}
D_\mu \Phi_{ww} + V^\nu F_{\nu \mu} =0 \,,
\end{equation}
\begin{equation} \label{Phiww-Phivw}
i  [   \Phi_{ww}   , \Phi_{vw}    ]  + V^\mu D_\mu  \Phi_{vw}   =0 \,,
\end{equation}
\begin{equation} \label{PhiwwPhivv}
 i  [   \Phi_{ww}, \Phi_{vv}       ]  +V^\mu   D_\mu \Phi_{vv} =0 \,.
\end{equation}

For the hypermultiplet we define
\begin{equation}
\boldsymbol{q}_\alpha := M_{\alpha a} q^a \,, 
\quad
\widetilde{\boldsymbol{q}}{}_\alpha := M_{\alpha a} \widetilde{q}\,^a \,,
\end{equation}
\begin{equation}
\boldsymbol{G}_\alpha := N_{\alpha}{}^{\check a} G_{\check a} \,, 
\quad
\widetilde{\boldsymbol{G}}_\alpha := N_{\alpha}{}^{\check a} \widetilde G_{\check a} \,, 
\end{equation}
We obtain
\begin{equation}
 \gamma^\mu D_\mu \boldsymbol{q} + \Phi_{vw}\boldsymbol{q} +  \boldsymbol{G}=0 \,,
\end{equation}
\begin{equation}
 i \Phi_{ww}\boldsymbol{q} + V^\mu D_\mu \boldsymbol{q}  + \frac{i}{2} \epsilon \, n^\mu \gamma_\mu \boldsymbol{q} =0 \,.
\end{equation}
\begin{equation}
 \gamma^\mu D_\mu \widetilde{\boldsymbol{q}} -  \widetilde{\boldsymbol{q}} \, \Phi_{vw} +  \widetilde{\boldsymbol{G}}=0 \,,
\end{equation}
\begin{equation} \label{qtPhiww}
- i \widetilde{\boldsymbol{q}} \, \Phi_{ww} + V^\mu D_\mu \widetilde{\boldsymbol{q}}  + \frac{i}{2} \epsilon \, n^\mu \gamma_\mu\widetilde{\boldsymbol{q}} = 0 \,.
\end{equation}

When we impose the reality conditions~(\ref{v-reality-condition})-(\ref{reality-conditions-hyper}) we can decompose yet again the equations~(\ref{FDPhiD})-(\ref{PhiwwPhivv}) into the hermitian and anti-hermitian parts.
In particular the hermitian part of ~(\ref{FDPhiD}) gives the Bogomolny equations
\begin{equation}\label{eq:Bogomolny}
\frac{1}{2} \varepsilon_\mu{}^{\nu\rho} F_{\nu\rho}  + D_\mu \Phi_{vw}=0 \,.
\end{equation}
Equations~(\ref{DmuPhiww}) and (\ref{Phiww-Phivw}) imply that the path integral localizes to the $\delta_\Omega^2$-invariant sub-locus of the monopole moduli space.
In the localization locus the hypermultiplet fields must all vanish.
Even when we drop the reality conditions, we assume that the choice of functional integration contours is compatible with~(\ref{eq:Bogomolny}).

\subsection{On-shell action and one-loop determinants}

In the localization calculation of the monopole operator vev, perturbative contributions are given by the on-shell value of the classical action and the one-loop determinants for the fluctuations around the saddle point field configuration without monopole bubbling.

The on-shell value of the classical action, supplemented by boundary terms~\cite{Gomis:2011pf,Ito:2011ea,Dedushenko:2017avn},
can be obtained by dimensionally reducing the corresponding 4d result in~\cite{Ito:2011ea}:
\begin{equation} \label{eq:on-shell-action}
S_\text{cl} =  \frac{8\pi}{g^2} {\rm Tr}\left( \Phi_{vw}^{(\infty)} {\bm B} \right)\,.
\end{equation}
 Here $\Phi_{vw}^{(\infty)}$ is the asymptotic value of $\Phi_{vw}$ at $|x^1|^2+|x^2|^2
 \rightarrow \infty$.

The path integral decomposes into topological sectors labeled by the asymptotic ($|x^1|^2+|x^2|^2 \rightarrow \infty $) magnetic charges $\mathfrak{m} \in \Lambda_\text{cochar}(G)$, where $\Lambda_\text{cochar}(G)$ is the cocharacter lattice of~$G$.
We claim that we should define the path integral on $\mathbb{R}^3$ by including a weight
\begin{equation}\label{GNO-weight}
\exp ( i\mathfrak{m} \cdot \gamma) \,,
\end{equation}
where $\gamma \in \mathrm{Lie}(\mathbb{T}) $ and the dot denotes the Killing form.
For an abelian gauge group this arises automatically when one dualizes compact scalars (dual photons) to gauge fields (see for example~\cite{Seiberg:1996nz}) in a  2-form field background;
the quantity~$\gamma$ is identified with the vevs of the compact scalars.
The weight combines with the on-shell action to give
\begin{equation}
e^{i \mathrm{Tr} ({\bm B } \gamma)} e^{-S_\text{cl}} =e^{\mathrm{Tr} ({\bm B}  {\bm b} )}=e^{ {\bm  B} \cdot {\bm b} } \,,
\label{eq:saddlept}
\end{equation}
with%
\footnote{%
In~\cite{Ito:2011ea} the theta term induces a mixing between parameters $a$ and $b$.
Upon dimensional reduction $R\rightarrow 0$ the mixing disappears because $a=\mathcal{O}(R)$ and $b=\mathcal{O}(1)$.
}
\begin{equation}\label{b-def}
{\bm b}:= i \gamma  - \frac{8 \pi }{g^2} \Phi_{vw}^{(\infty)} \in \mathrm{Lie}(\mathbb{T}) \otimes \mathbb{C}  \,.
\end{equation}

For the one-loop determinants, too, we take the results  of~\cite{Ito:2011ea} in $S^1 \times \mathbb{R}^3$  and perform a dimensional reduction  along  the $S^1$.
The dictionary can be obtained by comparing the squares of the localization supercharge.
SUSY transformations and the Lagrangians with $\Omega$-deformation in Section~\ref{sec:flat-omega} were obtained by relating the twist parameter $\lambda$, the $S^1$ radius $R$, and the $\Omega$-deformation parameter $\epsilon$ as
\begin{equation}
 - \lim_{R\rightarrow 0} \frac{\lambda}{R}  \quad \longleftrightarrow \quad \epsilon \,.
\end{equation}
From~(3.8) of \cite{Ito:2011ea} combined with the shift $\partial_\tau \rightarrow \partial_\tau - (\lambda/R) \partial_\phi$, we see that the supercharge-squared $Q^2$ contains $ \epsilon\partial_\phi + i a$.%
\footnote{%
To be precise, in~\cite{Ito:2011ea} the sign in the definition of $J_3$ and the sign in front of $\lambda$ in (2.12) should be flipped.
The conclusion above remains valid.
}
By comparing with~(\ref{delta-Omega-squared}) we find the dictionary
\begin{equation}
- \frac{a}{R} \quad \longleftrightarrow \quad  \Phi_{ww}^{(\infty)} =:{\bm \varphi} \in \mathrm{Lie}(\mathbb{T}) \otimes \mathbb{C}  \,.
\label{eq:VM_scalar}
\end{equation}
Here~$\Phi_{ww}^{(\infty)}$ is the asymptotic value of $\Phi_{ww}$ in the limit $|x^1|^2+|x^2|^2 \rightarrow \infty$.

We can introduce a mass parameter $\bm{m}$ as the quantity corresponding to $\bm{\varphi}$ in the non-dynamical vector multiplet for the flavor symmetry that acts on hypermultiplets.
 Let  us consider  fields in a hypermultiplet $(q_a, \psi_{\dot{a}}, G_a)$ that transform in a representation $\mathcal{R}$ of  $G$ and  in a representation $\mathcal{F}$ of the flavor symmetry group. 
The one-loop determinants for the vector multiplet and  the hypermultiplets  are given as
\begin{align}
&Z^{\text{vm}}_{1\mathchar `-\text{loop}}({\bm \varphi},\bm{B}; \epsilon)=\prod_{\alpha:\text{root} } \prod_{k=0}^{|\alpha \cdot \bm{B}|-1} 
  \left ( \alpha \cdot {\bm \varphi} + \left( \frac{|\alpha \cdot \bm{B}| }{2}-k \right) \epsilon \right) ^{-\frac{1}{2}},
\label{eq:1loopvec}
\\
&Z^{\text{hm}}_{1 \mathchar `-\text{loop}}({\bm \varphi}, {\bm m}; \bm{B}; \epsilon)= \prod_{{\rm w} \in \mathcal{R}} \prod_{\mu \in \mathcal{F}} \prod_{k=0}^{|{\rm w} \cdot \bm{B}|-1} 
    \left ( {\rm w} \cdot {\bm \varphi} +\mu \cdot {\bm m} + \left( \frac{|{\rm w} \cdot \bm{B}| -1}{2}-k \right) \epsilon \right) ^{\frac{1}{2}}, 
\label{eq:1loophy}
\end{align}
where the notation $\prod_{{\rm w} \in \mathcal{R}}$ indicates a product over weights ${\rm w}$ in the representation~$\mathcal{R}$, and 
$\prod_{\mu \in \mathcal{F}}$ a product over weights $\mu$ in the representation~$\mathcal{F}$ of the flavor symmetry group.
The notation $x \cdot y$ denotes the canonical pairing between $x$ and $y$.
When  $|\alpha \cdot \bm{B}|=0$ or $| {\rm w} \cdot \bm{B}|=0$, the product is understood as $ \prod_{k=0}^{|\alpha \cdot \bm{B}|-1} (\cdots):=1$ or $\prod_{k=0}^{|{\rm w} \cdot \bm{B}|-1} (\cdots):=1$, respectively. 
These are the dimensional reductions of the one-loop determinants on $S^1 \times \mathbb{R}^3$ computed in~\cite{Ito:2011ea}.
The total one-loop determinant is $Z_{1 \mathchar `-\text{loop}}({\bm \varphi}, {\bm m}; \bm{B} ;  \epsilon):=Z^{\text{vm}}_{1\mathchar `-\text{loop}} ({\bm \varphi}; \bm{B} ;  \epsilon)Z^{\text{hm}}_{1 \mathchar `-\text{loop}}({\bm \varphi}, {\bm m}; \bm{B} ;  \epsilon)$.

\subsection{Monopole  bubbling and $Z_\text{mono}$}\label{sec:bubbling}

We now consider the contributions from non-perturbative saddle points of the localization action~(\ref{localization-action}).
The saddle points are given by the solutions of the equations $\delta_\Omega(\text{fermions})=0$, which in particular include the Bogomolny equations~(\ref{eq:Bogomolny}).
For generic values of $\bm{\varphi}$, one can show~\cite{Ito:2011ea} that the solutions are given by the fixed points in the monopole moduli space under the action of the maximal torus of $[SU(2)_\text{rot}\times SU(2)_H]_\text{diag}$ times the gauge group.%
\footnote{%
Both in instanton counting and here, the saddle points are assumed to be invariant under the action of the whole maximal torus of the gauge group, not just under the action of a particular element $\bm\varphi$ of the Lie algebra of the gauge group.
} 
These fixed points arise in the so-called monopole bubbling locus of the moduli space and are given by abelian configurations, {\it i.e.}, configurations of  bosonic fields with values in the Cartan subalgebra, where the magnetic charge of the Dirac monopole is reduced from $\bm{B} \in\Lambda_\text{cochar} (G)$ to
\begin{equation}
\mathfrak{m} \in \Lambda_\text{cr}(G)+\bm{B}\quad \text{ with } \quad ||\mathfrak{m}||< ||\bm{B}|| \,.
\end{equation}
Physically, such configurations correspond to the screening of the magnetic charge by smooth 't Hooft-Polyakov monopoles.
We note that the coroot lattice~$\Lambda_\text{cr} (G) $, the cocharacter lattice~$\Lambda_\text{cochar} (G)$, and the coweight lattice~$ \Lambda_\text{coweight} (G)$ are related as
\begin{equation}
 \Lambda_\text{cr} (G) \subset \Lambda_\text{cochar} (G)\subset \Lambda_\text{coweight} (G) \,.
\end{equation}
Smooth 't Hooft-Polyakov monopoles that screen the Dirac monopole charge~$\bm{B}$ carry charges in~$\Lambda_\text{cr} (G)$.

In the localization calculation of non-perturbative contributions, we need the on-shell value of the classical action in the saddle point with monopole bubbling.
This is obtained by simply replacing $\bm{B}$ with $\mathfrak{m}$ in~(\ref{eq:on-shell-action}).
We also need the fluctuation determinant.
We define the monopole bubbling contribution%
\footnote{%
In some references on 't Hooft line operators such as~\cite{Gomis:2011pf,Ito:2011ea,Hayashi:2019rpw}, $Z_\text{mono}$ is called a monopole {\it screening}, rather than bubbling, contribution.
This was motivated by the usage of the term ``bubbling contribution'' in~\cite{Kapustin:2006pk,Kapustin:2007wm} as the contribution from a lower charge line operator in the decomposition of a product of line operators into a sum of irreducible ones.
}
 $Z_\text{mono}$ so that the total fluctuation determinant at the saddle point is 
\begin{equation}
Z_{1 \mathchar `-\text{loop}}({\bm \varphi}, {\bm m}; \mathfrak{m} ;  \epsilon) Z_{\text{mono}} ({\bm \varphi}, {\bm m} ;  {\bm B},\mathfrak{m} ; \epsilon) \,. 
\end{equation}

Again it is possible to dimensionally reduce the results of~\cite{Ito:2011ea} to obtain the three-dimensional version of~$Z_\text{mono}$.
In this paper, however, we take a different approach.
In Section~\ref{sec:matrix-branes} we will explain how to compute~$Z_\text{mono}$ using matrix models that can be read off from brane configurations that realize monopole screening.
In later sections we will compute $Z_\text{mono}$ explicitly for concrete examples.

\subsection{General results for SUSY localization}

We now put together the results of Sections~\ref{sec:locus}-\ref{sec:bubbling} to write down the localization formulas for the vevs of Coulomb branch operators.

We begin with Coulomb branch operators given by gauge invariant polynomials of the scalar $\Phi_{w w}$.
In the absence of a Dirac monopole singularity ($\bm{B}=0$), the one-loop determinants~\eqref{eq:1loopvec} and~\eqref{eq:1loophy} are trivial,  and the saddle point value is fixed to the asymptotic value~$\Phi^{(\infty)}_{ww}$.
The vev of 
$P(\Phi_{w w})$, where $P$ is a gauge invariant polynomial, 
 is simply
\begin{align}
\langle  P(\Phi_{w w}) \rangle  =  P(\Phi^{(\infty)}_{ww}) \,.
\label{eq:exsca}
\end{align}

For a bare monopole operator ${V}_{\bm B}$, the vev is given by
\begin{align}
 \langle {V}_{{\bm B}} \rangle =\sum_{\mathfrak{m} \in \Lambda_{\mathrm{cr}}(G)+{\bm B} \atop ||\mathfrak{m}|| \le ||{\bm B}|| }
 e^{\mathfrak{m} \cdot {\bm b} }  Z_{1 \mathchar `-\text{loop}}({\bm \varphi}, {\bm m}; \mathfrak{m} ;  \epsilon) Z_{\text{mono}} ({\bm \varphi}, {\bm m} ;  {\bm B},\mathfrak{m} ; \epsilon)\,.
\label{eq:exbaremono}
\end{align}
The one-loop determinant $Z_{1 \mathchar `-\text{loop}}({\bm \varphi}, {\bm m}; \mathfrak{m} ;  \epsilon):=Z^{\text{vm}}_{1\mathchar `-\text{loop}} ({\bm \varphi}; \mathfrak{m} ;  \epsilon)Z^{\text{hm}}_{1 \mathchar `-\text{loop}}({\bm \varphi}, {\bm m}; \mathfrak{m} ;  \epsilon)$ is given by the formulas in~(\ref{eq:1loopvec}) and~(\ref{eq:1loophy}) with $\bm{B}$ replaced by $\mathfrak{m}$. 
The symbol $\Lambda_{\mathrm{cr}}(G)$ denotes the coroot lattice of $G$, and $||\bullet||$ the norm.
We have $Z_{\text{mono}} ({\bm \varphi}, {\bm m} ;  {\bm B},\mathfrak{m} ; \epsilon)=1$ for $||\mathfrak{m}||=||{\bm B}||$.

For a massless theory on $\mathbb{R}^3$ without $\Omega$-deformation, the R-charge of a bare monopole operator,
originally computed in~\cite{Borokhov:2002cg} (see also~\cite{Gaiotto:2008ak}), can be read off from the one-loop determinants~(\ref{eq:1loopvec}) and~(\ref{eq:1loophy}).
We recall that the R-symmetry of the $\mathcal{N}=4$ theory is $\mathfrak{su}(2)_H \oplus \mathfrak{su}(2)_C$.
The boundary condition~(\ref{boundary-condition-Phi}) picks out $\Phi_{vw}$ as a special linear combination of three real scalars, and breaks $\mathfrak{su}(2)_C$ to a subalgebra $\mathfrak{u}(1)_C$,  which is part of the R-symmetry of an $\mathcal{N}=2$ subalgebra.
The R-charge of interest is the weight with respect to~$\mathfrak{u}(1)_C$.%
\footnote{%
The subalgebra $\mathfrak{u}(1)_C$ is generated by $-(v^{\dot a}w_{\dot b} + w^{\dot a} v_{\dot b})/2 \in \mathfrak{su}(2)_C$.
}
The R-charge of $\Phi_{ww}$, and hence of $\bm\varphi$, is $+1$.
The one-loop determinants~(\ref{eq:1loopvec}) and~(\ref{eq:1loophy}) imply that the R-charge of the monopole operator ${V}_{\bm B}$ is
\begin{equation}\label{eq:UV-R-charge}
\frac{1}{2}\left(
\sum_{ {\rm w} \in \mathcal{R}} | {\rm w} \cdot \bm{B}| - \sum_{\alpha:\text{root}} |\alpha\cdot \bm{B}|
\right).
\end{equation}
For the total expression \eqref{eq:exbaremono} to have a definite R-charge in the massless and $\epsilon\rightarrow 0$ limit, $Z_\text{mono}$ must also have an appropriate weight under the $\mathfrak{u}(1)_C$ transformation.

Next we consider a dressed monopole operator~${V}_{{\bm B}, f}$ with magnetic charge~${\bm B}$ and a dressing factor $f(\Phi_{ww})$, which is a polynomial invariant under the stabilizer $ G_{ \bm B}$ of $\bm B$.
Let~$W_G$ be the Weyl group of $G$.
A cocharacter~$\bm{B}$ is called minuscule when, in the irreducible representation of the Langlands dual of~$G$ corresponding to $\bm{B}$, all weights are in a single Weyl orbit.
When~${\bm B}$ is minuscule
 there is no monopole bubbling and the expectation value is given by inserting functions of~$\varphi$ to the right hand side of \eqref{eq:exbaremono}:
\begin{align}
\langle {V}_{{\bm B}, f} \rangle &=
\sum_{\mathfrak{m}\in W_G \cdot \bm{B}}  f_{\mathfrak{m}} (\bm\varphi)  e^{\mathfrak{m} \cdot {\bm b} }  Z_{1 \mathchar `-\text{loop}}({\bm \varphi}, {\bm m}; \mathfrak{m} ;  \epsilon) \,,
\label{eq:exdremono}
\end{align}
where $W_G\cdot \bm{B}$ is the Weyl orbit of $\bm{B}$, and $f_{\mathfrak{m}}$ is defined so that
$f_{\mathfrak{m}} (\bm\varphi) =f(\sigma^{-1}(\bm\varphi))$, $\mathfrak{m}=\sigma(\bm B)$ and $\sigma\in W_G$.%
\footnote{%
There is a subtlety in the precise value of the insertion because $\Phi_{ww}$ is not well-defined at the location of the Dirac monopole, as can be seen from~(\ref{theta-dependence-of-Phiww}).  Here we simply insert $f_{\mathfrak{m}} (\bm\varphi)$.
In~(\ref{eq:Uldressmono}) we use a different prescription that we will explain there.
}
In later sections we will see  in several examples that our localization formulas~\eqref{eq:exsca}, \eqref{eq:exbaremono}, and \eqref{eq:exdremono} reproduce  the known expressions for
the generators of  (quantized) Coulomb branch chiral rings.

Even for non-minuscule~${\bm B}$, the one-loop contributions from saddle points labeled $\mathfrak{m} \in \Lambda_{cr}(G)+{\bm B}$ are the same as those for bare monopole operators.
Thus~$\langle {V}_{{\bm B}, f} \rangle$ takes the form
\begin{align}
\langle {V}_{{\bm B}, f} \rangle 
&=
\sum_{\mathfrak{m}\in W_G \cdot \bm{B}} f_{\mathfrak{m}} (\bm\varphi)
 e^{\mathfrak{m} \cdot {\bm b} }  Z_{1 \mathchar `-\text{loop}}({\bm \varphi}, {\bm m}; \mathfrak{m} ;  \epsilon)
 \nonumber \\
&\qquad +\sum_{\mathfrak{m} \in \Lambda_{\mathrm{cr}}(G)+{\bm B} \atop ||\mathfrak{m}|| < ||{\bm B}|| } 
 e^{ {\bm b} \cdot \mathfrak{m} }  Z_{1 \mathchar `-\text{loop}}({\bm \varphi}, {\bm m}; \mathfrak{m} ;  \epsilon) Z_{\text{d.mono}} (f; {\bm \varphi}, {\bm m} ;  {\bm B},\mathfrak{m} ; \epsilon) .
\label{eq:exdremono2}
\end{align}
The first line is the same as in the minuscule case~\eqref{eq:exdremono}. 
The quantities~$Z_{\text{d.mono}} (f;{\bm \varphi}, {\bm m} ;  {\bm B},\mathfrak{m} ; \epsilon)$ in the second line represent bubbling contributions for a dressed monopole operator.%
\footnote{%
Though in 4d bubbling contributions for dyonic line operators are in general different from those for pure 't Hooft operators (see for example Appendix~E of~\cite{Ito:2011ea}), in 3d it seems likely that we can simply take $Z_{\text{d.mono}} (f;{\bm \varphi}, {\bm m} ;  {\bm B},\mathfrak{m} ; \epsilon)=Z_{\text{mono}}({\bm \varphi}, {\bm m} ;  {\bm B},\mathfrak{m} ; \epsilon)f(\sigma^{-1}(\bm\varphi))$ with $\sigma(\bm B)=\mathfrak{m}$, up to a possible shift of $\bm{\varphi}$ in $f$ by $\epsilon$ multiplied by a constant.
It would be interesting to clarify this point.
}

\section{Coulomb branch and its deformation quantization}\label{sec:quantization}


In Section~\ref{sec:localization} we obtained the localization formulas~\eqref{eq:exsca}, \eqref{eq:exbaremono}, and \eqref{eq:exdremono} for the vevs of Coulomb branch operators.
In this section we study how the localization formulas are related to the classical and quantized Coulomb branch chiral rings~\cite{Bullimore:2015lsa, Braverman:2016wma}.%
\footnote{%
The chiral ring for $\epsilon\neq0$ is called ``quantized'' in the sense that it is deformed to a non-commutative ring, with the deformation parameter~$\epsilon$ playing the role of the Planck constant~$\hbar$.
The commutative ring for $\epsilon=0$ is called classical, but from the QFT point of view it includes one-loop and non-perturbative corrections.} 

\subsection{Moyal product, wall-crossing, and operator ordering}
\label{sec:moyalorder}

In~\cite{Ito:2011ea} it was found for 4d $\mathcal{N}=2$ theories on $S^1\times \mathbb{R}^3$ with $\Omega$-deformation that the expectation value of a product of line operators equals the Moyal product of the vevs of the operators.
We now derive a similar relation for 3d $\mathcal{N}=4$ theories (with a Lagrangian) on $\mathbb{R}^3$ with $\Omega$-deformation; namely the expectation value of a product of Coulomb branch operators equals the Moyal product of the vevs of the operators.

Let us begin with a $U(1)$ gauge theory.
We consider the product of the polynomial~$P(\Phi_{ww})$ in the vector multiplet scalar~$\Phi_{ww}$ inserted at $x^\mu=(0,0,s)$ and a bare monopole operator ${V}_n$ inserted at $x^\mu=(0,0,0)$.

Upon localization the expectation value of the two operators is simply obtained by evaluating~$P( \Phi_{ww})$ in the unique localization saddle point of the path integral with the boundary condition that defines the monopole operator~${V}_n$.
The saddle point configuration is given by the solution to the localization equations~(\ref{FDPhiD})-(\ref{PhiwwPhivv}) for the vector multiplet.
Recall from~(\ref{eq:VM_scalar}) that~$\varphi=  \Phi_{ww}^{(\infty)} \equiv \lim_{|x^1|^2+|x^2|^2\rightarrow\infty} \Phi_{ww}(x^1,x^2,x^3)$.
The equations~(\ref{DmuPhiww}) and~(\ref{boundary-condition-vector}) imply that~$\Phi_{ww}$ in the saddle point configuration is a non-trivial function of $\theta$ given as
\begin{equation} \label{theta-dependence-of-Phiww}
\Phi_{ww} = - \epsilon \frac{n}{2} \cos\theta { +} \varphi .
\end{equation}
In particular we have
\begin{equation}\label{eq:Phiww00s}
\Phi_{ww}(0,0,s)= \left\{\
\begin{array}{ll}
\displaystyle
 \varphi - \epsilon \frac{n}{2}  \vspace{1mm}& \text{ for } s>0 , \\
\displaystyle
\varphi + \epsilon \frac{n}{2}   & \text{ for } s<0 .
\end{array}
\right.
\end{equation}
This implies that
\begin{equation} \label{eq:fV-abelian}
\left\langle P(\Phi_{ww})(s) {V}_n(0) \right\rangle =
\left\{\
\begin{array}{ll}
\displaystyle
P\left(\varphi - \epsilon \frac{n}{2}\right)  { \langle {V}_n  \rangle} \vspace{1mm} & \text{ for } s > 0 ,
\\ 
\displaystyle
P\left(\varphi + \epsilon \frac{n}{2}\right)  { \langle {V}_n  \rangle } & \text{ for } s<0 .
\end{array}
\right.
\end{equation}
The right hand side depends on the sign of $s$ but not on its magnitude.

For general functions $F$ and $G$ of $(\varphi,b)$ let us define the Moyal product $F * G$ by
\begin{align} \label{eq:Moyal-def-two-variables}
(F * G) (\varphi,b):= \exp \left[  \frac{\epsilon}{2} \left( \partial_{b} \partial_{\varphi^{\prime}}-\partial_{\varphi} \partial_{b^{\prime}} \right)  \right]
 F (\varphi,b) G (\varphi',b') |_{\varphi^{\prime} = \varphi,  b^{\prime}=b} \,.
\end{align}
We can write the relation~(\ref{eq:fV-abelian}) as
\begin{equation} 
\left\langle P(\Phi_{ww})(s) {V}_n(0) \right\rangle =
\left\{\
\begin{array}{ll}
\displaystyle
 \langle P(\Phi_{ww})\rangle * \langle  {V}_n \rangle & \text{ for } s > 0 ,
\\ 
\displaystyle
\langle  {V}_n \rangle * \langle P(\Phi_{ww})\rangle  & \text{ for } s<0 .
\end{array}
\right.
\end{equation}

We now consider an $\mathcal{N}=4$ theory with a general gauge group of rank $N$.
Following~\cite{Bullimore:2015lsa} let $\{\chi^i\}_{i=1}^N$ be a basis of the cocharacter lattice~$\Lambda_\text{cochar}(G)$ and expand
\begin{equation} \label{eq:expand-varphi-b}
\bm{\varphi}=\varphi_i \chi^i \,,
\quad
\bm{b} = b_i \chi^i \,.
\end{equation}
Let us define the matrix~$\kappa^{ij}:= \chi^i \cdot \chi^j$ that represents the Killing form and denote its inverse by $\kappa_{ij}$.
Let $F({\bm \varphi}, {\bm b})$ and $G({\bm \varphi}, {\bm b})$ be functions of~$\bm{\varphi}$ and $\bm{b}$.
The Moyal product $F * G$~\cite{Moyal:1949sk} is defined by
\begin{align} \label{eq:Moyal-def}
(F * G) ({\bm \varphi}, {\bm b}):= \exp \left[  \frac{\epsilon}{2} \sum_{i,j=1}^N \kappa_{ij}{  \left( \partial_{b_i} \partial_{\varphi^{\prime}_j}-\partial_{\varphi_i} \partial_{b^{\prime}_j} \right) }  \right]
 F ({\bm \varphi}, {\bm b}) G ({\bm \varphi}^{\prime}, {\bm b}^{\prime}) |_{{\bm \varphi}^{\prime} ={\bm \varphi}, {\bm b}^{\prime}={\bm b}} \,.
\end{align}
The sum is independent of the normalization of $\kappa_{ij}$ because it enters~$\bm{b}$ through~(\ref{b-def}).
 
We also define the Poisson bracket
\begin{align}\label{eq:Poisson-bracket}
  \{F ,G \}:=  \sum_{i,j=1}^N \kappa_{ij}  \left( \partial_{b_i} F \partial_{\varphi_j} G-\partial_{\varphi_i} F \partial_{b_j} G \right) 
\end{align}
that corresponds to the holomorphic symplectic form%
\footnote{%
HyperK\"ahler geometry implies that in equation (4.10) of~\cite{Bullimore:2015lsa}, for a general gauge group, $\kappa^{ab}$ (in their notation) should be included.
Then our~(\ref{eq:holo-symp}) is consistent with their (4.10).
}
\begin{equation}\label{eq:holo-symp}
\Omega = \sum_{i,j=1}^N \kappa^{ij} d\varphi_i \wedge d b_j \,.
\end{equation}
For small~$\epsilon$
the Moyal product reduces to the ordinary product, with the first order correction proportional to the Poisson bracket:
\begin{align}
  F *G  =F G {+} \frac{\epsilon}{2}  \{F ,G \} + O(\epsilon^2)  \,.
\end{align}

As discussed in Section~\ref{sec:bubbling} we assume, as in instanton counting~\cite{Nekrasov:2002qd}, that the localization saddle points are given by abelian field configurations where fields in the adjoint representation  take values in the Cartan subalgebra.
Then the argument above for the $U(1)$ theory implies that we have the  relation
\begin{equation} \label{op-prod-moyal-2-pt}
\langle \mathcal{O}_1(s_1)  \mathcal{O}_2(s_2) \rangle
= \left\{
\begin{array}{ll}
\langle \mathcal{O}_1  \rangle *\langle \mathcal{O}_2 \rangle & \text{ for } s_1 > s_2 , \\
\langle \mathcal{O}_2  \rangle *\langle \mathcal{O}_1 \rangle & \text{ for } s_2 > s_1 
\end{array}
\right.
\end{equation}
for two operators $\mathcal{O}_a$ ($a=1,2$) inserted at $(x^1,x^2,x^3)=(0,0, s_a)$.
This relation can be generalized further to the $\ell$-point function of Coulomb branch operators on the $x^3$-axis.
Let us consider $\ell$ Coulomb branch operators $ \mathcal{O}_a  (s_a)$ inserted at $(x^1,x^2,x^3)=(0,0, s_a)$ with $a=1,\ldots,\ell$.
We can specify the ordering of the $\ell$ operators by a permutation $\sigma\in S_\ell$:
\begin{equation} \label{s-sigma-indequalities}
s_{\sigma(1)} > s_{\sigma(2)} > \ldots > s_{\sigma(\ell)} .
\end{equation}
Then we have the relation 
\begin{align}
\langle \mathcal{O}_1  (s_1)  \mathcal{O}_2 (s_2) \cdots \mathcal{O}_\ell (s_{\ell}) \rangle
  =\langle \mathcal{O}_{\sigma(1)}  \rangle * \langle  \mathcal{O}_{\sigma(2)} \rangle * \cdots * \langle  \mathcal{O}_{\sigma(\ell)} \rangle.
\label{eq:lptmoyal}
\end{align}
Thus the $\ell$-point function~$\langle \mathcal{O}_1  (s_1)  \mathcal{O}_2 (s_2) \cdots \mathcal{O}_\ell (s_{\ell}) \rangle$ is piecewise constant, and depends only on the ordering of $s_a$'s.
We note that the Moyal product is associative: $F*(G*H)=(F*G)*H$, so that the right hand side of~(\ref{eq:lptmoyal}) is well-defined.
This demonstrates that the 3d theory reduces to a 1d topological field theory.\footnote{
 This is mirror dual to a similar reduction of the Rozansky-Witten theory with $\Omega$-deformation~\cite{Yagi:2014toa}.
}

As we will explain in Section~\ref{sec:matrix-branes}, we can sometimes realize bare monopole operators~${V}_{\bm{B}_a}$ by a brane construction.
In such a case, the locations of the operators $s_a$'s are related to the FI parameters of certain matrix models as $\zeta_a=s_{a+1}-s_a$.
We define as in~\cite{Hayashi:2019rpw}
\begin{equation}\label{eq:FIchamber}
\begin{aligned}
&
\hspace{-.5cm}
\text{{\it FI-chamber} specified by $\sigma\in S_\ell$}
\\
&:=  \text{chamber in the space $\mathbb{R}^{\ell-1}$ of $\bm{\zeta}=(\zeta_a=s_{a+1}-s_a)$ determined by (\ref{s-sigma-indequalities}).}
\end{aligned}
\end{equation}
For two bare monopole operators $\mathcal{O}_a  (s_a)$ and $\mathcal{O}_b  (s_b)$ in $\langle \mathcal{O}_1  (s_1)  \mathcal{O}_2 (s_2) \cdots \mathcal{O}_\ell (s_{\ell}) \rangle$, when their ordering ($s_a > s_b$ or $s_a< s_b$) along the $x^3$-axis changes, a discrete change may or may not occur in the $\ell$-point function.
If there is a discrete change, we say that a {\it wall-crossing} phenomenon occurs.%
\footnote{%
In a different terminology found in the literature, wall-crossing refers to a mere change of parameters across a wall, and may or may not be accompanied by a discrete change in a quantity of interest.
}

Let us consider a non-commutative ring generated by non-commutative variables $\widehat{\bm{\varphi}}=\widehat{\varphi}_i \chi^i$ and $\widehat{\bm{b}} =  \widehat{b}_i \chi^i$ satisfying the relations
\begin{equation}
[\widehat{b}_i, \widehat{\varphi}_j]  =  \epsilon \kappa_{ij}.
\end{equation}
Given a function $F$ of commutative variables $\bm{\varphi}$ and $\bm{b}$, its {\it Weyl transform}~$\widehat F$ is given by%
\begin{align}\label{eq:Moyal_Wey}
\widehat{F}
: =
\exp \left( \frac{\epsilon}{2}  \sum_{i,j=1}^N \kappa_{ij} \partial_{b_i} \partial_{\varphi_j} \right) F (\bm{\varphi},\bm{b})\Bigg |_{\bm{\varphi}\rightarrow \widehat {\bm \varphi}, \bm{b}\rightarrow \widehat{\bm b}}.
\end{align}
Here the substitution $\bm{\varphi}\rightarrow \widehat {\bm \varphi}, \bm{b}\rightarrow \widehat{\bm b}$ should be done after expanding the exponential, performing differentiations, and then writing all $\varphi_i$'s to the left and all $b_i$'s to the right, {\it i.e.}, a monomial before substitution should read $\varphi_1^{m_1}\ldots  \varphi_N^{m_N}b_1^{n_1}\ldots  b_N^{n_N}$~\cite{Moyal:1949sk}.%
\footnote{%
For $N=1$ and $\kappa_{11}=1$ we have $\widehat{\varphi b} =\widehat{\varphi}\widehat{b} + \frac{\epsilon}{2}=(\widehat{\varphi}\widehat{b}+\widehat{b}\widehat{\varphi})/2$,
$\widehat{\varphi*b} =\widehat{\varphi b - \frac{\epsilon}{2}}=\widehat{\varphi}\widehat{b}  $, $\widehat{b*\varphi} =\widehat{\varphi b + \frac{\epsilon}{2}}=\widehat{b}  \widehat{\varphi}$.
}
The Moyal product is then mapped to the non-commutative product in the ring:
\begin{equation}
\widehat{F *G} = \widehat{F}  \, \widehat{G} .
\end{equation}
Thus we can  read off the commutation relation of operators $\widehat{F}$ and $\widehat{G} $ from the Weyl transform of $F* G-G*F$ as
\begin{align}
[\widehat{F}, \widehat{G}]=\widehat{F* G}-\widehat{G*F}.
\end{align}
In the limit $\epsilon \to 0$,  the commutation relation reduces to the Poisson  bracket of $F$ and $G$;
\begin{align}
\{F,G \}=  \lim_{\epsilon \to 0} \frac{1}{\epsilon}(F* G-G*F)=   \lim_{\epsilon \to 0} \frac{1}{\epsilon}[\widehat{F}, \widehat{G}]  \Big|_{\widehat{\bm\varphi}\rightarrow \bm{\varphi},\widehat{\bm b}\rightarrow \bm{b}}
.
\label{eq:moyalpoisson}
\end{align}


\subsection{Abelianized  Coulomb branch  from SUSY localization}
\label{sec:abelCB}

We now show that the abelianized version of the quantum Coulomb branch chiral ring in~\cite{Bullimore:2015lsa}, obtained indirectly by the use of mirror symmetry, naturally arises from the Weyl transform of the vevs of the Coulomb branch operators computed directly by localization.

\subsubsection{Abelianization of $\mathcal{N}=4$  gauge theories (review)}

We begin by recalling the  construction of the Coulomb branch chiral ring in~\cite{Bullimore:2015lsa}.
In an $\mathcal{N}=4$ gauge theory with gauge group $G$ (abelian or non-abelian), the unbroken part of gauge symmetry is the maximal torus $\mathbb{T}=U(1)^{N}$ at a generic point of the Coulomb branch, where $N$ is the rank of $G$.
The vev of the complex scalar~$\bm{\varphi}$ in~\eqref{eq:VM_scalar} takes values in the complexified Cartan subalgebra $ \text{Lie} (\mathbb{T}) \otimes_{\mathbb{R}} {\mathbb{C}}$ and is expanded as $\bm{\varphi}=\varphi_i \chi^i$.
See~(\ref{eq:expand-varphi-b}).

To construct the classical version of {\it abelianized Coulomb branch chiral ring}, one introduces formal commutative variables  ${v}^{\prime}_{\bm A}$ labeled by $\bm{A} = A_i \chi^i \in \Lambda_\text{cochar}(G)$.
The variable ${v}^{\prime}_{\bm A}$ is meant to represent the vev of a bare monopole operator of the low-energy effective abelian gauge theory on $\mathbb{R}^3$.
Though we will later identify ${v}^{\prime}_{\bm A}$ with quantities $v_{\bm A}$ that appear in localization formulas, we denote them with a prime to make the logic clearer.
The vev of a bare monopole operator defined in the UV gauge theory should be a Weyl invariant combination of~${v}^{\prime}_{\bm A}$.

Mirror symmetry and one-loop corrections to the Coulomb branch metric suggest that one postulates the relations
\begin{align}
{v}^{\prime}_{{\bm A}} {v}^{\prime}_{{\bm B}}= {v}^{\prime}_{{\bm A}+{\bm B}} \frac{P^{\text{hyper}}_{{\bm A}, {\bm B}}({\bm \varphi}, {\bm m})}{P^{\text{W}}_{{\bm A}, {\bm B}} ({\bm \varphi}) }\,,
\label{eq:nonAB_CB_relation}
\end{align}
where the hypermultiplets give rise to the factors
\begin{align}
P^{\text{hyper}}_{{\bm A}, {\bm B}}(\varphi,m)  :=
\prod_{i} ({\rm w}_i \cdot {\bm \varphi}+ \mu_i \cdot {\bm m} )^{({\rm w}_i \cdot {\bm A})_+ + ({\rm w}_i \cdot {\bm B})_+ - ({\rm w}_i \cdot ({\bm A} + {\bm B} ))_+} \,,
\label{eq:Ab_gauge_hyper}
\end{align}
and the W-bosons 
\begin{align}
& P^{\text{W}}_{{\bm A}, {\bm B}} ({\bm \varphi})= 
\prod_{\alpha:\text{root}}
 (\alpha \cdot {\bm \varphi} )^{(\alpha \cdot {\bm A})_+ + (\alpha \cdot {\bm B})_+ - (\alpha \cdot ({\bm A} + {\bm B} ))_+} .
\end{align}
Here we denoted by ${\rm w}_i$ (resp. $\mu_i$) a weight of the gauge (resp. flavor) group for the $i$-th hypermultiplet, and by $\alpha$ a root of $G$.
The quantity~$(a)_+$ is defined to be $a$ for $a>0$ and $0$ for $a \le 0$. 
The relations~(\ref{eq:nonAB_CB_relation}) are compatible with the R-charge assignments.
The abelianized Coulomb branch chiral ring is defined as
\begin{align}
\mathbb{C}[\mathcal{M}^{\text{abel}}_{C}] : =\mathbb{C} [\{ {v}^{\prime}_{\bm A} \}_{{\bm A} \in \Lambda_{\text{cochar}}(G)}, \{ \varphi_j \}_{j=1}^{N}, \{ (\alpha \cdot {\bm \varphi})^{-1} \}_
{\alpha:\text{root}}
] \,/   \text{relations} \, \eqref{eq:nonAB_CB_relation}.
\label{eq:abelianizedCBrel}
\end{align}
It has been demonstrated by mirror symmetry in many examples that the true Coulomb branch chiral ring~$\mathbb{C}[\mathcal{M}_{C}]$ can be realized as a subring of the abelianized ring~$\mathbb{C}[\mathcal{M}^{\text{abel}}_{C}] $.

To construct the quantum version of the abelianized Coulomb branch chiral ring in~\cite{Bullimore:2015lsa}, one promotes the classical variables $\varphi_i$ and $v'_{\bm A}$ to the quantum variables $\hat{\varphi}_i$ and $\hat{v}^{\prime}_{\bm A}$ and postulate the commutation relations%
\footnote{%
The $\Omega$-deformation parameter~$\epsilon$ in our convention corresponds to $ - \epsilon$ in \cite{Bullimore:2015lsa}.} 
\begin{align}
[\hat{\varphi}_j, \hat{v}^{\prime}_{\bm A}]= - \epsilon A_j   \hat{v}^{\prime}_{\bm A} \,.
\label{eq:commphiv}
\end{align}
For the product of $\hat{v}^{\prime}_{\bm A}$ and $\hat{v}^{\prime}_{\bm B}$, one postulates the quantum versions of the relations~(\ref{eq:nonAB_CB_relation})
\begin{equation}
\begin{aligned}
 \hat{v}^{\prime}_{{\bm A}} \hat{v}^{\prime}_{{\bm B}}
 \
 &=  \
\frac{ {\displaystyle \prod_{\begin{subarray}{c} i : |{\rm w}_i \cdot {\bm A}| \le  |{\rm w}_i \cdot {\bm B}|  \\  ({\rm w}_i \cdot {\bm A}) ({\rm w}_i \cdot {\bm B} ) <0 \end{subarray} } }
[{\rm w}_i \cdot \widehat{\bm \varphi} + \mu_i \cdot {\bm m} -\frac{\epsilon}{2} ]^{- ({\rm w}_i \cdot {\bm A})} } 
 { {\displaystyle \prod_{\begin{subarray}{c} \alpha : |\alpha \cdot {\bm A}| \le  |\alpha \cdot {\bm B}|  \\  (\alpha \cdot {\bm A}) (\alpha \cdot {\bm B} ) <0 \end{subarray}} }
 [\alpha \cdot \widehat{\bm \varphi}  ]^{- ( \alpha \cdot {\bm A} )} } 
\hat{v}^{\prime}_{{\bm A}+{\bm B}}
\\
&
\qquad\qquad\qquad
\times
 \frac{ {\displaystyle \prod_{\begin{subarray}{c} i : |{\rm w}_i \cdot {\bm A}| >  |{\rm w}_i \cdot {\bm B}|  \\  ({\rm w}_i \cdot {\bm A}) ({\rm w}_i \cdot {\bm B} ) <0 \end{subarray} } }
[{\rm w}_i \cdot \widehat{\bm \varphi} + \mu_i \cdot {\bm m} -\frac{\epsilon}{2}  ]^{ ({\rm w}_i \cdot {\bm B})} } 
 { {\displaystyle \prod_{\begin{subarray}{c} \alpha : |\alpha \cdot {\bm A}| >  |\alpha \cdot {\bm B}|  \\  (\alpha \cdot {\bm A}) (\alpha \cdot {\bm B} ) <0 \end{subarray}} }
 [\alpha \cdot \widehat{\bm \varphi}  ]^{ ( \alpha \cdot {\bm B})} }  \,,
\label{eq:nonABQ_CB_relation}
\end{aligned}
\end{equation}
where we use the special notation
\begin{align}\label{eq:sqbracket}
[x]^b:=
\left\{
\begin{array}{cc}
 \prod_{a=0}^{b-1} (x-a \epsilon )
 &  \, \, b > 0, \\
 \prod_{a=1}^{|b|} (x+a \epsilon )
 &  \, \, b < 0, \\
 1
 &  \, \, b = 0. \\
\end{array}
\right.
\end{align}
 Then the quantum version of the abelianized Coulomb branch chiral ring is defined as
\begin{equation}
\begin{aligned}
& \mathbb{C}_{\epsilon}[\mathcal{M}^{\text{abel}}_{C}]
\\
&:= \mathbb{C} [\{ \hat{v}^{\prime}_{\bm A} \}_{{\bm A} \in \Lambda_{\text{cochar}}(G)} , \{ \hat{\varphi}_j \}_{j=1}^{N}, 
 {\displaystyle \{ (\alpha \cdot \widehat{\bm \varphi} +n \epsilon)^{-1} \}_{\begin{subarray}{c} n \in \mathbb{Z}  \\ {\alpha: \text{root}}  \end{subarray}} }
] \,/   \text{relations} \, \eqref{eq:commphiv} \text{ and } \eqref{eq:nonABQ_CB_relation}.
\label{eq:abelianizedQCB}
\end{aligned}
\end{equation} 
The quantization~$\mathbb{C}_\epsilon[\mathcal{M}_{C}]$ of the true Coulomb branch chiral ring~$\mathbb{C}[\mathcal{M}_{C}]$ is a subalgebra of~$\mathbb{C}_\epsilon[\mathcal{M}^\text{abel}_{C}]$.
 When $\epsilon=0$, the quantum rings reduce to the classical rings: $\mathbb{C}_{\epsilon=0}[\mathcal{M}^{\text{abel}}_{C}]=\mathbb{C}[\mathcal{M}^{\text{abel}}_{C}]$, $\mathbb{C}_{\epsilon=0}[\mathcal{M}_{C}]=\mathbb{C}[\mathcal{M}_{C}]$.
 
Given the above description of the abelianized ring~$ \mathbb{C}_{\epsilon}[\mathcal{M}^{\text{abel}}_{C}]$, the remaining step in the determination of~$\mathbb{C}_\epsilon[\mathcal{M}_{C}]$ is the identification of its generators in~$ \mathbb{C}_{\epsilon}[\mathcal{M}^{\text{abel}}_{C}]$.

\subsubsection{Abelianized Coulomb branch chiral rings from SUSY localization}

We now show that the relations 
 \eqref{eq:nonAB_CB_relation} and \eqref{eq:nonABQ_CB_relation} in the abelianized Coulomb branch chiral rings are satisfied by the Weyl transforms~(\ref{eq:Moyal_Wey}) of the bare monopole operator vevs given by the localization formula~\eqref{eq:exbaremono}.

We define the quantity $v_{\bm{B}}$ as the product of the classical contribution~\eqref{eq:saddlept} 
and the one-loop determinant~$Z_{1 \mathchar `-\text{loop}}({\bm \varphi}, {\bm m};\bm{B} ;  \epsilon):=Z^{\text{vm}}_{1\mathchar `-\text{loop}} ({\bm \varphi}; \bm{B} ;  \epsilon)Z^{\text{hm}}_{1 \mathchar `-\text{loop}}({\bm \varphi}, {\bm m}; \bm{B};  \epsilon)$ given by~\eqref{eq:1loopvec} and~\eqref{eq:1loophy} for $\bm{B} \in \Lambda_{\text{cochar}} (G)$:
\begin{align}
 v_{\bm{B}} := e^{ \bm{B} \cdot \bm{b}} Z_{1 \mathchar `-\text{loop}}({\bm \varphi}, {\bm m}; \bm{B} ;  \epsilon) \,.
\end{align}
Let us introduce the functions
\begin{equation}
\begin{aligned}
&  {\sf F}_a ( {\rm w}_i, {\bm m}; {\bm A}, {\bm B}) 
\\
:=  &
\prod_{k=0}^{|{\rm w}_i \cdot {\bm A} |-1} \left( {\rm w}_i \cdot {\bm \varphi}  + \mu_i \cdot m+ \left( \frac{ | {\rm w}_i \cdot {\bm A} |- {\rm w} \cdot {\bm B} -a}{2} - k\right) \epsilon \right)^{\frac{1}{2}} 
  \\
&  \times \prod_{l=0}^{|{\rm w}_i \cdot {\bm B} |-1}  \left( {\rm w}_i \cdot  {\bm \varphi}  
+ \mu_i \cdot m+ \left( \frac{ | {\rm w}_i \cdot {\bm B} |+{\rm w}_i \cdot {\bm A}  -a}{2} - l\right) \epsilon \right)^{\frac{1}{2}}
 \\
& \qquad \times \prod_{n=0}^{|{\rm w}_i \cdot ( {\bm A}+ {\bm B} ) |-1}  \left( {\rm w}_i \cdot  {\bm \varphi}  + \mu_i \cdot m+ \left( \frac{ | {\rm w}_i \cdot ({\bm A} +{\bm B} )| -a}{2} - n\right) \epsilon \right)^{-\frac{1}{2}} 
\label{eq:moyalabmonopole}
\end{aligned}
\end{equation}
for $a=0,1$.
The Moyal product of $v_{\bm A}$ and $v_{\bm B}$ is given by
\begin{align}
 v_{\bm A} * v_{\bm B} &=   v_{{\bm A}+{\bm B}}
\frac{ \prod_{i} {\sf F}_1 ({\rm w}_i, {\bm m} ; {\bm A}, {\bm B}) }{ \prod_{\alpha : \text{root} } {\sf F}_0 ( \alpha, {\bm m}=0  ; {\bm A}, {\bm B}) }\, .
\label{eq:moyalabmonople}
\end{align}
For the rather complicated function~(\ref{eq:moyalabmonopole}), we can show, by the formula~(\ref{eq:Moyal_Wey}), the following simple result:
\begin{align}
& \text{Weyl transform of} \, \, e^{  ( {\bm A}+ {\bm B} ) \cdot {\bm b} } {\sf F}_a ( {\rm w}_i, {\bm m}; {\bm A}, {\bm B})  \nonumber \\
&=
\left\{
\begin{array}{cc}
 [{\rm w}_i \cdot  \widehat{\bm \varphi}  + \mu_i \cdot {\bm m}  { -} \frac{a}{2} \epsilon]^{- {\rm w}_i \cdot {\bm A} } \, e^{ ({\bm A}+ {\bm B}) \cdot \hat{{\bm b} } }
 & \text{for }  ( {\rm w}_i \cdot {\bm A} )( {\rm w}_i \cdot {\bm B} ) <0  \, \text{and} \, |{\rm w}_i \cdot {\bm A} | \le |{\rm w}_i \cdot {\bm B} | \,,\\
e^{ ({\bm A}+ {\bm B}) \cdot \hat{{\bm b} } } [{\rm w}_i \cdot  \widehat{\bm \varphi}  + \mu_i \cdot {\bm m} { -} \frac{a}{2} \epsilon]^{ {\rm w}_i \cdot {\bm B} }
 & \text{for }  ( {\rm w}_i \cdot {\bm A} )( {\rm w}_i \cdot {\bm B} ) <0  \, \text{and} \, |{\rm w}_i \cdot {\bm A} | > |{\rm w}_i \cdot {\bm B} | \,,\\
  e^{ ({\bm A}+ {\bm B}) \cdot \hat{{\bm b} } }
 &  \,   \text{otherwise.} \\
\end{array}
\right.
\label{eq:WWtrans1} 
\end{align}
The derivation of~(\ref{eq:WWtrans1}), which we detail in Appendix~\ref{app:QACB}, involves the consideration of several separate cases.
The same consideration applied to
 \eqref{eq:moyalabmonople} gives
\begin{equation}
\begin{aligned}
 \hat{v}_{\bm A}  \hat{v}_{\bm B} & \, =    \,
\frac{ {\displaystyle \prod_{\begin{subarray}{c} i : |{\rm w}_i \cdot {\bm A}| \le  |{\rm w}_i \cdot {\bm B}|  \\  ({\rm w}_i \cdot {\bm A}) ({\rm w}_i \cdot {\bm B} ) <0 \end{subarray} } }
[{\rm w}_i \cdot \widehat{\bm \varphi} + \mu_i \cdot {\bm m} -\frac{\epsilon}{2} ]^{- ({\rm w}_i \cdot {\bm A})} } 
 { {\displaystyle \prod_{\begin{subarray}{c} \alpha : |\alpha \cdot {\bm A}| \le  |\alpha \cdot {\bm B}|  \\  (\alpha \cdot {\bm A}) (\alpha \cdot {\bm B} ) <0 \end{subarray}} }
 [\alpha \cdot \widehat{\bm \varphi}  ]^{- ( \alpha \cdot {\bm A} )} } 
\hat{v}_{{\bm A}+{\bm B}}
\\
&\qquad \qquad \qquad 
\times
 \frac{ {\displaystyle \prod_{\begin{subarray}{c} i : |{\rm w}_i \cdot {\bm A}| >  |{\rm w}_i \cdot {\bm B}|  \\  ({\rm w}_i \cdot {\bm A}) ({\rm w}_i \cdot {\bm B} ) <0 \end{subarray} } }
[{\rm w}_i \cdot \widehat{\bm \varphi} + \mu_i \cdot {\bm m} -\frac{\epsilon}{2}  ]^{ ({\rm w}_i \cdot {\bm B})} } 
 { {\displaystyle \prod_{\begin{subarray}{c} \alpha : |\alpha \cdot {\bm A}| >  |\alpha \cdot {\bm B}|  \\  (\alpha \cdot {\bm A}) (\alpha \cdot {\bm B} ) <0 \end{subarray}} }
 [\alpha \cdot \widehat{\bm \varphi}  ]^{ ( \alpha \cdot {\bm B})} }  \,.
\label{eq:relationQab}
\end{aligned}
\end{equation}
We find that~$\hat{v}_{\bm B}$ satisfy the same relations as~$\hat{v}^{\prime}_{\bm B}$. 
We can thus identify $\hat{v}_{\bm B}^{\prime}$ with  $\hat{v}_{\bm B}$.

In~\cite{Bullimore:2015lsa} it was proposed that the classical (quantized) Coulomb branch chiral ring $\mathbb{C}[\mathcal{M}_{C}]$ (resp. $\mathbb{C}_{\epsilon}[\mathcal{M}_{C}]$)
is embedded as a subalgebra of $\mathbb{C}[\mathcal{M}^{\text{abel}}_{C}]$ (resp. $\mathbb{C}_{\epsilon}[\mathcal{M}^{\text{abel}}_{C}]$).

 Although the definitions~\eqref{eq:abelianizedCBrel} and \eqref{eq:abelianizedQCB} involve negative powers of~$\prod_{ \alpha:\text{root} }\alpha \cdot {\bm \varphi}$ and
 $\prod_{\alpha:\text{root}  }(\alpha \cdot \widehat {\bm \varphi} +n \epsilon )$, it should be possible to find an appropriate set of generators so that any element of $\mathbb{C}[\mathcal{M}_{C}]$ or $\mathbb{C}_{\epsilon}[\mathcal{M}_{C}]$ can be expressed in terms of positive powers of generators.

If the gauge group is abelian the embedding should be an isomorphism, {\it i.e.}, 
\begin{equation}
\mathbb{C}[\mathcal{M}_{C}] \simeq \mathbb{C}[\mathcal{M}^{\text{abel}}_{C}] \,,
\quad
\mathbb{C}_{\epsilon}[\mathcal{M}_{C}] \simeq \mathbb{C}_{\epsilon}[\mathcal{M}^{\text{abel}}_{C}] \,.
\end{equation}
Thus for abelian gauge groups we demonstrated, by a direct localization calculation, the validity of the (quantized) Coulomb branch chiral ring as described in~\cite{Bullimore:2015lsa}.

\section{Matrix models for $Z_\text{mono}$ from branes}\label{sec:matrix-branes}


In this section we explain an approach to the computation of the monopole  bubbling contribution~$Z_\text{mono}$ in~(\ref{eq:exbaremono}).
The approach makes use of matrix models, and is closely related to the approach based on supersymmetric quantum mechanics to computation of the 4d version of the $Z_\text{mono}$'s that appear in the 't Hooft operator vev.

Let us briefly explain the 3d-4d relation.
One may obtain monopole operators of a 3d theory by dimensionally reducing 't Hooft operators of a 4d theory.
In the early study of 't Hooft operator vevs on $S^4$~\cite{Gomis:2011pf} and $S^1\times\mathbb{R}^3$~\cite{Ito:2011ea}, the non-perturbative contributions from monopole screening were derived using Kronheimer's relation~\cite{Kronheimer:MTh} between instantons on the Taub-NUT space and singular monopoles.
One can use Kronheimer's relation to compute monopole operator vevs in 3d.
Recently a new approach to the screening contributions in 4d has been advanced based on supersymmetric quantum mechanics on~$S^1$~\cite{Brennan:2018yuj,Brennan:2018rcn,Assel:2019iae,Hayashi:2019rpw}.
In this section we study the matrix models that result from the dimensional reduction of the supersymmetric quantum mechanics on~$S^1$.

We will first review the standard type IIB brane construction of  the 3d $\mathcal{N}=4$ $U(N)$ gauge theory with $N_F$ hypermultiplets in the fundamental representation (SQCD with $N_F$ flavors).
Then we will realize monopole operators and monopole screening using branes.
We will explain how to read off the matrix models that capture  monopole bubbling contributions, and also how to use the matrix models to compute the monopole  bubbling contributions~$Z_\text{mono}$ to monopole operator vevs.
Finally we will generalize the study from SQCD to linear and circular quiver theories.

\subsection{Brane realization of $U(N)$ SQCD and monopole operators}\label{sec:brane-SQCD}

\begin{figure}[t]
\centering
\subfigure[]{\label{subfig:SQCD-branes}
\includegraphics[width=4cm]{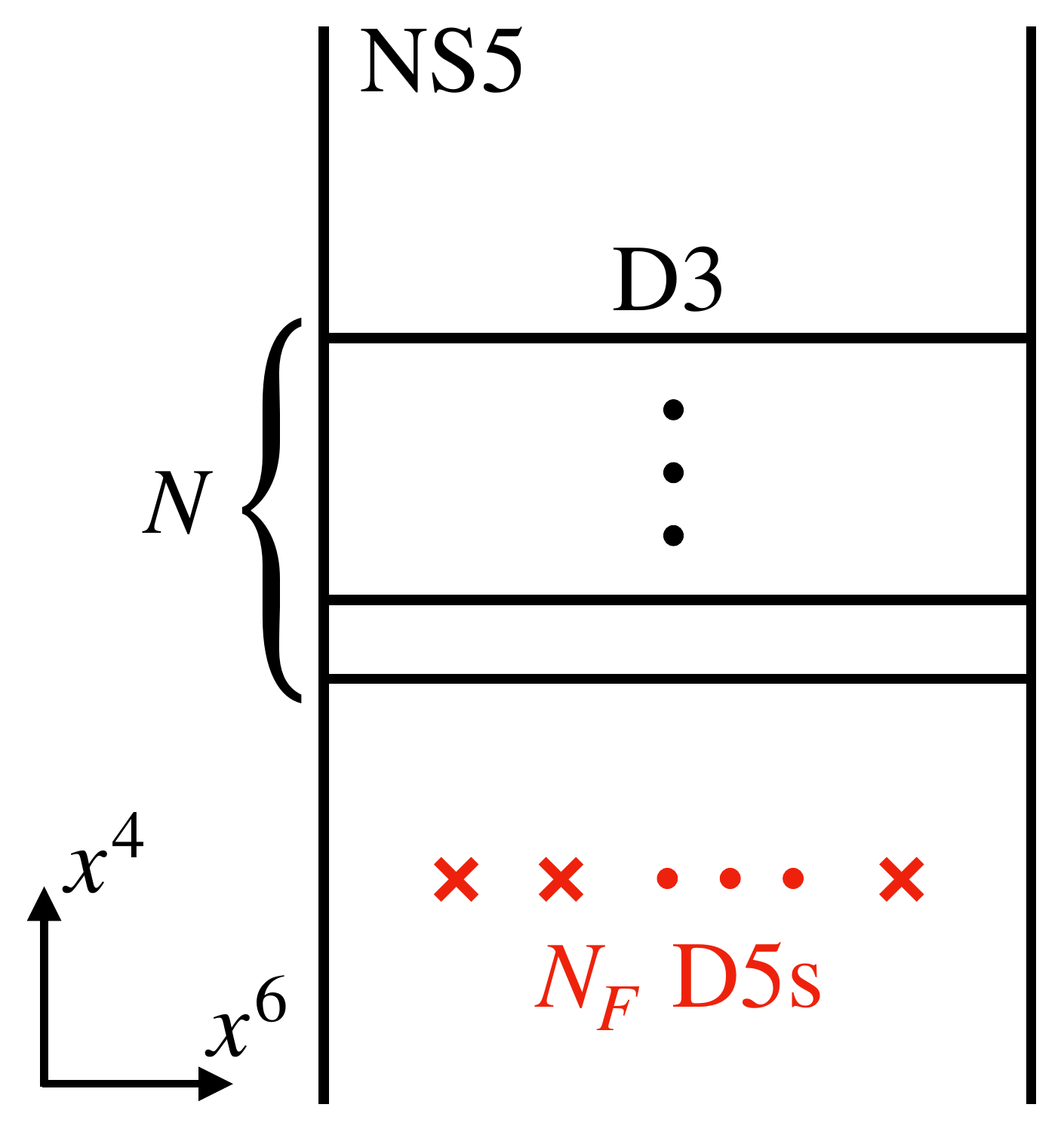}}
\hspace{0.5cm}
\subfigure[]{\label{subfig:SQCD-quiver}
\raisebox{.5cm}{\includegraphics[width=1cm]{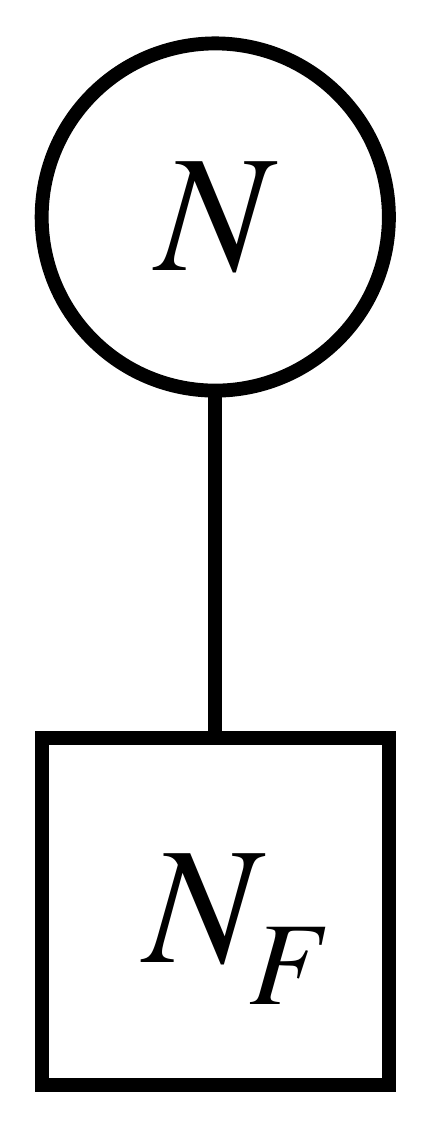}}
\hspace{.15cm}}
\hspace{0.35cm}
\subfigure[]{\label{subfig:SQCD-branes-3d}
\includegraphics[width=4.5cm]{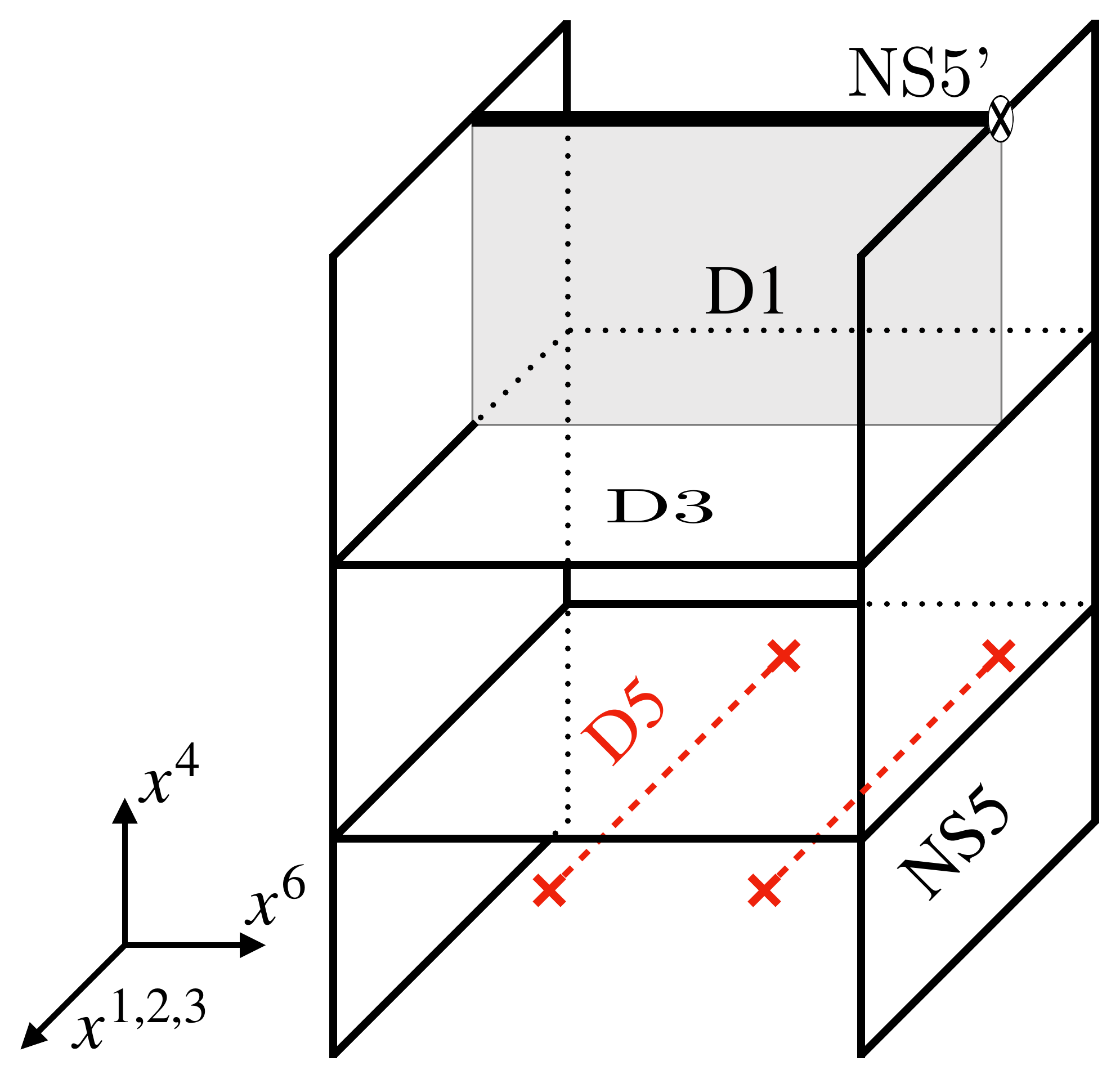}}
\hspace{.5cm}
\subfigure[]{\label{subfig:SQCD-branes-1234}
\raisebox{.5cm}{\includegraphics[width=4cm]{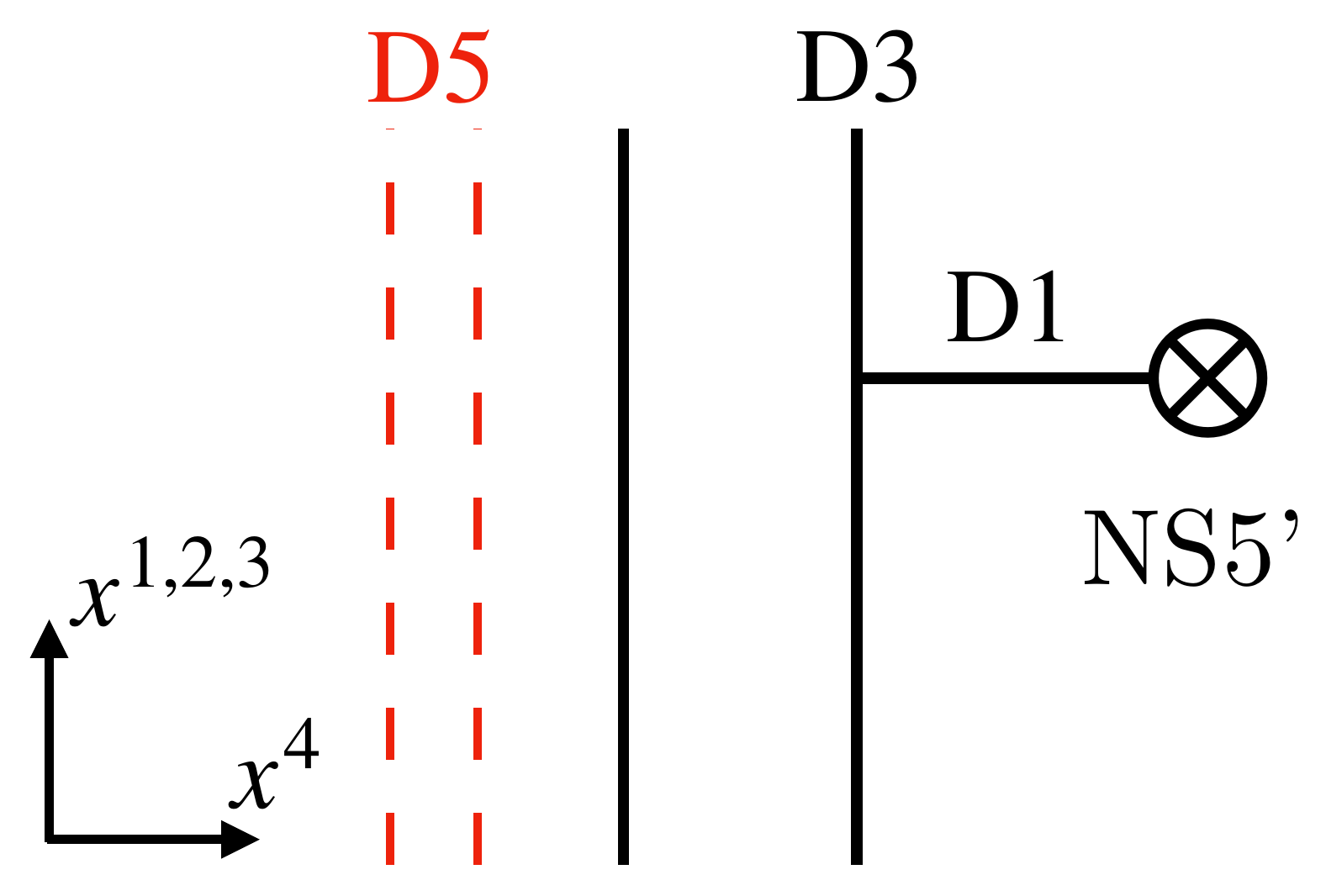}}}
\caption{(a): The brane configuration for the 3d $\mathcal{N}=4$ SQCD.
The $(x^4,x^6)$-directions are shown explicitly.
(b): The 3d $\mathcal{N}=4$ quiver diagram corresponding to~(a).
(c): The brane configuration for the SQCD with $N=N_F=2$, with a D1-brane and an NS5'-brane realizing a monopole operator.
In addition to the $(x^4,x^6)$-directions of (a), another direction that collectively represents $(x^1,x^2,x^3)$ is added to the figure.
(d): The projection of the same brane system to the $x^4$-direction and another direction that collectively represents $(x^1,x^2,x^3)$.
The NS5'-brane and the D1-brane inserts a monopole operator with charge $\bm{B}=\bm{e}_2$.
}
\label{fig:SQCD}
\end{figure}

Let us recall the brane engineering of the $U(N)$ SQCD~\cite{Hanany:1996ie}, with a slight change so that we take the spacetime to be the 10d Euclidean space rather than the Minkowski space.
Let $x^\mu$ ($\mu=0,1,\ldots,9$) be the coordinates on~$\mathbb{R}^{10}$.
We realize the gauge theory on the world-volume of $N$ D3-branes that extend infinitely in the $(x^1,x^2,x^3)$-directions and stretch in the $x^6$-direction between two NS5-branes.
The D3-branes are localized in the $(x^0,x^4,x^5,x^7,x^8,x^9)$-space.
The NS5-branes extend in the $(x^0,x^1,x^2,x^3,x^4,x^5)$-directions; they are at two different values of the $x^6$-coordinate, but are located at the same point in the $(x^7,x^8,x^9)$-space.
We also introduce $N_F$ D5-branes that extend in the $(x^1,x^2,x^3,x^7,x^8,x^9)$-directions.
They are localized in the $(x^0,x^4,x^5,x^6)$-space, and in particular sit at values of $x^6$ between those of the two NS5-branes.
On the world-volume of D3-branes we obtain a 3d $\mathcal{N}=4$ gauge theory.
The open strings with two ends on D3-branes give rise to a $U(N)$ vector multiplet.
The open strings with one end on a D3-brane and the other on a D5-brane  give rise to $N_F$ hypermultiplets in the fundamental representation of $U(N)$.
The locations of the D3-branes in the $(x^0,x^4,x^5)$-space are linear combinations of the three real scalars $\Phi_{\dot a\dot b}=\Phi_{\dot b\dot a}$ ($\dot a,\dot b \in\{1,2\}$) in the vector multiplet. 
Similarly the locations of the D5-branes  are the mass parameters for the hypermultiplets.
See Figures~\ref{subfig:SQCD-branes} and~\ref{subfig:SQCD-quiver} for the brane configuration and the quiver diagram for the 3d theory.

We can realize monopole operators in the SQCD by introducing D1-branes each of which has one end on a D3-brane and the other on what we call an NS5'-brane; we use a prime to distinguish an NS5-brane extended in the $(x^0,x^5,x^6,x^7,x^8,x^9)$-directions from the NS5-branes extended in the $(x^0,x^1,x^2,x^3,x^4,x^5)$-directions.
An NS5'-brane is localized in the $(x^1,x^2,x^3,x^4)$-space.
A D1-brane that ends on the NS5'-brane and on a D3-brane is localized in the $(x^1,x^2,x^3,x^5)$-space.
We can also consider D1-branes that have both ends on D3-branes; these D1-branes are free to move in the $(x^1,x^2,x^3)$-space and realize the smooth monopoles~\cite{Diaconescu:1996rk} that play the role of instantons in the 3d theory~\cite{Polyakov:1976fu}.
Whether a D1-brane ends on an NS5'-brane or not, it has a finite world-volume in the $(x^4,x^6)$-space.
To study the magnetic charges induced in the 3d theory, let us label the $N$ D3-branes by their values of the coordinate $x^4$ in the ascending order%
\footnote{%
We assume that the D3-brane are located at generic positions in the $(x^0,x^4,x^5)$-space.
}:
$x^4_1 < x^4_2 < \ldots < x^4_N$.
Suppose that $k^+_i$ ($k^-_i$) D1-branes that end on the $i$-th D3-brane have a world-volume extended in the direction $x^4\geq x^4_i$ ($x^4\leq x^4_i$).
We choose a convention such that the induced magnetic charge on the D3-branes world-volume is
\begin{equation}
\text{magnetic charge} = \sum_{i=1}^N (k_i^+ - k_i^-) \bm{e}_i \in \Lambda_\text{cochar} (U(N)) \,,
\end{equation}
where the cocharacter lattice for $U(N)$ was given in~(\ref{eq:cochar-UN}).
Figures~\ref{subfig:SQCD-branes-3d} and~\ref{subfig:SQCD-branes-1234} illustrate the brane configuration for the monopole operator with charge~$\bm{B}=\bm{e}_1$ in the $U(2)$ SQCD with $N_F=2$ flavors.
Table~\ref{table:brane-directions} summarizes the directions in which various branes extend.
\begin{table}[t]
\begin{center}
\begin{tabular}{c|c|ccc|cc|c|ccc}
&0&1&2&3&4&5&6&7&8&9
\\
\hline
D3&&$\times$&$\times$&$\times$&&&$\times$ 
\\
D5&&$\times$&$\times$&$\times$&&&&$\times$&$\times$&$\times$
\\
NS5 &$\times$&$\times$&$\times$&$\times$&$\times$&$\times$&&
\\
\hline
D1 &&&&&$\times$&&$\times$
\\
NS5'&$\times$&&&&&$\times$&$\times$&$\times$&$\times$&$\times$
\end{tabular}
\end{center}
\caption{The symbol $\times$ indicates the directions in which branes extend.}
\label{table:brane-directions}
\end{table}

 In order to compare our brane construction with those considered in~\cite{Brennan:2018yuj,Brennan:2018moe,Brennan:2018rcn,Assel:2019iae,Hayashi:2019rpw},
let us consider another brane configuration that realizes the SQCD in a limit.
We let the branes extend in the directions indicated in Table~\ref{table:brane-directions}.
We compactify the $x^6$-direction to a circle and place a single NS5-brane.
We fiber the $(x^7,x^8,x^9)$-space over the circle, {\it i.e.}, when we go around the $x^6$-circle we shift $(x^7,x^8,x^9)$ by some constant vector.
Then we let $N$ D3-branes end on the NS5-brane; the shift forces each D3-brane to end on the NS5-brane at two different points in the $(x^7,x^8,x^9)$-space.
We also introduce $N_F$ D5-branes.
The D3 world-volume theory has a $U(N)$ vector multiplet, $N_F$ hypermultiplets in the fundamental representation, and a hypermultiplet in the adjoint representation whose mass parameters are proportional to the shift in the $(x^7,x^8,x^9)$-space.
If one wishes the shift can be made large so that the adjoint hypermultiplet is integrated out, realizing the SQCD in the limit.

Starting with the configuration in the previous paragraph with a finite shift, we may consider a T-duality in the $x^6$-direction.
The NS5-brane at a point in the $(x^7,x^8,x^9)$-space turns into a deformation, by some background fields corresponding to the shift, of the single-center Taub-NUT space (topologically the same as $\mathbb{R}^4$) in the $(x^6_\text{dual},x^7,x^8,x^9)$-directions.
The T-duality turns D3-branes into D2-branes, and D5-branes into D6-branes.
A circle compactification and another T-duality along the Euclidean $x^0$-direction gives the brane set-up for the 4d $\mathcal{N}=2$ theory considered in~\cite{Brennan:2018rcn,Hayashi:2019rpw}.

\subsection{Quiver matrix models for  $U(N)$ SQCD from branes}\label{sec:matrix-SQCD-branes}

Next let us consider monopole screening~\cite{Kapustin:2006pk}, which is a phenomenon where a smooth monopole screens, partially or completely, the magnetic charges of the Dirac singularities that correspond to monopole operators.
Terms labeled by $\mathfrak{m}$ with $||\mathfrak{m}|| <||{\bm B}||$ in~(\ref{eq:exbaremono-intro}) are the contributions from such field configurations.
Correspondingly D1-branes that stretch between D3-branes reconnect with the D1-branes that stretch between an NS5'-brane a D3-brane.
The reconnection is possible only when the position of the D1-brane in the $(x^1,x^2,x^3)$-space coincides with that of the NS5'-brane.

Let us consider a configuration such as Figure~\ref{subfig:SQCD-branes-1234}.
When D3-branes and D5-branes move in the $x^4$-direction, they can pass each other without actually colliding in 10d, because they are point-like in the $(x^0,x^4,x^5)$-space as can be seen from Table~\ref{table:brane-directions}.
On the other hand we also see from the same table that when an NS5'-brane moves in the $x^4$-direction it cannot pass a D3-, or D5-brane without a collision in 10d.
In particular when an NS5'-brane and a D3-brane pass each other a D1-brane is created or annihilated via a {\it Hanany-Witten} (HW) {\it transition}~\cite{Hanany:1996ie}.
See~\cite{Hanany:1996ie} for the precise rules that govern brane creation/annihilation.

After introducing D1-branes stretched between D3-branes, the coefficients ${\rm m}_i$ in $\mathfrak{m}=\sum_i {\rm m}_i \bm{e}_i$ may not be non-decreasing. 
In such a situation, we can move the $N$ D3-branes in the $(x^0,x^4,x^5)$-space so that after the permutation we have ${\rm m}_1\leq {\rm m}_2\leq\ldots\leq {\rm m}_N$~\cite{Hayashi:2019rpw}.

\begin{figure}[t]
\centering
\subfigure[]{\label{subfig:screening1}
\includegraphics[scale=.25]{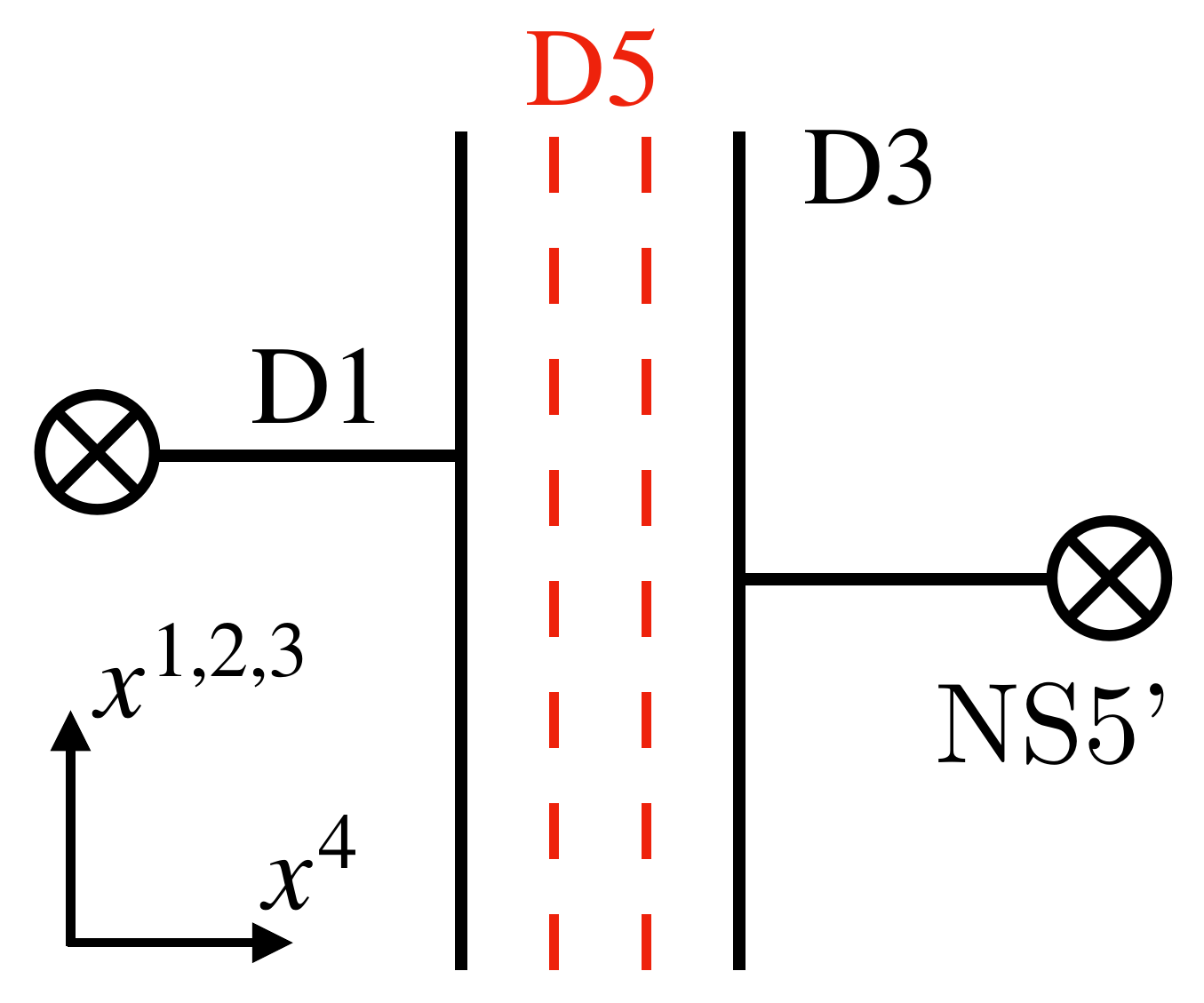}}
\hspace{0cm}
\subfigure[]{\label{subfig:screening2}
\includegraphics[scale=.25]{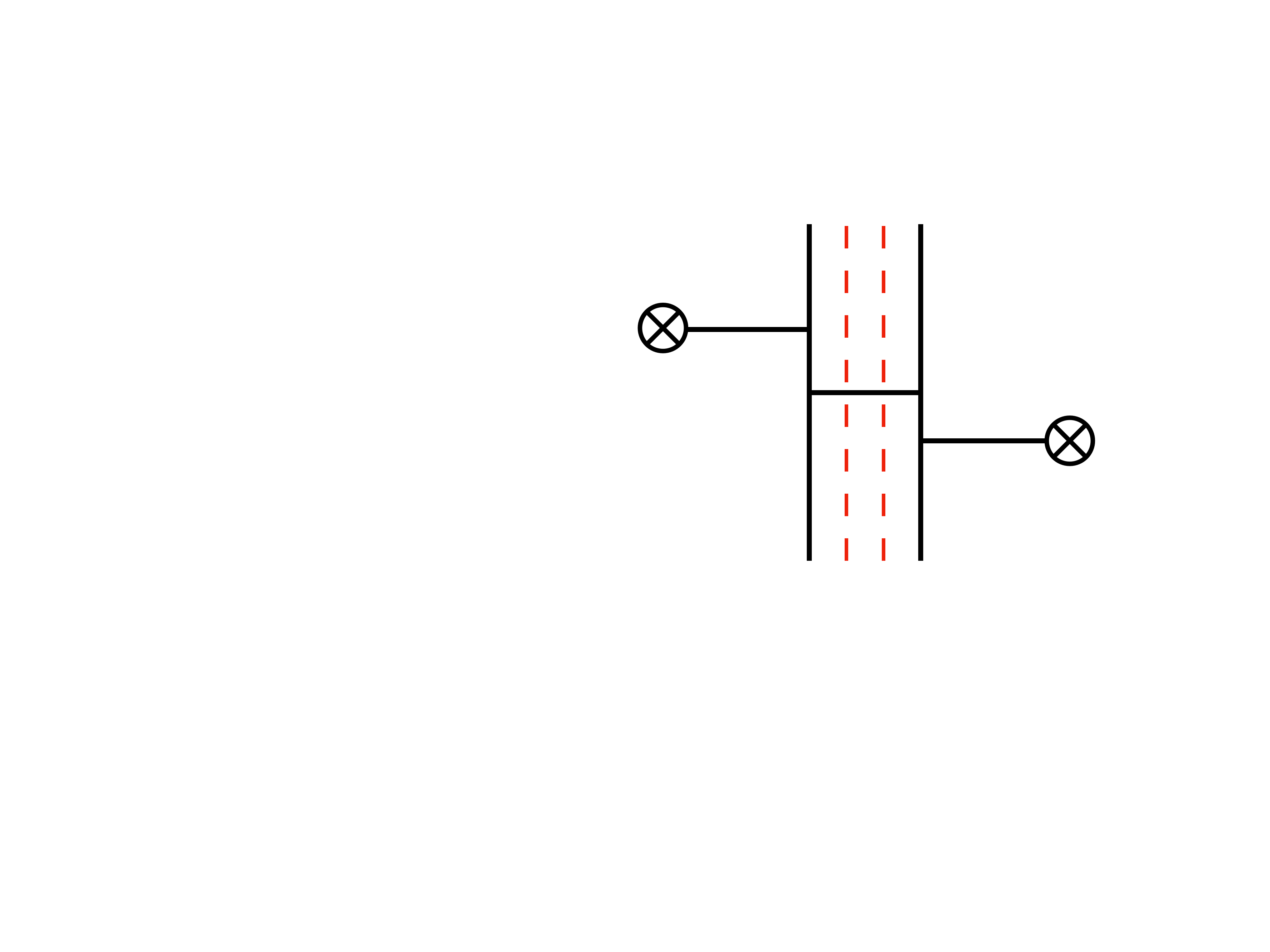}}
\hspace{0cm}
\subfigure[]{\label{subfig:screening3}
\includegraphics[scale=.25]{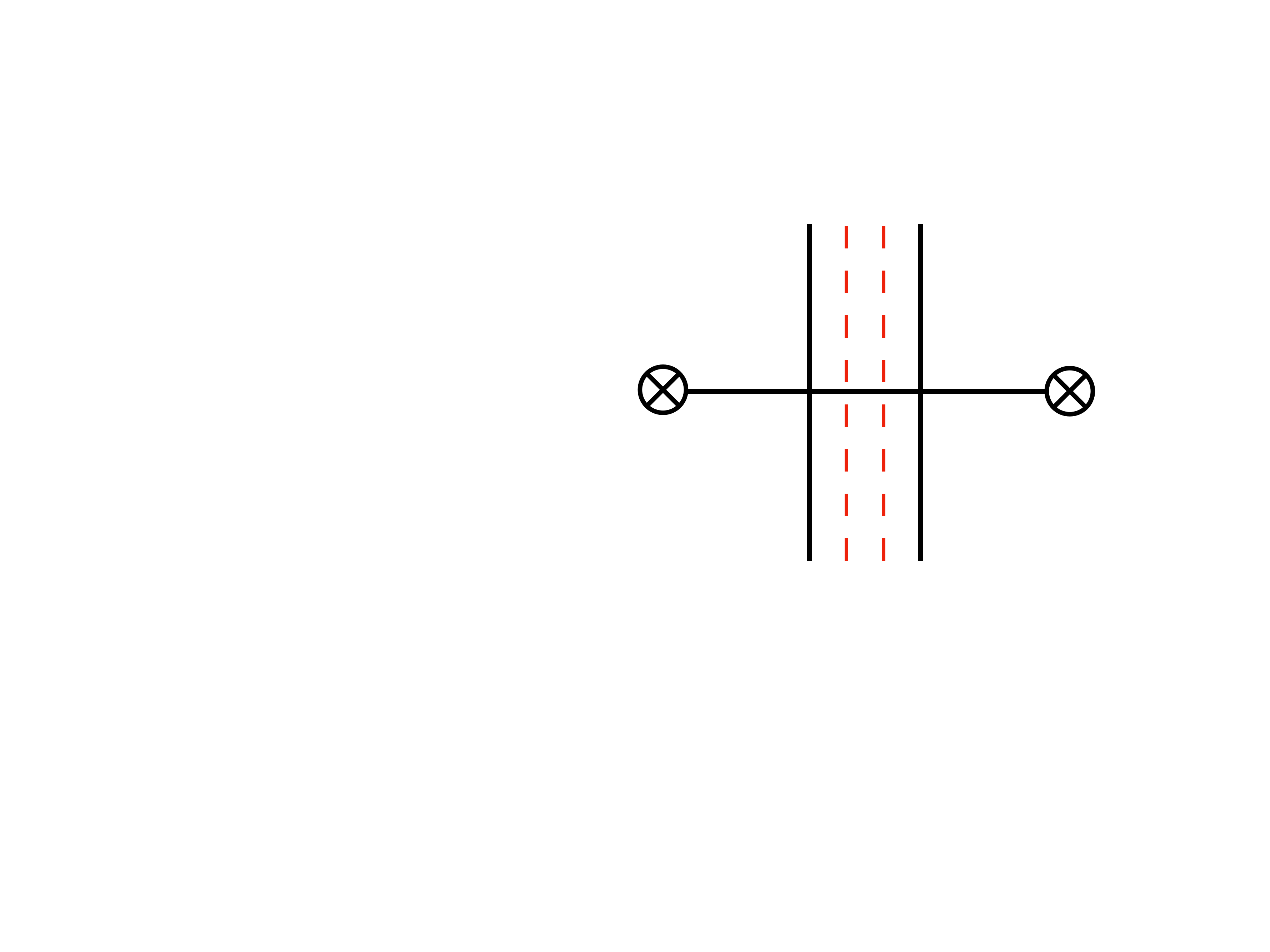}}
\hspace{1cm}
\subfigure[]{\label{subfig:screening4}
\includegraphics[width=.6cm]{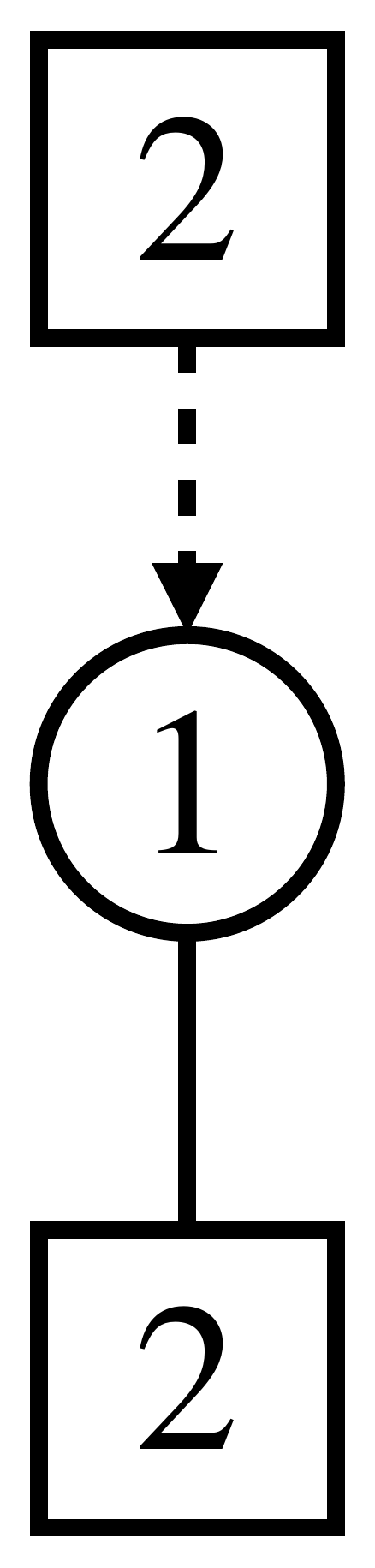}
\hspace{.1mm}
}
\caption{(a): The NS5'-branes and the D1-branes insert a product monopole operator with total charge $\bm{B}=-\bm{e}_1 + \bm{e}_2$.
(b): A finite D1-brane representing a smooth monopole is added to the system.
(c): The finite D1-brane reconnects with the other D1-branes to form a single D1-brane that only end on NS5'-branes.
The system describes the sector with $\mathfrak{m}=0$; the magnetic charge is completely screened.
(d): The  $\mathcal{N}=(0,4)$ quiver diagram for the matrix model that lives on the D1-brane in (c).
}
\label{fig:SQCD-screening}
\end{figure}

At this point we can follow the procedure in Section~3.3 of~\cite{Brennan:2018yuj} to read off  the matter content of a quiver matrix model.
 We will not spell out the precise interactions as they only serve to constrain fugacities for global symmetries.
This involves moving NS5'-branes via a sequence of HW transitions~\cite{Hanany:1996ie}.
The matrix model is read off from the HW frame where the D1-branes end only on NS5'-branes (and on NS5-branes), and not on any D3-brane.
The Higgs branch of the matrix model is a component of the monopole moduli space.
The partition function is an equivariant integral that can be identified with $Z_\text{mono}({\bm B}, \mathfrak{m})$.
The brane realization of monopole screening and the procedure are illustrated in Figure~\ref{fig:SQCD-screening}.

\begin{figure}[t]
\centering
\subfigure[]{\label{subfig:U(N)-vector}
\includegraphics[scale=.28]{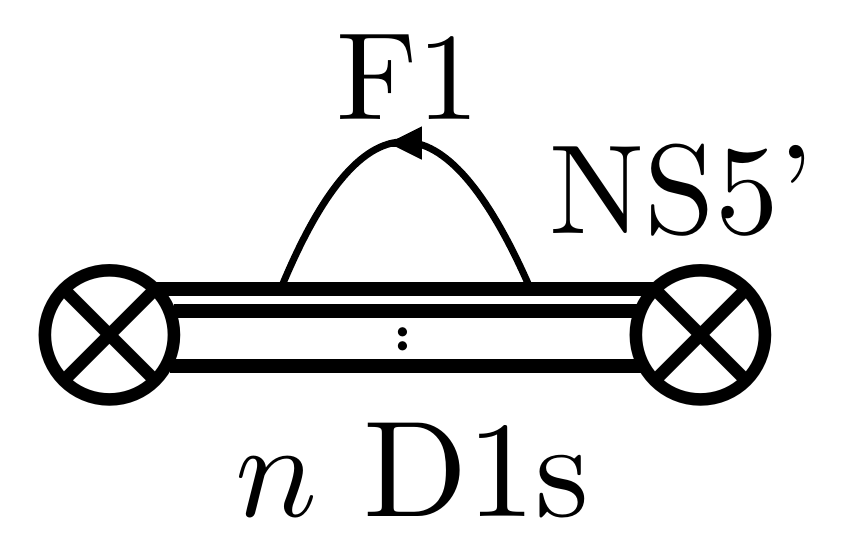}}
\hspace{1cm}
\subfigure[]{\label{subfig:bifundamenta-hyper}
\includegraphics[scale=.3]{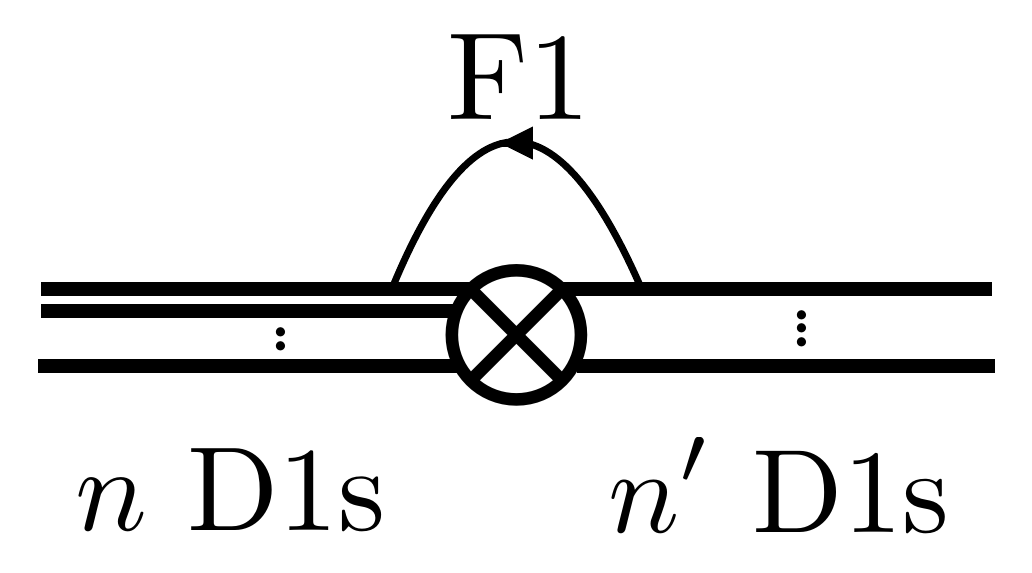}}
\hspace{1cm}
\subfigure[]{\label{subfig:fundamental-hyper}
\includegraphics[scale=.3]{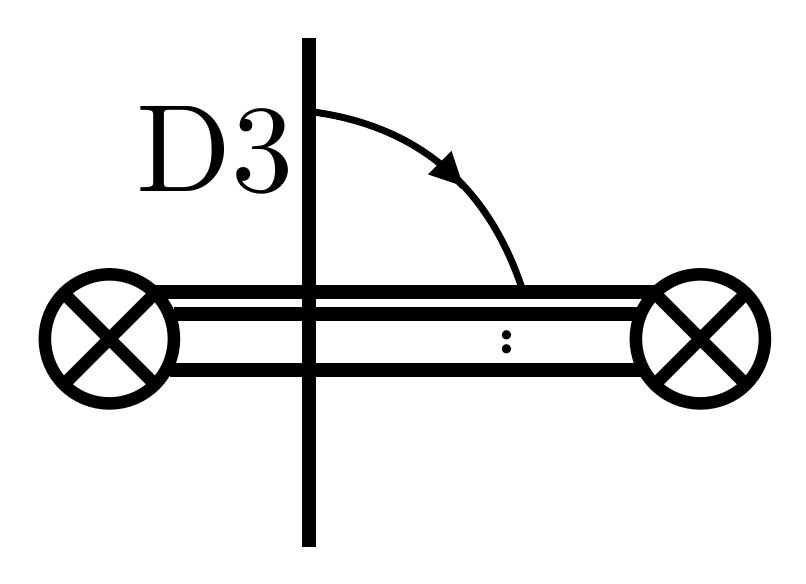}}
\hspace{1cm}
\subfigure[]{\label{subfig:short-Fermi}
\includegraphics[scale=.3]{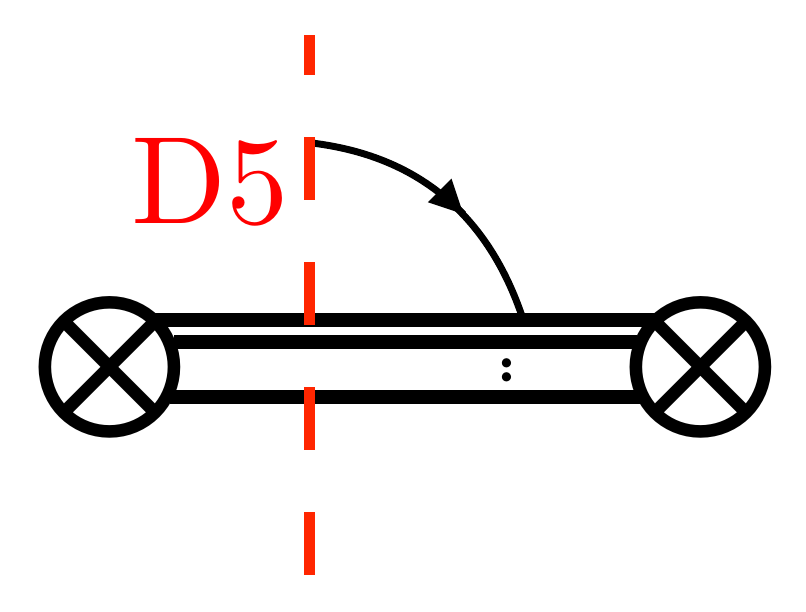}}
\caption{(a): $\mathcal{N}=(0,4)$ $U(n)$ vector multiplet.
(b): $\mathcal{N}=(0,4)$ hypermultiplet in the bifundamental representation of $U(n)\times U(n')$.
(c): $\mathcal{N}=(0,4)$ hypermultiplet in the fundamental representation of $U(n)$.
(d): $\mathcal{N}=(0,4)$ short Fermi multiplet in the fundamental representation of $U(n)$.
}
\label{fig:SQCD-screening-multiplets}
\end{figure}

The matrix model is the reduction to zero dimensions of a two-dimensional $\mathcal{N}=(0,4)$ supersymmetric gauge theory, possibly with parameters that break the SUSY to $\mathcal{N}=(0,2)$. 
D1-D1 strings on a single stack of $n$ D1-branes give rise to an $\mathcal{N}=(0,4)$ $U(n)$ vector multiplet, as illustrated in Figure~\ref{subfig:U(N)-vector}.
D1-D1 strings that connect a stack of $n$ D1-branes with another stack of $n'$ D1-branes separated by an NS5'-brane give an $\mathcal{N}=(0,4)$ hypermultiplet in the bifundamental representation of $U(n)\times U(n')$.
See Figure~\ref{subfig:bifundamenta-hyper}.
D1-D3 strings on $n$ D1-branes and a D3-brane yield an $\mathcal{N} = (0,4)$ hypermultiplet in the fundamental representation of~$U(n)$, if the value of $x^4$ for the D3 is in the range spanned by the D1-brane, as shown in Figure~\ref{subfig:bifundamenta-hyper}.
Similarly D1-D5 strings on D1-branes and a D5-brane give rise to a short $\mathcal{N}=(0,4)$ Fermi multiplet in the fundamental representation of $U(n)$, if the value of $x^4$ for the D5 is in the range spanned by the D1-brane.  See Figure~\ref{subfig:short-Fermi}.

The $\mathcal{N}=(0,4)$ $U(n)$ vector multiplet from D1-D1 strings discussed above come with three real FI-parameters $(\zeta^1,\zeta^2,\zeta^3)$.
They are related to the relative positions of the monopole operators and the NS'5-branes in the $(x^1,x^2,x^3)$-space.
Let us consider the brane construction of $\ell$ monopole operators.
Let $(x^1,x^2,x^3)=(s^1_a,s^2_a,s^3_a)$ be the location of the $a$-th monopole operator.
Correspondingly, the NS5'-brane that realizes this monopole operator via a D1-brane attached to it is also located at $(x^1,x^2,x^3)=(s^1_a,s^2_a,s^3_a)$.
The FI-parameters and the locations of the monopole operators and the NS5'-branes in the $(x^1,x^2,x^3)$-space are related, up to an overall rescaling, as 
\begin{equation} \label{eq:zeta-sa}
(\zeta^1_a,\zeta^2_a,\zeta^3_a) = (s^1_{a+1}-s^1_a,s^2_{a+1}-s^2_a,s^3_{a+1}-s^3_a) , \quad a = 1,\ldots, \ell-1 .
\end{equation}
When the $\Omega$-deformation is introduced to the 3d theory, two of the coordinates are restricted to zero~$s^1_a=s^2_a=0$ for each $a$.
Correspondingly FI-parameters also get restricted: $\zeta^1_a=\zeta^2_a=0$.
The remaining FI-parameters will be denoted as $\bm{\zeta}\equiv(\zeta_a)_{a=1}^{\ell-1}\equiv(\zeta^3_a)_{a=1}^{\ell-1}$.

\subsection{$Z_\text{mono}$ for $U(N)$ SQCD from quiver matrix models}\label{sec:SQCD-matrix}

The monopole bubbling contribution $Z_{\text{mono}} $ can be computed as the partition function of a supersymmetric quiver matrix model, which is the dimensional reduction of the gauged quantum mechanics 
 in \cite{Brennan:2018yuj,Brennan:2018rcn}.
By analogy with the localization computations of the elliptic genus~\cite{Benini:2013xpa} and the Witten index~\cite{Hwang:2014uwa,Cordova:2014oxa,Hori:2014tda}, we propose%
\footnote{%
 The proposal is supported by our explicit comparisons between the Moyal products and the matrix model partition functions.
It would be interesting to prove the proposal by an analysis similar to~\cite{Benini:2013xpa} and~\cite{Hori:2014tda}.
} that the partition function is given as the matrix integral according to the Jeffrey-Kirwan (JK) residue prescription~\cite{MR1318878}:
\begin{align}
Z^{({\bm \zeta})}_{\text{mono}}({\bm \varphi},  {\bm m} ; {\bm B},\mathfrak{m} ; \epsilon)= 
\frac{1}{|W_{G^{\prime}}|}
\oint_{\mathrm{JK}({\bm \zeta})} \prod_{{ a}=1}^{\mathrm{rank}(G^{\prime})} \frac{d u_{a}}{2\pi i} \prod Z^{\text{0d}}_{\mathrm{vec}} \prod Z^{\text{0d}}_{\mathrm{hyper}}
 \prod Z^{\text{0d}}_{\mathrm{Fermi}} .
\label{eq:monosc}
\end{align}
Here $Z^{\text{0d}}_{\mathrm{vec}}$, $Z^{\text{0d}}_{\mathrm{hyper}}$, and $Z^{\text{0d}}_{\mathrm{Fermi}}$ are the one-loop determinants of the dimensional reduction of 2d $\mathcal{N}=(0,4)$ vector, hyper and Fermi multiplets to zero dimensions.%
\footnote{We often  refer to zero-dimensional multiplets simply  as $\mathcal{N}=(0,4)$ vector, hyper-, and Fermi multiplets. } 
When a 3d $U(N)$ adjoint hypermultiplet is present, $Z^{\text{0d}}_{\mathrm{vec}}$ and $Z^{\text{0d}}_{\mathrm{hyper}}$ contain contributions that can be regarded as those of $\mathcal{N}=(4,4)$ vector and hypermultiplets.
The JK residue prescription for the matrix integral is summarized in Appendix~\ref{app:JK}.
The quantity $|W_{G^{\prime}}|$ is the order of the Weyl group $W_{G'}$ of the matrix model gauge group $G^{\prime}$.%
\footnote{
While there are no gauge fields in 0d, the $G^{\prime}$ gauge symmetry associated with the vector multiplets restricts the physical observables to be the combinations of the matrix model variables invariant under $G^{\prime}$.
}
In general, $Z^{({\bm \zeta})}_{\text{mono}}$ is a piecewise constant function of ${\bm \zeta}$.

For $G'=U(n)$ the one-loop determinants for the vector multiplet is 
\begin{align}
Z^{\text{0d}}_{\mathrm{vec}} 
=\left\{\
\begin{array}{ll}
\displaystyle
\prod_{  a \neq b}^n   ( u_{a} -u_{b}  ) \prod_{  a , b =1 }^n ( u_{a} -u_{b} -\epsilon ) & \text{ for } \mathcal{N}=(0,4) , \\
\displaystyle
\prod_{  a \neq b}^n   ( u_{a} -u_{b}  ) \prod_{  a , b =1 }^n \frac{ u_{a} -u_{b} -\epsilon } 
{\prod_{s=\pm1} (u_{a} -u_{b}-{m}_{\text{ad}} -\frac{s}{2}\epsilon)}   & \text{ for } \mathcal{N}=(4,4).
\end{array}
\right. 
\label{one-loop-0d-vec} 
\end{align}
The parameter~$m_\text{ad}$ is the mass of the 3d adjoint hypermultiplet.
The one-loop determinant for a hypermultiplet in the bifundamental representation of  $U(n) \times U(n^{\prime})$ is 
\begin{align}
&Z^{\text{0d}}_{\mathrm{hyp}}
=\left\{\
\begin{array}{ll}
\displaystyle
\prod_{s=\pm1} \prod_{a =1}^n \prod_{b=1}^{n^{\prime}}  \frac{1}{ s ( u_a -u^{\prime}_b) -\frac{1}{2}  \epsilon } & \text{ for } \mathcal{N}=(0,4) , \\
\displaystyle
\prod_{s=\pm1} \prod_{a =1}^n \prod_{b=1}^{n^{\prime}}  \frac{s ( u_a -u^{\prime}_b)-m_{\text{ad}} -\frac{1}{2}  \epsilon}{ s ( u_a -u^{\prime}_b) -\frac{1}{2}  \epsilon }   & \text{ for } \mathcal{N}=(4,4).
\end{array}
\right. 
\label{one-loop-0d-hyp}
\end{align}
When $U(n')$ is a flavor symmetry of the matrix model and corresponds to a 3d gauge group, 
the integration variables $\{u^{\prime}_b \}_{b=1}^{n^{\prime}}$ in \eqref{one-loop-0d-hyp} are replaced by the coefficients $\varphi_i$ in the expansion of the 3d scalar $\bm{\varphi}=\sum_{i=1}^N \varphi_i \bm{e}_i$.
The one-loop determinant for $l$ short Fermi multiplets  in the fundamental representation of $U(n)$ is 
\begin{align}
&Z^{\text{0d}}_{\mathrm{Fermi}}=\prod_{a=1 }^n \prod_{f=1}^{l}   ( u_a-m_f). 
\label{one-loop-0d-Fermi}
\end{align}
Here $m_f$ are identified with the masses of the 3d fundamental hypermultiplets.
The 0d gauge group and the representation of the 0d hypermultiplets depend on  the 3d gauge group and the matter content, and also on the magnetic charges $\mathfrak{m}$ and ${\bm B}$ of the monopole operators with bubbling.

\subsection{Generalization to 3d linear and circular quiver gauge theories}\label{sec:brane-quiver}
 
In Sections~\ref{sec:brane-SQCD}-\ref{sec:SQCD-matrix} we considered the 3d $U(N)$ SQCD and its monopole operators.
In this subsection we generalize the brane construction and the matrix models to 3d (linear as well as circular) quiver gauge theories.

\begin{figure}[t]
\centering
\subfigure[]{\label{subfig:3d-quiver-branes-A}
\includegraphics[scale=.3]{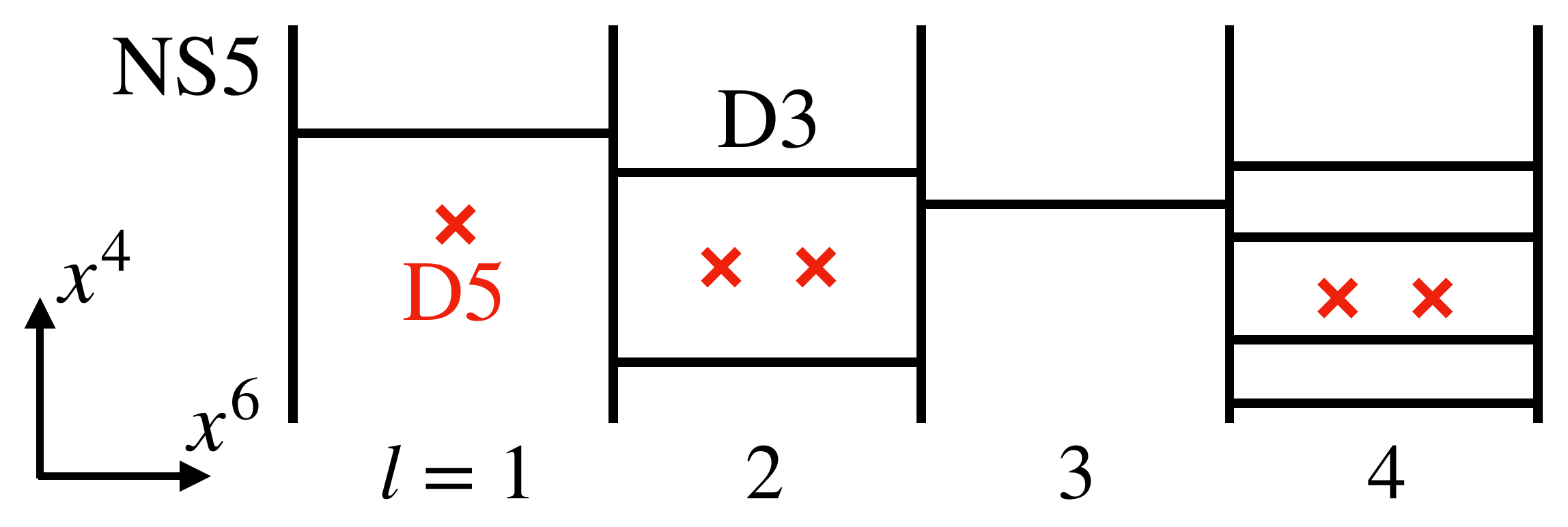}}
\hspace{1cm}
\subfigure[]{\label{subfig:3d-quiver-A}
\includegraphics[width=5cm]{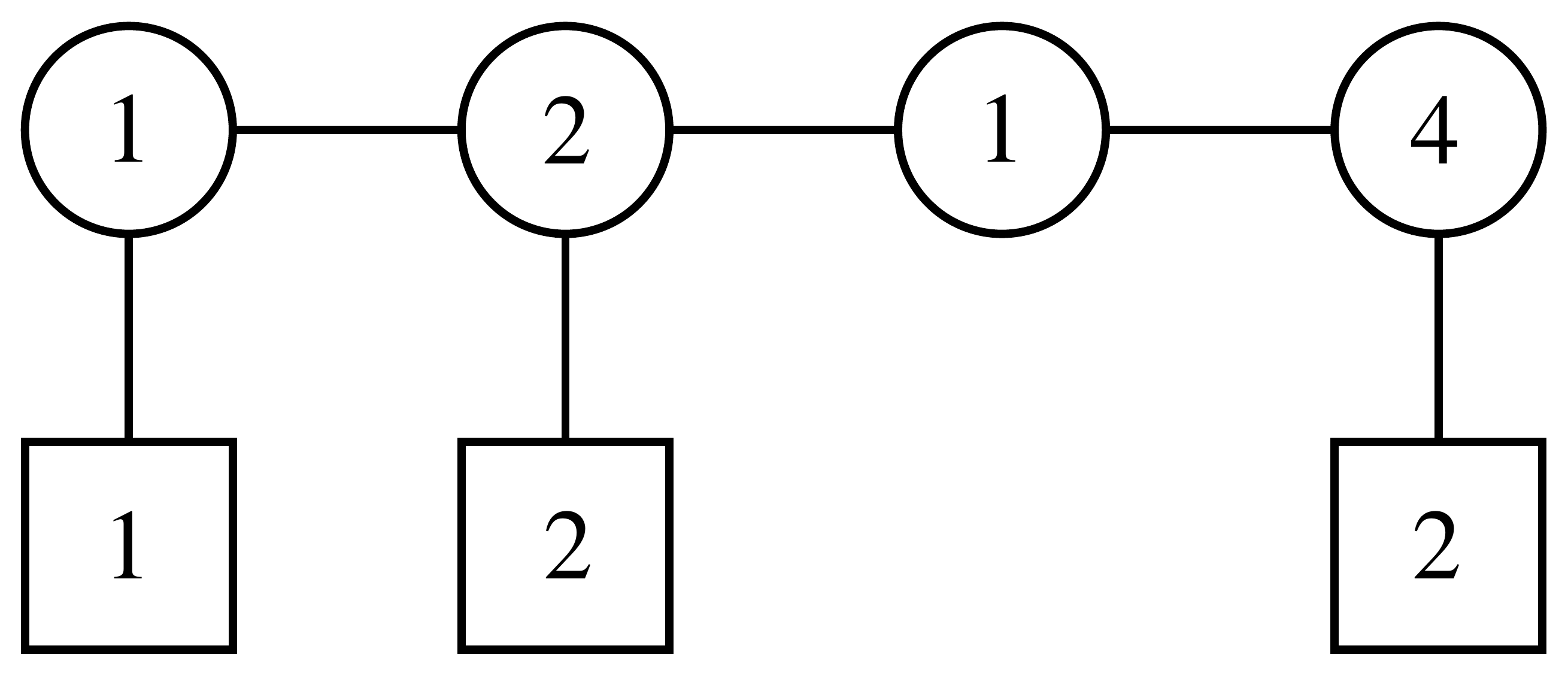}}
\hspace{0cm}
\subfigure[]{\label{subfig:3d-quiver-branes-monopoles-A}
\includegraphics[width=12cm]{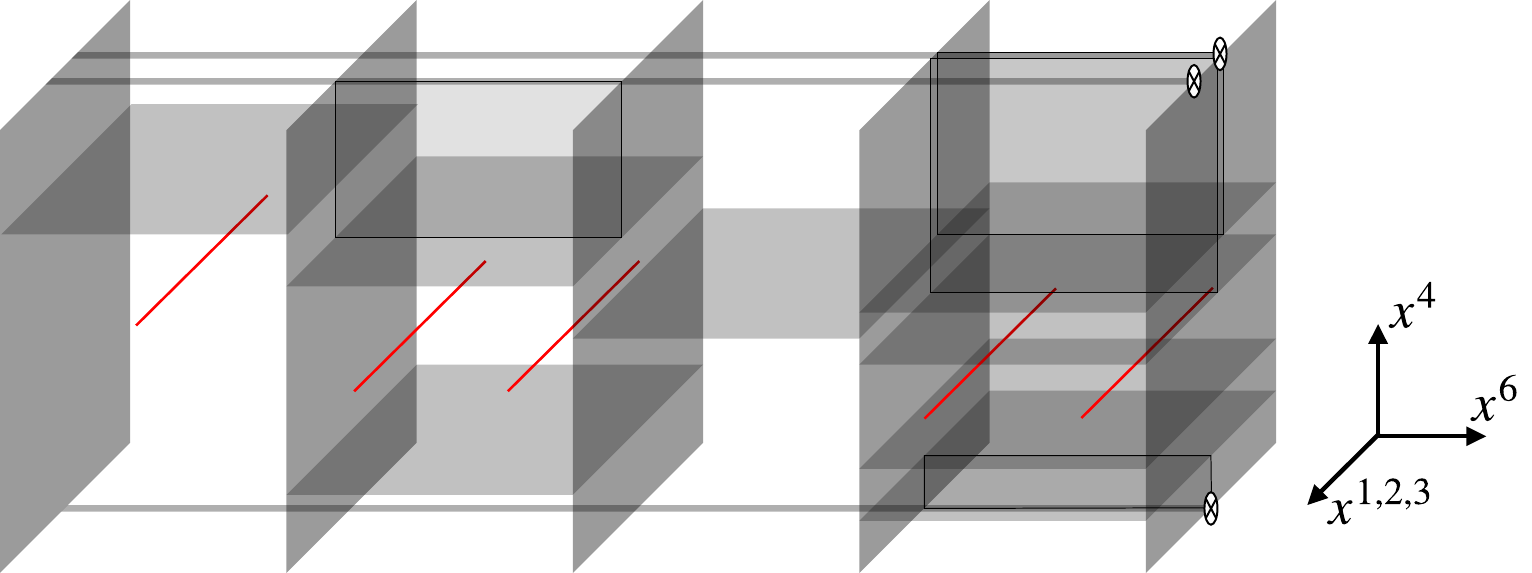}}
\caption{(a): The brane configuration for a 3d $\mathcal{N}=4$ quiver gauge theory.
The $(x^4,x^6)$-directions are shown explicitly.
(b): The 3d $\mathcal{N}=4$ quiver diagram corresponding to~(a).
(c): The brane configuration of (a) together with D1-  and NS5'-branes that realize monopole operators, shown in a three-dimensional figure.
The monopole operators realized are $V^{(2)}_{\bm{B}_1}$, $V^{(4)}_{\bm{B}_2}$, and $V^{(4)}_{\bm{B}_3}$ with charges $\bm{B}_1={\bm e}_2^{(2)}$, $\bm{B}_2=
{\bm e}_3^{(4)}+ {\bm e}_4^{(4)}
$, and $\bm{B}_3=
-{\bm e}_1^{(4)}
$, where $V^{(l)}_{\bm{B}}$ denotes a monopole operator with charge $\bm{B}$ for the $l$-th gauge group factor.
See~(\ref{cochar-quiver}) for the cocharacter notation.
}
\label{fig:3d-quiver}
\end{figure}

To construct a 3d quiver gauge theory, we consider several NS5-branes separated in the $x^4$-direction, which we take to be either a non-compact $\mathbb{R}$ or a compact $S^1$.  
(The latter will be necessary for a circular quiver theory.)
The $x^4$-direction is then divided into several segments that we label by $l=1,2,\ldots,L$; in the $l$-th segment we have $N_l$ D3-branes. 
The $N_l$ D3-branes in the $l$-th segment correspond to the $l$-th gauge node in the quiver diagram for a 3d quiver gauge theory.
An open string between the $l$-th stack of $N_l$ D3-branes and the $(l+1)$-th stack of $N_{l+1}$ D3-branes gives rise to a 3d hypermultiplet in the bifundamental representation of $U(N_l)\times U(N_{l+1})$.
See Figures~\ref{subfig:3d-quiver-branes-A} and~\ref{subfig:3d-quiver-A} for an illustration.

As in the case of SQCD we can introduce an NS5'-brane D1-branes with one end on the NS5'-brane and the other on the $l$-th stack of D3-branes.
This induces a monopole operator for the $U(N_l)$ gauge group corresponding to the $l$-th gauge node.
See Figures~\ref{subfig:3d-quiver-branes-monopoles-A}.

To read off the matrix model that describes monopole screening, we follow the same procedure as for the $U(N)$ SQCD described in Section~\ref{sec:matrix-SQCD-branes}.
We move the D3-branes relative to the NS5'-branes along the $x^4$-direction.
When a D3-brane crosses an NS5'-brane, a Hanany-Witten transition occurs where a D1-brane is created or destroyed~\cite{Hanany:1996ie}.
Applying the same procedure as in the 4d $\mathcal{N}=2$ theory~\cite{Brennan:2018yuj,Brennan:2018rcn}, we can reach a Hanany-Witten frame where D1-branes end on NS5'-branes (and NS5-branes), but not on D3-branes.

For a general 3d quiver gauge theory and its monopole operator, this brane configuration is the dimensional reduction of the brane box models~\cite{Hanany:1998it,Hanany:2018hlz}.
Here we focus on 0d multiplets that arise from open strings that end on D-branes located on different sides of an NS5-brane. 

Open strings ending on $n_l$ D1-branes in the $l$-th segment and $n_{l+1}$ D1-branes  in the $(l+1)$-th segment  give rise to $\mathcal{N}=(0,4)$  hypermultiplets in the bi-fundamental representation  of $U(n_1) \times U(n_2)$.  

On the other hand, open strings ending on $n_l$ D1-branes in the $l$-th segment and $n_{l-1}$ D3-branes  in the $(l-1)$-th segment give rise to 
$n_{l-1}$ $\mathcal{N}=(0,4)$ short Fermi multiplets in the fundamental representation of $U(n_l)$.
To see this, let us consider a configuration of $n_l$ D1-branes and  $n_{l-1}$ D5-branes in the $l$-th segment.
D1-D5 open stings give rise to $n_{l-1}$ $\mathcal{N}=(0,4)$ fundamental short Fermi multiplets.  
If we move the  $n_{l-1}$ D5-branes to the $(l-1)$-th segment, they pass an NS5-brane and a HW transition occurs.
We obtain $n_{l-1}$ D3-branes suspended between the D5-branes and the NS5-brane  in the $(l-1)$-th segment.  
Now $n_{l-1}$ $\mathcal{N}=(0,4)$ $U(n_l)$ short Fermi multiplets must arise from the open strings ending on  the D1-branes in the $l$-th segment  and the D3 branes in the $(l-1)$-th segment because the low energy world-volume theory on D1-branes is expected to be invariant in the transition.


\section{Rank one gauge theories}\label{sec:rank-one}

We now turn to concrete examples of 3d gauge theories.
In this section we study the classical and quanatized Coulomb branch chiral rings for rank one gauge theories.
\subsection{$U(1)$ with $N_F$ fundamental hypermultiplets}

In Section~\ref{sec:abelCB} we showed for abelian gauge theories that the vevs of the Coulomb branch operators in the $\Omega$-background together with Moyal multiplication
give the quantized  Coulomb branch chiral ring~$\mathbb{C}_{\epsilon}[\mathcal{M}_{C}]$ as described in~\cite{Bullimore:2015lsa}. 
Here we explicitly write down the generators and the relations for the  3d $\mathcal{N}=4$ $U(1)$ gauge theory with $N_F$ hypermultiplets with unit gauge charge. 
As explained in~\eqref{eq:exsca}, the expectation value of a polynomial~$P( {\Phi}_{ww})$ in the Coulomb branch scalar~${\Phi}_{ww}$ is a position independent constant
\begin{align}
\langle P({\Phi}_{ww} )\rangle = P( \varphi) \,.
\end{align}
Since there is no monopole bubbling  in an abelian  gauge theory,
 the expectation values of monopole operators are determined by the one-loop determinants for hypermultiplets and the classical part of the action.
For a magnetic charge ${\bm B}=p \in \mathbb{Z}$, the expectation value of the monopole operator ${V}_p$ is given by 
\begin{align}
v_{p}:=\langle {V}_{p} \rangle=e^{ p b }
\prod_{f=1}^{N_F}  \prod_{k=0}^{|p|-1}  
    \left (  \varphi -m_f+ \left( \frac{|p| -1}{2}-k \right) \epsilon \right) ^{\frac{1}{2}} .
\end{align}
We first consider the Moyal products of the vevs of identical  minimal monopole operators.
The vev of the monopole operator with magnetic charge $ |p|$ (resp. $-|p|$) is the Moyal product of $|p|$ copies of the minimal monopole operator $v_{+1}$ (resp. $v_{-1}$) 
\begin{align}
v_{\pm |p|}=\underbrace{v_{\pm 1}* \cdots * v_{\pm 1}}_{|p|\text{-times}}.
\label{eq:p_prod_mono}
\end{align}
Applying the Weyl transform \eqref{eq:Moyal_Wey} to 
\eqref{eq:p_prod_mono}, we obtain relations between $\hat{v}_{\pm|p|}$ and  $\hat{v}_{\pm 1}$ in the quantized chiral ring
\begin{align}
\hat{v}_{\pm |p|}= (\hat{v}_{\pm 1} )^{|p|} \,,
\label{eq:p_products}
\end{align}
which mean  that $\hat{v}_{\pm1}$  generate monopole operators with higher charges.

Next we evaluate the Moyal products of $\varphi$, $v_+$, and $ v_-$. Then we obtain 
\begin{align}
& {v}_{ +1}  *  {v}_{ -1} 
=\prod_{f=1}^{N_f}   \left( \varphi + \frac{\epsilon}{2}  -m_f   \right), \quad 
{v}_{ -1}  * {v}_{ +1} 
=\prod_{f=1}^{N_F}   \left( \varphi - \frac{\epsilon}{2}  -m_f   \right) , 
\label{eq:QAn1}
\\
&{\varphi} * {v}_{\pm 1}
= \left( \varphi \mp \frac{\epsilon}{2} \right) v_{\pm1} 
,  \quad 
{v}_{\pm 1} * {\varphi}
= \left( \varphi \pm \frac{\epsilon}{2} \right) v_{\pm1} .
\label{eq:QAn2}
\end{align}
Again, applying the Weyl transform to \eqref{eq:QAn1} and \eqref{eq:QAn2},
 we obtain the following relation between $\hat{\varphi}$, $\hat{v}_+$, and $ \hat{v}_-$ in the quantized Coulomb branch chiral ring :
\begin{align}
&\hat{v}_{+1} \hat{v}_{-1}=\prod_{f=1}^{N_F}   \left( \hat{\varphi} + \frac{\epsilon}{2}  -m_f   \right), \quad 
\hat{v}_{-1} \hat{v}_{+1}=\prod_{f=1}^{N_F}   \left( \hat{\varphi} - \frac{\epsilon}{2}  -m_f   \right), 
\label{eq:Qrelation1}
  \\
& [ \hat{\varphi}, \hat{v}_{\pm 1}]  = \mp \epsilon  \hat{v}_{\pm1} . 
\label{eq:Qrelation2}
\end{align}
\eqref{eq:p_products}, \eqref{eq:Qrelation1}, and \eqref{eq:Qrelation2} show that the quantized Coulomb branch chiral ring is generated by $\hat{\varphi}, \hat{v}_{+1}$, and $\hat{v}_{-1}$. 

The classical Coulomb branch relation  can be read off  by setting $\epsilon = 0$.
In this limit all the Coulomb branch operators commute and 
the quantized relations  \eqref{eq:Qrelation1} or equivalently the Moyal products \eqref{eq:QAn1}
 reduces to  the following relation 
\begin{align}
{v}_{+1} {v}_{-1} =\prod_{f=1}^{N_F}   \left( {\varphi}   -m_f   \right) .
\label{eq:classical_SQED}
\end{align}
The relation~\eqref{eq:classical_SQED} is  a deformation of the relation for $\mathbb{C}^2/ \mathbb{Z}_{N_F}$ by the parameters $m_f$,  and correctly reproduces the classical Coulomb branch of $N_F$-flavor SQED~\cite{Seiberg:1996bs}.  
Since the expectation values of arbitrary monopole operators are  given by products of  $v_{\pm1}$, the  classical Coulomb branch chiral ring is 
\begin{align}
\mathbb{C}[\mathcal{M}_{C}]&:=\mathbb{C}[ \{ v_{p} \}_{p \in \mathbb{Z}}, \varphi] / \left( v_{\pm n } =v_{\pm 1}^{n}\text{ for } n \in \mathbb{N} \, ,\,\, {v}_{+1} {v}_{-1} 
=\prod_{f=1}^{N_F} \left( \varphi -m_f   \right)\right )
\nonumber \\
&=\mathbb{C}[v_{+1}, v_{-1}, \varphi]/ \left( {v}_{+1} {v}_{-1} -\prod_{f=1}^{N_F}  \left( \varphi   -m_f   \right)\right ),   
\label{eq:classical_SQED_CB}
\end{align}
which is the well-known coordinate ring of a deformation of $\mathbb{C}^2/ \mathbb{Z}_{N_F}$.
The relations \eqref{eq:Qrelation1} and \eqref{eq:Qrelation2} imply that  the Poisson brackets between Coulomb branch operators are given by
\begin{align}
\{v_{+1}, v_{-1}  \}= \partial_{ \varphi} \prod_{f=1}^{N_F}   \left( \varphi   -m_f   \right), \quad 
\{\varphi, v_{\pm1}  \}=\mp  {v}_{\pm1}.
\end{align}

\subsection{$PSU(2)$  with $N_F$ adjoint hypermultiplets}
It is known that the Coulomb branch of the theory with gauge group~$PSU(2)\simeq SO(3)$ and $N_F$ adjoint hypermultiplets is  isomorphic to the orbifold $\mathbb{C}^2/D_{ N_F+1}$ where $D_{N_F+1}$ is a dihedral group~\cite{Braverman:2016wma}. 
We study the quantization   Coulomb branch in terms of localization formula and show classical Coulomb branch is reproduced by
 taking the limit $\epsilon \to 0$. The GNO magnetic (or equivalently Langlands dual) group of $PSU(2)$ is $SU(2)$ and the minimal magnetic charge, corresponding to the doublet, is ${\bm B}=\text{diag} (\frac{1}{2}, -\frac{1}{2})$.
Here we identify the Cartan subalgebra of $SU(2)$ with the space of traceless diagonal $2\times 2$ matrices, use the trace as the Killing form.
Let $\mathrm{diag} (\varphi, -\varphi)$ be  the  saddle point value of the scalar ${\Phi}_{ww}$,
${V}_{(1/2,-1/2)}$ be the minimal monopole operator, and  ${V}_{ {(1/2,-1/2)},\varphi}$ be a dressed monopole operator with magnetic charge $\text{diag} (\frac{1}{2}, -\frac{1}{2})$ and dressed by $\mathrm{diag} (\varphi, -\varphi)$. 
We define $X, Y$, and $Z$ by
\begin{align}
{X}&:=  -\frac{1}{2} \langle  \mathrm{Tr} \, {\Phi}_{w w}^2 \rangle=- \varphi^2 ,    \nonumber  \\
{Y}&:=\frac{1}{2^{N_F}}\langle {V}_{(1/2,-1/2)} \rangle= \frac{1}{2^{N_F}} \left(  v_+ + v_- \right), 
\label{eq:CopSU2ad} \\
{Z}&:=\frac{1}{2^{N_F}} \langle {V}_{{(1/2,-1/2)},\varphi}   \rangle= \frac{\varphi}{2^{N_F}} ( v_+ -v_-).  \nonumber 
\end{align}
Here   $v_{\pm}$ are defined by
\begin{align}
v_{\pm}:= e^{\pm  b} \frac{\prod_{f=1}^{N_F} \left(2   \varphi -m_f \right)^{\frac{1}{2}}\left( - 2 \varphi -m_f \right)^{\frac{1}{2}}}
{\left(  2 \varphi  + \frac{\epsilon}{2} \right)^{\frac{1}{2}}\left( - 2 \varphi +\frac{\epsilon}{2} \right)^{\frac{1}{2}}}.  
\end{align}
First we consider the classical case $\epsilon=0$. We find that $X, Y,$ and $Z$  satisfy 
the  relation
\begin{align}
Z^2+X Y^2= \prod_{f=1}^{N_F} \left( X + \frac{m_f^2}{4} \right)\,,
\label{eq:ALEDn}
\end{align}
 which
correctly reproduces  the algebraic description of the orbifold~$\mathbb{C}^2/D_{N_F+1}$   with complex deformations by the mass parameters.
Next we evaluate the commutation relations of $\widehat{X}, \widehat{Y}, \widehat{Z}$ in the  quantized Coulomb branch. 
For $PSU(2)$,  the Moyal product \eqref{eq:Moyal-def} and the Weyl transform \eqref{eq:Moyal_Wey} are given by
\begin{align} 
(F * G) ({\varphi}, { b})&= \exp \left[  \frac{\epsilon}{4} {  \left( \partial_{b} \partial_{\varphi^{\prime}}-\partial_{\varphi} \partial_{b^{\prime}} \right) }  \right]
 F ( \varphi, { b}) G ({ \varphi}^{\prime}, { b}^{\prime}) |_{{ \varphi}^{\prime} ={ \varphi}, { b}^{\prime}={ b}} \,, \\
\widehat{F}
& =
\exp \left({\epsilon}   \partial_{b} \partial_{\varphi} \right) F ({\varphi},{b})\Bigg |_{{\varphi}\rightarrow \widehat { \varphi}, b \rightarrow \widehat{ b}} \,.
\end{align}

We calculate the Moyal products between
$X$, $Y$,  and $Z$,  and then take their Weyl transforms.   We obtain  
\begin{align}
[ \widehat{X} ,  \widehat{Y}]& = \epsilon \widehat{Z}, \\
[\widehat{X}  , \widehat{Z}] &=  \epsilon \left( -\widehat{X }\widehat{Y} + \frac{\epsilon}{2}  \widehat{Z} -\frac{\epsilon^2}{16} \widehat{Y} \right), \\
[\widehat{Y}  ,  \widehat{Z}]&=\frac{\epsilon}{2} \widehat{Y}^2 +\frac{\prod_{f=1}^{N_F} \left(   \sqrt{-\widehat{X}}+\frac{\epsilon}{4} -\frac{m_f}{2} \right)\left(  - \sqrt{-\widehat{X}} -\frac{\epsilon}{4} -\frac{m_f}{2} \right) }{  2\sqrt{-\widehat{X}}} 
\nonumber \\
&\qquad \qquad -\frac{\prod_{f=1}^{N_F} \left(     \sqrt{-\widehat{X}}  -\frac{\epsilon}{4} -\frac{m_f}{2} \right)\left(  -  \sqrt{-\widehat{X}} +\frac{\epsilon}{4} -\frac{m_f}{2} \right) }{ 2 \sqrt{-\widehat{X}}} .
\end{align}
The square-roots are understood to be combined into powers with non-negative integer exponents when the expressions are expanded out.
The ring relation \eqref{eq:ALEDn} is deformed to
\begin{align}
\widehat{Z}^2+\widehat{Y} \widehat{X} \widehat{Y}=-\frac{\epsilon}{16} \widehat{Y}^2 +L(\widehat{X}, \epsilon) ,
\end{align}
where $L(\widehat{X}, \epsilon)$ is defined by 
\begin{align}
L(\widehat{X}, \epsilon) &=   \frac{\left( \sqrt{-\widehat{X}}+\frac{\epsilon}{4}  \right) \prod_{f=1}^{N_F}  \left(     \sqrt{-\widehat{X}} +\frac{\epsilon}{4} -\frac{m_f}{2} \right) \left(    - \sqrt{-\widehat{X}} -\frac{\epsilon}{4} -\frac{m_f}{2} \right)}
 { 2 \sqrt{-\widehat{X}}   } \nonumber \\
& \qquad +
  \frac{ \left( \sqrt{-\widehat{X}}-\frac{\epsilon}{4}  \right) \prod_{f=1}^{N_F}  \left(    \sqrt{-\widehat{X}} -\frac{\epsilon}{4} -\frac{m_f}{2} \right)\left( -   \sqrt{-\widehat{X}} +\frac{\epsilon}{4} -\frac{m_f}{2} \right) }
 { 2\sqrt{-\widehat{X}}  }. 
\end{align}

The Poisson brackets between classical Coulomb branch elements $X$, $Y$, and $Z$  are obtained by setting $\epsilon$ to zero:
\begin{align}
\{ X, Y \}= Z,  \quad \{ Z, X \}=- X Y, \quad \{ Y, Z \}= \frac{Y^2}{2}+S(X).
\end{align}
Here $S(X)$ is defined by 
\begin{align}
S(X)&=\lim_{\epsilon \to 0} \frac{1}{\epsilon} \left( 
\frac{\prod_{f=1}^{N_F} \left(   \sqrt{-\widehat{X}}+\frac{\epsilon}{4} -\frac{m_f}{2} \right)\left(  - \sqrt{-\widehat{X}} -\frac{\epsilon}{4} -\frac{m_f}{2} \right) }{  2\sqrt{-\widehat{X}}} 
\right. \nonumber \\
&\qquad \qquad \left. -\frac{\prod_{f=1}^{N_F} \left(     \sqrt{-\widehat{X}}  -\frac{\epsilon}{4} -\frac{m_f}{2} \right)\left(  -  \sqrt{-\widehat{X}} +\frac{\epsilon}{4} -\frac{m_f}{2} \right) }{ 2 \sqrt{-\widehat{X}}}
\right) .
\end{align}


\section{$U(N)$ with $N_F$ fundamental hypermultiplets: operator ordering and wall-crossing}
\label{sec:U(N)-SQCD-NF-flavors}

In this section we study
bare monopole operators 
 in the $\Omega$-deformed $U(N)$ SQCD, {\it i.e.}, the gauge theory with $N_F$ hypermultiplets  in the fundamental representation.
 We will study the situations where several minimal operators, {\it i.e.}, copies of  $  V_{\bm{e}_N} $ and  $  V_{-\bm{e}_1} $, are inserted at distinct points on the $x^3$-axis of the spacetime $\mathbb{R}^3$.
We will evaluate the Moyal products of the vevs of two and three minimal monopole operators, and relate them to the vevs of higher charge monopole operators with monopole bubbling. 
We will find that the ordering of  operators in a product is closely related to wall-crossing in the matrix models that compute monopole bubbling contributions.

The $U(N)$ SQCD treated in this section is a  special case of the linear quiver gauge theory with gauge group $\prod_{l=1}^n U(N_l)$ that we will study in Section~\ref{sec:linear-quiver}.
 In general the Weyl transforms of the vevs of monopole operators can be regarded as elements of the quantum deformation~$\mathbb{C}_\epsilon[\mathcal{M}_C]$ of the Coulomb branch chiral ring $\mathbb{C}[\mathcal{M}_C] \subset  \mathbb{C}[\mathcal{M}^{\text{abel}}_C]$.
We will study the relation between  the localization formulas for the Coulomb branch operators and the  quantized Coulomb branch in Section  \ref{sec:qCBlinear}. 

In this section we  focus on the relation between wall-crossing and the ordering of bare monopole operators.
 In  the $\mathcal{N}=2$ $U(N)$  gauge theory with $2N$ fundamental hypermultiplets  in four dimensions,  a similar relation between wall-crossing and the ordering of  minimal 't Hooft operators with opposite charges was studied in \cite{Assel:2019iae, Hayashi:2019rpw}.
 We refer the reader to~\cite{Hayashi:2019rpw} for more details on the relation between wall-crossing and the ordering of  operators.

Let us compute the vevs of minimal monopole operators.
For  $\mathfrak{m} ={\bm e}_k$ the one-loop determinants~\eqref{eq:1loopvec} and~\eqref{eq:1loophy}, with $\bm{B}$ replaced by $\mathfrak{m}$, read
\begin{align}
Z^{\text{vm}}_{1\mathchar `-\text{loop}}({ \bm \varphi} ; \mathfrak{m}= \pm {\bm e}_k)&=\prod_{j =1\atop j \neq k}^N  
   \left ( \varphi_{k j}  +  \frac{\epsilon}{2} \right) ^{-\frac{1}{2}} \left ( \varphi_{j k} +  \frac{\epsilon}{2} \right) ^{-\frac{1}{2}} \,,
 \\
Z^{\text{hm}}_{1\mathchar `-\text{loop}}({\bm \varphi}, {\bm m} ; \mathfrak{m}=\pm {\bm e}_k)&=\prod_{f=1}^{N_F}    
  \left(   \varphi_{k} -m_f   \right)^{\frac{1}{2}} \,,
\end{align}
where we defined $\varphi_{i j}:=\varphi_{i}-\varphi_{j}$. 
The vevs of minimal monopole operators are determined  unambiguously by the classical action and the one-loop determinants because the monopole bubbling effect is absent for ${\bm B}=\bm{e}_N$ and ${\bm B}=-\bm{e}_1$. 
Then the localization formula~(\ref{eq:exbaremono}) or~(\ref{eq:exdremono}) gives the expectation values of the minimal  monopole operators as
\begin{align}
\langle  V_{\bm{e}_N} \rangle
&=\sum_{k=1}^N e^{ b_{k}} \frac{\prod_{f=1}^{N_F}  
  \left(   \varphi_{k} -m_f   \right)^{\frac{1}{2}}}{\prod_{j=1 \atop j \neq k}^N  
   \left ( \varphi_{k j} +  \frac{\epsilon}{2} \right) ^{\frac{1}{2}}  \left ( - \varphi_{k j} +  \frac{\epsilon}{2} \right) ^{\frac{1}{2}} }  \,,
\label{eq:minimalmonople1}
\\
\langle  V_{-\bm{e}_1} \rangle
&=\sum_{k=1}^N e^{- b_{k}} \frac{\prod_{f=1}^{N_F}  
  \left(   \varphi_{k} -m_f   \right)^{\frac{1}{2}}}{\prod_{j=1 \atop j \neq k}^N  
   \left ( \varphi_{k j} +  \frac{\epsilon}{2} \right) ^{\frac{1}{2}}  \left ( - \varphi_{k j} +  \frac{\epsilon}{2} \right) ^{\frac{1}{2}} }  \,.
\end{align}
We will use these vevs as basic building blocks in the rest of the section.

\begin{figure}[thb]
\centering
\subfigure[]{\label{fig:adjmonopole1}
\includegraphics[width=7cm]{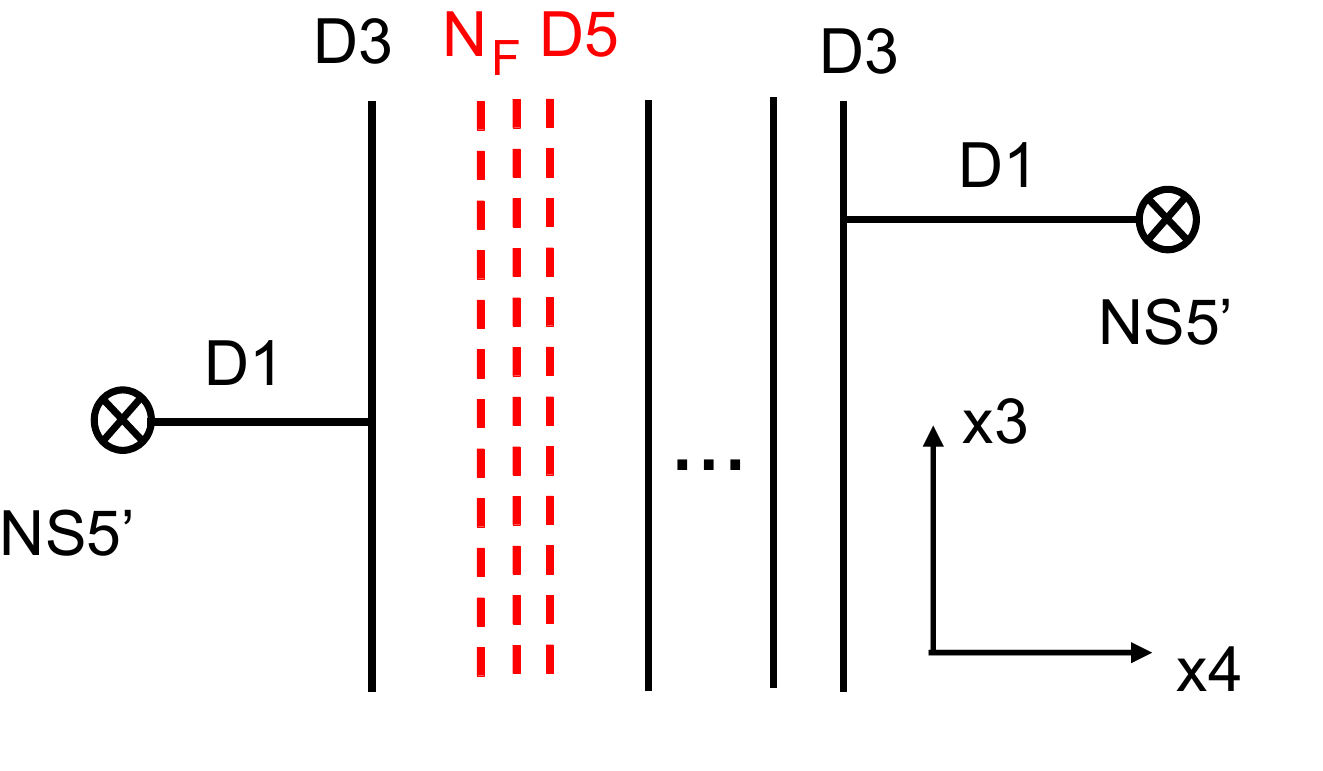}}
\hspace{1cm}
\subfigure[]{\label{fig:adjmonopole2}
\includegraphics[width=7cm]{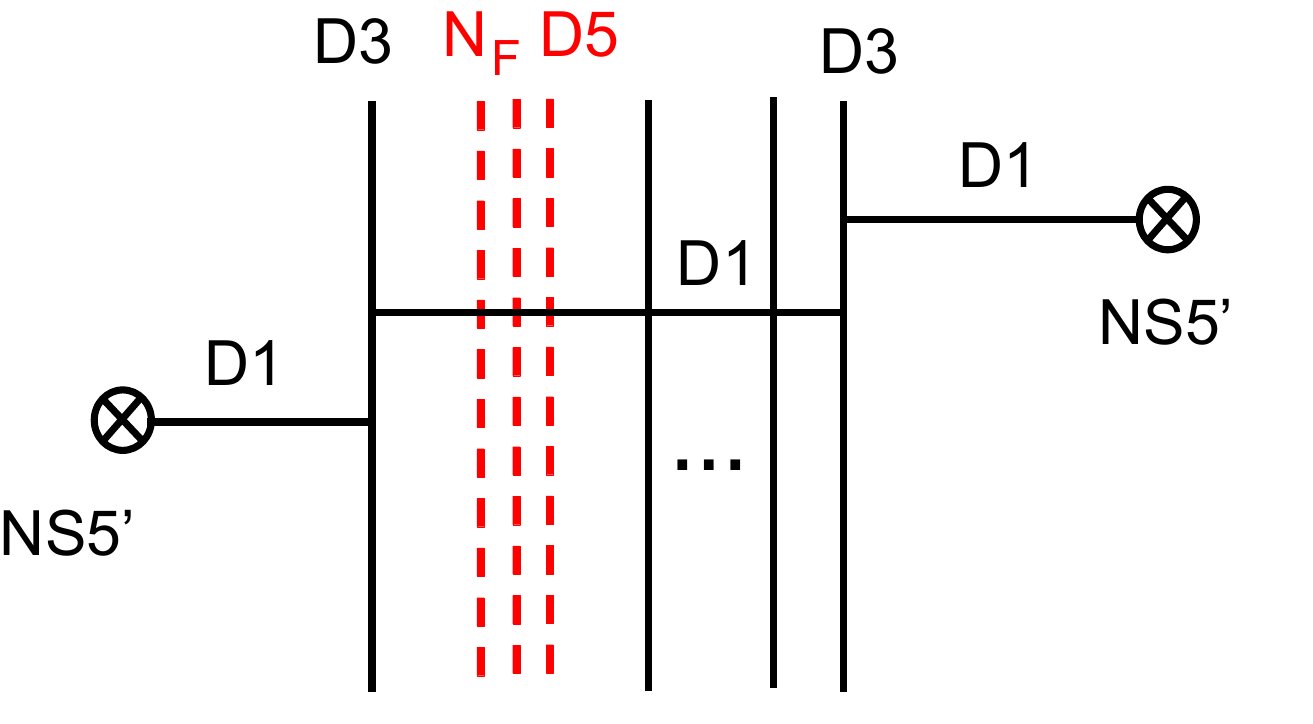}}
\subfigure[]{\label{fig:adjmonopole3}
\includegraphics[width=6cm]{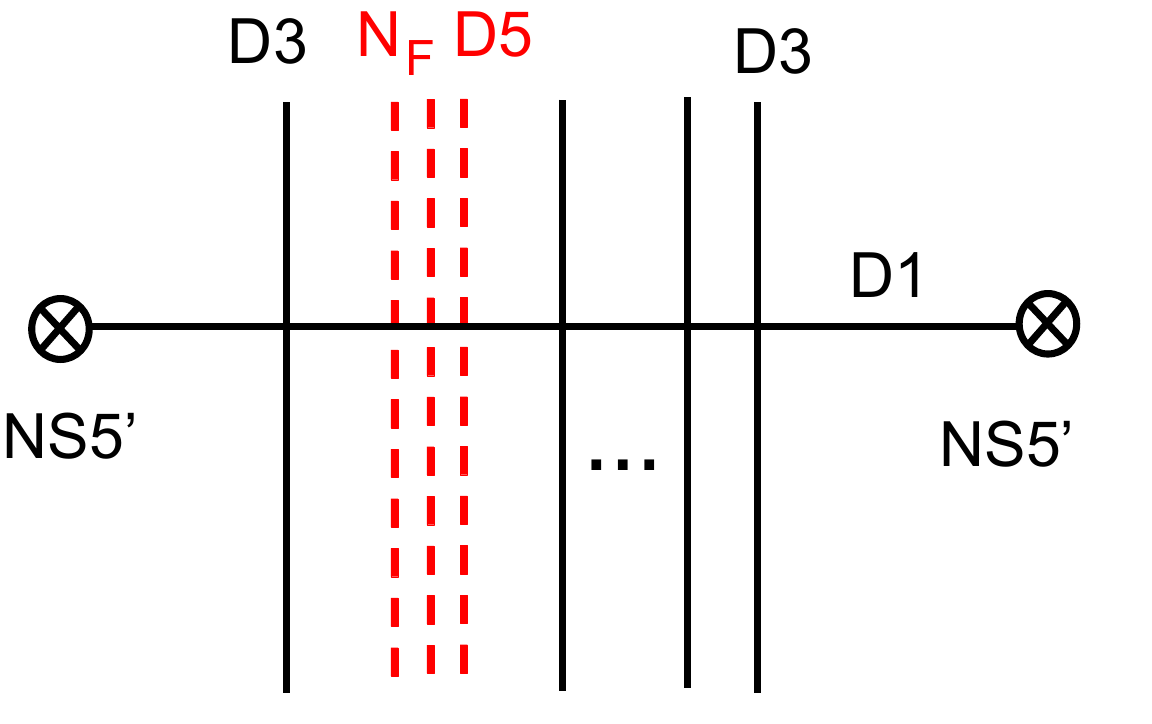}}
\hspace{1cm}
\subfigure[]{\label{fig:D0quiver1}
\hspace{-5mm}
\includegraphics[width=1cm]{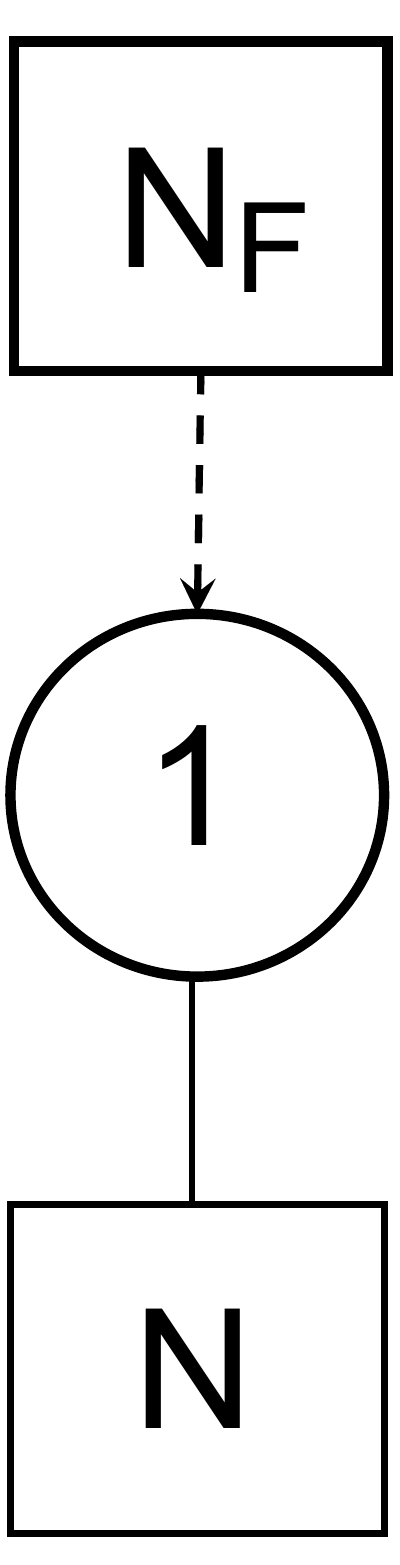}}
\caption{(a): The brane configuration for the (product) monopole operator with ${\bm B}=-\bm{e}_1 + \bm{e}_N$. (b) A D1-brane suspended between the leftmost D3-brane and the rightmost D3-brane corresponds to a smooth 't Hooft-Polyakov monopole with a magnetic charge $ \bm{e}_1 - \bm{e}_N$. 
(c): When  the   values of $x^3$ for the three D1-branes in Figure~(b) coincide,  the D1-branes  reconnect to form a single D1-brane and monopole bubbling occurs. 
The brane configuration corresponds to the bubbling sector $\mathfrak{m}={\bm 0}$.  (d) The quiver diagram  for the matrix model  associated with the world-volume theory on the D1-brane in Figure~(c).
  }
\label{fig:monopole3}
\end{figure}

\subsection{ Basic case: product of \texorpdfstring{$ V_{\bm{e}_N}$ and $ V_{-\bm{e}_1}$}{Basic case: product of ... and ...}}

We begin by studying the relation between wall-crossing in  $\langle  V_{-\bm{e}_1 + \bm{e}_N} \rangle$ and the ordering in the Moyal product of  $\langle  V_{\bm{e}_N} \rangle$ and $\langle  V_{-\bm{e}_1} \rangle$.  
We assume that $N\geq 2$.

The localization formula~(\ref{eq:exbaremono}) gives $\langle  V_{-\bm{e}_1 + \bm{e}_N} \rangle$ in the form
\begin{align}
\langle  V_{-\bm{e}_1 + \bm{e}_N} \rangle^{(\zeta)} 
&=  \sum_{1 \le k  \neq  l \le N} e^{  b_k - b_l } Z_{1\mathchar `-\text{loop}}( \mathfrak{m}= {\bm e}_k- {\bm e}_l) \nonumber \\
& \qquad  \qquad +Z^{(\zeta)}_{\mathrm{mono}} ( {\bm B}=-\bm{e}_1 + \bm{e}_N , \mathfrak{m}={\bm 0}). 
\label{eq:vevmonopolem1p1}
\end{align}
Here  $Z^{(\zeta)}_{\mathrm{mono}} ( {\bm B}=-\bm{e}_1 + \bm{e}_N, \mathfrak{m}={\bm 0} )$ is the monopole bubbling contribution that depends on the sign of the FI parameter $\zeta$   in the matrix model.  
 We will show that $\langle  V_{-\bm{e}_1 + \bm{e}_N} \rangle^{(\zeta)} $, where the superscript indicates that it  can depend on  (the sign of) $\zeta$,  coincides with the Moyal product between $\langle  V_{-\bm{e}_1} \rangle $ and $ \langle  V_{\bm{e}_N} \rangle$  in an appropriate ordering.

\subsubsection{ Moyal products of \texorpdfstring{$\langle  V_{-\bm{e}_1}\rangle$ and $ \langle  V_{\bm{e}_N}\rangle$}{... and ...}}

The Moyal products in two orderings  can be readily computed.
We write the results as
\begin{align}
\langle  V_{\bm{e}_N} \rangle *  \langle  V_{-\bm{e}_1} \rangle 
&=\sum_{1 \le k  \neq  l \le N}  e^{  b_k - b_l } Z_{1\mathchar `-\text{loop}}( \mathfrak{m}= {\bm e}_k- {\bm e}_l) \nonumber \\
& \qquad +\sum_{ k =1}^N  \frac{ \prod_{f=1}^{N_F} (\varphi_k -m_f+\frac{\epsilon}{2} )}{\prod_{j \neq k}\varphi_{k j} (\varphi_{j k}- \epsilon)} ,
\label{eq:moyalplusminus}
 \\
\langle  V_{-\bm{e}_1} \rangle *  \langle  V_{\bm{e}_N} \rangle 
&=\sum_{1 \le k  \neq  l \le N}  e^{  b_k - b_l } Z_{1\mathchar `-\text{loop}}( \mathfrak{m}= {\bm e}_k- {\bm e}_l) \nonumber \\
& \qquad +\sum_{ k =1}^N  \frac{ \prod_{f=1}^{N_F} (\varphi_k -m_f-\frac{\epsilon}{2} )}{\prod_{j=1 \atop j \neq k}^N \varphi_{k j} (\varphi_{j k}+ \epsilon)} .
\label{eq:moyalminusplus}
\end{align}
Comparing \eqref{eq:moyalplusminus} and \eqref{eq:moyalminusplus} with  \eqref{eq:vevmonopolem1p1},
 we expect
 the second  term in \eqref{eq:moyalplusminus}  or in \eqref{eq:moyalminusplus}
  to  equal 
$Z^{(\zeta)}_{\mathrm{mono}} ( {\bm B}=-\bm{e}_1 + \bm{e}_N, \mathfrak{m}={\bm 0} )$ in~(\ref{eq:vevmonopolem1p1})
 with $\zeta>0$ or $\zeta<0$, respectively.

\subsubsection{ Matrix model for \texorpdfstring{$(\bm{B},  \mathfrak{m})=(-\bm{e}_1 + \bm{e}_N,0)$}{...}}

 We now compute the monopole bubbling contribution $Z^{(\zeta)}_{\mathrm{mono}}$ in \eqref{eq:vevmonopolem1p1}  by applying the JK residue prescription in Appendix~\ref{app:JK} to the matrix model obtained by the brane construction in Section~\ref{sec:matrix-branes}.

The set-up with two minimal monopole operators, with total magnetic charge ${\bm B}=-\bm{e}_1 + \bm{e}_N$, is realized by the brane configuration in Figure~\ref{fig:adjmonopole1}. 
The $\mathfrak{m}={\bm 0}$ sector with complete screening of the charge is realized, as  shown in Figures~\ref{fig:adjmonopole2} and~\ref{fig:adjmonopole3}, by adding  a  D1-brane suspended between two D3-branes.
The contribution due to monopole bubbling is given as the partition function of the matrix model realized as the low energy world-volume theory on the Euclidean D1-brane in  Figure~\ref{fig:adjmonopole3}.
The matter content of the supersymmetric matrix model is  encapsulated in the quiver diagram shown in  Figure~\ref{fig:D0quiver1}.
By applying  the localization formula summarized in Appendix~\ref{app:JK} to the matrix model, we obtain  a contour integral expression  for the bubbling contribution:
\begin{align}
Z^{(\zeta)}_{\mathrm{mono}} ( {\bm B}=-\bm{e}_1 + \bm{e}_N, \mathfrak{m}={\bm 0} ) 
=
\oint_{\mathrm{JK} (\zeta)} \frac{d u}{2 \pi i} 
\frac{ (- \epsilon) \prod_{f=1}^{N_F}(u-m_f)  }{\prod_{s=\pm1} \prod_{i=1}^N \left( s(u-\varphi_i) -\frac{\epsilon}{2} \right)} .
\label{eq:contourint1}
\end{align}
Here the integration contour $\mathrm{JK} (\zeta)$ in \eqref{eq:contourint1} is determined by
 the sign of the FI parameter~$\zeta$;
the residues are evaluated at $u=\varphi_i+\frac{\epsilon}{2}$ ($i=1, \cdots, N$) for $\zeta >0$, and at $u=\varphi_i-\frac{\epsilon}{2}$ ($i=1, \cdots, N$) for $\zeta <0$, respectively. 
We find
\begin{align}
Z^{(\zeta>0)}_{\mathrm{mono}} ( {\bm B}=-\bm{e}_1 + \bm{e}_N, \mathfrak{m}={\bm 0} ) &= \sum_{ k =1}^N  \frac{ \prod_{f=1}^{N_F} (\varphi_k -m_f+\frac{\epsilon}{2} )}{\prod_{j \neq k}\varphi_{k j} (\varphi_{j k}- \epsilon)},  
\label{eq:screeningplus}\\
 Z^{(\zeta<0)}_{\mathrm{mono}} ( {\bm B}=-\bm{e}_1 + \bm{e}_N, \mathfrak{m}={\bm 0} ) &=\sum_{ k =1}^N \frac{ \prod_{f=1}^{N_F} (\varphi_k -m_f-\frac{\epsilon}{2} )}{\prod_{j \neq k}\varphi_{k j} (\varphi_{j k}+ \epsilon)} .
\label{eq:screeningminus}
\end{align}
We see that there is a relation
\begin{align}
Z^{(\zeta>0)}_{\mathrm{mono}} ( {\bm B}=-\bm{e}_1 + \bm{e}_N, \mathfrak{m}={\bm 0} ; \epsilon ) =
 Z^{(\zeta<0)}_{\mathrm{mono}} ( {\bm B}=-\bm{e}_1 + \bm{e}_N, \mathfrak{m}={\bm 0} ; -\epsilon ) .
\end{align}

\subsubsection{Operator ordering and wall-crossing}

 The expressions \eqref{eq:screeningplus} and \eqref{eq:screeningminus}   coincide with the monopole bubbling  contributions anticipated from the Moyal products \eqref{eq:moyalplusminus} and \eqref{eq:moyalminusplus}. 
Therefore  the signs of $\zeta$ for $\langle  V_{-\bm{e}_1 + \bm{e}_N} \rangle^{ (\zeta)}$
are in a  one-to-one correspondence with the  orderings in the Moyal product:
\begin{align}
\langle  V_{\bm{e}_N} \rangle *  \langle  V_{-\bm{e}_1} \rangle 
&=\langle  V_{-\bm{e}_1 + \bm{e}_N} \rangle^{(\zeta >0)}, 
\label{eq:moyalvpvm}\\
 \langle  V_{-\bm{e}_1} \rangle * \langle  V_{\bm{e}_N} \rangle 
&=\langle  V_{-\bm{e}_1 + \bm{e}_N} \rangle^{(\zeta < 0)}.
\label{eq:moyalvmvp}
\end{align}
 We  saw
 in Section~\ref{sec:moyalorder}  that the orderings 
in the Moyal product are same as the orderings of the operators  along the $x^3$-direction. 
 We can therefore write
\begin{align}
\langle  V_{\bm{e}_N} (s_2)  V_{-\bm{e}_1} (s_1) \rangle=
\left\{
\begin{array}{cc}
\langle  V_{-\bm{e}_1 + \bm{e}_N} \rangle^{(\zeta >0)}  & (s_1 < s_2 ),  \\
\langle  V_{-\bm{e}_1 + \bm{e}_N} \rangle^{(\zeta < 0)}  & (s_1 > s_2), \\
\end{array}
\right.
\label{eq:ordermonopole}
\end{align}
where $s_1$ and $s_2$ are the positions  of $ V_{-\bm{e}_1} $ and $ V_{\bm{e}_N}$ in the $x^3$-direction, respectively, and are related to the FI parameter as $\zeta=s_2-s_1$.

When the $\Omega$-deformation is turned off we have
\begin{equation}
\langle  V_{-\bm{e}_1 + \bm{e}_N} \rangle 
=
\langle  V_{\bm{e}_N} \rangle   \langle  V_{-\bm{e}_1} \rangle
\quad
\text{ for } \epsilon=0.
\label{eq:omegazero}
\end{equation}
The correlation function~\eqref{eq:omegazero}  is completely independent of the positions of the operators;  it is independent of not only the distance but also the ordering.
This is as expected because our monopole operators are observables of a 
  topological field theory as explained in Section~\ref{sec:twist-Coulomb}.

 The two expressions in~\eqref{eq:screeningplus} and~\eqref{eq:screeningminus} appear rather different.
For low values of $N_F$, however, their values in fact coincide and there is no wall-crossing.
Their difference is the residue of~\eqref{eq:contourint1} at  $u =\infty$.
We find%
\footnote{%
 In the special case~$N_F=2N$, the result~(\ref{eq:gooduglybad}) can also be obtained from a mathematical result, the one-instanton case of Theorem~3.6 in~\cite{Ohkawa:2015dam}, on wall-crossing in the instanton counting in $\mathbb{C}^2$ by applying the $U(1)_K$ projection procedure described in Section~5.3 of~\cite{Ito:2013kpa}.
 }
\begin{equation}
\begin{aligned}
& \langle  V_{\bm{e}_N} \rangle *  \langle  V_{-\bm{e}_1} \rangle  - {  \langle  V_{-\bm{e}_1} \rangle * \langle  V_{\bm{e}_N} \rangle}   \\
=&Z^{(\zeta>0)}_{\mathrm{mono}} ( {\bm B}=-\bm{e}_1 + \bm{e}_N, \mathfrak{m}={\bm 0} )-Z^{(\zeta<0)}_{\mathrm{mono}} ( {\bm B}=-\bm{e}_1 + \bm{e}_N, \mathfrak{m}={\bm 0} )  \\
=&\oint_{u= \infty} \frac{d u}{2 \pi i} \frac { (-\epsilon) \prod_{f=1}^{N_F}(u-m_f)}{\prod_{s=\pm1} \prod_{i=1}^N \left( s(u-\varphi_i) -\frac{\epsilon}{2} \right)}  \\
=&\left\{
\begin{array}{cll}
0  & \text{for $N_F<2N-1$} & \hspace{-1mm} \text{(bad)}, \\
 (-1)^{N-1} \epsilon  & \text{for $N_F=2N-1$} & \hspace{-1mm}\text{(ugly),}\\
\hspace{-2.5mm}
 (-1)^{N-1}\epsilon \Big( 2 \sum_{i=1}^N \varphi_i 
-\sum_{f=1}^{2N} m_f  \Big)
\hspace{-2mm}   &\text{for $N_F=2N$} & \hspace{-1mm}\text{(good and balanced),}\\
 \epsilon A ({\bm \varphi}, {\bm m}, \epsilon) & \text{for $N_F>2N$}  &\hspace{-1mm} \text{(good but not balanced)}
\end{array}
\right.
\label{eq:gooduglybad}
\end{aligned}
\end{equation}
with 
\begin{align}
A ({\bm \varphi}, {\bm m}, \epsilon):= \frac{1}{(N_F+1-2N)!} \left( \frac{d}{ d w } \right)^{ N_F+1-2N}   \frac{\prod_{f=1}^{N_F}(1 -m_f w)}
{ \prod_{i=1}^N \left( 1 - \left( \varphi_i + \frac{\epsilon}{2} \right) w  \right) \left( -1 + \left( \varphi_i - \frac{\epsilon}{2} \right) w  \right) } \Bigg |_{w=0} .
\end{align}
The Possion bracket  between $\langle V_{\bm{e}_N}\rangle$ and $ \langle V _{-\bm{e}_1}\rangle$, obtained by applying the relation~\eqref{eq:moyalpoisson}
to~\eqref{eq:gooduglybad}, is
\begin{equation}\label{SQCD-Poisson-bracket}
\begin{aligned}
& \left\{ \langle V_{\bm{e}_N} \rangle, \langle V _{-\bm{e}_1}\rangle \right\}
\\
=&
\left\{
\begin{array}{cll}
0 & \text{for $N_F<2N-1$}   & \text{(bad)}, \\
 (-1)^{N-1}   &  \text{for $N_F=2N-1$} \hspace{-1mm} & \text{(ugly),}\\
\displaystyle  
 (-1)^{N-1}   (2 \,{ \sum_{i=1}^N \varphi_i}-\sum_{f=1}^{2N} m_f) &  \text{for $N_F=2N$} & \text{(good and balanced),}\\
A ({\bm \varphi}, {\bm m}, \epsilon=0)&  \text{for $N_F>2N$}  & \text{(good but not balanced).}
\end{array}
\right.
\end{aligned}
\end{equation}

The properties ``bad'', ``ugly'', ``good'', and ``balanced'' were introduced in~\cite{Gaiotto:2008ak}, and we review them here.
They concern the violation of the unitarity bound by monopole operators in the massless limit.
Let $q_R$ denote the UV $\mathcal{N}=2$ R-charge given by the formula~(\ref{eq:UV-R-charge}).
A {\it bad} theory has at least one monopole operator with $q_R\leq 0$, and cannot flow to an $\mathcal{N}=4$ superconformal field theory (SCFT) with the R-symmetry given by that in the UV.
For $N_F<2N-1$ the minimal monopole operators have~$q_R\leq 0$.
In an {\it ugly} theory the smallest value of $q_R$ is $1/2$; the monopole operators with this R-charge decouple in the IR and become the scalar components of free twisted hypermultiplets.
For $N_F=2N-1$ the two minimal operators have $q_R=1/2$ and therefore must decouple in the IR to reside in a single free twisted hypermultiplet; the Moyal commutator in~(\ref{eq:gooduglybad}) and the Poisson bracket in~(\ref{SQCD-Poisson-bracket}) in this case are indeed those of such a free theory.
A {\it good} theory flows to an~$\mathcal{N}=4$ SCFT with the R-symmetry given by that in the UV.
A {\it balanced} theory has monopole operators with $q_R=1$; such monopole operators are the bottom components of $\mathcal{N}=2$ current multiplets that are responsible for enhancing $\mathfrak{u}(1)_C$ to $\mathfrak{su}(2)_C$.
For $N_F=2N$ the minimal operators do have $q_R=1$ and are part of the $\mathcal{N}=4$ stress tensor multiplet.

Thus we find  that the behavior of  $Z_{\text{mono}}$ under wall-crossing is closely related to  the division of the $\mathcal{N}=4$ gauge theories into the categories ``bad'', ``ugly'', ``good'', and ``balanced''.

\begin{figure}[htb]
\centering
\subfigure[]{\label{fig:symmonopole1}
\includegraphics[width=5cm]{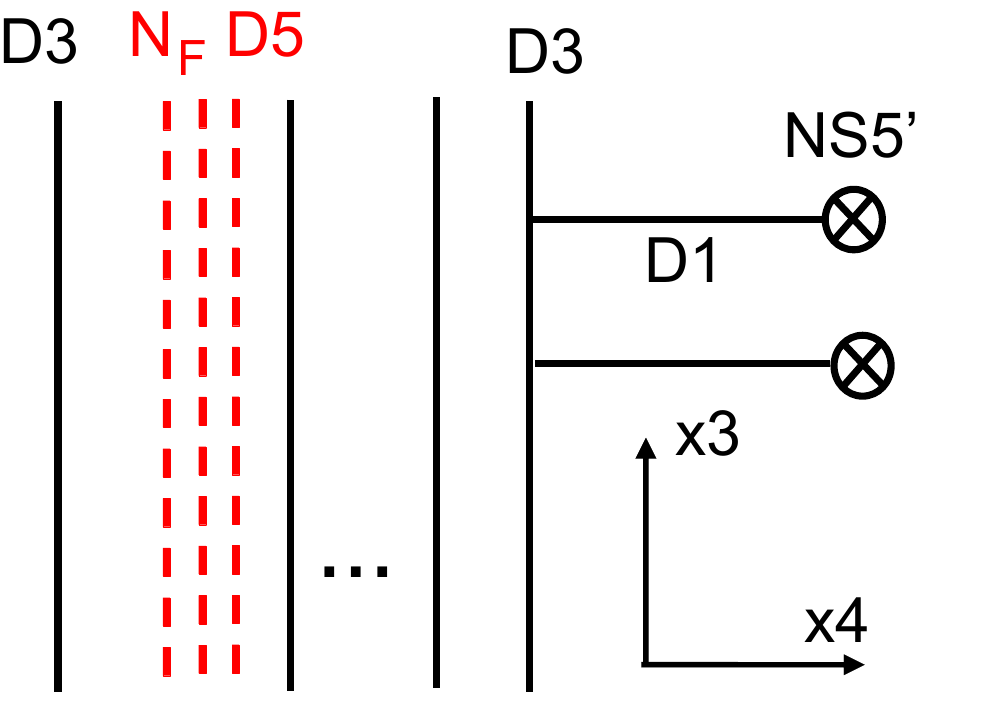}}
\subfigure[]{\label{fig:symmonopole2}
\includegraphics[width=5cm]{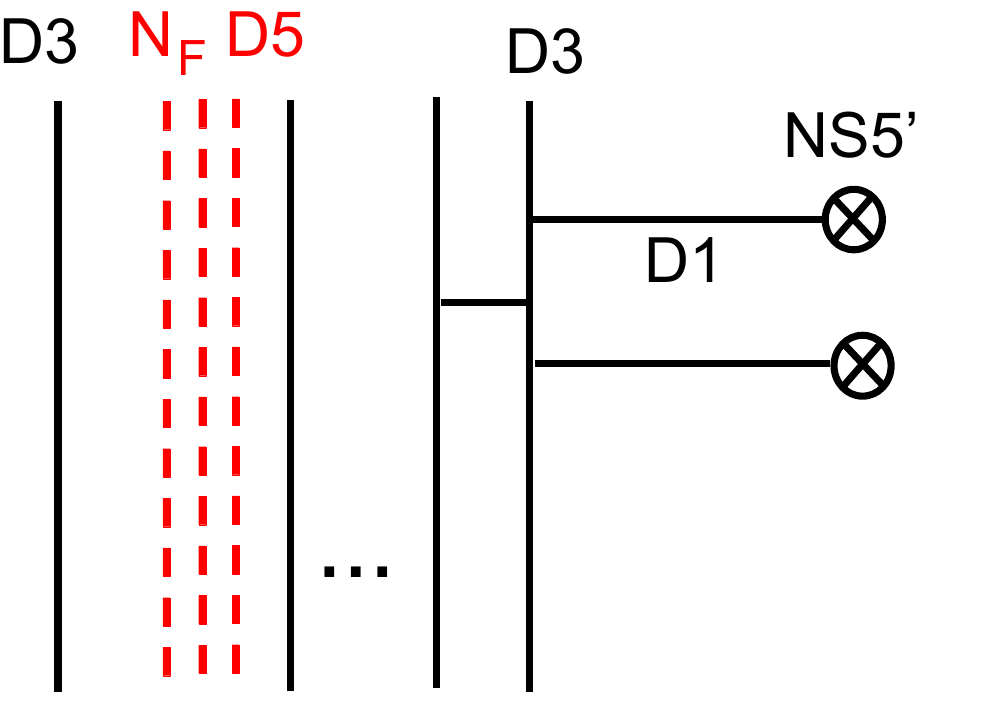}}
\subfigure[]{\label{fig:symmonopole3}
\includegraphics[width=4.5cm]{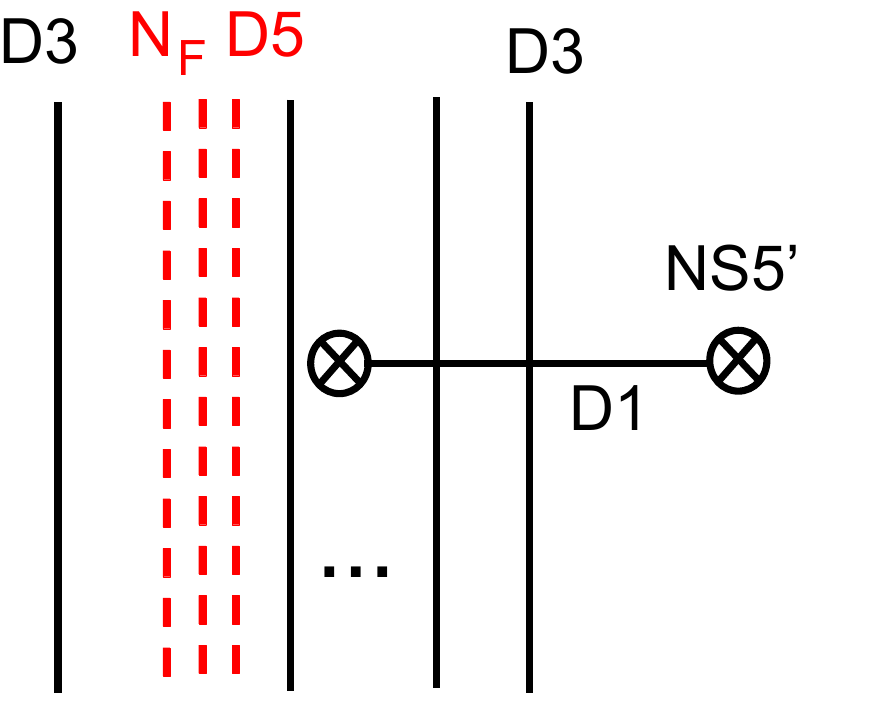}}
\caption{(a): The brane configuration for monopole with ${\bm B}=
2 \bm{e}_N
$.
(b): A  D1-brane describing a smooth monopole with  magnetic charge $
\bm{e}_{N-1} - \bm{e}_N
$ is introduced. (c) When the coordinates of two D1-branes along $x^3$-direction coincide,  
the magnetic charge is partially screened to $\mathfrak{m}=
\bm{e}_{N-1}+\bm{e}_{N}
$. }
\label{fig:monopole4}
\end{figure}

\begin{figure}[htb]
\begin{center}
\includegraphics[width=2cm]{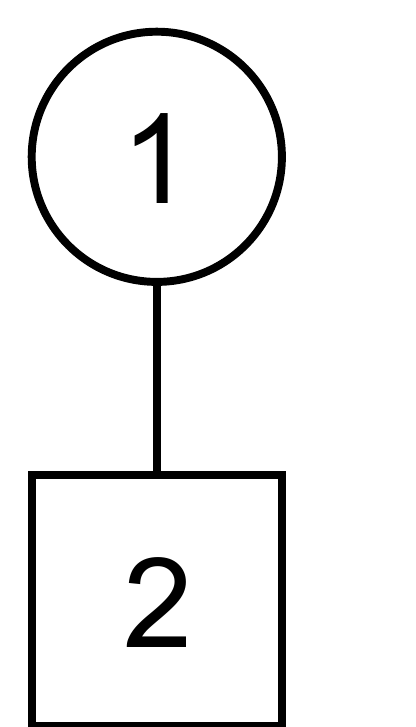}
\end{center}
\vspace{-0.5cm}
\caption{ The quiver diagram of the matrix model  for $(\bm{B}, \mathfrak{m})=(2 \bm{e}_N,   \bm{e}_k+\bm{e}_l)$ with $k<l$.
 }
\label{fig:D1quiversym}
\end{figure}

\subsection{ Product of two \texorpdfstring{$  V_{\bm{e}_N}$'s}{...}}

Next we will compute  the Moyal product of two $\langle  V_{\bm{e}_N} \rangle$ and compare it with  $\langle  V_{2 \bm{e}_N} \rangle$.

\subsubsection{ Moyal product of two \texorpdfstring{$\langle  V_{\bm{e}_N}\rangle$'s}{...}}

 We can readily compute the Moyal product of { two} $\langle  V_{\bm{e}_N} \rangle$'s  in~\eqref{eq:minimalmonople1}. 
We write the result as
\begin{align} \label{eq:MoyalVeNVeN}
\langle  V_{\bm{e}_N} \rangle * \langle  V_{\bm{e}_N} \rangle
&=\sum_{k=1 }^N  e^{ 2 b_k  } Z_{1\mathchar `-\text{loop}}( \mathfrak{m}=2 {\bm e}_k)  \nonumber \\
& \qquad \qquad + \sum_{1 \le k < l \le N}  e^{  b_{k }+b_{l }}  Z_{1\mathchar `-\text{loop}}( \mathfrak{m}= {\bm e}_k+{\bm e}_l)   \frac{2}{(\varphi_{kl}+\epsilon) (\varphi_{lk}  +\epsilon)}.
\end{align}
Then 
\begin{align}
{
\Big( Z_{\text{mono}}(\mathfrak{m}= {\bm e}_k+{\bm e}_l) \, \, \text{read off from} \, \, \langle V_{\bm{e}_N} \rangle * \langle V_{\bm{e}_N} \rangle \Big )
}
\ =  \
   \frac{2}{(\varphi_{kl}+\epsilon) (\varphi_{lk}+\epsilon)} .
\label{eq:zmono20}
\end{align}
For identical operators there is no ordering ambiguity; we expect no wall-crossing for the corresponding matrix model.

\subsubsection{ Matrix model for \texorpdfstring{$(\bm{B},  \mathfrak{m})=(2 \bm{e}_N,{\bm e}_k+{\bm e}_l )$ with $k<l$}{...}}

Next we evaluate the monopole bubbling  contribution from  the sector $ \mathfrak{m}={\bm e}_k+{\bm e}_l $ with $1\leq k < l\leq N$.
 As noted in Section~\ref{sec:matrix-SQCD-branes} (see also Figure~3 of~\cite{Hayashi:2019rpw}), we can permute the D3-branes.
The brane configuration for the case $(k,l)=(N-1,N)$, which one can reach by a permutation, is shown in Figure~\ref{fig:monopole4}. 
The matter content of the matrix model is  that specified by the quiver diagram in Figure~\ref{fig:D1quiversym}.
The partition function, identified with the bubbling contribution, can be written  as a contour integral
\begin{align}
Z^{(\zeta)}_{\mathrm{mono}} ( {\bm B}=2 \bm{e}_N, \mathfrak{m}={\bm e}_k+{\bm e}_l ) 
=\oint_{\mathrm{JK} (\zeta)} \frac{d u}{2 \pi i} \frac{-\epsilon }{\prod_{s=\pm1} \prod_{i=k,l} \left( s(u-\varphi_i) -\frac{\epsilon}{2} \right)} .
\label{eq:contourint2}
\end{align}
 According to the JK residue prescription in Appendix~\ref{app:JK}, we should pick poles at $u=\varphi_i+\frac{\epsilon}{2}$ ($i=k, l$) for $\zeta >0$, and at $u=\varphi_i-\frac{\epsilon}{2}$ ($i=k,l$) for $\zeta <0$.
Even before evaluating the residues, we can see that $Z^{(\zeta>0)}_{\mathrm{mono}}=Z^{(\zeta<0)}_{\mathrm{mono}}$ because there is  no  pole at $u= \infty$. 
 There is no wall-crossing, and for either sign of $\zeta$ the JK residues sum up to
\begin{align}
  Z^{(\zeta)}_{\mathrm{mono}} ( {\bm B}=2 \bm{e}_N, \mathfrak{m}={\bm e}_k+{\bm e}_l ) =\frac{2}{(\varphi_{kl}+\epsilon) (\varphi_{lk}+\epsilon)} .
\label{eq:Zmonoekel}
\end{align}
This coincides with \eqref{eq:zmono20}.

With this result, we see that the whole expression~(\ref{eq:exbaremono}) for $\langle  V_{2 \bm{e}_N} \rangle$ coincides with~(\ref{eq:MoyalVeNVeN}):
\begin{align}
\langle  V_{2 \bm{e}_N} \rangle&=\langle  V_{\bm{e}_N} \rangle * \langle  V_{\bm{e}_N} \rangle.
\end{align}

\subsection{ Product of \texorpdfstring{one $  V_{-\bm{e}_1} $ and two $  V_{\bm{e}_N} $'s}{...}}
\begin{figure}[htb]
\centering
\subfigure[]{\label{fig:21monopole1}
\includegraphics[width=7cm]{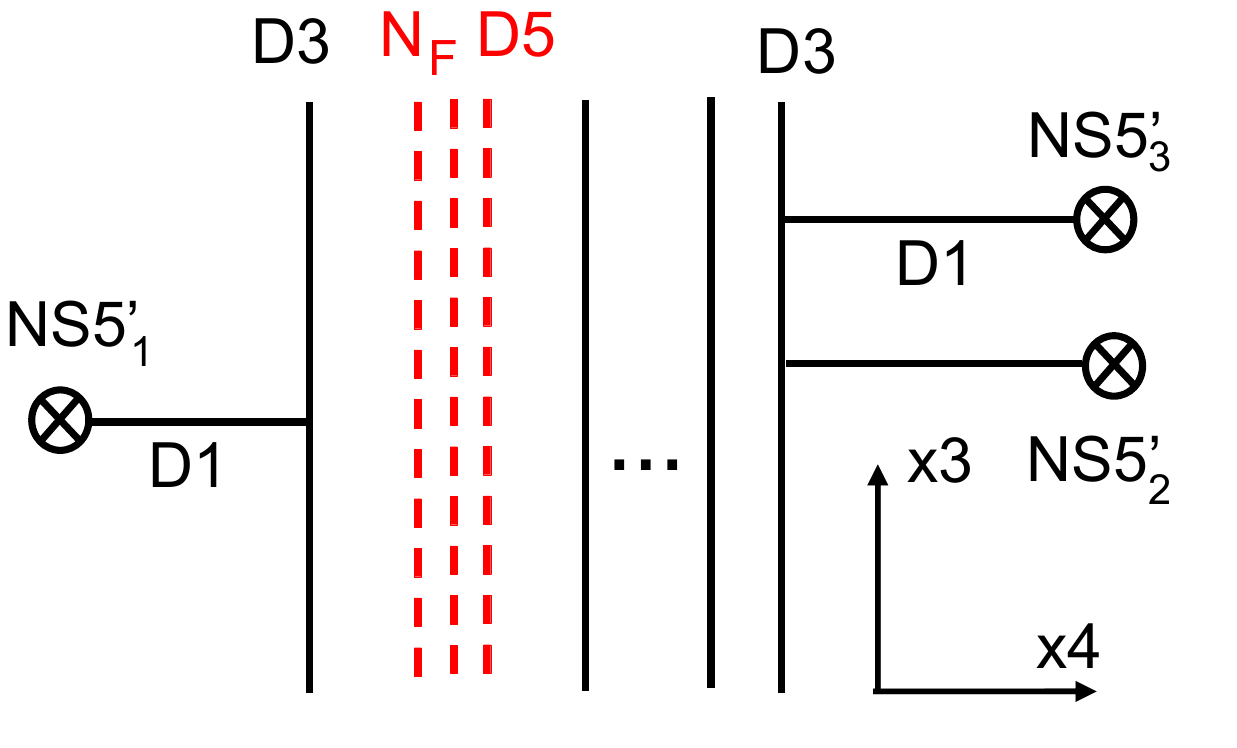}}
\hspace{.5cm}
\subfigure[]{\label{fig:21monopole2}
\includegraphics[width=6cm]{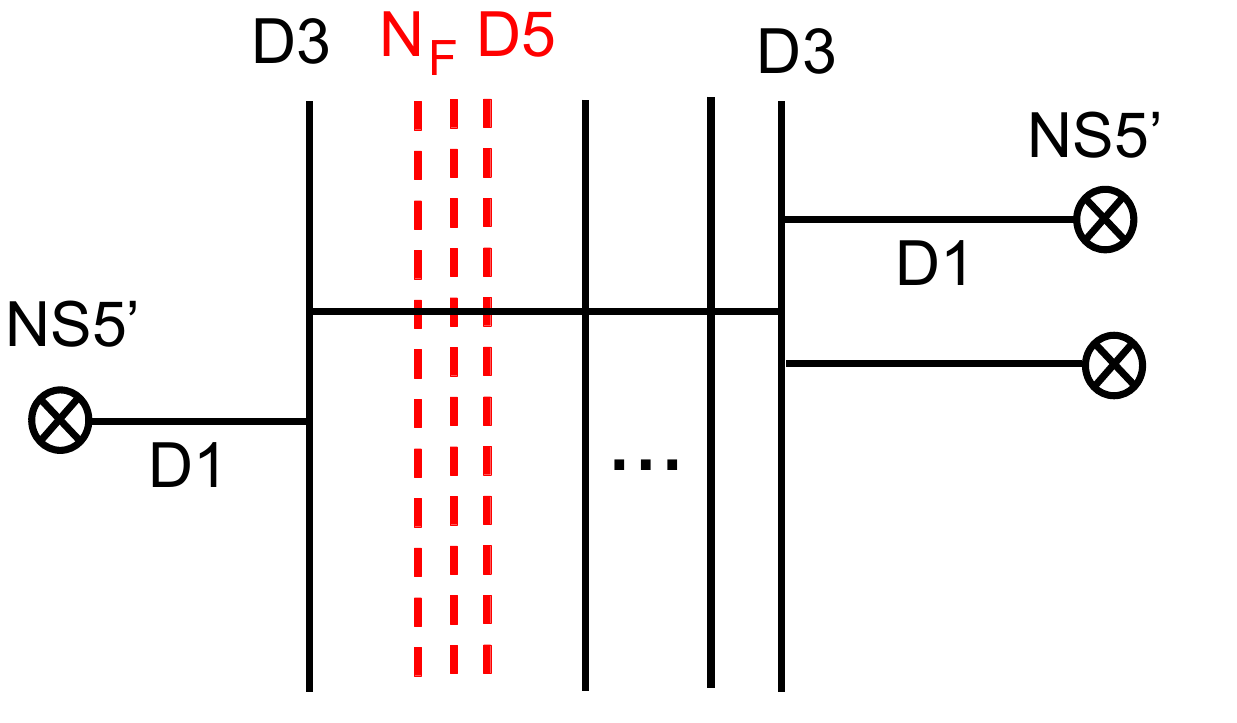}}
\subfigure[]{\label{fig:21monopole3}
\includegraphics[width=6cm]{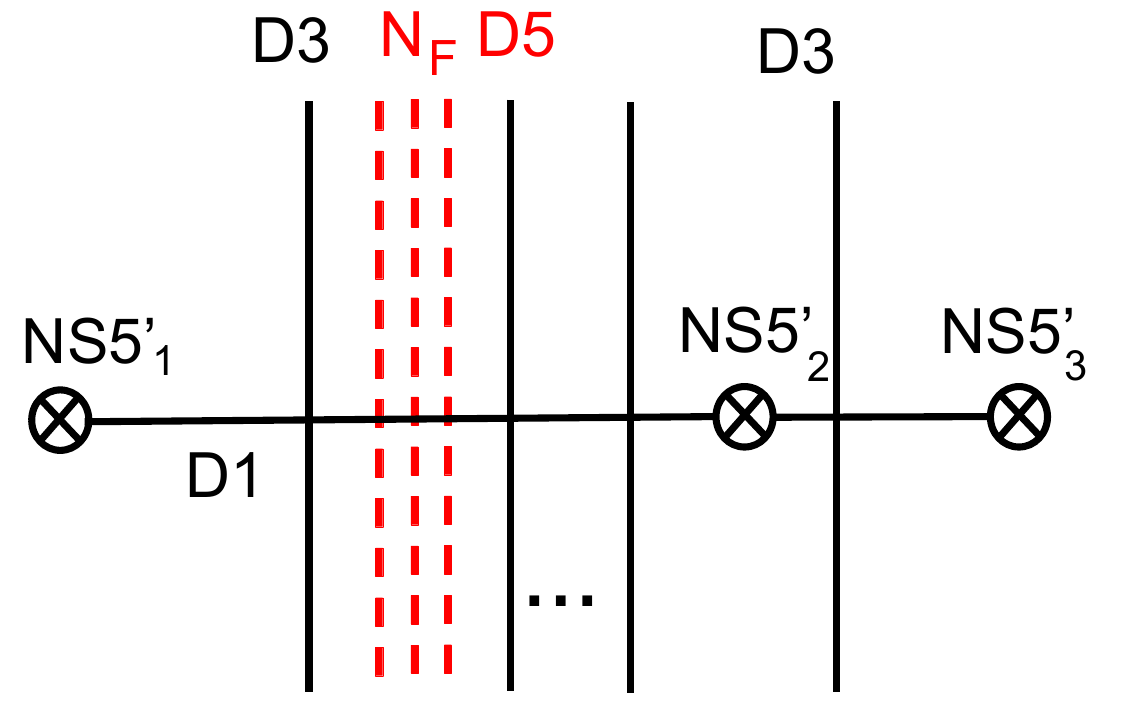}}
\hspace{1cm}
\subfigure[]{\label{fig:21quiverMM}
\includegraphics[width=3cm]{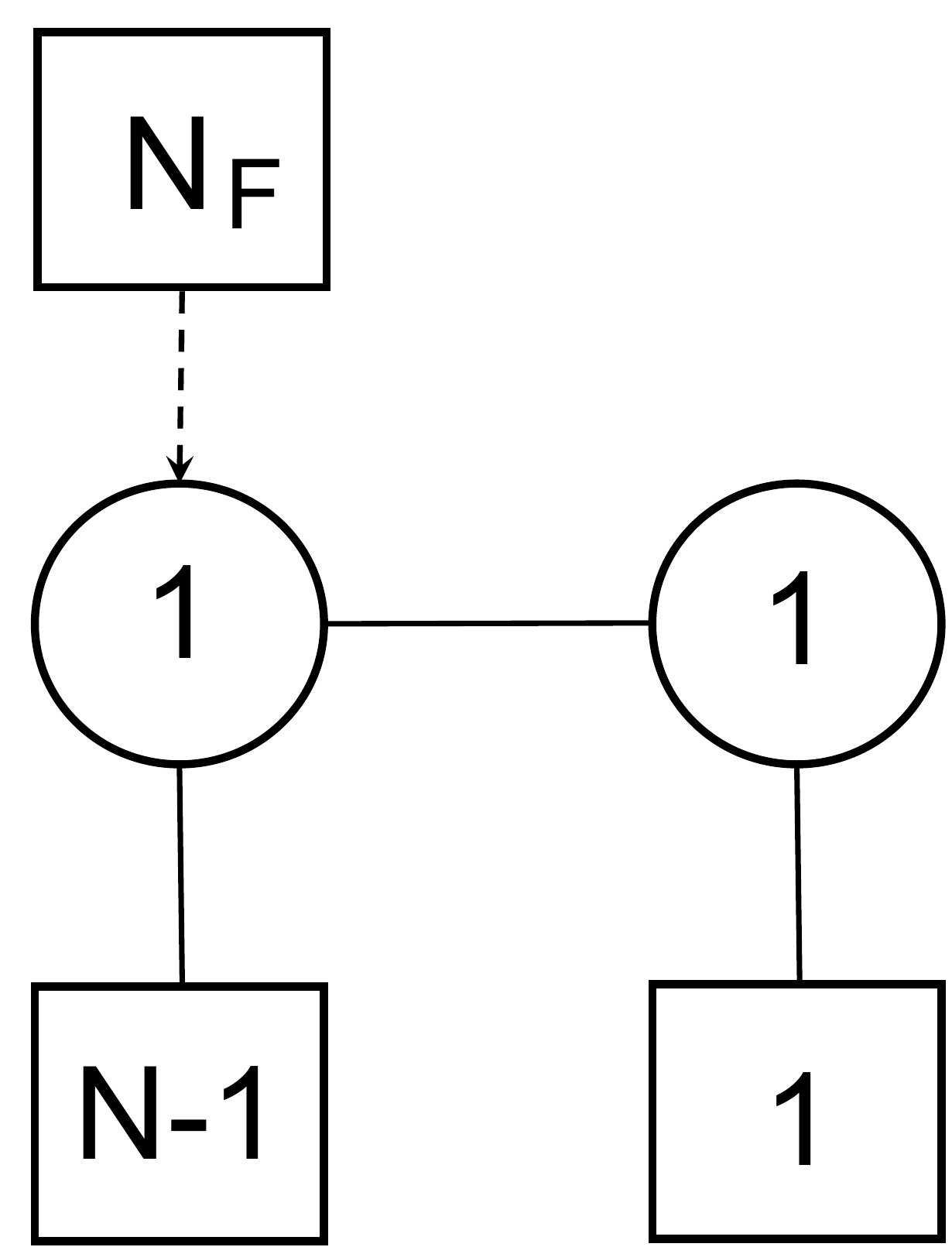}}
\caption{(a): The brane configuration for monopole with ${\bm B}=-\bm{e}_1 + 2\bm{e}_N$.
(c): The brane configuration corresponding to the monopole bubbling sector $\mathfrak{m}=\bm{e}_N$. 
(d) The quiver diagram of the matrix model for monopole bubbling sector $\mathfrak{m}={\bm e}_{N}$ in ${\bm B}=-\bm{e}_1 + 2\bm{e}_N$.
 }
\label{fig:monopole5}
\end{figure}
We now extend the correspondence between operator ordering and  the dependence of $Z_{\text{mono}}$ on the FI-parameter, or more precisely on the FI-chamber defined in~(\ref{eq:FIchamber}), to the products of  three minimal monopole operators.
We assume that $N\geq 3$.
 We begin with the product of one $  V_{-\bm{e}_1} $ and two $  V_{\bm{e}_N} $'s.

\subsubsection{ Moyal product of  \texorpdfstring{one $ \langle V_{-\bm{e}_1} \rangle$ and two $ \langle  V_{\bm{e}_N}\rangle $'s}{one ... and two ...'s}}
There are three  distinct orderings, for each of which the Moyal product takes the form 
\begin{equation}\label{eq:3moyalprodut}
\begin{aligned}
&
\quad
\langle  V_{\bm{e}_N} \rangle *\langle  V_{\bm{e}_N} \rangle *\langle  V_{-\bm{e}_1} \rangle
\ \text{ or } \
\langle  V_{\bm{e}_N} \rangle*\langle  V_{-\bm{e}_1} \rangle *\langle  V_{\bm{e}_N} \rangle
\ \text{ or } \
\langle  V_{-\bm{e}_1} \rangle*\langle  V_{\bm{e}_N} \rangle *\langle  V_{\bm{e}_N} \rangle
  \\
&= \sum_{1 \le k \neq l \le N} e^{2 b_k -b_l} Z_{1\mathchar `-\text{loop}}( \mathfrak{m}=2 {\bm e}_k -{\bm e}_l)
+\sum_{k=1}^N e^{ b_k} Z_{1\mathchar `-\text{loop}}( \mathfrak{m}= {\bm e}_k)  
Z_{\text{mono}}( \mathfrak{m}= {\bm e}_k)  \\
&\qquad
+\sum_{1 \le k < l \le N} \sum_{n=1 \atop n \neq k, l}^N e^{ b_k+b_l- b_n} Z_{1\mathchar `-\text{loop}}( \mathfrak{m}= {\bm e}_k+ {\bm e}_l- {\bm e}_n)  
Z_{\text{mono}}( \mathfrak{m}={\bm e}_k+ {\bm e}_l- {\bm e}_n) .
\end{aligned}
\end{equation}
Here the one-loop determinants~$Z_{1\mathchar `-\text{loop}}$
 are independent of the ordering  in the Moyal product  and are
given by \eqref{eq:1loopvec} and \eqref{eq:1loophy}.
 It turns out that
$Z_{\text{mono}}( \mathfrak{m}={\bm e}_k+ {\bm e}_l- {\bm e}_n)$ obtained from the three Moyal products  is also independent of the ordering and is given as%
\footnote{%
We note
that \eqref{eq:zmonokln} is  the same as \eqref{eq:zmono20}; see Section~\ref{sec:Bme1p2eNekelmen} for an explanation using branes.
}
\begin{align}
 Z_{\text{mono}}(  {\bm B}=-\bm{e}_1 + 2\bm{e}_N, \mathfrak{m}={\bm e}_k+ {\bm e}_l- {\bm e}_n) 
&= \frac{2}{(\varphi_{l k}+  \epsilon) (\varphi_{k l  } +  \epsilon) } .
\label{eq:zmonokln}
\end{align}
On the other hand, the  bubbling
  contribution
 $Z_{\text{mono}}( \mathfrak{m}= {\bm e}_k)$ read off from the  Moyal products in three orderings takes different expressions:
\begin{align}
& Z_{\text{mono}}( \mathfrak{m}= {\bm e}_k) \, \, \text{from} \, \, 
\langle  V_{\bm{e}_N} \rangle *\langle  V_{\bm{e}_N} \rangle *\langle  V_{-\bm{e}_1} \rangle
\nonumber \\
&=\sum_{l=1 \atop l \neq k}^N \frac{2 \prod_{f=1}^{N_F} \left (\varphi_l-m_f +\frac{\epsilon}{2} \right) }
{ \left(\varphi_{k l }+\frac{\epsilon}{2} \right)  \left(\varphi_{ l k}+\frac{3}{2} \epsilon \right) \prod_{j =1 \atop j \ne k, l} ^{N}  \varphi_{j l} \left(\varphi_{l j}+\epsilon \right) }
+
\frac{\prod_{f=1}^{N_F} \left (\varphi_k-m_f +\epsilon\right) }{\prod_{j =1 \atop j \ne k} ^{N} \left(\varphi_{k j}+\frac{\epsilon}{2} \right)  \left(\varphi_{j k}-\frac{3 \epsilon}{2} \right)},
\label{eq:ekzmono} \\
& Z_{\text{mono}}( \mathfrak{m}= {\bm e}_k) \, \, \text{from} \, \, 
\langle  V_{\bm{e}_N} \rangle *\langle  V_{-\bm{e}_1} \rangle *\langle  V_{\bm{e}_N} \rangle 
\nonumber \\
&=\sum_{s=\pm 1} \sum_{l=1 \atop l \neq k}^N 
\frac{ \prod_{f=1}^{N_F} \left (\varphi_l-m_f +s \frac{\epsilon}{2} \right) }
{ \left(\varphi_{  l k }+s \frac{\epsilon}{2} \right)  \left(\varphi_{   k l}- s \frac{3}{2} \epsilon \right) \prod_{j =1 \atop j \ne k, l} ^{N}  \varphi_{ l j} \left(\varphi_{ j l}-s \epsilon \right) } 
+ 
\frac{\prod_{f=1}^{N_F} \left (\varphi_k-m_f \right) }{\prod_{j =1 \atop j \ne k} ^{N} \left(\varphi_{k j}-\frac{\epsilon}{2} \right)  \left(\varphi_{j k}-\frac{ \epsilon}{2} \right)},
\label{eq:} \\
& Z_{\text{mono}}( \mathfrak{m}= {\bm e}_k) \, \, \text{from} \, \, 
\langle  V_{-\bm{e}_1} \rangle *\langle  V_{\bm{e}_N} \rangle *\langle  V_{\bm{e}_N} \rangle 
\nonumber \\
&=\sum_{l=1 \atop l \neq k}^N \frac{2 \prod_{f=1}^{N_F} \left (\varphi_l-m_f -\frac{\epsilon}{2} \right) }
{ \left(\varphi_{k l }-\frac{\epsilon}{2} \right)  \left(\varphi_{ l k}-\frac{3}{2} \epsilon \right) \prod_{j =1 \atop j \ne k, l} ^{N} \varphi_{j l}  \left(\varphi_{l j}-\epsilon \right)  }
+
\frac{\prod_{f=1}^{N_F} \left (\varphi_k-m_f -\epsilon\right) }{\prod_{j =1 \atop j \ne k} ^{N} \left(\varphi_{k j}-\frac{\epsilon}{2} \right)  \left(\varphi_{j k}+\frac{3 \epsilon}{2} \right)}.
\end{align}

\begin{figure}[thb]
\begin{center}
\subfigure[]{\label{fig:sixFIregion}
\includegraphics[width=6cm]{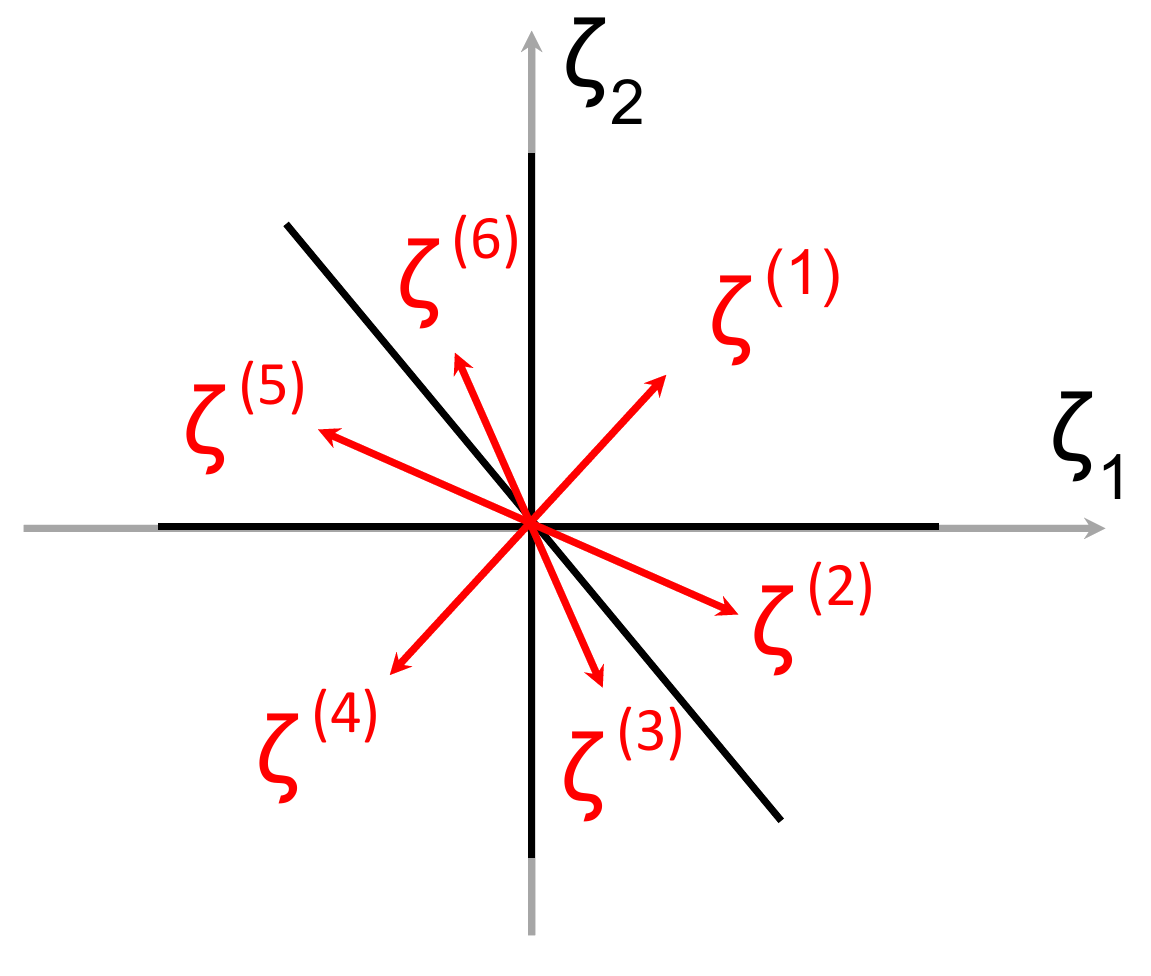}}
\subfigure[]{\label{fig:wall3monopole}
\includegraphics[width=6cm]{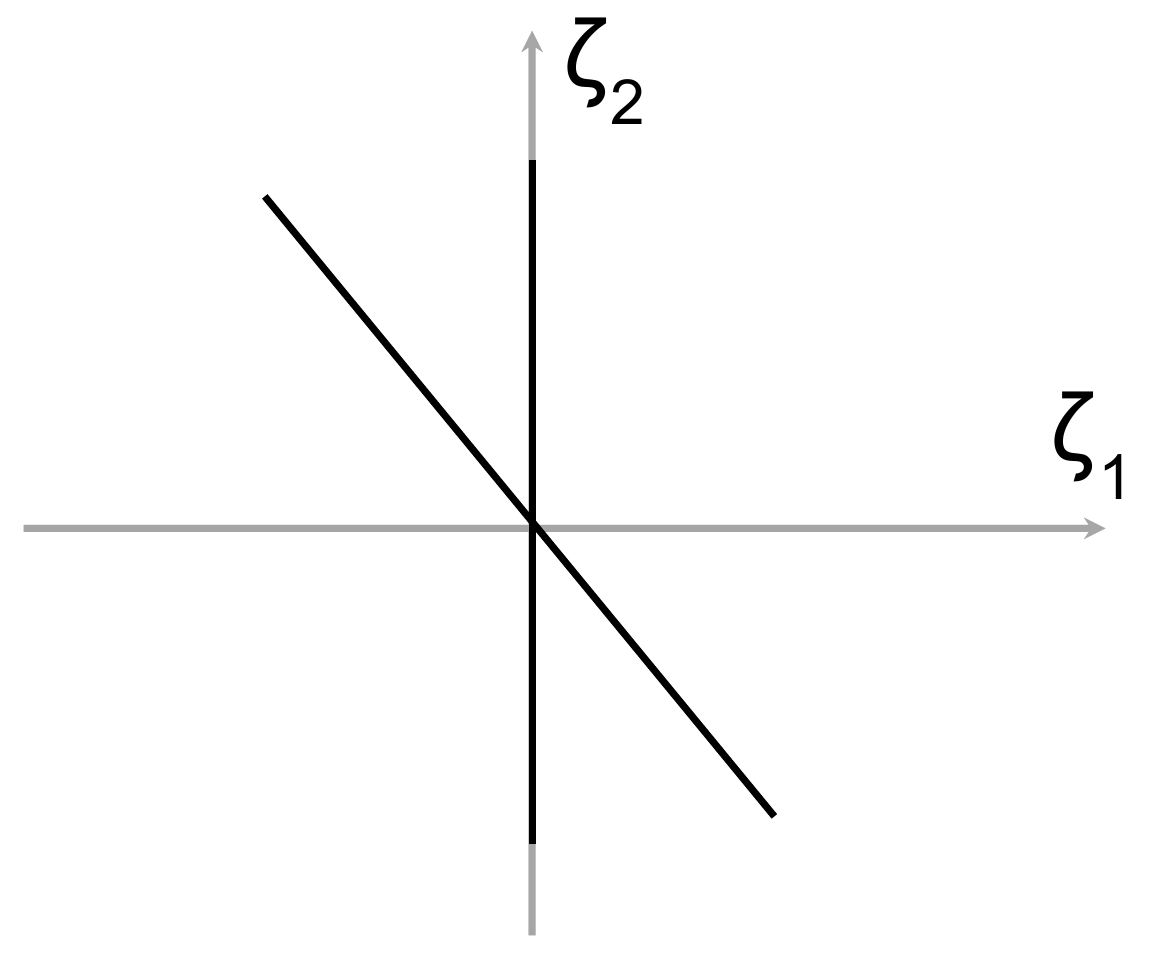}}
\end{center}
\vspace{-0.5cm}
\caption{(a)  Representatives~${\bm \zeta}^{(i)}$ of the FI-cone $\mathcal{C}^{(i)}$ for $i=1, \cdots, 6$ are indicated by red arrows. The six black  lines indicate  the boundaries  between $\mathcal{C}^{(i)}$ and $\mathcal{C}^{(i+1)}$ for $i=1, \cdots, 6$. 
(We set $\mathcal{C}^{(7)}:=\mathcal{C}^{(1)}$.)
 (b) The four boundaries on which wall-crossing actually occurs  for $N_F\geq 2N-1$ are  indicated by black lines.  }
\end{figure}

\subsubsection{ Matrix model for \texorpdfstring{$(\bm{B},  \mathfrak{m})=(-\bm{e}_1 + 2\bm{e}_N,{\bm e}_k)$}{...}}

 We now evaluate the bubbling  contributions for $\mathfrak{m}={\bm e}_k$ for $k=1, \cdots, N$  and ${\bm B}=-\bm{e}_1 + 2\bm{e}_N$  using the matrix model   (unique up to permutation of $k$) realized as the low-energy world-volume theories on  Euclidean D1-branes.

The brane construction for the sector $(\bm{B},  \mathfrak{m})=(-\bm{e}_1 + 2\bm{e}_N,{\bm e}_N)$ is illustrated in Figure~\ref{fig:monopole5}. 
For convenience  we refer to the two  NS5'-branes for ${\bm B}=\bm{e}_N$  
 as $\text{NS5'}_2$ and $\text{NS5'}_3$,
 and the NS5'-brane for ${\bm B}=-\bm{e}_1$ 
as NS5'${}_{1}$. 
 As in Section~\ref{sec:matrix-SQCD-branes} we denote the value of the coordinate $x^3$ for the location of  $\text{NS5'}_i, i=1,2,3$ 
  by $s_i$. 
Let  ${\bm \zeta}:=(\zeta_1, \zeta_2)$ be the FI parameters for $U(1)_1 \times U(1)_2$, where  $U(1)_1$  and $U(1)_2$ respectively denote the $U(1)$  gauge groups on the left and the right in  Figure~\ref{fig:21quiverMM}. 
 Recall from~(\ref{eq:zeta-sa}) that $\zeta_1=s_2-s_1$ and $\zeta_2=s_3-s_2$.

The  bubbling contribution for $\mathfrak{m}={\bm e}_N$ and ${\bm B}=-\bm{e}_1 + 2\bm{e}_N$ is given by
\begin{align}
&Z^{({\bm \zeta})}_{\mathrm{mono}} ( {\bm B}=-\bm{e}_1 + 2\bm{e}_N, \mathfrak{m}={\bm e}_N ) 
{
= \oint_{\text{JK}(\bm{\zeta})} \omega^{(1)}
}
=\sum_{{\bm u}_{ *}} \mathop{\text{JK-Res}}_{{\bm u}_{*}}({\bm Q}_*, {\bm \zeta}) \, \omega^{(1)}, 
 \label{eq:contourint4}
\end{align}
with
\begin{align}
\omega^{(1)}=\frac{\epsilon^2 \prod_{f=1}^{N_F}(u_1-m_f) d u_1  \wedge d u_2 }{\prod_{s=\pm1} 
\left( s(u_1-u_2) -\frac{\epsilon}{2} \right) \left( s(u_2-\varphi_N) -\frac{\epsilon}{2} \right)  \prod_{i=1}^{N-1} \left( s(u_1-\varphi_i) -\frac{\epsilon}{2} \right) }  .
\end{align}

There are six distinct FI-chambers as defined in~(\ref{eq:FIchamber}):
\begin{equation} \label{eq:regions}
{
\begin{array}{llllll}
\mathcal{C}^{(1)} =  \{ \, (\zeta_1, \zeta_2) \, | \, \zeta_1 >0, \, \zeta_2 >0   \} & \longleftrightarrow & s_1<s_2<s_3, \\
\mathcal{C}^{(2)} =  \{ \, (\zeta_1, \zeta_2) \, | \, \zeta_1+\zeta_2 >0, \, \zeta_2 < 0   \} & \longleftrightarrow & s_1<s_3 < s_2, \\
\mathcal{C}^{(3)} =  \{ \, (\zeta_1, \zeta_2) \, | \, \zeta_1 > 0, \, \zeta_1+\zeta_2  < 0    \} & \longleftrightarrow &  s_3 < s_1 < s_2,\\
\mathcal{C}^{(4)} =  \{ \, (\zeta_1, \zeta_2) \, | \, \zeta_1 <  0, \, \zeta_2 < 0    \} & \longleftrightarrow & s_3 < s_2 < s_1,\\
\mathcal{C}^{(5)} =  \{ \, (\zeta_1, \zeta_2) \, | \,  \zeta_1+\zeta_2 < 0, \, \zeta_2 > 0    \} & \longleftrightarrow &  s_2 < s_3 < s_1,\\
\mathcal{C}^{(6)} =  \{ \, (\zeta_1, \zeta_2) \, | \,  \zeta_1 < 0, \, \zeta_1+\zeta_2 > 0    \} & \longleftrightarrow &  s_2 < s_1 < s_3 . 
\end{array}
}
\end{equation}
We  exhibit representatives  ${\bm \zeta}^{(i)} \in \mathcal{C}^{(i)}$ for $i=1,\cdots, 6$ in Figure~\ref{fig:sixFIregion}.
  For different FI-chambers the singular hyperplanes  relevant to the JK residue operation, and hence the JK residues themselves, are possibly different.    
 For each value of $i=1,\ldots,6$ we evaluate the matrix model partition function with FI parameter $ {\bm \zeta}^{(i)}$.

Let us first consider ${\bm \zeta}^{(1)} \in \mathcal{C}^{(1)}$.  The JK residues   are evaluated by setting ${\bm \eta}={ \bm{\zeta}^{(1)}}$ in the definition of the JK residue for a non-degenerate intersection~\eqref{JK-def-non-degenerate}.
The non-zero contributions come from  the intersection points 
{ ${\bm u}^{(a)}_{ k}$, $ {\bm u}^{(b)}_{ k}$}, and
 ${\bm u}^{(c)}$ ($k=1,\cdots, N-1 $)  in the ${\bm u}$-plane defined by
\begin{align}
&
{ {\bm u}^{(a)}_{ k}}
:=\left\{ u_1- \varphi_k -\frac{\epsilon}{2} =0 \right\} \cap \left\{ u_2- \varphi_N -\frac{\epsilon}{2} =0 \right\},    
\label{eq:chargecone1} \\
&
{ {\bm u}^{(b)}_{ k}}
:=\left\{ u_1- \varphi_k -\frac{\epsilon}{2} =0 \right\} \cap \left\{ u_2- u_1 -\frac{\epsilon}{2} =0 \right\}, 
\label{eq:chargecone2}  \\
&
{ \bm{u}^{(c)}}
:=\left\{ u_1-u_2 -\frac{\epsilon}{2} =0 \right\} \cap \left\{ u_2- \varphi_N -\frac{\epsilon}{2} =0 \right\}.
\label{eq:chargecone3}
\end{align}
Let ${\bm Q}^{(i)}_*=\{ {\bm Q}^{(i)}_1, {\bm Q}^{(i)}_2 \}$ be the set of gauge charges  associated with singular hyperplanes~\eqref{eq:chargecone1} for $i=a$, \eqref{eq:chargecone2} for $i=b$ and \eqref{eq:chargecone3} for $i=c$, respectively,  given as
\begin{equation}
\begin{array}{lll}
{\bm Q}^{(a)}_1:=(1,0) ,& \quad {\bm Q}^{(a)}_2:=(0,1), \\
{\bm Q}^{(b)}_1:=(1,0)  ,&\quad {\bm Q}^{(b)}_2:=(-1,1), \\
{\bm Q}^{(c)}_1:=(1,-1) ,&  \quad {\bm Q}^{(c)}_2:=(0,1). 
\end{array}
\end{equation}
Note that ${\bm \zeta}^{(1)} \in \text{Cone}[{\bm Q}^{(i)}_1, {\bm Q}^{(i)}_2]$ for $i=a, b, c$. 
The JK residues at these points are given by
\begin{align}
\sum_{k=1}^{N-1}  \mathop{\text{JK-Res}}_{{\bm u}_{ *}= { {\bm u}^{(a)}_{ k}} } ({\bm Q}^{(a)}_*, {\bm \zeta}^{(1)}) \, \omega^{(1)}  
&=\sum_{k=1}^{N-1} \frac{\prod_{f=1}^{N_F} \left (\varphi_k-m_f +\frac{\epsilon}{2}\right) }{ \left(\varphi_{k N}-\frac{\epsilon}{2} \right)  \left(\varphi_{N k}-\frac{\epsilon}{2} \right) \prod_{j =1 \atop j \neq k} ^{N-1} \varphi_{k j} (\varphi_{j k} -\epsilon) },
\label{eq:JKres_a} \\
\sum_{k=1}^{N-1}  \mathop{\text{JK-Res}}_{{\bm u}_{ *}= { {\bm u}^{(b)}_{ k}}} ({\bm Q}^{(b)}_*, {\bm \zeta}^{(1)}) \, \omega^{(1)}  
&=\sum_{k=1}^{N-1} \frac{\prod_{f=1}^{N_F} \left (\varphi_k-m_f +\frac{\epsilon}{2}\right) }{ \left(\varphi_{k N}+\frac{\epsilon}{2} \right)  \left(\varphi_{N k}-\frac{3 \epsilon}{2} \right)
 \prod_{j =1 \atop j \neq k} ^{N-1} \varphi_{k j} (\varphi_{j k} -\epsilon) },
\label{eq:JKres_b} \\
 \mathop{\text{JK-Res}}_{ {\bm u}_{ *}= { {\bm u}^{(c)}} } ({\bm Q}^{(c)}_*, {\bm \zeta}^{(1)} ) \, \omega^{(1)}  
&= \frac{\prod_{f=1}^{N_F} \left (\varphi_N-m_f +\epsilon\right) }{\prod_{j =1} ^{N-1} \left(\varphi_{N j}+\frac{\epsilon}{2} \right)  \left(\varphi_{j N}-\frac{3 \epsilon}{2} \right)} .
\label{eq:JKres_c}
\end{align}
 The bubbling contribution corresponding to $\bm{\zeta}={\bm \zeta}^{(1)}$ is  the sum of \eqref{eq:JKres_a}, \eqref{eq:JKres_b} and \eqref{eq:JKres_c}, which combine into
\begin{align}
&Z^{({\bm \zeta}^{(1)})}_{\mathrm{mono}} ( {\bm B}=-\bm{e}_1 + 2\bm{e}_N, \mathfrak{m}={\bm e}_N ) 
\nonumber \\
&=
\sum_{ l\neq N}
\frac{   2 \prod_{f=1}^{N_F} \left (\varphi_l-m_f +\frac{\epsilon}{2} \right) }
{  \left( \varphi_{N  l}+\frac{\epsilon}{2} \right)  \left(\varphi_{l N }+\frac{3 \epsilon}{2} \right)
 \prod_{ j\neq l, N}
 {\varphi_{ j l}}
  (\varphi_{ l j} +\epsilon)}
+
\frac{\prod_{f=1}^{N_F} \left (\varphi_N-m_f +\epsilon\right) }{
\prod_{ j\neq N}
 \left(\varphi_{N j}+\frac{\epsilon}{2} \right)  \left(\varphi_{j N}-\frac{3 \epsilon}{2} \right)} .
\label{eq:zemono_zeta1}
\end{align}

We find that  \eqref{eq:zemono_zeta1}  reproduces a monopole bubbling effect \eqref{eq:ekzmono} for $\mathfrak{m}={\bm e}_N$ evaluated from the Moyal product $\langle  V_{\bm{e}_N} \rangle * \langle  V_{\bm{e}_N} \rangle * \langle  V_{-\bm{e}_1} \rangle$.
The monopole bubbling effect for $\mathfrak{m}={\bm e}_k$  is obtained by  replacing the  symbol $N$ by $k$ in~\eqref{eq:zemono_zeta1}.

Similarly we evaluate the JK residues with  the JK parameter $\bm{\eta}$ set to $\bm{\zeta}^{(i)}\in\mathcal{C}^{(i)}$  for $i=2, \cdots, 6$.
 We obtain the following results for the bubbling contributions corresponding to $\mathfrak{m}={\bm e}_k$ and ${\bm B}=-\bm{e}_1 + 2\bm{e}_N$:
\begin{align}
&Z^{({\bm \zeta}^{(1)})}_{\mathrm{mono}} = Z^{({\bm \zeta}^{(2)})}_{\mathrm{mono}} =
\Big(
 Z_{\mathrm{mono}} \, \, \text{from} \, \,  \langle  V_{\bm{e}_N} \rangle *\langle  V_{\bm{e}_N} \rangle *\langle  V_{-\bm{e}_1} \rangle \Big),  
\label{eq:noWC1}  \\
&Z^{({\bm \zeta}^{(3)})}_{\mathrm{mono}} =Z^{({\bm \zeta}^{(6)})}_{\mathrm{mono}} =
 \Big(Z_{\mathrm{mono}} \, \, \text{from} \, \,  \langle  V_{\bm{e}_N} \rangle *\langle  V_{-\bm{e}_1} \rangle*\langle  V_{\bm{e}_N} \rangle\Big), 
\label{eq:noWC2}  \\
&Z^{({\bm \zeta}^{(4)})}_{\mathrm{mono}} = Z^{({\bm \zeta}^{(5)})}_{\mathrm{mono}}=
\Big( Z_{\mathrm{mono}} \, \, \text{from} \, \,  \langle  V_{\bm{e}_N} \rangle *\langle  V_{\bm{e}_N} \rangle *\langle  V_{-\bm{e}_1} \rangle\Big).   
\label{eq:noWC3} 
\end{align}
Therefore we have  a perfect agreement between  the Moyal products and the matrix model partition functions.

 The equality $Z^{({\bm \zeta}^{(1)})}_{\mathrm{mono}} = Z^{({\bm \zeta}^{(2)})}_{\mathrm{mono}}$  in~(\ref{eq:noWC1}) implies the absence of wall-crossing between $\mathcal{C}^{(1)}$ and $\mathcal{C}^{(2)}$.
This is expected because the boundary between $\mathcal{C}^{(1)}$ and $\mathcal{C}^{(2)}$ is where the locations $s_2$ and $s_2$ of two identical operators $V_{\bm{e}_N}$ coincide.
The equalities in~(\ref{eq:noWC2}) and~(\ref{eq:noWC3}) are interpreted similarly.
The four boundaries on which the ordering of $ V_{\bm{e}_N} $ and $  V_{-\bm{e}_1} $  changes  are indicated by black  lines in Figure~\ref{fig:wall3monopole}.
When $N_F \ge 2 N-1$, the values of $Z_\text{mono}^{({\bm \zeta}_i)}$'s  in the four regions separated by the black lines are actually different, and wall-crossing occurs.

\begin{figure}[htb]
\centering
\subfigure[]{\label{fig:111monopole1}
\includegraphics[width=6cm]{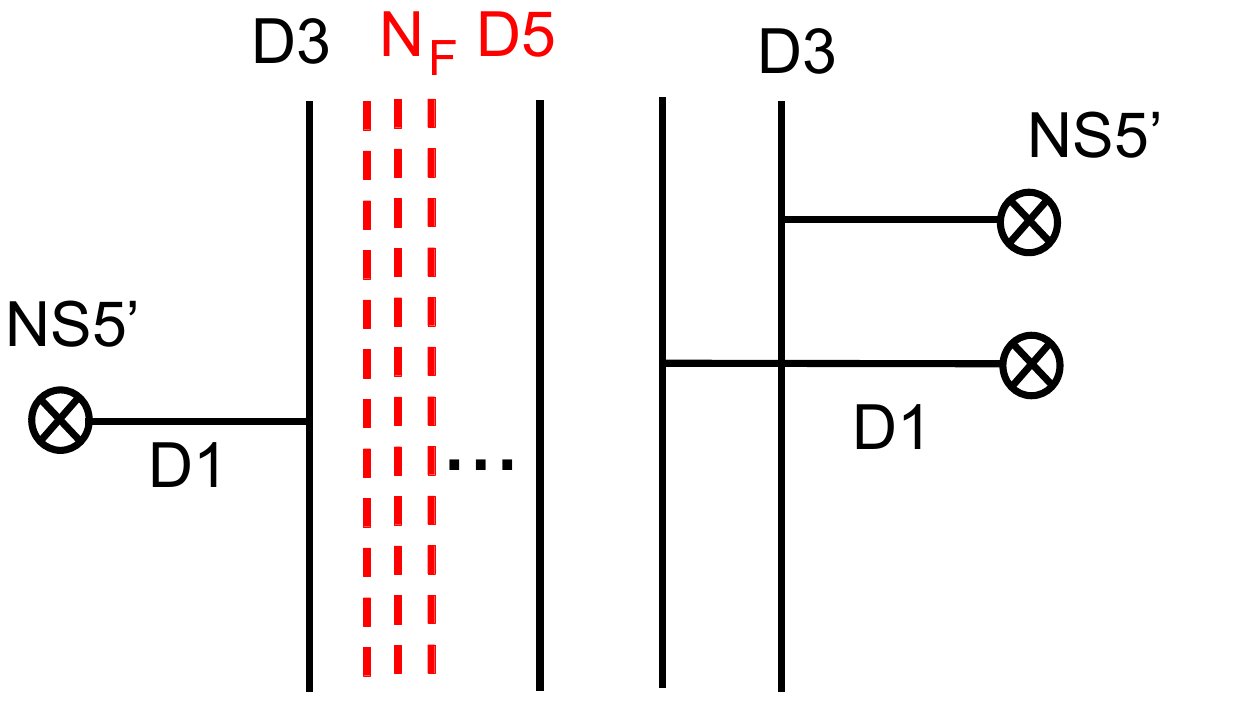}}
\subfigure[]{\label{fig:111monopole2}
\includegraphics[width=6cm]{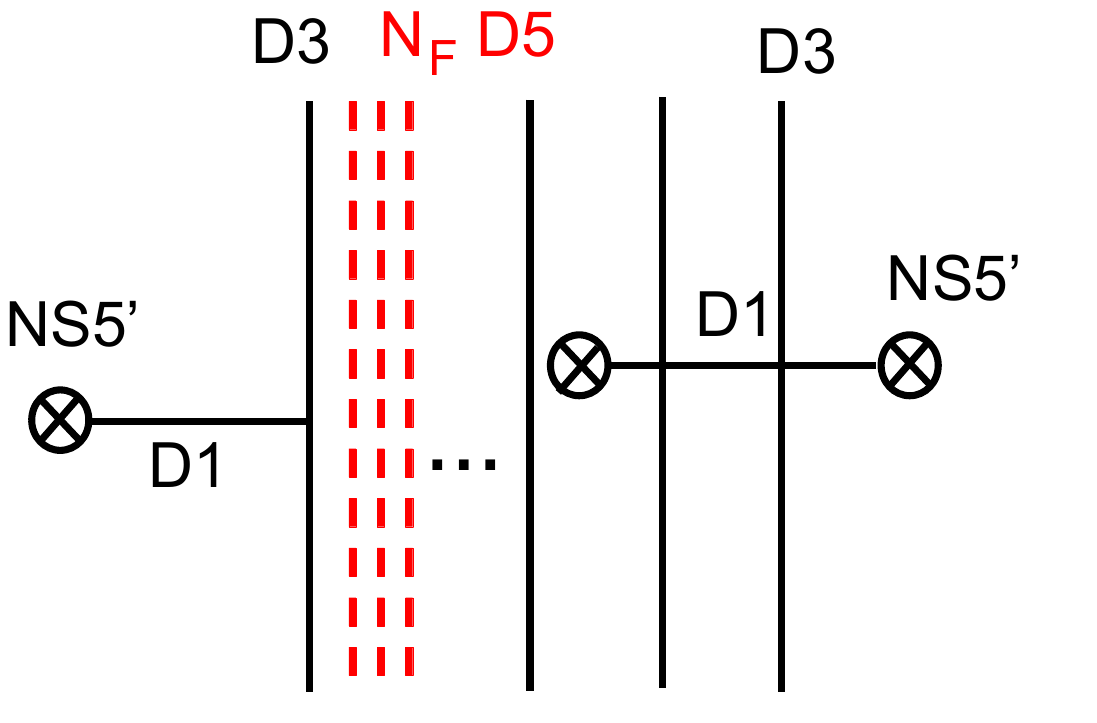}}
\caption{(a):  We added a D1-brane representing a smooth monopole with charge $\bm{e}_{N-1}-\bm{e}_N$ to Figure~\ref{fig:21monopole1}.
(b)  We moved  an NS5'-brane to left. 
When the NS5'-brane crosses D3-branes Hanany-Witten transitions occur,  and we obtain  this brane configuration.
 }
\label{fig:monopole6}
\end{figure}

\subsubsection{Matrix model for \texorpdfstring{$(\bm{B},  \mathfrak{m})=(-\bm{e}_1 + 2\bm{e}_N,{\bm e}_{k}+{\bm e}_{l}-{\bm e}_{n})$ with
 { $k<l$ and $n\neq k,l$}
  }{...with...and...}}
\label{sec:Bme1p2eNekelmen}

Next we compute  the monopole bubbling contribution from the sector specified by 
$(\bm{B},  \mathfrak{m})=(-\bm{e}_1 + 2\bm{e}_N,{\bm e}_{k}+{\bm e}_{l}-{\bm e}_{n})$ 
with 
 { $k<l$ and $n\neq k,l$}.
The brane construction is  illustrated in Figure~\ref{fig:monopole6}.  
 The matrix model can be read off from the configuration in Figure~\ref{fig:111monopole2}, and is found to be the
same as 
that for $\mathfrak{m}={\bm e}_{k}+{\bm e}_{l}$ and ${\bm B}=-\bm{e}_1 + 2\bm{e}_N$. 
The partition function is given by \eqref{eq:Zmonoekel}, which does not exhibit wall-crossing.

\subsubsection{ Operator ordering and wall crossing}

To summarize, we have found the following relations:
\begin{align}
& \langle  V_{-\bm{e}_1 + 2\bm{e}_N} \rangle^{( {\bm \zeta}^{(1)}) }  = \langle  V_{-\bm{e}_1 + 2\bm{e}_N} \rangle^{( {\bm \zeta}^{(2)}) }=
\langle  V_{\bm{e}_N} \rangle * \langle  V_{\bm{e}_N} \rangle * \langle  V_{-\bm{e}_1} \rangle ,
\label{eq:JKand3moyal1}   \\
& \langle  V_{-\bm{e}_1 + 2\bm{e}_N} \rangle^{( {\bm \zeta}^{(3)}) }  =  \langle  V_{-\bm{e}_1 + 2\bm{e}_N} \rangle^{( {\bm \zeta}^{(6)}) } =
\langle  V_{\bm{e}_N} \rangle *  \langle  V_{-\bm{e}_1} \rangle *\langle  V_{\bm{e}_N} \rangle   ,
\label{eq:JKand3moyal2} \\
& \langle  V_{-\bm{e}_1 + 2\bm{e}_N} \rangle^{( {\bm \zeta}^{ (4)} ) }  = \langle  V_{-\bm{e}_1 + 2\bm{e}_N} \rangle^{( {\bm \zeta}^{ (5)} ) } =
\langle  V_{-\bm{e}_1} \rangle * \langle  V_{\bm{e}_N} \rangle *  \langle  V_{\bm{e}_N} \rangle   .
\label{eq:JKand3moyal3}
\end{align}
The only bubbling contribution that exhibits wall-crossing is $Z^{({\bm \zeta})}_{\mathrm{mono}} ( {\bm B}=-\bm{e}_1 + 2\bm{e}_N, \mathfrak{m}={\bm e}_N ) $, for which we already discussed the chamber structure in the paragraph below~(\ref{eq:noWC1})-(\ref{eq:noWC3}).
When $\epsilon=0$ the ${\bm \zeta}$-dependence of $Z_{\text{mono}}^{({\bm \zeta})}$ disappears, and the non-commutative Moyal products reduce to the ordinary commutative products.
From~\eqref{eq:JKand3moyal1}-\eqref{eq:JKand3moyal3} we obtain   
\begin{align}
 \langle  V_{-\bm{e}_1 + 2\bm{e}_N} \rangle  = \langle  V_{-\bm{e}_1} \rangle \langle  V_{\bm{e}_N} \rangle^2 \quad \text{ for }\epsilon=0.
\end{align}


\subsection{ Product of \texorpdfstring{three $  V_{\bm{e}_N} $'s }{...}}

Finally we study the relation between $\langle  V_{3\bm{e}_N} \rangle$ and 
 the Moyal product $\langle  V_{\bm{e}_N} \rangle * \langle  V_{\bm{e}_N} \rangle * \langle  V_{\bm{e}_N} \rangle $.
 We expect no wall-crossing because the ordering is unique for the product of identical operators.

\subsubsection{ Moyal product of  \texorpdfstring{three $ \langle  V_{\bm{e}_N}\rangle $'s}{...}}
 
The Moyal product has following form
\begin{align}
&\langle  V_{\bm{e}_N} \rangle * \langle  V_{\bm{e}_N} \rangle * \langle  V_{\bm{e}_N} \rangle
=\sum_{k=1}^N e^{ 3 b_k} Z_{1\mathchar `-\text{loop}}( \mathfrak{m}=3 {\bm e}_k) 
\nonumber \\
&+\sum_{1 \le k  \neq  l \le N} e^{2 b_k +b_l} Z_{1\mathchar `-\text{loop}}( \mathfrak{m}=2 {\bm e}_k +{\bm e}_l) Z_{\text{mono}}( \mathfrak{m}=2 {\bm e}_k+ {\bm e}_l)
 \nonumber \\
&+\sum_{1 \le k < l < n \le N}  e^{ b_k+b_l+ b_n} Z_{1\mathchar `-\text{loop}}( \mathfrak{m}= {\bm e}_k+ {\bm e}_l+ {\bm e}_n)  
Z_{\text{mono}}( \mathfrak{m}={\bm e}_k+ {\bm e}_l+ {\bm e}_n) .
\end{align}
The monopole bubbling  contributions read off from the Moyal product are
\begin{align}
 Z_{\text{mono}}( \mathfrak{m}=2 {\bm e}_k+ {\bm e}_l)&=\frac{3}{\left( \varphi_{k l}+\frac{3}{2}\epsilon \right) \left( \varphi_{l k}+\frac{3}{2}\epsilon \right) }, 
\label{eq:zmono12in3} \\
Z_{\text{mono}}( \mathfrak{m}={\bm e}_k+ {\bm e}_l+ {\bm e}_n)&
= \frac{6 }{\prod_{ i, j =N-2,i \neq j }^{N}   (\varphi_{i j }+\epsilon)(\varphi_{ j i }+\epsilon)  }. 
\label{eq:zmono111in3}
\end{align}
We will reproduce  these expressions as matrix model partition functions.
The brane configuration for the monopole charge $\bm{B}
=3\bm{e}_N$ is depicted in Figure~\ref{fig:3monopole1}. 

\subsubsection{ Matrix model for \texorpdfstring{$(\bm{B},  \mathfrak{m})=(3\bm{e}_N, {\bm e}_k+{\bm e}_l+ {\bm e}_n)$ with $k<l<n$ }{...}}

\begin{figure}[htb]
\centering
\subfigure[]{\label{fig:3monopole1}
\includegraphics[width=6cm]{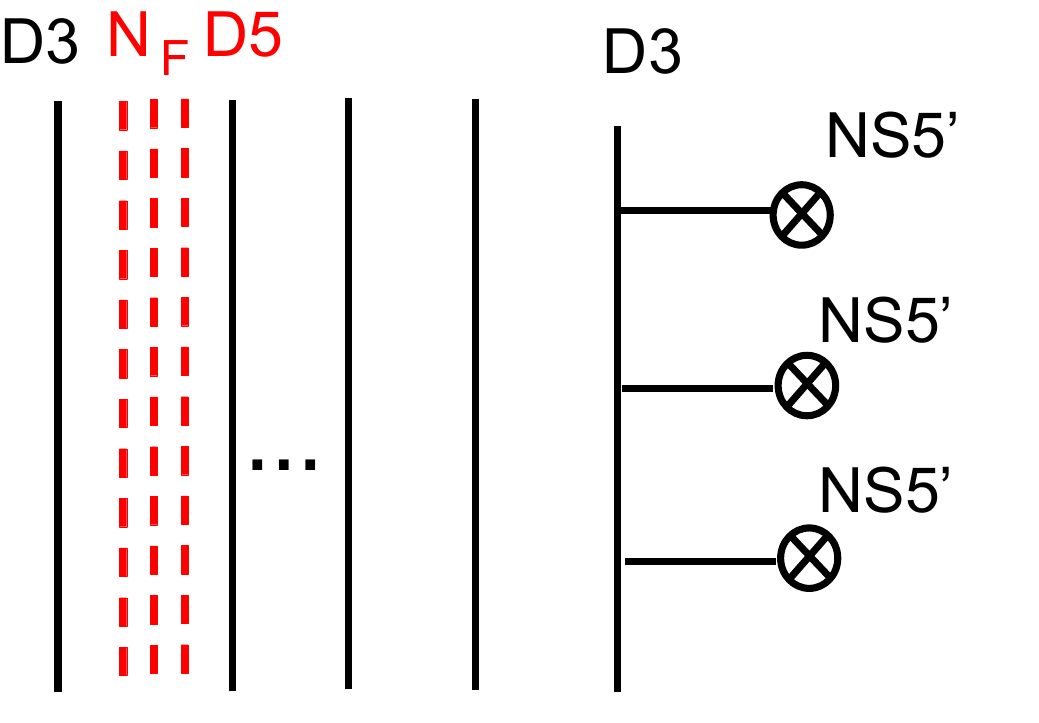}}
\subfigure[]{\label{fig:3monopole2}
\includegraphics[width=6cm]{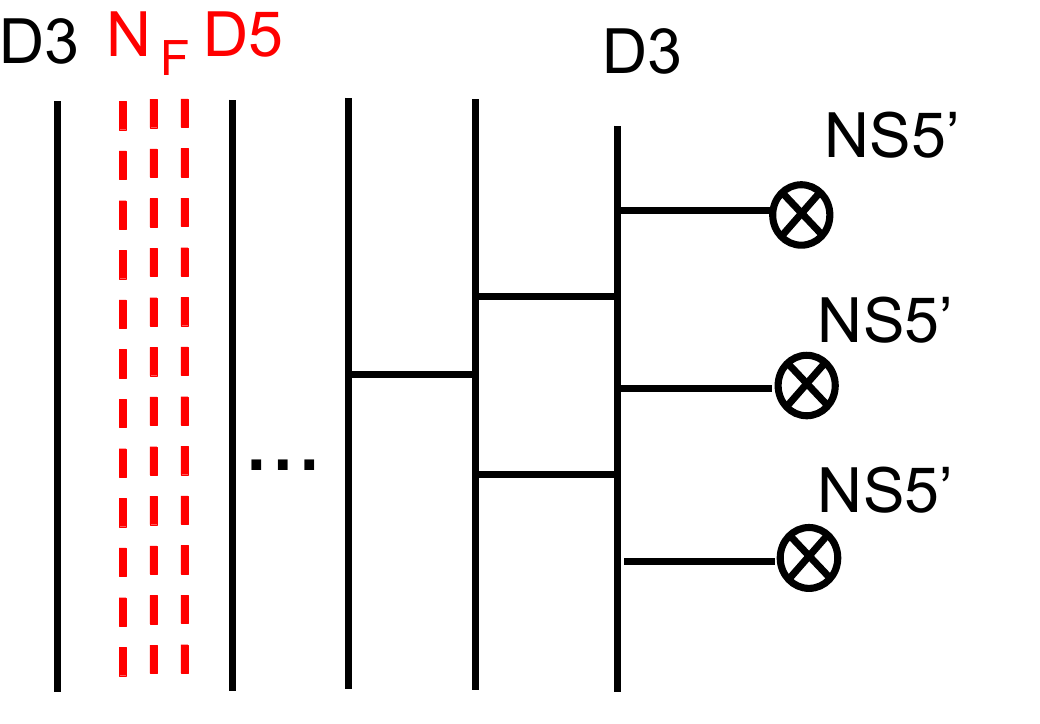}}
\subfigure[]{\label{fig:3monopole3}
\includegraphics[width=6cm]{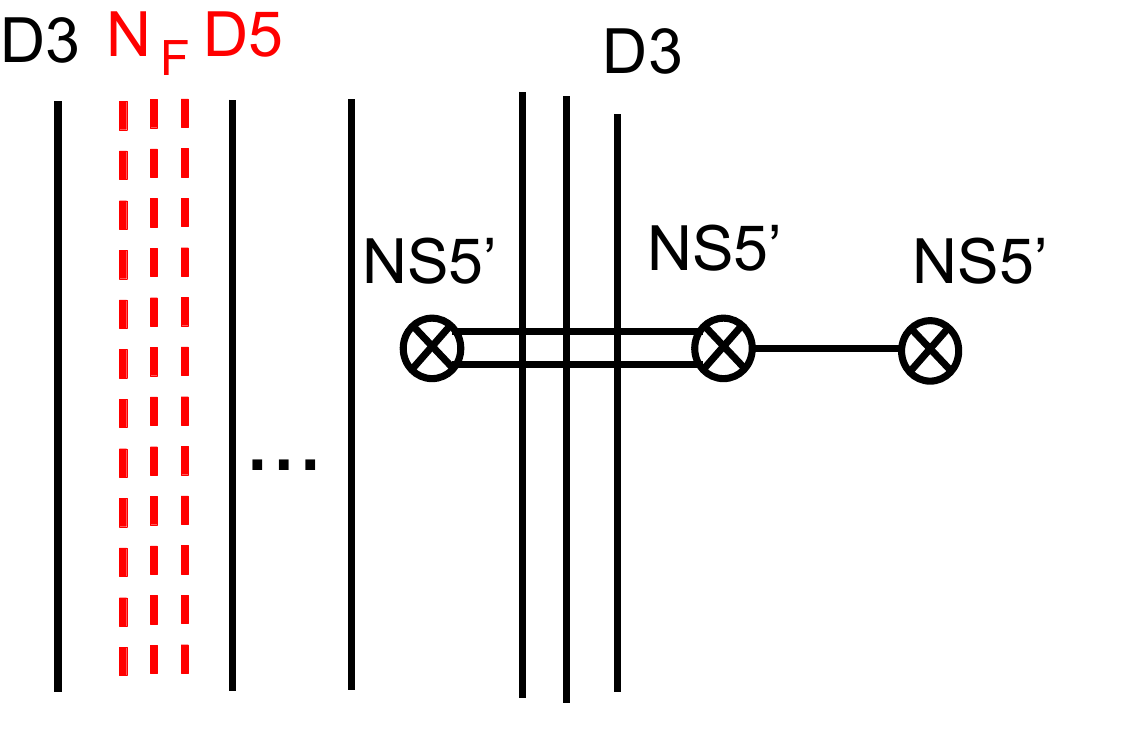}}
\subfigure[]{\label{fig:21quiverMM321}
\includegraphics[width=1cm]{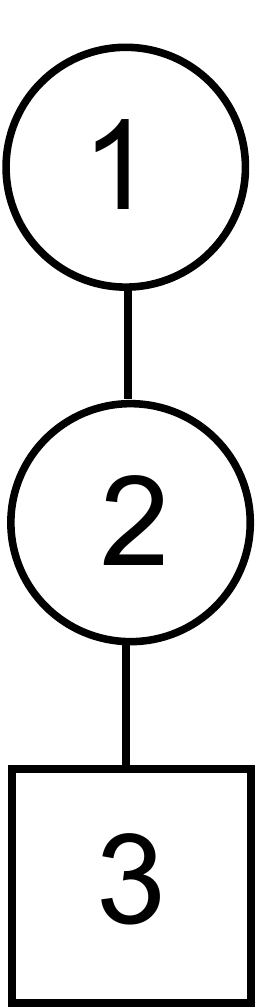}}
\caption{(a): The brane configuration for  the monopole operator with ${\bm B}=
3 \bm{e}_N
$.
(b): We add three D1-branes to realize the bubbling sector for  $\mathfrak{m}=
 \bm{e}_{N-2}+ \bm{e}_{N-1}+ \bm{e}_N
$. 
(c): After a sequence of Hanany-Witten transitions,  we obtain a brane configuration  where D1-branes end only on NS5'-branes, from which the matter content of  the matrix model can be read off. 
(d) The  $\mathcal{N}=(0,4)$ quiver diagram for the world-volume theory  on the D1-branes in (c).
 }
\label{fig:monopole5new}
\end{figure}

We first consider  the monopole bubbling sector  with 
$\mathfrak{m}= {\bm e}_{N-2}+{\bm e}_{N-1}+ {\bm e}_{N}$.
 (Other charges $\mathfrak{m}= {\bm e}_k+{\bm e}_l+ {\bm e}_n$ with $1\leq k <l < n \leq N$ can be obtained by permutation.)
 The sector is realized by adding 
 three D1-branes suspended between D3-branes
as  in Figure \ref{fig:3monopole2}. 
We  then obtain the brane configuration of  Figure~\ref{fig:3monopole3}  via Hanany-Witten transitions.
The matter content of  the matrix model for $Z_{\text{mono}}(\mathfrak{m}=
{ {\bm e}_{N-2}+{\bm e}_{N-1}+ {\bm e}_{N}}
)$ read off 
from  the configuration
is  summarized as the $\mathcal{N}=(0,4)$ quiver diagram in Figure~\ref{fig:21quiverMM321}.

The bubbling contribution
 is given as
\begin{align}
 Z^{({\bm \zeta})}_{\text{mono}}( {\bm B}=
{ 3\bm{e}_N},  \mathfrak{m}= {\bm e}_{N-2}+{\bm e}_{N-1}+ {\bm e}_{N} )&
=
{
\int_{{\rm JK}(\bm{\zeta})} 
}
\omega^{(2)}, 
 \label{eq:contourint5}
\end{align}
with
\begin{align}
\omega^{(2)}= \frac{1}{2}  \frac{ (-\epsilon)^3 \prod_{1 \le i \neq j \le 2}(u_i- u_j)(u_i- u_j-\epsilon)  d u_1  \wedge d u_2 \wedge d u^{\prime}  }{ \prod_{i=1}^2 \prod_{s=\pm1} 
\left( s(u_i- u^{\prime} ) -\frac{\epsilon}{2} \right) \prod_{k=N-2}^{N} \left( s(u_i-\varphi_k) -\frac{\epsilon}{2} \right)}  .
\end{align}
 The right hand side of~(\ref{eq:contourint5}) is to be evaluated according to the prescription in Appendix~\ref{app:JK}.

For  the choice $\bm{\zeta} = \bm{\zeta}^{(1)}=(1,1) \in \mathcal{C}^{(1)}$, see~(\ref{eq:regions}), 
 the cones of gauge charges  that contain $\widetilde {\bm \zeta}^{ (1)}=(1,1,1)$ are
\begin{equation}\label{eq:conesnondeg}
\begin{array}{ll}
\text{Cone}[(1,0,0), (0,1,0), (-1,0,1)] ,& \quad \text{Cone}[(1,0,0), (0,1,0), (0,-1, 1)],  \\
\text{Cone}[(1,0,0), (0,1,-1), (-1,0, 1)], & \quad \text{Cone}[(0, 1, 0), (1, 0,-1), (0, -1, 1)] .
\end{array}
\end{equation}
 For example, let us consider $\text{Cone}[(1,0,0), (0,1,0), (-1,0,1)]$.
The intersections ${\bm u}_*$ of singular hyperplanes  for which the ordered set of associated gauge charges $\bm{Q}_*({\bm u}_*)$ equals  ${\bm Q}^{(0)}_*:=\{ (1,0,0), (0,1,0), (-1,0,1) \}$ are 
\begin{equation}
 {\bm u}_* = \, {\bm u}_{i, j}:=\left\{ u_1- \varphi_i -\frac{\epsilon}{2} =0 \right\} \cap \left\{ u_2- \varphi_j -\frac{\epsilon}{2} =0 \right\} \cap \left\{ -u_1 + u^{\prime} -\frac{\epsilon}{2} =0 \right\} ,
\end{equation}
where $i, j  \in \{N-2, N-1 ,N\}  $ with $ i \neq j$.
 Since the number of hyperplanes that intersect equals the number of integration variables, these intersections are non-degenerate for which the JK residues are computed according to~(\ref{JK-def-non-degenerate}).
The residues at these points  sum up to
\begin{align}
 \sum_{ i, j  =N-2 \atop  i \neq j }^N \mathop{\text{JK-Res}}_{{\bm u}_{i ,j} } (  {\bm Q}^{(0)}_*
  ,
\widetilde{ \bm{\zeta}}^{(1)}
 ) \, \omega^{(2)}
= \frac{3 }{\prod_{ i, j =N-2,i \neq j }^{N}  (\varphi_{i j }+\epsilon)(\varphi_{ j i }+\epsilon)  }. 
\label{eq:JKCone1}
\end{align}
In  a similar manner, we evaluate the JK residues at the other intersection points associated with the charge cones \eqref{eq:conesnondeg}.
We find that the JK residues associated with $\text{Cone}[(1,0,0),$ $ (0,1,0), (0,-1, 1)]$  again sum up to the right hand side of  \eqref{eq:JKCone1}. 
The JK residues associated with $\text{Cone}[(1,0,0), (0,1,-1), (-1,0, 1)] $ and $\text{Cone}[(0, 1, 0), (1, 0,-1), (0, -1, 1)] $ separately vanish.
There are also degenerate intersections of four singular hyperplanes. For example, a  degenerate intersection is given by
\begin{equation}
\begin{aligned}
& \left\{ u_1- \varphi_i -\frac{\epsilon}{2} =0 \right\} \cap \left\{ u_2- \varphi_i -\frac{\epsilon}{2} =0 \right\}
\\
&\qquad\qquad
 \cap \left\{ -u_1 + u^{\prime} -\frac{\epsilon}{2} =0 \right\} 
\cap \left\{ -u_2 + u^{\prime} -\frac{\epsilon}{2} =0 \right\} 
\end{aligned}
\end{equation}
with $i\in  \{N-2, N-1 ,N\} $.
But 
 by applying the prescription for degenerate intersections summarized in Appendix~\ref{app:JK}%
\footnote{
In particular we slightly shift~$\widetilde{\zeta}^{(1)}$ to~$\widetilde{\zeta}^{\prime(1)}$ so that $\bm{\eta}=\widetilde{\zeta}^{\prime(1)}$ satisfies the strong regularity condition~\eqref{equation:strongref}. 
The residues do not depend on the  choice of such a shift.
}
we find that the contribution from each degenerate point  actually vanishes.
We conclude that  $Z^{({\bm \zeta}^{ (1)})}_{\text{mono}}$
is twice~\eqref{eq:JKCone1} and agrees with \eqref{eq:zmono111in3} with $k=N-2, l=N-1, n=N$.  
We also evaluated~\eqref{eq:contourint5} in the other FI-chambers and obtained the same value;  there is no wall-crossing as expected for the product of identical operators.

\subsubsection{ Matrix model for \texorpdfstring{$(\bm{B},  \mathfrak{m})=(3\bm{e}_N, \bm{e}_k + 2 \bm{e}_l )$ with $k\neq l$ }{...}}

\begin{figure}[htb]
\centering
\subfigure[]{\label{fig:21in3monopole1}
\includegraphics[scale=.5]{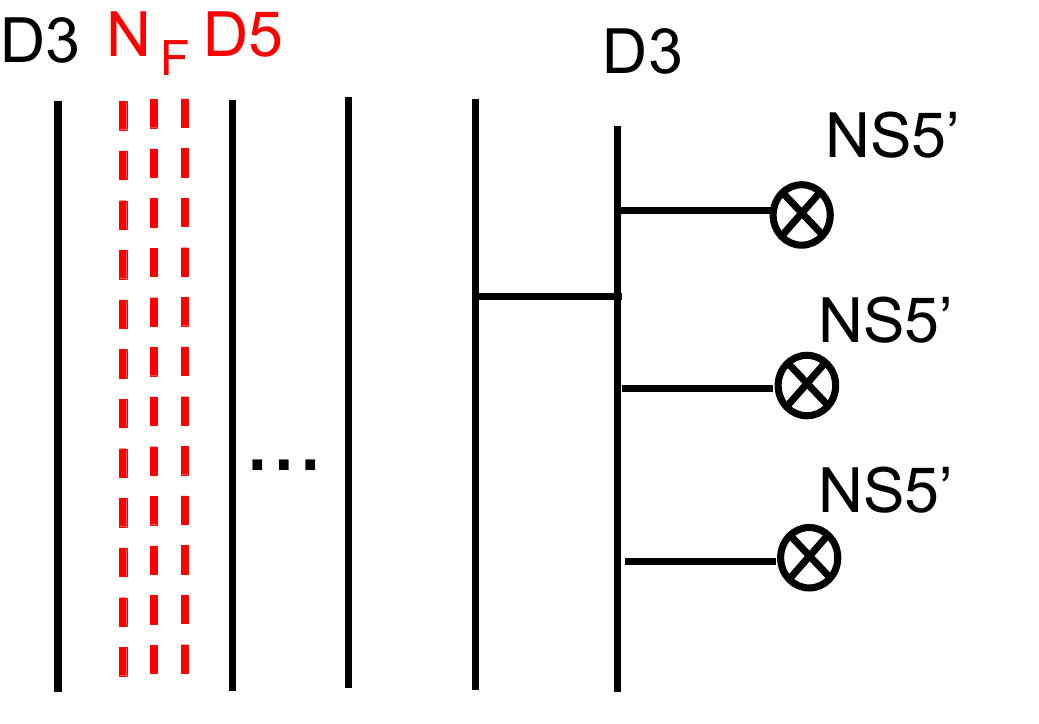}}
\hspace{1cm}
\subfigure[]{\label{fig:21in3monopole2}
\includegraphics[scale=.5]{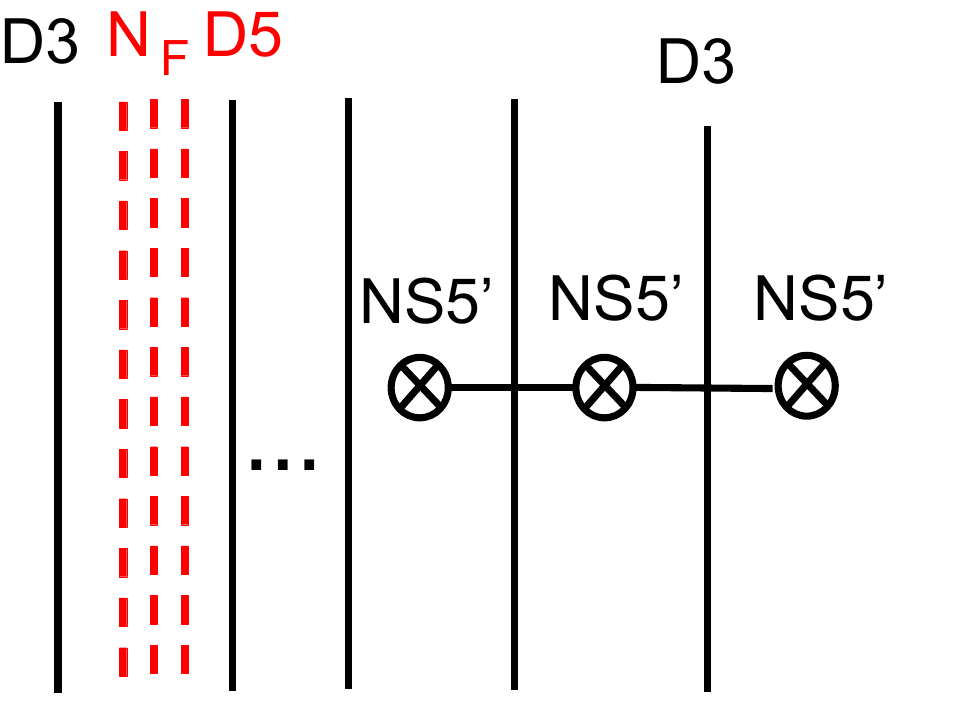}}
\caption{(a): 
  We add a finite D1-branes in Figure~\ref{fig:3monopole1}  to describe monopole bubbling effect for   $\mathfrak{m}=
\bm{e}_{N-1}+2\bm{e}_N
  $ in ${\bm B}=
  3\bm{e}_N
  $.
(b): After a sequence of Hanany-Witten transitions,  we obtain a brane configuration from which the matter content of matrix model can be read off. }
\label{fig:monopole52}
\end{figure}

The brane construction for monopole bubbling  with $\mathfrak{m}=
\bm{e}_{N-1} + 2 \bm{e}_N
$ and ${\bm B}=3\bm{e}_N$ is   illustrated in Figure \ref{fig:monopole52}.
The quiver diagram  for the matrix model  is the special case of the one in  Figure~\ref{fig:21quiverMM} with $N_F=0$ and $N=2$. 
Wall-crossing does not occur and  we obtain
\begin{align}
 Z^{({\bm \zeta}^{\prime})}_{\text{mono}}(  {\bm B}=3\bm{e}_N, \mathfrak{m}= {\bm e}_{N-1}+ 2 {\bm e}_{N} )&
=\frac{3}{\left( \varphi_{N-1, N}+\frac{3}{2}\epsilon \right) \left( \varphi_{N, N-1}+\frac{3}{2}\epsilon \right) }.
\label{}
\end{align}
This reproduces \eqref{eq:zmono12in3} with $k=N, l=N-1$. 
 The other values of $(k,l)$ can be reached by permutation.

Finally we obtain 
\begin{align}
\langle  V_{3\bm{e}_N} \rangle=\langle  V_{\bm{e}_N} \rangle * \langle  V_{\bm{e}_N} \rangle * \langle  V_{\bm{e}_N} \rangle.
\end{align}

\section{$U(N)$ gauge theory with $N_F$ fundamentals and an adjoint: Jordan quiver}

\begin{figure}[htb]
\vspace{5mm}
\centering
\includegraphics[width=6cm]{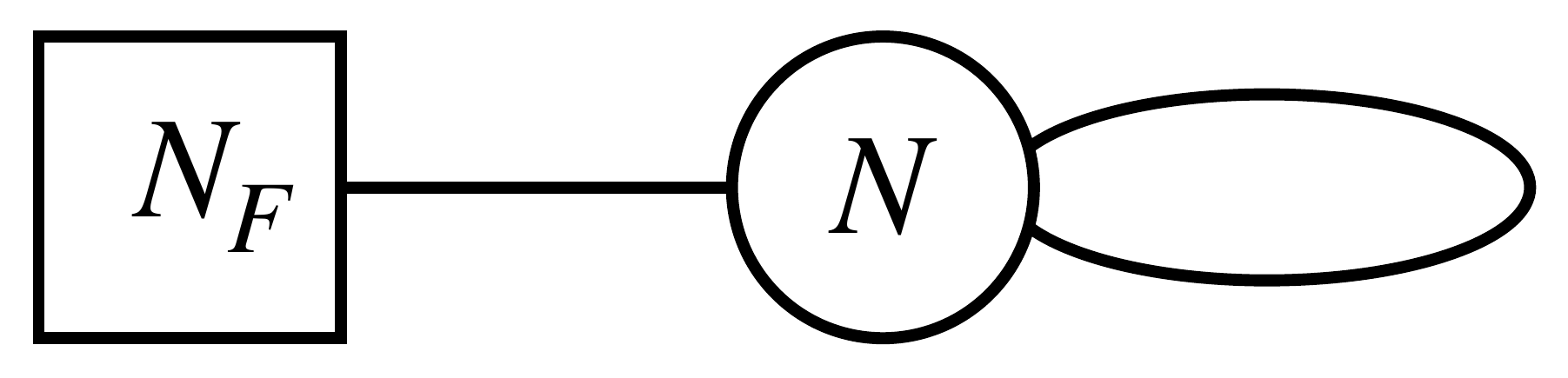}
\caption{The Jordan quiver.}
\label{fig:quiver-adjoint-SQCD}
\end{figure}

In this section we study the $U(N)$ gauge theory with $N_F$ hypermultiplets in the fundamental representation and a hypermultiplet in the adjoint representation. 
The corresponding quiver diagram, shown in Figure~\ref{fig:quiver-adjoint-SQCD}, is known as the Jordan quiver. 
The Higgs branch is isomorphic to the moduli space of $N$-instantons on $\mathbb{R}^4$ for gauge group $SU(N_F)$; the equations that define the Higgs branch in the gauge theory are precisely those of the ADHM construction.

\subsection{Quantized Coulomb branch chiral ring}
\label{sec:CBjordan}
The abelianization procedure of~\cite{Bullimore:2015lsa} was motivated by what one would obtain by a localization calculation in the $\Omega$-background.
In~\cite{Braverman:2016wma}
 the equivariant localization applied to the mathematical definition~\cite{Nakajima:2015txa,Braverman:2016wma} of the Coulomb branch was related to the abelianization procedure  for some $\mathcal{N}=4$ theories.
  For the $\mathcal{N}=4$ theory we study in this section, \cite{Braverman:2016aa} constructed explicitly the generators of the quantized Coulomb branch as difference operators which correspond to the dressed monopole operators.
The magnetic charge of such a dressed operator corresponds to an exterior power of the fundamental or anti-fundamental representation of the Langlands dual group $U(N)$ of the gauge group.
Here we confirm that the actual SUSY localization formula \eqref{eq:exdremono} for dressed monopole operators indeed reproduces the difference operators.

We consider dressed monopole operators with magnetic charges
\begin{equation}
{\bm B}_n :=
\bm{e}_{N-n+1} + \ldots + \bm{e}_N
\quad \text{ and } \quad
{\bm B}'_n :=
-\bm{e}_1 -\ldots - \bm{e}_n
\end{equation}
 for $n=0,1,\cdots, N$. 
We can apply the localization formula \eqref{eq:exdremono} to evaluate the vevs of the dressed monopole operators, since
these coweights are minuscule 
and   monopole bubbling terms are absent. 
The expectation values of the  dressed monopole operators 
$V_{{B_n}, g}$
 and 
 $V_{{B'_n}, g}$ 
 with a dressing factor $g(\varphi_I)$  are given  by
\begin{align}
&\langle
V_{{B_n}, g}
 \rangle
=\sum_{I \subset \{1, \cdots, N \} \atop |I|=n} g (\varphi_I) \prod_{i \in I  \atop j \notin I} 
\frac{\left( \varphi_{i j}-m_{\mathrm{ad}} \right)^{\frac{1}{2}} \left( \varphi_{j i} -m_{\mathrm{ad}} \right)^{\frac{1}{2}}}{\left( \varphi_{i j}+\frac{\epsilon}{2} \right)^{\frac{1}{2}} \left( \varphi_{j i}+\frac{\epsilon}{2} \right)^{\frac{1}{2}} } \prod_{i \in I} \left( \prod_{f=1}^{N_F}(\varphi_i -m_f)^{\frac{1}{2}}  \right) e^{  b_i}, 
\label{eq:jordandress1} \\
&\langle 
V_{{B'_n}, g}
 \rangle
=\sum_{I \subset \{1, \cdots, N \} \atop |I|=n} g (\varphi_I) \prod_{i \in I \atop j \notin I} 
\frac{\left( \varphi_{i j}-m_{\mathrm{ad}} \right)^{\frac{1}{2}} \left( \varphi_{j i} -m_{\mathrm{ad}} \right)^{\frac{1}{2}}}{\left( \varphi_{i j}+\frac{\epsilon}{2} \right)^{\frac{1}{2}} \left( \varphi_{j i}+\frac{\epsilon}{2} \right)^{\frac{1}{2}} } \prod_{i \in I} \left( \prod_{f=1}^{N_F}(\varphi_i -m_f)^{\frac{1}{2}}  \right) e^{-  b_i}.
\label{eq:jordandress2}
\end{align}
Here $I$ with $|I|=n$ denotes a subset of $\{1, \cdots, N \}$ with $n$ elements. $g(\varphi_I)$ is a symmetric polynomial of $\{\varphi_i\}_{i \in I}$. 
In order to compare the mathematical literature we introduce $u_I$ defined as
\begin{align}
u_I:= \prod_{i \in I} \left( \prod_{ j \notin I} 
\frac{\left( \varphi_{j i}+\frac{\epsilon}{2} \right) \left( \varphi_{i j}-m_{\mathrm{ad}} \right)  }
{\left( \varphi_{i j}+\frac{\epsilon}{2} \right)  \left( \varphi_{j i} -m_{\mathrm{ad}} \right) } \right)^{\frac{1}{2}}
 \left( \prod_{f=1}^{N_F}(\varphi_i -m_f)^{\frac{1}{2}}  \right) e^{  b_i} .
\end{align}
In terms of $\varphi_i$ and $u_I$ we can write \eqref{eq:jordandress1} and \eqref{eq:jordandress2} as
\begin{align}
\langle V_{{B_n}, g}\rangle
&=\sum_{I \subset \{1, \cdots, N \} \atop |I|=n} g (\varphi_I) \prod_{i \in I \atop j \notin I} 
\frac{ \varphi_{j i }-m_{\mathrm{ad}}  }{  \varphi_{j i}+\frac{\epsilon}{2} } u_I, \\
\langle V_{{B'_n}, g}\rangle
&=\sum_{I \subset \{1, \cdots, N \} \atop |I|=n} g (\varphi_I) \prod_{i \in I \atop j \notin I} 
\frac{ \varphi_{i j }-m_{\mathrm{ad}}  }{  \varphi_{i j}+\frac{\epsilon}{2} }  
\left( \prod_{f=1}^{N_F} (\varphi_i -m_f)   \right) u^{-1}_I .
\end{align}
The Weyl transform of the dressed monopole operators in anti-symmetric representation is given by
\begin{align}
&
\widehat{V}_{{B_n}, g}
=\sum_{I \subset \{1, \cdots, N \} \atop |I|=n} g \left(\hat{\varphi}_I+\frac{\epsilon}{2} \right) \prod_{i \in I , j \notin I} 
\frac{ \hat{\varphi}_{j i }-\frac{\epsilon}{2}-m_{\mathrm{ad}}  }{  \hat{\varphi}_{j i} }   \hat{u}_I \\
& 
\widehat{V}_{{B'_n}, g}
=\sum_{I \subset \{1, \cdots, N \} \atop |I|=n} g \left(\hat{\varphi}_I-\frac{\epsilon}{2} \right) \prod_{i \in I , j \notin I} 
\frac{ \hat{\varphi}_{i j }-\frac{\epsilon}{2}-m_{\mathrm{ad}}  }{  \hat{\varphi}_{i j} }   \prod_{i \in I} \left( \prod_{f=1}^{N_F} (\hat{\varphi}_i-\frac{\epsilon}{2} -m_f)   \right) \hat{u}^{-1}_I .
\end{align}
Here  $\hat{\varphi}_i, \hat{u}^{\pm1}_{I}$ are defined by the Weyl transform of  ${\varphi}_i, {u}^{\pm 1}_{I}$ and satisfy the following relations
\begin{align}
\hat{u}^{\pm 1}_I  \hat{\varphi}_i=
\left\{
\begin{array}{cl}
( \hat{\varphi}_i \pm \epsilon ) \hat{u}^{\pm 1}_I & \text{ for } i \in I, \\
\hat{\varphi}_i \hat{u}^{\pm 1}_I & \text{ for } i \notin I ,
\end{array}
\right.
\label{eq:rel_phi_u}
\end{align}
and
\begin{align}
&\hat{u}_I  \hat{u}^{-1 }_J= \hat{u}_{I \cap J^{c} }  \hat{u}^{-1 }_{I^c \cap J}, \quad 
\hat{u}^{-1}_I  \hat{u}_J= \hat{u}^{-1}_{I \cap J^{c} }  \hat{u}_{I^c \cap J},     \nonumber \\
&\hat{u}_I  \hat{u}_J =\hat{u}_J  \hat{u}_I, \quad \hat{u}^{-1}_I  \hat{u}^{-1}_J =\hat{u}^{-1}_J  \hat{u}^{-1}_I .
\label{eq:rel_u_u}
\end{align}
Here $I^{c}$ is the compliment of $I$ in $\{1, \cdots, N \}$. If we introduce operators $\{ \hat{\sf u}^{\pm 1}_i \}_{i=1}^N$ subject to the relations
\begin{align} \label{rm-u-relations}
[ \hat{\sf u}^{\pm 1}_i, \hat{\varphi}_j] = \pm \epsilon \delta_{i j}  \hat{\sf u}^{\pm 1}_i , \quad [\hat{\sf u}_i, \hat{\sf u}_j]=[\hat{\sf u}^{-1}_i, \hat{\sf u}^{-1}_j]=0, \quad \hat{\sf u}_i \hat{\sf u}^{-1}_i =\hat{\sf u}^{-1}_i \hat{\sf u}_i=1,
\end{align}
we find that  $\prod_{i \in I} \hat{\sf u}^{\pm 1}_i$ satisfies the same relations \eqref{eq:rel_phi_u} and \eqref{eq:rel_u_u} of $\hat{u}^{\pm 1}_I$ and can be  identified as $\prod_{i \in I}  \hat{\sf u}^{\pm 1}_i  \equiv \hat{u}_I^{\pm 1}$.
We  also redefine   $f(\varphi_I):=g \left(\varphi_I+\frac{\epsilon}{2} \right)$ and $t:=-\frac{\epsilon}{2}-m_{\mathrm{ad}}$ and $z_f:=-\frac{\epsilon}{2} +m_f$. Then 
the Weyl transform of the dressed monopoles operators  are expressed as
\begin{align}
&
\widehat{V}_{{B_n}, g}
=\sum_{I \subset \{1, \cdots, N \} \atop |I|=n} f \left(\hat{\varphi}_I\right) \prod_{i \in I} \left( \prod_{ j \notin I} 
\frac{ \hat{\varphi}_{ i } -\hat{\varphi}_{ j } -t   }{ \hat{\varphi}_{ i } -\hat{\varphi}_{ j } } \right)  
\hat{\sf u}_i,
\label{eq:Enf}\\
&
\widehat{V}_{{B'_n}, g}
=\sum_{I \subset \{1, \cdots, N \} \atop |I|=n} f \left(\hat{\varphi}_I-\epsilon \right) \prod_{i \in I } \left(\prod_{ j \notin I}  
\frac{ \hat{\varphi}_{ i } -\hat{\varphi}_{ j } +t   }{ \hat{\varphi}_{ i } -\hat{\varphi}_{ j } }    \right)  \left( \prod_{f=1}^{N_F} (\hat{\varphi}_i-z_f-\epsilon )   \right) 
\hat{\sf u}^{-1}_i . 
\label{eq:Fnf}
\end{align}
The operators \eqref{eq:Enf} and \eqref{eq:Fnf} are the same as   $E_n[f], F_n[f]$ defined in (A.5) of~\cite{Braverman:2016aa} if we identify $ \{ \hat{\sf u}^{\pm 1}_i \}_{i=1}^N$ with the corresponding difference operators obeying~(\ref{rm-u-relations}).
It was shown in~\cite{Braverman:2016aa} that quantized Coulomb branch chiral ring is generated by $E_n[f], F_n[f]$ with $f$ taken as the $n$-th symmetric polynomial of $\{ \varphi_i \}_{i \in I} $ for $n=0, 1, 2, \cdots ,N$. 
The authors of \cite{Kodera:2016faj} showed that the quantized Coulomb branch chiral ring is isomorphic to certain well known algebras: the spherical trigonometric  Double Affine Hecke algebra (DAHA) of type $\mathfrak{gl}(N)$  for $N_F=0$, and the spherical cyclotomic rational Cherednik algebra for $N_F >0$.

\subsection{K-theoretic Coulomb branch, line operators in 4d and the spherical DAHA}
The spherical trigonometric  DAHA  is a degeneration of  the  spherical DAHA  \cite{MR2133033}.
The K-theoretic version \cite{Braverman:2016wma, Braverman:2016aa} of the quantized  Coulomb branch (K-theoretic Coulomb branch) for  $N_F=0$ is isomorphic to the  spherical DAHA  of type $\mathfrak{gl}(N)$ \cite{Braverman2016ef}.
It is natural to expect that this degeneration is related  to the dimensional reduction of line operators in $G=U(N)$ $\mathcal{N}=2^*$ theory  on $S^1 \times \mathbb{R}^2_{\epsilon} \times \mathbb{R}$ and the algebra of line operators on the $\Omega$-background 
 is related to the quantized K-theoretic  Coulomb branch. 
In this subsection we point out that the expressions that follow from the localization formula of~\cite{Ito:2011ea} for Wilson-'t Hooft line operators with minuscule coweights as magnetic charges indeed coincide with those for the generators of the  spherical DAHA  of type $\mathfrak{gl}(N)$  in the functional representation~\cite{Francesco2017aa}.%
\footnote{%
We understand, through talks and conversations, that the connection between line operators in 4d $\mathcal{N}=2^*$ theories and DAHA has been discussed for many years by various people, though we were not able to find a paper that explicitly studies the connection.
See, for example,~\cite{Nawata-KITP-talk} for an online talk.
For other connections between SUSY gauge theories and DAHA or Macdonald operators, see for example~\cite{Razamat:2013jxa,Bullimore:2014nla}.
}

We can write the vevs of Wilson-'t Hooft operators  with magnetic charges 
${\bm B}_n$
and 
${\bm B}'_n$
using the SUSY localization formula.
 When the 4d theory on  $S^1 \times \mathbb{R}^3$ is dimensionally reduced along the $S^1$, the 4d parameters $\varphi_i$ and $b_i$ that we use below become small and proportional to the corresponding 3d parameters.
Let us introduce the shorthand notation
\begin{align}
\mathrm{sh} (z):=2 \sinh \left( \frac{z}{2} \right) .
\end{align}
Then the vevs for the operators with no bubbling contributions, {\it i.e.}, those with minuscule magnetic charges~$\bm{B}_n$ and $\bm{B}'_n$, are given as
\begin{align}
&\langle
 T_{{\bm B}_n, g_1}
  \rangle
=\sum_{I \subset \{1, \cdots, N \} \atop |I|=n} g_1 (e^{\varphi_I}) \left( \prod_{i \in I , j \notin I} 
\frac{ \mathrm{sh} \left( \varphi_{i j}-m_{\mathrm{ad}} \right)  \mathrm{sh}  \left( \varphi_{j i} -m_{\mathrm{ad}} \right)}{\mathrm{sh} \left( \varphi_{i j}+\frac{\epsilon}{2} \right)  \mathrm{sh} \left( \varphi_{j i}+\frac{\epsilon}{2} \right) } \right)^{\frac{1}{2}} \prod_{i \in I}  e^{  b_i}, 
\label{eq:jordandyonic1} \\
&\langle 
T_{{\bm B}'_n, g_2}
 \rangle
=\sum_{I \subset \{1, \cdots, N \} \atop |I|=n} g_2 (e^{\varphi_I})
\left( \prod_{i \in I , j \notin I} 
\frac{ \mathrm{sh} \left( \varphi_{i j}-m_{\mathrm{ad}} \right)  \mathrm{sh}  \left( \varphi_{j i} -m_{\mathrm{ad}} \right)}{\mathrm{sh} \left( \varphi_{i j}+\frac{\epsilon}{2} \right)  \mathrm{sh} \left( \varphi_{j i}+\frac{\epsilon}{2} \right) } \right)^{\frac{1}{2}}  \prod_{i \in I}  e^{-  b_i}.
\label{eq:jordandyonic2}
\end{align}
Functions $g_1$ and $g_2$  are  symmetric Laurent polynomials in $\{ x_i \}_{i \in I}$ with $x_i:= e^{- \varphi_i}, i=1, \cdots, N$ which are  Wilson line operators for a stabilizer of $U(N)$ for an $n$-th anti-symmetric cocharacters.
We define 
\begin{align}
&\theta:=e^{- \frac{m_{\text{ad}}}{2}- \frac{\epsilon}{4} }, \quad q:=e^{-\epsilon},  \\
&\Gamma_I:= \prod_{i \in I} \left(  \prod_{ j \notin I} 
\frac{ \mathrm{sh} \left( \varphi_{j i}+\frac{\epsilon}{2} \right) \mathrm{sh}  \left( \varphi_{i j}-m_{\mathrm{ad}} \right)  }
{\mathrm{sh}  \left( \varphi_{i j}+\frac{\epsilon}{2} \right) \mathrm{sh}  \left( \varphi_{j i} -m_{\mathrm{ad}} \right)} \right)^{\frac{1}{2}} e^{  b_i} .
\end{align}
Then the Weyl transform of \eqref{eq:jordandyonic1} and \eqref{eq:jordandyonic2} with respect to \eqref{eq:Moyal_Wey} can be written as
\begin{align}
&
 \widehat{T}_{{\bm B}_n, g_1}
=\sum_{I \subset \{1, \cdots, N \} \atop |I|=n} g_1 (\hat{x}_I q^{-\frac{1}{2}}) \left( \prod_{i \in I , j \notin I} 
\frac{\hat{x}_i \theta -\hat{x}_j \theta^{-1}}{\hat{x}_i -\hat{x}_j} \right) \hat{\Gamma}_I , 
\label{eq:jordandyonic3} \\
&
\widehat{T}_{{\bm B}'_n, g_2}
=\sum_{I \subset \{1, \cdots, N \} \atop |I|=n} g_2 (\hat{x}_I q^{\frac{1}{2}})
\left( \prod_{i \in I , j \notin I} \frac{\hat{x}_i \theta^{-1} -\hat{x}_j \theta}{\hat{x}_i -\hat{x}_j}  \right)  \hat{\Gamma}_I^{-1}.
\label{eq:jordandyonic4}
\end{align}
where $\hat{x}_i$ and $\hat{\Gamma}_I$ are the Weyl transform of ${x}_i$ and ${\Gamma}_I$. The commutation relation of  $\hat{\Gamma}^{\pm 1}_I$ and $\hat{x}_{i}$ are  
\begin{align}
\hat{\Gamma}^{\pm 1}_I  \hat{x}_i=
\left\{
\begin{array}{ll}
q^{\pm 1} \hat{x}_i \hat{\Gamma}^{\pm 1}_I  & i \in I, \\
\hat{x}_i \hat{\Gamma}^{\pm 1}_I  & i \notin I .
\end{array}
\right.
\end{align}
which means $\hat{\Gamma}_I$ acts on functions of $x_i$ as a $q$-difference operator.  
When we take $g_1 (\hat{x}_I q^{-\frac{1}{2}})=g_2 (\hat{x}_I q^{\frac{1}{2}}) =\prod_{i \in I } \hat{x}^k_i$ for $k=0,1,2, \cdots$, 
\eqref{eq:jordandyonic3} and \eqref{eq:jordandyonic4} are  same as generalized Macdonald operators $\mathcal{D}^{(q,t)}_{\alpha, k}$ and $\mathcal{D}^{(q^{-1},t^{-1})}_{\alpha, k}$ with $\alpha=n$ in \cite{Francesco2017aa}.  When $n=0$ the Wilson-'t Hooft line operators becomes 't Hooft line operators in the anti-symmetric representation, our computation show that 't Hooft line operators in the anti-symmetric representations agree with the Macdonald operators $\mathcal{D}^{(q,t)}_{\alpha, 0}$ and $\mathcal{D}^{(q^{-1},t^{-1})}_{\alpha, 0}$.

It was shown in \cite{Francesco2017aa} that  the generalized Macdonald operators  generate  the spherical DAHA. Therefore the algebra of Wilson, 't Hooft, and dyonic line operators generated by \eqref{eq:jordandyonic3} and \eqref{eq:jordandyonic4} which is a deformation quantization of Coulomb branch of $G=U(N)$ $\mathcal{N}=2^*$ super Yang-Mills theory and coincides with the spherical DAHA of type $\mathfrak{gl} (N)$. 

\subsection{Operator ordering and wall-crossing}
\begin{figure}[htb]
\begin{center}
\includegraphics[width=5cm]{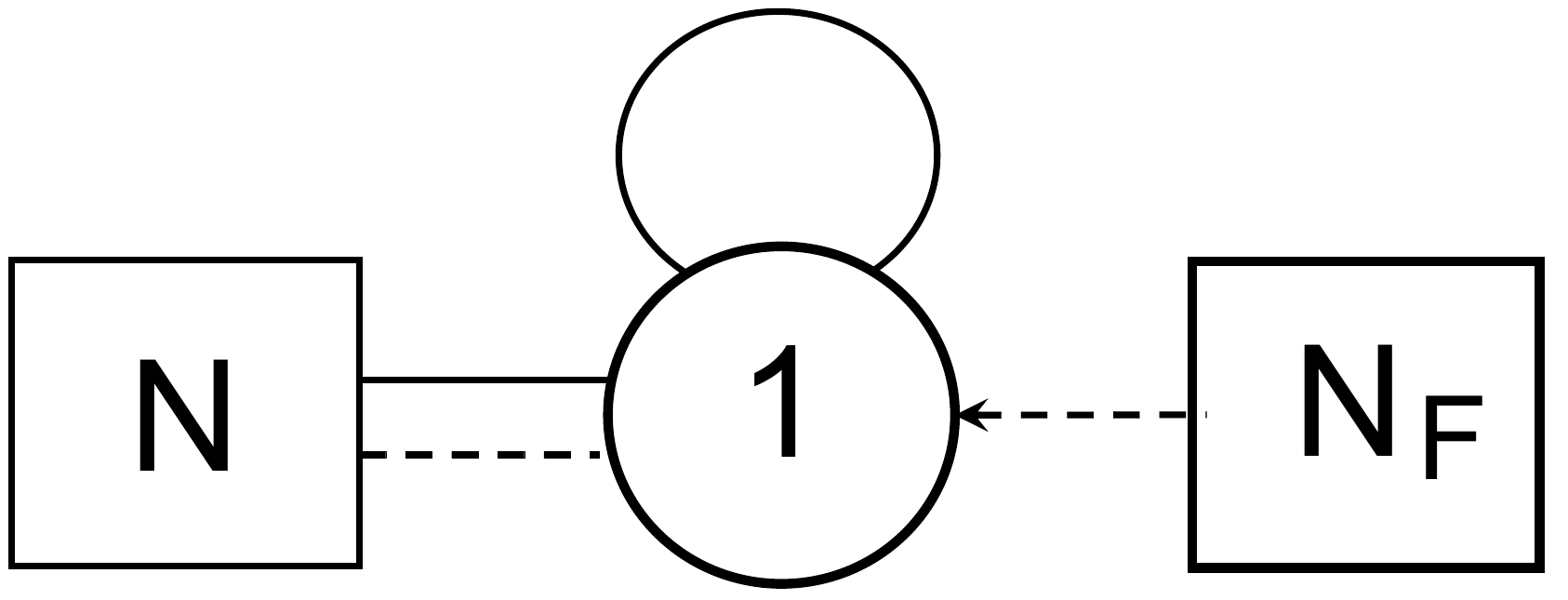}
\end{center}
\vspace{-0.5cm}
\caption{The quiver diagram for the matrix model that computes $Z_{\text{mono}} (\mathfrak{m}=0, {\bm B}=
\bm{e}_N-\bm{e}_1
) $ in the 3d $\mathcal{N}=4$ $U(N)$  gauge theory with an adjoint hypermultiplet and $N_F$ fundamental hypermultiplets. 
The dashed line without an arrow represents $\mathcal{N}=(0,4)$ long Fermi multiplets. The curve connected to the circle with $1$ represents an adjoint $\mathcal{N}=(0,4)$  twisted hypermultiplet.}
\label{fig:quiverMMjordan}
\end{figure}
We study  the ordering of  Moyal products and wall-crossing in the partition functions of  matrix models. 
We assume that $N\geq 2$.
The computation of Moyal products and JK residues are parallel to that  in the $U(N)$ with $N_F$  fundamental hypermultiplets case.
 So we do not repeat all of the computations done in Section~\ref{sec:U(N)-SQCD-NF-flavors}, but focus on  new  features in the brane picture and  the matrix model arising from an adjoint hypermultiplet.

To realize the 3d gauge theory on the world-volume of D3-branes~\cite{deBoer:1996ck} we compactify the $x^6$-direction to a circle and place one, rather than two, NS5-brane.  We thus have $N$ D3-branes, one NS5-brane, and $N_F$ D5-branes extended in the directions as specified in Table~\ref{table:brane-directions}.
The D3-D3 open string across the NS5-brane gives rise to the adjoint hypermultiplet.

The matrix models for monopole bubbling for the 3d theory with an adjoint are almost the same as for the 3d theory without an adjoint  hypermultiplet.
They are obtained by replacing $\mathcal{N}=(0,4)$ vector multiplets and hypermultiplets by $\mathcal{N}=(4,4)$ vector multiplets and hypermultiplets, respectively.
In the brane set-up above, the extra fields correspond to D1-D1 and D1-D3 open strings whose ends are on different sides of the NS5-brane.

For example, let us consider the matrix model that computes the monopole bubbling contribution $Z^{(\zeta)}_{\mathrm{mono}} (\mathfrak{m}={\bm 0} )$ in  $\langle  V_{\bm{e}_N-\bm{e}_1} \rangle$.  
The brane configurations for monopole bubbling, projected onto the $(x^3,x^4)$-plane, look the same as  Figures~\ref{fig:adjmonopole1}, \ref{fig:adjmonopole2} and \ref{fig:adjmonopole3},
 but we get as extra matter fields an $\mathcal{N}=(0,4)$ 
 hypermultiplet in the adjoint representation (actually neutral because the gauge group is abelian), and $N$ long Fermi multiplets with charge $1$.
Thus the quiver diagram changes from Figure~\ref{fig:D0quiver1} to Figure~\ref{fig:quiverMMjordan}. 
The partition function of the matrix model is given by
\begin{align}
Z^{(\zeta)}_{\mathrm{mono}} (\mathfrak{m}={\bm 0} ) 
=\oint_{\mathrm{JK} (\zeta)} \frac{d u}{2 \pi i} \frac{ (- \epsilon) \prod_{f=1}^{N_F}(u-m_f)\prod_{s=\pm1} \prod_{i=1}^N \left( s(u-\varphi_i) -m_{\text{ad}}  \right) }
{ \prod_{s=\pm1} \left(  -m_{\text{ad}}+s \frac{\epsilon}{2}  \right) \prod_{i=1}^N \left( s(u-\varphi_i) -\frac{\epsilon}{2} \right)} .
\label{eq:contourjordan}
\end{align}
The residues are again evaluated at $u=\varphi_i+\frac{\epsilon}{2}$, $i=1, \cdots, N$ for $\zeta >0$ and at $u=\varphi_i-\frac{\epsilon}{2}$, $i=1, \cdots, N$ for $\zeta <0$.
 On the other hand,   $Z_{\text{mono}}( \mathfrak{m}={\bm 0} )$ in $\langle  V_{\bm{e}_N-\bm{e}_1} \rangle$ 
from the Moyal product of  $\langle  V_{-\bm{e}_1} \rangle$ and $\langle  V_{\bm{e}_N} \rangle$ are easily 
calculated as
\begin{align}
& Z_{\text{mono}} (\mathfrak{m}={\bm 0} )  \, \text{from} \,  \langle  V_{
{ \bm{e}_N}
} \rangle *  \langle  V_{-\bm{e}_1} \rangle
\nonumber \\ 
&\qquad =\sum_{ k =1}^N  
   \frac{\prod_{f=1}^{N_F} \left( \varphi_k -m_f+\frac{\epsilon}{2} \right) \prod_{j = 1 \atop j \neq k }^N \left(\varphi_{k j} - m_{\text{ad}}+\frac{\epsilon}{2} \right) 
 \left(\varphi_{  j k}- m_{\text{ad}}- \frac{\epsilon}{2} \right)}
{\prod_{j = 1 \atop j \ne k}^N \varphi_{ k j } (\varphi_{ j k}- \epsilon)}, 
\label{eq:moyalplusminus_ad}
 \\
& Z_{\text{mono}} (\mathfrak{m}={\bm 0} )  \, \text{from} \,  \langle  V_{-\bm{e}_1} \rangle *  \langle  V_{\bm{e}_N} \rangle
\nonumber \\
&\qquad = \sum_{ k =1}^N  
   \frac{\prod_{f=1}^{N_F} \left( \varphi_k -m_f-\frac{\epsilon}{2} \right) \prod_{j = 1 \atop j \neq k }^N \left(\varphi_{k j} - m_{\text{ad}}-\frac{\epsilon}{2} \right) 
 \left(\varphi_{  j k}- m_{\text{ad}}+ \frac{\epsilon}{2} \right)}
{\prod_{j = 1 \atop j \ne k}^N \varphi_{ k j } (\varphi_{ j k}+ \epsilon)}. 
\label{eq:moyalminusplus_ad}
\end{align}
We have agreement between the Moyal product and the matrix model computation;
\begin{align}
Z^{(\zeta>0)}_{\mathrm{mono}} ( {\bm B}=\bm{e}_N-\bm{e}_1, \mathfrak{m}={\bm 0} ) &= (
Z_{\text{mono}} (\mathfrak{m}={\bm 0} )  \, \text{from} \,  \langle  V_{
\bm{e}_N} \rangle *  \langle  V_{-\bm{e}_1} \rangle ) ,
\label{eq:screeningplus_ad}\\
 Z^{(\zeta<0)}_{\mathrm{mono}} ( {\bm B}=\bm{e}_N-\bm{e}_1, \mathfrak{m}={\bm 0} ) &=
( Z_{\text{mono}} (\mathfrak{m}={\bm 0} )  \, \text{from} \,  \langle  V_{-\bm{e}_1} \rangle *  \langle  V_{\bm{e}_N} \rangle ) . 
\label{eq:screeningminus_ad}
\end{align}

The non-commutativity of the Moyal product and wall-crossing behavior of the matrix model are again evaluated as
\begin{align}
&\langle  V_{\bm{e}_N} \rangle *  \langle  V_{-\bm{e}_1} \rangle
-
\langle  V_{-\bm{e}_1} \rangle *  \langle  V_{\bm{e}_N} \rangle 
\nonumber \\
&\qquad=
\oint_{u=\infty} \frac{d u}{2 \pi i} \frac{ (- \epsilon) \prod_{f=1}^{N_F}(u-m_f)\prod_{s=\pm1} \prod_{i=1}^N \left( s(u-\varphi_i) -m_{\text{ad}}  \right) }
{ \prod_{s=\pm1} \left(  -m_{\text{ad}}+s \frac{\epsilon}{2}  \right) \prod_{i=1}^N \left( s(u-\varphi_i) -\frac{\epsilon}{2} \right)}  \nonumber \\
&\qquad=\left\{
\begin{array}{cll}
0  & \text{ for } N_F=0 & \text{(bad),} \\
N \epsilon  & \text{ for } N_F=1 &\text{(ugly),}  \\
\epsilon  \left( 2 \sum_{i=1}^N \varphi_i- N \sum_{f=1}^{2} m_f   \right) 
  & \text{ for } N_F=2 & \text{(good and balanced),} \\
 \epsilon \tilde{A} ({\bm \varphi}, {\bm m}, \epsilon) &  \text{ for } N_F>2 & \text{(good but not balanced)}\\
\end{array}
\right.
\label{eq:gooduglybad3}
\end{align}
with
\begin{align}
\tilde{A} ({\bm \varphi}, {\bm m}, \epsilon):= { \frac{1}{(N_F+1)!}} \left( \frac{d}{ d w } \right)^{ N_F+1}  \frac{\prod_{f=1}^{N_F}(1 -m_f w) \prod_{s=\pm 1}  \prod_{i=1}^N \left( 1 - \left( \varphi_i + s \frac{\epsilon}{2} \right) w  \right) }
{ \prod_{s=\pm 1} \left(-m_{\text{ad}} +s \frac{\epsilon}{2} \right) \prod_{i=1}^N \left( 1 - \left( \varphi_i + s \frac{\epsilon}{2} \right) w  \right)  } \Bigg |_{w=0} .
\end{align}


\section{Linear quiver gauge theories}
\label{sec:linear-quiver}

\begin{figure}[tb]
\begin{center}
\includegraphics[width=5cm]{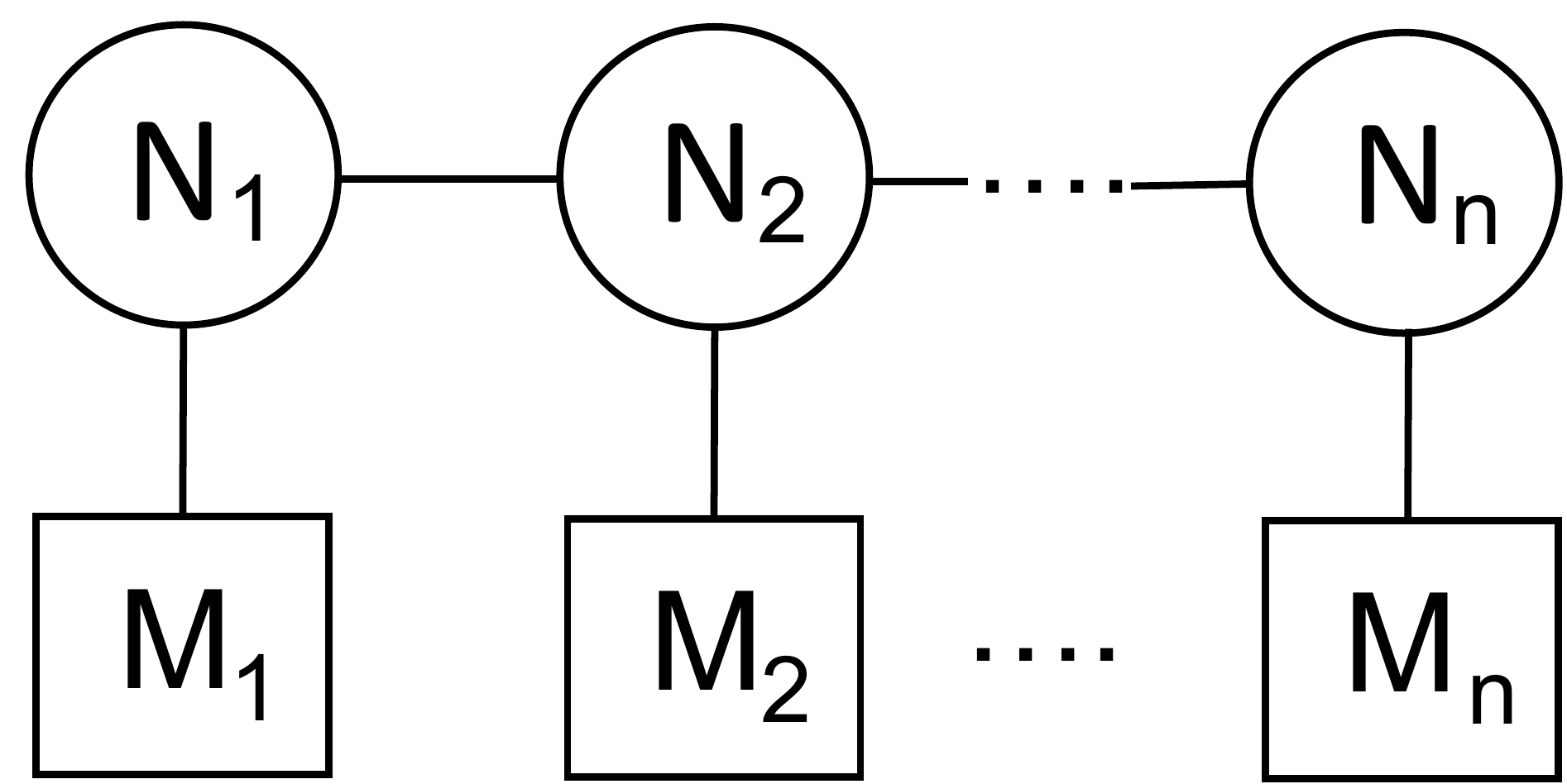}
\end{center}
\vspace{-0.5cm}
\caption{The quiver diagram for a 3d $\mathcal{N}=4$  linear quiver gauge theory with gauge group $G=\prod_{l=1}^n U(N_l)$.
A circle with $N_l$ represents a vector multiplet for a gauge group factor $U(N_l)$ and a line connecting a circle with $N_l$ and a box with $M_l$ represents $M_l$ hypermultiplets in the fundamental representation of $U(N_l)$. The solid line connecting two circles with $N_l$ and $N_{l+1}$ represents a $U(N_l) \times U(N_{l+1})$ bi-fundamental hypermultiplet.}
\label{fig:quiverAlinear}
\end{figure}
In this section we consider a 3d $\mathcal{N}=4$ linear quiver gauge theory  with gauge group $G=\prod_{l=1}^n U(N_l)$ specified by the quiver diagram depicted in  Figure~\ref{fig:quiverAlinear}.

We recall from~(\ref{cochar-quiver}) that the cocharacter lattice and the Cartan subalgebra of the gauge group is generated by $\bm{e}_i^{(l)}$ ($i=1,\ldots, N_l$, $l=1,\ldots,n$).
We define $\varphi^{(l)}_i$ and ${ b}^{(l)}_i$ by
\begin{equation}
\bm{\varphi} = \sum_{l=1}^n \sum_{i=1}^{N_l} \varphi^{(l)}_i\bm{e}_i^{(l)} \,,
\qquad
\bm{b} = \sum_{l=1}^n \sum_{i=1}^{N_l} b^{(l)}_i\bm{e}_i^{(l)} 
\end{equation}
and define $m^{(l)}_{ f}$ ($f=1,\cdots, M_l$) as the mass of  the $f$-th hypermultiplet in the fundamental representation of $U(N_l)$.
Recall that we denote the generators of the cocharacter lattice~(\ref{cochar-quiver}) of the gauge group $G=\prod_{l=1}^n U(N_{l})$ by ${\bm e}^{(l)}_i$ ($i=1,\ldots, N_l$, $l=1,\ldots, n$).
We denote the bare monopole operators with the minimal magnetic charges for  $U(N_l)$ by ${V}^{(l)}_{+}:={V}_{ {\bm e}^{(l)}_{N_l} }$ and ${V}^{(l)}_{-}:={V}_{- {\bm e}^{(l)}_{1} }$.

\subsection{Quantized Coulomb branch chiral ring}
\label{sec:qCBlinear}
In this subsection we will see explicitly that the Weyl transforms of the localization formulas~\eqref{eq:exsca}, \eqref{eq:exdremono}, and \eqref{eq:exdremono2} are identified with the description
 of the quantized Coulomb branch chiral ring $ \mathbb{C}_{\epsilon}[\mathcal{M}_{C}]$ in~\cite{Bullimore:2015lsa}. 
We will also  see that  the Weyl transform of vevs of (dressed) monopole operators can be identified with 
difference operators describing  the quantized Coulomb branch in   \cite{Braverman:2016aa}.

We consider dressed monopole operators with the minimal magnetic charges.
We take the dressing factor to be 
$p_s(\Phi_{ww}): =((\Phi_{ww}^{(l)})_{N_lN_l})^s$
 ($s=0, 1, 2, \cdots$) for  $\bm{B}={\bm e}^{(l)}_{N_l  }$, and 
 $p_s(\Phi_{ww}): =((\Phi_{ww}^{(l)})_{11})^s$ for $\bm{B}= \bm{e}^{(l)}_1$.
 Here we regard $\Phi_{ww}^{(l)} \in {\rm Lie}(U(N_l))$ as an $N_l\times N_l$ hermitian matrix.
From the localization formula \eqref{eq:exdremono2}, the vevs of dressed  monopole operators $\langle {V}^{(l)}_{+, s} \rangle :=\langle V_{ {\bm e}^{(l)}_{N_l  }, p_s } \rangle $
 and  $\langle {V}^{(l)}_{-, s} \rangle:=\langle V_{ -{\bm e}^{(l)}_{1  }, p_s } \rangle $
are  given by
\begin{align}
\langle {V}^{(l)}_{\pm,s} \rangle = \sum_{i=1}^{N_l} \left(\varphi^{(l)}_i-\frac{\epsilon}{2} \right)^s v^{ \pm}_{ {l, i}} \,,
\label{eq:Uldressmono}
\end{align}
where  we defined $v_{l, i}^{\pm}$ (without a hat) by
\begin{align}
 v_{l, i}^{\pm} &:= v_{ \pm {\bm e}^{(l)}_i} \nonumber \\
&=e^{ \pm { b}^{(l)}_i } \left(  \frac{ \prod_{f=1}^{M_l} (\varphi^{(l)}_i-m^{(l)}_{ f}) 
\prod_{k=1}^{N_{l-1}} (\varphi^{(l)}_i -\varphi^{(l-1)}_{k} ) 
\prod_{k^{\prime}=1}^{N_{l+1}} (\varphi^{(l+1)}_{k^{\prime}}-\varphi^{(l)}_{i} )  }
{\prod_{j=1 \atop j \neq i}^{N_l} \left(\varphi^{(l)}_{i j}  +\frac{\epsilon}{2} \right)\left(\varphi^{(l)}_{j i}  +\frac{\epsilon}{2} \right) } \right)^{\frac{1}{2}} 
\label{eq:vli}
\end{align}
with   $\varphi^{(l)}_{i j}:=\varphi^{(l)}_{i}-\varphi^{(l)}_{j}$  ($l=1, \cdots, n$, $i=1, \cdots, N_l $).

The shift of $\varphi_i^{(l)}$ by $-\epsilon/2$ in (\ref{eq:Uldressmono}) is to be understood as a generalization of (\ref{eq:fV-abelian});  when we try to construct the dressed monopole operator is by inserting $p_s(\Phi_{ww})$ at the location of the bare monopole operator, the value of~$\Phi_{ww}$ is ambiguous as we can see in~(\ref{theta-dependence-of-Phiww}).  Here we choose to remove the ambiguity by inserting $p_s(\Phi_{ww})$ slightly above (resp. below) the monopole operator $V^{(l)}_{\bm{e}^{(l)}_{N_l}}$ (resp. $V^{(l)}_{-\bm{e}^{(l)}_{1}}$) along the $x^3$-axis.

 We define $E_l(z)$ and $F_l(z)$ as the generating functions of the Weyl transforms of $ \langle {V}^{(l)}_{\pm , s}  \rangle$ given by 
\begin{align}
E_l(z)&:= (-1)^{N_{l+1}+1} \sum_{i=1}^{N_l} \sum_{s=0}^{\infty} \widehat{V}^{(l)}_{-,s}  z^{-s-1} = (-1)^{N_{l+1}+1}  \sum_{i=1}^{N_l} \frac{1}{(z-\hat{\varphi}^{(l)}_i)} \hat{v}^{ -}_{ {l, i}}\,,  \\
F_l(z)&:= (-1)^{N_{l}+1}  \sum_{i=1}^{N_l} \sum_{s=0}^{\infty} \widehat{V}^{(l)}_{+, s}  z^{-s-1}=  (-1)^{N_{l}+1}  \sum_{i=1}^{N_l} \frac{1}{(z-\hat{\varphi}^{(l)}_i -\epsilon)} \hat{v}^+_{l, i}\,.
\end{align}
Here we include the overall conventional sign factors $(-1)^{N_{l+1}+1}$ and $(-1)^{N_{l}+1 }$.
From the Weyl transform of the   Moyal product, we can show that $\hat{v}^{\pm}_{ {l, i}}$ and $\hat{\varphi}^{(l)}_i$ satisfy the following relations in $\mathbb{C}_{\epsilon}[\mathcal{M}^{\mathrm{abel}}_{C}]$:
\begin{align}
&[\hat{\varphi}^{(l)}_i, \hat{\varphi}^{(k)}_j ] =0, \quad
[\hat{\varphi}^{(l)}_i, \hat{v}^{\pm}_{ {k, j}} ]=\mp \epsilon \delta_{l k} \delta_{i j} \hat{v}^{ \pm}_{ {k, j}}, 
\label{eq:phiphivv} \\
&\hat{v}^{+}_{ k, j } \hat{v}^{-}_{ k, j } = (-1)^{N_{l}+N_{l-1}-1}
\frac{ Z_l ( \hat{\varphi}^{(l)}_i+\epsilon)  W_{l+1}(\hat{\varphi}^{(l)}_{i}+\frac{\epsilon}{2} )
W_{l-1}(\hat{\varphi}^{(l)}_{i}+\frac{\epsilon}{2} ) }
{ W_{l, i} (\hat{\varphi}^{(l)}_{i }  +\epsilon )W_{l, i} (\hat{\varphi}^{(l)}_{i }  ) }, \\
&\hat{v}^{-}_{ k, j } \hat{v}^{+}_{ k, j } = (-1)^{N_{l}+N_{l-1}-1}
\frac{ Z_l ( \hat{\varphi}^{(l)}_i)  W_{l+1}(\hat{\varphi}^{(l)}_{i}-\frac{\epsilon}{2} )
W_{l-1}(\hat{\varphi}^{(l)}_{i}-\frac{\epsilon}{2} ) }
{ W_{l, i} (\hat{\varphi}^{(l)}_{i }  - \epsilon )W_{l, i} (\hat{\varphi}^{(l)}_{i }  ) } .
\label{eq:vmvplinear}
\end{align}
Here we defined
\begin{align}
Z_l (z) := \prod_{f=1}^{M_l} \left(z-m^{(l)}_f-\frac{\epsilon}{2}\right) , \quad W_{l} (z):=\prod_{i=1}^{N_l} (z-\hat{\varphi}^{(l)}_{i}), \quad W_{l,i} (z):=\prod_{j=1 \atop j \neq i}^{N_l} (z-\hat{\varphi}^{(l)}_{j}) .
\end{align}
We also define
\begin{align}
H_l(z)
&=  Z_l (z) \frac{  W_{l+1} (z-\frac{\epsilon}{2}) W_{l-1} (z-\frac{\epsilon}{2}) }
{W_{l}(z) W_{l}(z-\epsilon) } .
\label{eq:Hlz}
\end{align}
The coefficient of $z^{-j}$  with $j > 0$ 
in $H_l(z)$ is a symmetric polynomial of $\{ \hat{\varphi}_{i} \}_{i=1}^{N_k}$ for $k=l-1,l ,l+1$. 
This means that  the coefficient of $z^{-j}$ is described by  vevs of gauge invariant function of Coulomb branch scalar of $\Phi_{ww}$ for $U(N_k)$ for $k=l-1, l, l+1$  and mass parameters.
Then we find $E_l(z)$, $F_l(z)$, $H_l(z)$ with the relations \eqref{eq:phiphivv}-\eqref{eq:vmvplinear}
  are the same as  $(-1)^{N_{l+1}} E_l(z)$, $ (-1)^{N_{l}+1} F_l(z)$, $z^{ M_{l}+N_{l-1}+N_{l+1}-2 N_l} H_l(z)$ in \cite{Bullimore:2015lsa} which give  the quantized Coulomb branch. 
Therefore the Weyl transform of vevs of Coulomb branch operators generates the quantized Coulomb branch in \cite{Bullimore:2015lsa}.

In \cite{Braverman:2016aa} the  quantized Coulomb branches of ADE quiver gauge theories were constructed in terms of difference operators.
Let us regard $E_l(z)$ and $F_l(z)$ above as difference operators, just like $E_n[f]$ and $F_n[f]$ in Section~\ref{sec:CBjordan}.
If we define (without a hat)
\begin{align}
{\sf u }_{l,i}:=e^{ { b}^{(l)}_i }  \left( \frac{ \prod_{j=1 \atop j \neq i}^{N_l} \left(\varphi^{(l)}_{j i}  +\frac{\epsilon}{2} \right) \prod_{f=1}^{M_l} (\varphi^{(l)}_i-m^{(l)}_{ f})  
\prod_{k^{\prime}=1}^{N_{l+1}} (\varphi^{(l+1)}_{k^{\prime}}-\varphi^{(l)}_{i} )  }
{\prod_{j=1 \atop j \neq i}^{N_l} \left(\varphi^{(l)}_{i j}  +\frac{\epsilon}{2} \right) \prod_{k=1}^{N_{l-1}} (\varphi^{(l)}_i -\varphi^{(l-1)}_{k}  ) } \right)^{\frac{1}{2}}  \,,
\end{align}
we can express $E_l(z)$ and $F_l(z)$ as
\begin{align}
&E_l(z)= - \sum_{i=1}^{N_l} \frac{Z_l (\hat{\varphi}^{(l)}_i) W_{l+1} (\hat{\varphi}^{(l)}_i -\frac{\epsilon}{2}) } 
{(z-\hat{\varphi}^{(l)}_i) W_{l,i} (\hat{\varphi}^{(l)}_i ) }  \hat{\sf u}^{-1}_{l,i},
\label{eq:Eldiff} \\
&F_l(z)= \sum_{i=1}^{N_l} \frac{W_{l-1} (\hat{\varphi}^{(l)}_i +\frac{\epsilon}{2})}{(z-\hat{\varphi}^{(l)}_i-\epsilon) W_{l,i} (\hat{\varphi}^{(l)}_i )}  \hat{\sf u}_{l,i} .
\label{eq:Fldiff}
\end{align}
Here $\hat{\sf u}_{l,i}$ and $\hat{\varphi}^{(l)}_{i}$ are defined as the Weyl transforms of ${\sf u}_{l,i}$ and ${\varphi}^{(l)}_{i}$, and  satisfy the commutation relations
\begin{align}
[ \hat{\sf u}^{\pm1}_{l,i}, \widehat\varphi^{(k)}_{j}] =\pm \epsilon \delta_{l,k} \delta_{i,j} \hat{\sf u}^{\pm1}_{l,i} .
\end{align}
Then  we see that \eqref{eq:Hlz}, \eqref{eq:Eldiff} and \eqref{eq:Fldiff}  coincide with the difference operators $H_l(z), E_l(z)$ and $F_l (z)$ in Appendix~B of~\cite{Braverman:2016aa}  for an A-type linear quiver gauge theory.
It is shown in \cite{Braverman:2016aa} that the coefficients of $z^{-j}$ in $H_l(z), E_l(z)$ and $F_l (z)$  with the commutation relations give 
 the quantized Coulomb branch chiral rings of ADE quiver theories.
 The rings are isomorphic to the so-called truncated shifted Yangians.

\subsection{Operator ordering and wall-crossing}\label{quiver-wall-crossing}
We now evaluate some bubbling contributions $Z_{\text{mono}}$ from the Moyal products of minimal monopole operator vevs
and compared with the partition functions of the matrix models obtained by brane construction.
As written in \eqref{eq:Uldressmono} and \eqref{eq:vli},  
the expectation values of minimal bare monopole operators with non-zero magnetic charge of $U(N_l)$ are given by
\begin{align}
\langle {V}^{(l)}_{\pm} \rangle = 
\sum_{i=1}^{N_l} e^{ \pm { b}^{(l)}_i } \left(  \frac{ \prod_{f=1}^{M_l} (\varphi^{(l)}_i-m^{(l)}_{ f}) \prod_{n=1}^{N_{l+1}} (\varphi^{(l+1)}_n-\varphi^{(l)}_{i} ) \prod_{k=1}^{N_{l-1}} (\varphi^{(l)}_i-\varphi^{(l-1)}_{k} ) }
{\prod_{j=1 \atop j \neq i}^{N_l} \left(\varphi^{(l)}_{i j}  +\frac{\epsilon}{2} \right)\left(\varphi^{(l)}_{j i}  +\frac{\epsilon}{2} \right) } \right)^{\frac{1}{2}} .
\end{align}

\begin{figure}[tb]
\begin{center}
\includegraphics[width=4cm]{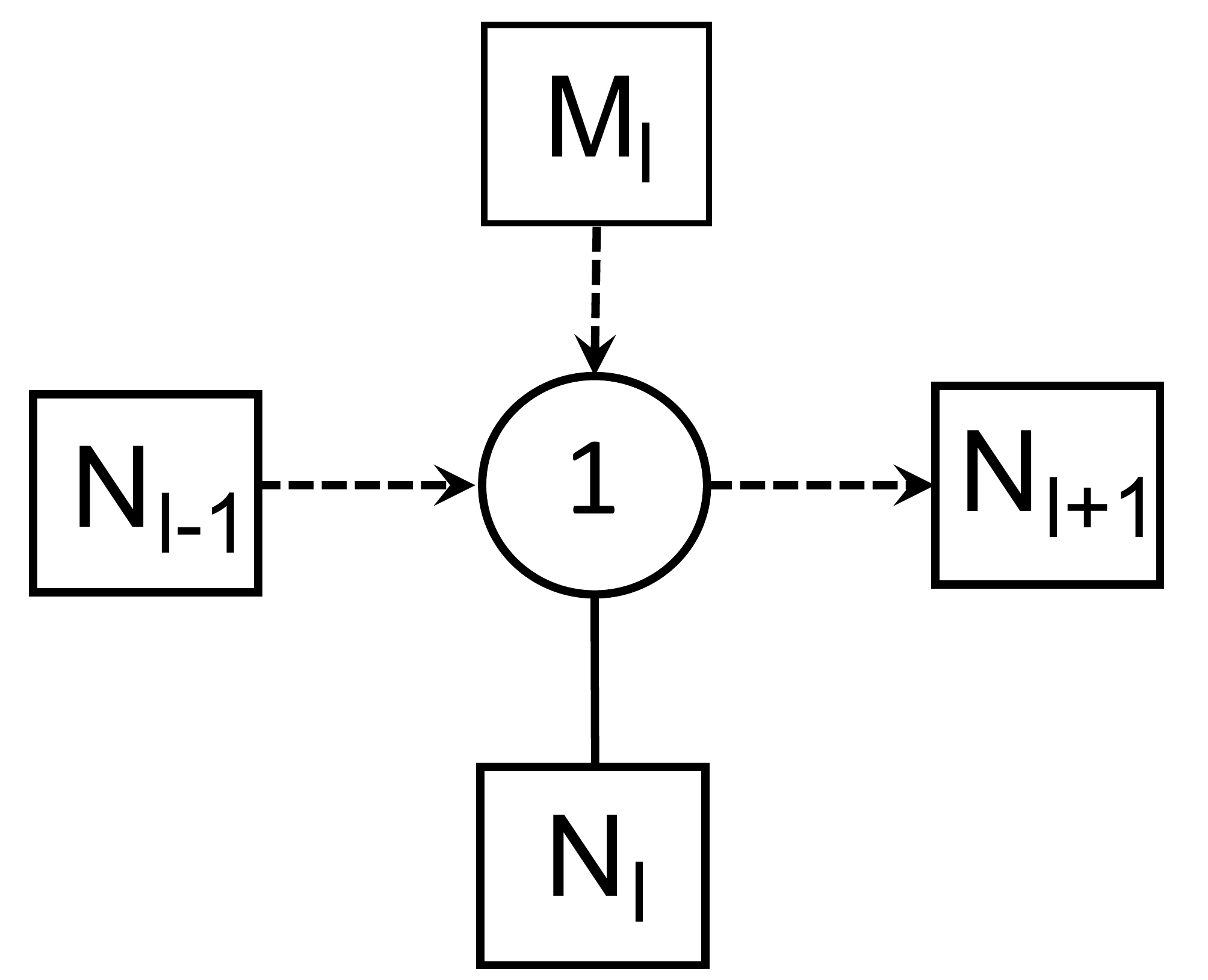}
\end{center}
\vspace{-0.5cm}
\caption{The quiver diagram represents the matter content of the matrix model for monopole bubbling with~$\mathfrak{m}={\bm 0}$ and ${\bm B}={\bm e}^{(l)}_{N_l}- {\bm e}^{(l)}_{1}$.
 }
\label{fig:quiverMMlinear}
\end{figure}

Let us consider the monopole bubbling contribution for $\mathfrak{m}=0
$ in $\langle {V}_{{\bm e}^{(l)}_{N_l}- {\bm e}^{(l)}_{1} } \rangle$. 
The functions $Z_{\text{mono}}$ read off from the Moyal product of $\langle {V}^{(l)}_{+} \rangle$ and $\langle {V}^{(l)}_{-} \rangle$  are 
\begin{align}
& Z_{\text{mono}} (\mathfrak{m}={\bm 0}) \text{ from } \langle {V}^{(l)}_{+} \rangle * \langle {V}^{(l)}_{-} \rangle   \nonumber \\
&=\sum_{i=1}^{N_l}   \frac{ \prod_{f=1}^{M_l} (\varphi^{(l)}_i-m^{(l)}_{f}+\frac{\epsilon}{2}) \prod_{n=1}^{N_{l+1}} (\varphi^{(l+1)}_n-\varphi^{(l)}_{i}-\frac{\epsilon}{2} ) \prod_{k=1}^{N_{l-1}} (\varphi^{(l)}_i-\varphi^{(l-1)}_{k}+\frac{\epsilon}{2} ) }
{\prod_{j=1 \atop j \neq i}^{N_l} \left(\varphi^{(l)}_{i j}  +\epsilon \right)\left(\varphi^{(l)}_{j i}  \right) } ,
\label{eq:zmonoquiver1}
\\
& Z_{\text{mono}} (\mathfrak{m}={\bm 0}) \text{ from } \langle {V}^{(l)}_{-} \rangle * \langle {V}^{(l)}_{+} \rangle   \nonumber \\
&=\sum_{i=1}^{N_l}   \frac{ \prod_{f=1}^{M_l} (\varphi^{(l)}_i-m^{(l)}_{f}-\frac{\epsilon}{2}) \prod_{n=1}^{N_{l+1}} (\varphi^{(l+1)}_n-\varphi^{(l)}_{i}+\frac{\epsilon}{2} ) \prod_{k=1}^{N_{l-1}} (\varphi^{(l)}_i-\varphi^{(l-1)}_{k}-\frac{\epsilon}{2} ) }
{\prod_{j=1 \atop j \neq i}^{N_l} \left(\varphi^{(l)}_{i j}  -\epsilon \right)\left(\varphi^{(l)}_{j i}  \right) } .
\label{eq:zmonoquiver2}
\end{align}
We can show that \eqref{eq:zmonoquiver1} and \eqref{eq:zmonoquiver2} coincide with the values of the partition function  of a matrix model specified by the quiver diagram in Figure~\ref{fig:quiverMMlinear} evaluated in two FI-chambers.
The matrix model formula for $Z_{\text{mono}}$ is
\begin{align}
&Z_{\text{mono}}^{(\zeta)} ( {\bm B}={\bm e}^{(l)}_{N_l}- {\bm e}^{(l)}_{1}, \mathfrak{m}={\bm 0}) \nonumber \\
&= \oint_{\mathrm{JK}(\zeta)} \frac{du}{2\pi i} 
\frac{ (-\epsilon)  \prod_{f=1}^{M_l}(u- m^{(l)}_{f})\prod_{n=1}^{N_{l+1}} (\varphi^{(l+1)}_n-u ) \prod_{k=1}^{N_{l-1}} (u-\varphi^{(l-1)}_{k} )}
{\prod_{s=\pm 1}\prod_{i=1}^{N_l} \left( s(u-\varphi^{(l)}_{ i})-\frac{\epsilon}{2} \right)} .
\label{eq:zmonoquiver3}
\end{align}
The JK residues are evaluated at  poles $u=\varphi^{(l)}_{ i}+\frac{\epsilon}{2}$ with $i=1,\cdots, N_l$ for 
$\zeta >0$ and $u=\varphi^{(l)}_{ i}-\frac{\epsilon}{2}$ with $i=1,\cdots, N_l$ for 
$\zeta <0$.
The jump in $Z_\text{mono}$ can be again obtained by evaluating the residue at $u=\infty$:
\begin{align}
&\langle  {V}^{(l)}_{+}  \rangle *  \langle {V}^{(l)}_{-}  \rangle-\langle {V}^{(l)}_{-} \rangle *  \langle {V}^{(l)}_{+}  \rangle \nonumber \\
&\qquad=\left\{
\begin{array}{cll}
0  & \text{for } \Delta_l<-1 &  \text{(bad),} \\
 (- 1)^{N_l+N_{l+1}-1} \epsilon  & \text{for }  \Delta_l=-1 &  \text{(ugly),}\\
&&\\
\hspace{-5mm}
\begin{matrix}
(- 1)^{N_l+N_{l+1}-1} \epsilon \left(2 \sum_{i=1}^{N_l} \varphi^{(l)}_i \right.
\\
\left. \qquad\qquad
- \sum_{k=l-1}^{l+1} \sum_{f=1}^{N_k} m^{(k)}_f  \right) 
\end{matrix}  
  & \text{for } \Delta_l=0 & \text{(good and balanced),}\\
&&\\
 \epsilon A^{\prime} ({\bm \varphi}, {\bm m}, \epsilon) & \text{for } \Delta_l> 0 & \text{(good but not balanced),}\\
\end{array}
\right.
\label{eq:gublinear}
\end{align}
with $\Delta_l:=M_l +N_{l-1}+N_{l+1} -2N_l$ and 
\begin{align}
{A}^{\prime} ({\bm \varphi}, {\bm m}, \epsilon)&:= { \frac{1}{(\Delta_l+1)!} \left( \frac{d}{ d w } \right)^{ \Delta_l+1} }   \nonumber \\
&\times
\frac{  \prod_{f=1}^{M_l}(1- m^{(l)}_{f} w)\prod_{n=1}^{N_{l+1}} (w \varphi^{(l+1)}_n-1 ) \prod_{k=1}^{N_{l-1}} (1-\varphi^{(l-1)}_{k} w )}
{\prod_{s=\pm 1}\prod_{i=1}^{N_l} \left( s(1-\varphi^{(l)}_{ i} w)-\frac{\epsilon}{2} w \right)} 
\Bigg |_{w=0} .
\end{align}
Again we find that the behaviors in wall-crossing are related to the division into the categories ``bad'', ``ugly'', ``good'', and ``balanced'' in~\cite{Gaiotto:2008ak}.

\begin{figure}[htb]
\centering
\subfigure[]{\label{subfig:monoquiver1}
\includegraphics[width=6cm]{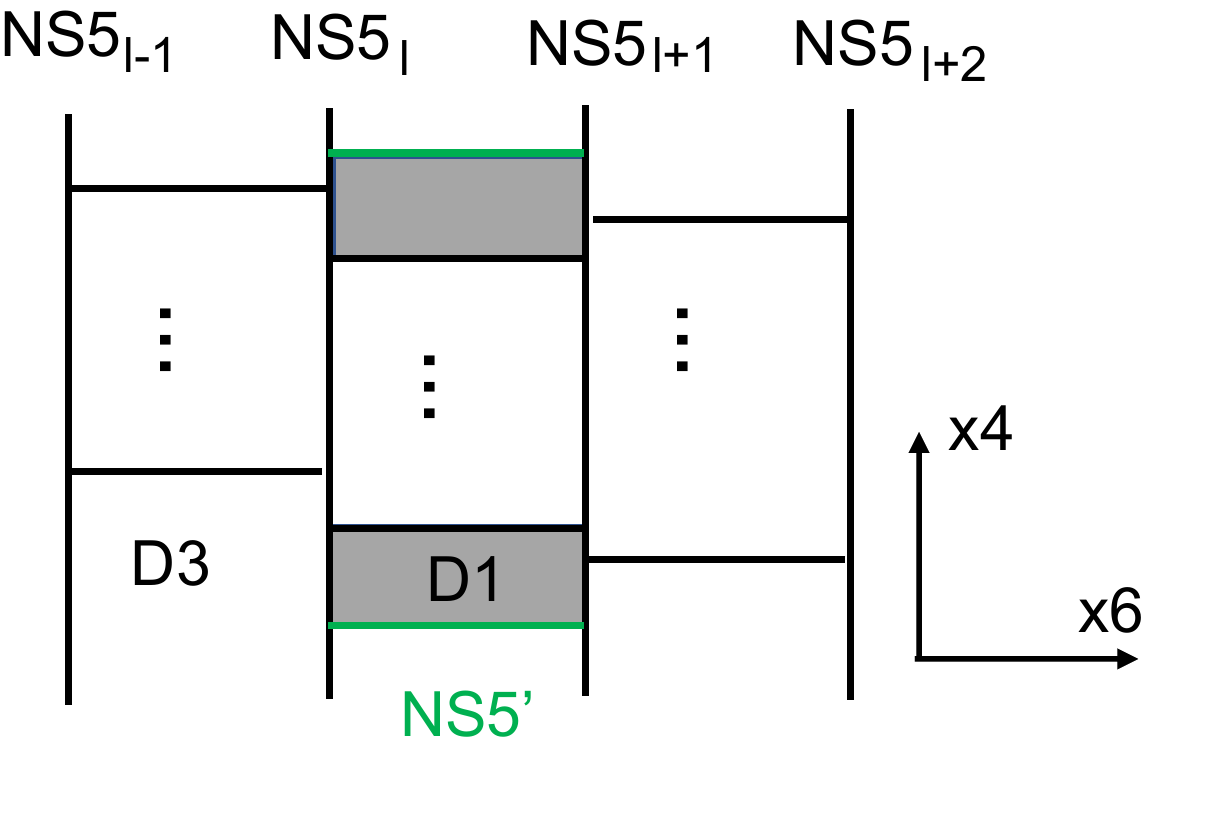}}
\hspace{0cm}
\subfigure[]{\label{subfig:monoquiver2}
\includegraphics[width=5.5cm]{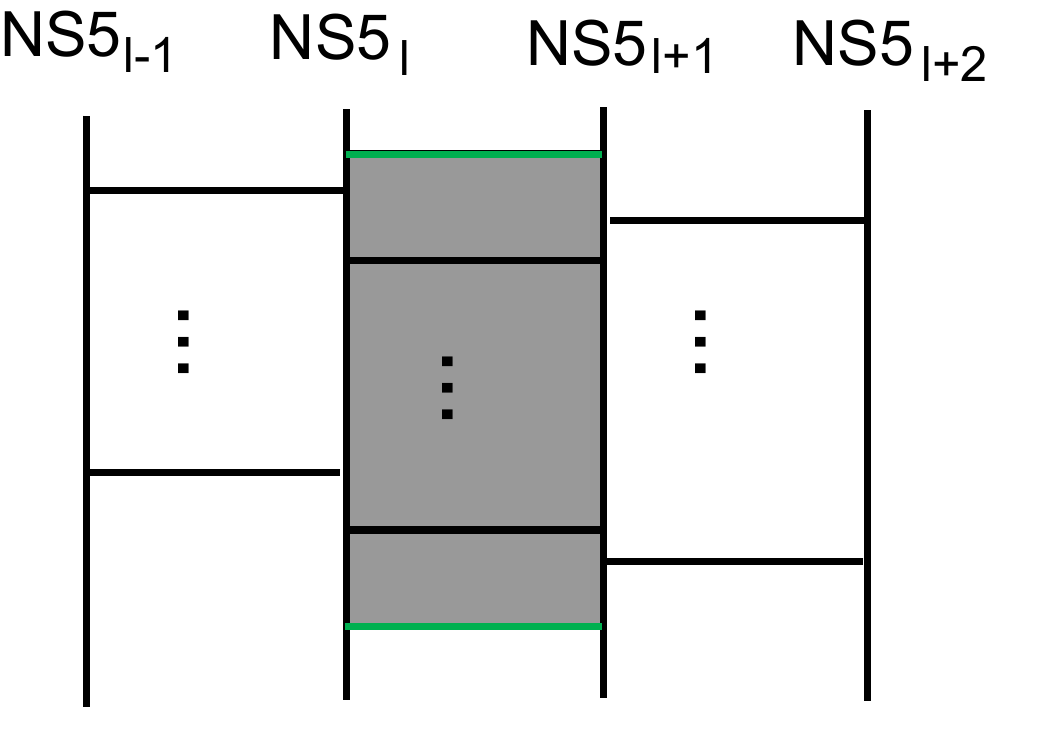}}
\caption{(a): The brane configuration for monopole operator with a magnetic charge ${\bm B}={\bm e}^{(l)}_{N_l} -{\bm e}^{(l)}_{1}$. 
The vertical black lines express NS5-branes. The $N_l$ horizontal black lines between the $l$-th NS5-brane (denoted as NS5${}_{l}$) and the $(l+1)$-th 
(denoted as NS5${}_{l+1}$) gives a $U(N_l)$ 3d $\mathcal{N}=4$ vector multiplet. The horizontal breen lines express NS5'-brane. The gray regions express D1-branes.
The upper D1-brane between a D3-brane and a NS5'-brane gives monopole charge ${\bm e}^{(l)}_{N_l}$ 
and the lower D1-brane $-{\bm e}^{(l)}_{1}$
(b): The brane configuration for monopole bubbling effect $\mathfrak{m}={\bm 0}$ in ${\bm B}={\bm e}^{(l)}_{N_l} -{\bm e}^{(l)}_{1}$.
The D1-brane end only on NS5-branes and NS5'-branes.
 }
\label{fig:linearquiverbubb1}
\end{figure}

The matrix model specified by the quiver diagram in Figure~\ref{fig:quiverMMlinear} indeed arises from a brane construction explained in Section~\ref{sec:brane-quiver}.  
The monopole operator with~${\bm B}={\bm e}^{(l)}_{N_l}- {\bm e}^{(l)}_{1}$, or more precisely the product of monopole operators with charges~${\bm e}^{(l)}_{N_l}$ and $- {\bm e}^{(l)}_{1}$, is realized by the brane configuration in Figure~\ref{fig:linearquiverbubb1}(a).
Monopole bubbling with $\mathfrak{m}={\bm 0}$ and ${\bm B}={\bm e}^{(l)}_{N_l}- {\bm e}^{(l)}_{1}$ is realized by the configuration inFigure~\ref{fig:linearquiverbubb1}(b).
Compared with the theory with a single $U(N)$ factor in the gauge group, there appears an extra short Fermi multiplet from an open string ending on the a D1-brane and a D3-brane separated by an NS5-brane (not an NS5'-brane).   
We find that $N_{l+1}$ Fermi multiplets  arise from open strings ending on a D1-brane and $N_{l+1}$ D3-branes  separated by the NS5${}_{l+1}$-brane and $N_{l-1}$ Fermi multiplets  arise from  open strings ending on  a D1-brane and $N_{l-1}$ D3-branes  separated by  the NS5${}_{l+1}$-brane. 
The partition function of the matrix model obtained by brane construction is~\eqref{eq:zmonoquiver3}, which reproduces the $Z_\text{mono}$'s obtained from the Moyal products.

\section{Discussion}\label{sec:discussion}
\label{sec:disc-concl}


Let us discuss several issues and future directions.

For bare monopole operators in 3d $\mathcal{N}=4$ $U(N)$ gauge theories, 
we evaluated monopole bubbling contributions by two independent methods, {\it i.e.}, Moyal products and matrix models, and we have found the perfect agreement between them. 
The Moyal product method is applicable to other gauge groups and also to dressed monopole operators. 
On the other hand, matrix models were read off from brane constructions, and are so far limited to bare monopole operators in $U(N)$ gauge theories. 
It is desirable to extend brane construction to other gauge groups and also to dressed monopole operators and check agreement with Moyal products.
It is indeed possible to extend the matrix models for $Z_\text{mono}$ from $U(N)$ to $SO(N)$ and $USp(N)$ gauge groups by brane construction involving orientifolds.

For $SU(N)$ gauge theories, even in the presence of fundamental flavors, it is known that results for such theories can be obtained from the results for $U(N)$ gauge theories by gauging the topological symmetry usually called $U(1)_J$.
It is the symmetry whose current is the Hodge dual of ${\rm Tr}\, F$.
Gauging $U(1)_J$ amounts to the operation to introduce a BF coupling $\frac{1}{2\pi}\int B\wedge {\rm Tr}\,F$ (or more precisely, its $\mathcal{N}=4$ supersymmetric extension~\cite{Brooks:1994nn}) and integrate out the gauge field~$B$.
This restricts $F= dA -i A\wedge A$ to be traceless, and reduces the gauge group from~$U(N)$ to~$SU(N)$.
In the context of the Coulomb branch, this operation was performed for $N=2$ in~\cite{Dey:2017fqs,Assel:2018exy}.

In 4d the localization results for 't Hooft operators in $U(N)$ theories reduce, by imposing the tracelessness condition, to those in $SU(N)$ theories if the matter content  only consists of an adjoint hypermultiplet, but not when it includes fundamental hypermultiplets.
It was noticed in~\cite{Ito:2011ea} that the localization computations for 't Hooft operators in the $U(N)$ SQCD does not coincide, upon imposing the tracelessness condition, with a prediction of the AGT correspondence~\cite{Alday:2009fs,Drukker:2009id,Passerini:2010pr,Gomis:2010kv} for the~$SU(N)$ SQCD.
More work was required to compute 't Hooft operator vevs by localization in the $SU(N)$ SQCD~\cite{Brennan:2018rcn,Assel:2019iae}.
It is unclear whether the BF coupling method above, using the 4d $\mathcal{N}=2$ version of such a coupling~\cite{deWit:1982na}, allows one to obtain results for the $SU(N)$ SQCD from those of the $U(N)$ theory.
There are also other operations that relate gauge groups~$U(N)$, $SU(N)$, and $SU(N)/\mathbb{Z}_N$~\cite{Kapustin:2014gua} by gauging an electric $U(1)$ one-form symmetry~\cite{Gaiotto:2014kfa}, but they break down in the presence of fundamental matter.

We pointed out that Wilson--'t Hooft line operators in 4d  $\mathcal{N}=2^*$ $U(N)$ gauge theory  generate the spherical DAHA of $\mathfrak{gl}(N)$ which coincides with the K-theoretic Coulomb branch of the quiver with a loop.
Our observation for the algebra of line operators for  $G=U(N)$ $\mathcal{N}=2^*$ super Yang-Mills theory implies that  the algebra of line operators for 
4d $\mathcal{N}=2$ gauge theories correspond to the quantized K-theoretic Coulomb branch chiral rings. 
 In \cite{Braverman2016ef, Finkelberg:2017qpy} quantized K-theoretic Coulomb branch of (quiver) gauge theories are studied and it was shown that quantized K-theoretic Coulomb branches
 possess nice algebraic structures; shifted quantum affine algebras and cousins of spherical DAHAs.  
It would be interesting to give the precise identification between the algebra of line operators and the quantized K-theoretic Coulomb branch chiral ring for other  gauge theories.  
The mathematical approaches to 3d and 4d Coulomb branches amount to considering 1d and 2d ($S^1\times \mathbb{R}$) sigma models with an infinite dimensional target space, the total space of a vector bundle over the affine Grassmannian.
Coulomb branch operators in 3d and line operators in 4d can be viewed as domain walls.
It may be beneficial to combine the analysis in this paper with the study of domain walls in~\cite{Nekrasov:2009uh,Nekrasov:2009ui,Honda:2013uca}.

In~\cite{Dedushenko:2018icp}, a method was developed to determine monopole bubbling contributions up to $1/r$-dependent operator mixing, by the requirement that the product of generators of the chiral ring is a polynomial in the generators.  Here $r$ is the radius of $S^3$ and plays the role of an $\Omega$-deformation parameter.  Their analysis was done on $S^3$, but it seems straightforward and useful to adapt their method to $\mathbb{R}^3$ with $\Omega$-deformation.
It would also be interesting to incorporate boundaries~\cite{Bullimore:2016nji,Bullimore:2016hdc} and 3d line operators~\cite{Dimofte:2019zzj} into the explicit SUSY localization framework of this paper.

\section*{Acknowledgements}
We thank H.~Hayashi for collaboration on related projects.
We are also grateful to H.~Nakajima and Y.~Tachikawa for helpful discussions.
The work of T.O is supported in part by JSPS KAKENHI Grant Number JP16K05312.
The work of Y.Y. is supported in part by JSPS KAKENHI Grant Number JP16H06335 and also by World Premier International Research Center Initiative (WPI), MEXT Japan.

\appendix

\section{Jeffrey-Kirwan prescription}\label{app:JK}\


As explained in Section~\ref{sec:SQCD-matrix}, in this paper we use the Jeffrey-Kirwan (JK) residue prescription to compute the matrix model partition functions.
In this appendix (similar to one in~\cite{Hayashi:2019rpw}) we give a summary of the JK residue prescription that we apply to non-degenerate and degenerate poles.

The partition functions of interest take the form
\begin{equation} \label{eq:monosc-app}
Z
= 
\frac{1}{|W_{G^{\prime}}|}
\oint_{\mathrm{JK}({ {\bm \zeta}})} 
\prod_{{ a}=1}^{\mathrm{rank}(G^{\prime})} \frac{d u_{a}}{2\pi i} \prod Z^{\text{0d}}_{\mathrm{vec}} \prod Z^{\text{0d}}_{\mathrm{hyper}}
 \prod Z^{\text{0d}}_{\mathrm{Fermi}} ,
 \end{equation}
where $|W_{G^{\prime}}|$ is the order of the Weyl group $W_{G'}$ of the matrix model gauge group $G^{\prime}$ { and the {\it JK parameter} $\bm{\eta}$ will be determined by the FI parameter $\bm{\zeta}$ of the matrix model}.
We denote by $Z^{\text{0d}}_{\mathrm{vec}} $, $Z^{\text{0d}}_{\mathrm{hyper}}$, and $Z^{\text{0d}}_{\mathrm{Fermi}}$ the one-loop determinants of the 0d $\mathcal{N}=(0,4)$ vector, hyper, and short Fermi multiplets.  Sometimes the integrand is constructed from contributions of $\mathcal{N}=(4,4)$ vector and hypermultiplets. 
Such $\mathcal{N}=(4,4)$ multiplets are decomposed into $\mathcal{N}=(0,4)$ multiplets as follows.  
 An $\mathcal{N}=(4,4)$ vector multiplet is decomposed into a $\mathcal{N}=(0,4)$ vector multiplet and an adjoint $\mathcal{N}=(0,4)$  twisted hypermultiplet. 
 An $\mathcal{N}=(4,4)$ hypermultiplet is decomposed into an $\mathcal{N}=(0,4)$ hypermultiplet and two $\mathcal{N}=(0,4)$ short Fermi multiplets.

In this paper we are only interested in the case that $G^{\prime}$ is a product of unitary groups, for which the one-loop determinants are given in~(\ref{one-loop-0d-vec})-(\ref{one-loop-0d-Fermi}).

Let $\mathfrak{h}$ be the Cartan subalgebra of the Lie algebra of $G^{\prime}$, and $\mathfrak{h}^*$ its dual.
We define the {\it charge vectors} $\bm{Q}_i\in\mathfrak{h}^*$ to be the weights of the gauge group $G^{\prime}$ in the representations carried by the $\mathcal{N}=(0,2)$ chiral multiplets.
The charge vectors appear as $\prod Z^{\text{0d}}_{\mathrm{hyper}} = \prod_i (\bm{Q}_i(\bm u)-c_i)^{-1}$ in~(\ref{eq:monosc-app}).
The quantity $c_i$ is a linear combination of $\epsilon$ and $\varphi_j$, $\bm{u}\in\mathfrak{h}$ collectively denotes all $u_j$'s, and $\bm{Q}_i(\bm u)$ denotes the pairing of $\bm{Q}_i$ and $\bm{u}$.
The locus $\bm{Q}_i(\bm{u})-c_i=0$ in the $\bm{u}$-space for each value of $i$ is called a {\it singular hyperplane}.
Let us suppose that exactly $l$ hyperplanes given (after a $\bm{u}_*$-dependent relabeling of the $i$'s) as $\bm{Q}_1(\bm{u}-\bm{u}_*)=0, \ldots, \bm{Q}_l(\bm{u}-\bm{u}_*)=0$ intersect at a point $\bm{u}_*$.
Such $l$ has to satisfy the inequality $l\geq n :=\dim \mathfrak{h}$.
We set $\bm{Q}_*:=\{\bm{Q}_1,\ldots,\bm{Q}_l\}$, keeping in mind that $\bm{Q}_*$ depends on~$\bm{u}_*$.
We assume that $\bm{Q}_*$ is {\it projective}; this means that all the elements of $\bm{Q}_*$ are contained in a half space of $\mathfrak{h}^*$.
The poles for which we need to evaluate the residues in this paper all satisfy the projectivity condition.
We will distinguish between the case~$l=n$ and the case $l>n$.
In the former (resp. latter) case the intersection point $\bm{u}_*$ is called a {\it non-degenerate} (resp. {\it degenerate}) pole.

At a non-degenerate pole~$\bm{u}_*$ the JK residue is defined as
\begin{equation}\label{JK-def-non-degenerate}
\begin{aligned}
&\mathop{\text{JK-Res}}_{\bm u = \bm{u}_*} (\bm{Q}_*,\bm{\eta}) \frac{du_1\wedge\cdots\wedge du_n}{\bm{Q}_1(\bm{u}-\bm{u}_*)\cdots \bm{Q}_n(\bm{u}-\bm{u}_*)}
\\
& \qquad \quad
=
\left\{
\begin{array}{cl}
|\det(\bm{Q}_1,\ldots,\bm{Q}_n)|^{-1} & \text{if } \bm{\eta}\in \text{Cone}[\bm{Q}_1,\ldots,\bm{Q}_n], \\
0 & \text{otherwise,}
\end{array}
\right. 
\end{aligned}
\end{equation}
where  $\text{Cone}[{\bm Q}_1, \ldots, {\bm Q}_n]$ is defined as $ \{\sum_{i=1}^n y_i{\bm Q}_i | y_i > 0 \text{ for } i=1, \ldots, n\}$.

At a degenerate pole, we use the constructive ``definition''~\cite{Szenes-Vergne} of the JK residue that we review following~\cite{Benini:2013xpa}.
Let us set
\begin{align}
{\Sigma {\bm Q}_{\ast}  } := \Bigl\{ \sum_{i \in \pi } {\bm Q}_{i} \Big| \pi \subset \{1, \ldots, l \}  \Bigr\}
\end{align}
and define ${\rm Cone}_{\rm sing} [\Sigma {\bm Q}_{\ast}]$ to be the union of all the cones  spanned by $(n-1)$ elements of $\Sigma {\bm Q}_{\ast}$. 
We assume that $\bm{\eta}\in \mathfrak{h}^*$ satisfies the {\it strong regularity condition}
\begin{equation}\label{equation:strongref}
\bm{\eta}\notin \text{Cone}_\text{sing}[\Sigma\bm{Q}_*].
\end{equation}
Let  ${\cal F L} ({\bm Q}_{\ast})$ be the set of flags  
\begin{equation}
F=[\{0 \}=F_{0} \subset F_1 \subset \cdots \subset F_{n} = \mathfrak{h}^{\ast}], \quad \dim F_i = i 
\end{equation}
such that ${\bm Q}_{\ast}$ contains a basis of $F_i$ for $i=1,2, \ldots, n$.
We let $\mathfrak{B}(F)=\{ {\bm Q}_{j_1}, \cdots, {\bm Q}_{j_n} \}$
be the ordered set whose first $i$ elements  form a basis of $F_i$ for $i=1, \ldots, n$.
For each flag $F \in {\cal F L} ({\bm Q}_{\ast})$, the iterated residue ${\rm Res}_F$ of an $n$-form $ \omega$ is defined by 
\begin{align}
{\rm Res}_{F} \, \omega= \oint_{\tilde{u}_{j_n}=0} \frac{d \tilde{u}_{j_n}} {2 \pi i}  \cdots  \oint_{\tilde{u}_{j_1}=0}  \frac{d \tilde{u}_{j_1}}{2 \pi i}\tilde{\omega}_{j_1 \cdots j_n } , 
\end{align}
where $\tilde{u}_i= {\bm Q}_i ({\bm u}-{\bm u}_{\ast})$ and $\omega = \tilde{\omega}_{j_1 \cdots j_n } d \tilde{u}_{j_1} \wedge \cdots \wedge d \tilde{u}_{j_n}$.
For each flag $F \in {\cal F L} ({\bm Q}_{\ast})$, let us introduce the vectors
\begin{align}
 {\bm \kappa}^{F}_i=\sum_{k=1}^i  {\bm Q_{j_k}}
\end{align}
and the closed cone
\begin{align}
\mathfrak{s}^+ (F, {\bm Q}_{\ast}) := \sum_{i=1}^n \mathbb{R}_{\ge 0} {\bm \kappa}^{F}_{i}.
\end{align}
We then define
\begin{align}
{\cal F L}^+ ({\bm Q}_{\ast}, {\bm \eta}) :=  \{ F\in {\cal F L} ({\bm Q}_{\ast})| {\bm \eta} \in \mathfrak{s}^+ (F, {\bm Q}_{\ast})\}.
\end{align}
Finally, the JK residue at the pole ${\bm u}={\bm u}_*$ is defined by
\begin{align}
\mathop{{\rm JK} \mathchar `- {\rm Res} }_{{\bm u}={\bm u}_*} ({\bm Q}_{\ast}, {\bm \eta} )
 =\sum_{F \in {\cal F L}^+ ({\bm Q}_{\ast},{\bm \eta}) } \nu (F) {\rm Res}_F  ,
 \label{eq:consJKres}
\end{align}
where $\nu(F)= {\rm sign} \det ({\bm \kappa}^{F}_{1},\cdots {\bm \kappa}^{F}_{n})$ with ``sign'' defined so that ${\rm sign}\, x$ is $+1$ for $x>0$, 0 for $x=0$, and $-1$ for $x<0$.

For gauge group $U(n)$ the FI parameter $\zeta$ determines an element  $\widetilde{\bm{\zeta}}=\zeta \sum_{i=1}^n \bm{e}_i 
\in \mathbb{R}^n \simeq \mathfrak{h}^{\ast}$~\cite{Hori:2014tda}.
More generally for a product gauge group $U(n_1)\times U(n_2)\times\ldots\times U(n_L)$ the FI parameters $\bm{\zeta}=(\zeta_1,\zeta_2,\ldots,\zeta_L)$ determine an element $\widetilde{\bm \zeta}:=\sum_{a=1}^L \zeta_a 
( \bm{e}_{\tilde{n}_{a-1}+1} + \ldots +\bm{e}_{\tilde{n}_a} )
\in\mathfrak{h}^*$, 
where 
$\{\bm{e}_{\tilde{n}_{a-1}+1} ,\ldots,\bm{e}_{\tilde{n}_a}\}$
is an orthonormal basis of $\mathfrak{h}^*_{U(n_a)}\simeq \mathbb{R}^{n_a}$, $\tilde{n}_a = n_1+\ldots+n_{a}$, and $\tilde{n}_0=0$.

We propose that, when all the zero-dimensional intersections of the singular hyperplanes are non-degenerate, the matrix model partition is given as
\begin{align}
Z
&= \frac{1}{|W_{G^{\prime} }|}\oint_{{\rm JK}({\bm \zeta})}  \prod_{{ a}=1}^{\mathrm{rank}(G^{\prime}) }  \frac{d u_{a}}{2\pi i}  \prod Z_{\text{vec}}^\text{0d} \prod Z_{\text{hyp}}^\text{0d} \prod Z_{\text{Fermi}}^\text{0d} \label{partitionfunction} \\ 
&:=\frac{1}{|W_{G^{\prime}}|} \sum_{{\bm u}_* } \mathop{{\rm JK} \mathchar `- {\rm Res} }_{{\bm u}={\bm u}_*} ({\bm Q}_{\ast}, {\bm \eta} = { \widetilde{\bm \zeta}} ) 
 \prod Z_{\text{vec}}^\text{0d} \prod Z_{\text{hyp}}^\text{0d} \prod Z_{\text{Fermi}}^\text{0d} \,
 d u_1 \wedge \cdots \wedge  d u_{\mathrm{dim} \;\mathfrak{h} },   \nn
\end{align}
where the JK-residues are computed according to~(\ref{JK-def-non-degenerate}).

When some of the zero-dimensional intersections are degenerate and when they satisfy the strong regularity condition~(\ref{equation:strongref}), we propose that the matrix model partition function can be calculated by applying the constructive definition~(\ref{eq:consJKres}) to the degenerate poles.
We take the summation in  \eqref{partitionfunction} over all the degenerate poles ${\bm u}_*$ with ${\bm \eta} \in \mathfrak{s}^+ (F, {\bm Q}_{\ast} (\bm{u}_*))$ {for some $F\in {\cal F L}^+ ({\bm Q}_{\ast} (\bm{u}_*),{\bm \eta})$
and also over all the non-degenerate poles ${\bm u}_*$ with 
 ${\bm \eta} \in \text{Cone}[{\bm Q}_1, \cdots, {\bm Q}_n]$.

When some of the zero-dimensional intersections are degenerate and when some of them violate the strong regularity condition~(\ref{equation:strongref}), we compute $Z$ for~${\bm \zeta}$ in the interior of an FI-chamber \eqref{eq:FIchamber} as follows.
We use almost the same formula~(\ref{partitionfunction}) and apply the constructive definition~(\ref{eq:consJKres}) to the degenerate poles and sum the JK-residues as in the previous paragraph, but at a degenerate pole that violates the strong regularity condition~(\ref{equation:strongref}), we use as $\bm{\eta}$ not $\widetilde{\bm \zeta}$ itself but a vector $\widetilde{\bm\zeta}'$ that is obtained by infinitesimally shifting~$\widetilde{\bm \zeta}$ and that satisfies the strong regularity condition.

We use the terminology
\begin{align}
\text{{\it JK-chamber} } := \text{ a connected component of }\mathfrak{h}^{\ast} \backslash  \mathrm{Cone}_{\text{sing}} [\cup_{{\bm u}_*} {\bm Q}_{*}] \label{def-JK-chamber}
\end{align}
to distinguish it from an FI-chamber defined in~(\ref{eq:FIchamber}).
Here
$\cup_{{\bm u}_*} {\bm Q}_*$ is the union of the~${\bm Q}_*{ (\bm{u}_*)}$'s for all  the poles ${\bm u}_*$  
and 
$\mathrm{Cone}_{\text{sing}} [\cup_{{\bm u}_*} {\bm Q}_{*}] $  is the union of all the cones generated by subsets of $\cup_{{\bm u}_*} {\bm Q}_*$ with $n-1$ elements.
$\mathrm{Cone}_{\text{sing}} [\Sigma {\bm Q}_*]$ divides a JK-chamber into  subchambers.
The prescription above to shift $\widetilde{\bm \zeta}$ to $\widetilde{\bm \zeta}'$ is motivated by the fact that the expression
$$
\frac{1}{|W_{G^{\prime}}|} \sum_{{\bm u}_* } \mathop{{\rm JK} \mathchar `- {\rm Res} }_{{\bm u}={\bm u}_*} ({\bm Q}_{\ast}, {\bm \eta} ) 
\prod Z_{\text{vec}}^\text{0d} \prod Z_{\text{hyp}}^\text{0d} \prod Z_{\text{Fermi}}^\text{0d} \,
 d u_1 \wedge \cdots \wedge  d u_{\mathrm{dim} \;\mathfrak{h} }
 $$ 
 as a function of $\bm{\eta}$ is constant as long as~$\bm{\eta}$ stays within the same JK-chamber~\cite{Szenes-Vergne,Benini:2013xpa}.

The matches between the matrix model computations and the Moyal products in the main text are the evidence for the validity of our proposal and prescription.

\section{Derivation of~\eqref{eq:WWtrans1}}\label{app:QACB}\


In this appendix we explain the derivation of~\eqref{eq:WWtrans1}.
Equation~(\ref{eq:relationQab}), which is what we really need, can be derived in exactly the same way.

Let us consider the first line on the right hand side of~\eqref{eq:WWtrans1}, namely the case satisfying~$( {\rm w}_i \cdot {\bm A} )( {\rm w}_i \cdot {\bm B} ) <0 $ and $ |{\rm w}_i \cdot {\bm A} | \le |{\rm w}_i \cdot {\bm B} |$.  It is divided into the two subcases
\begin{align}
& 0 < - ({\rm w}_i \cdot {\bm A}) \le  {\rm w}_i \cdot {\bm B}
\label{eq:cond1} \,,\\  
&\qquad\qquad
\text{ or } \nonumber \\
& 0 <  {\rm w}_i \cdot {\bm A}  \le -({\rm w}_i \cdot {\bm B})  \,.
\label{eq:cond2}
\end{align}
The function $ {\sf F}_a ( {\rm w}_i, {\bm m}; {\bm A}, {\bm B}) $ defined in~(\ref{eq:moyalabmonopole}) can be expressed as
\begin{equation}
\begin{aligned}
  {\sf F}_a ( {\rm w}_i, {\bm m}; {\bm A}, {\bm B})  &=
\left\{
\begin{array}{cc}
{\displaystyle \prod_{l= \frac{{\rm w}_i \cdot ( {\bm A} + {\bm B})}{2}}^{\frac{{\rm w}_i \cdot (  {\bm B}- {\bm A} )}{2}-1} }
 \left( {\rm w}_i \cdot {\bm \varphi}-l \epsilon-\frac{a}{2} \epsilon \right)
 & \text{for }  0 < - ({\rm w}_i \cdot {\bm A}) \le  {\rm w}_i \cdot {\bm B} \,,\\
 &\\
{\displaystyle \prod_{l= \frac{{\rm w}_i \cdot (  {\bm B}-{\bm A} )}{2}}^{\frac{{\rm w}_i \cdot ( {\bm A} + {\bm B} )}{2}-1} }
\left( {\rm w}_i \cdot {\bm \varphi}-l \epsilon-\frac{a}{2} \epsilon \right)
 & \text{for } 0 <  {\rm w}_i \cdot {\bm A}  \le -({\rm w}_i \cdot {\bm B}) \,.
\end{array}
\right.
\end{aligned}
\end{equation}
In case~\eqref{eq:cond1}, the left hand side of \eqref{eq:WWtrans1} becomes
\begin{equation}
\begin{aligned}
& \text{Weyl transform of} \, \, e^{  ( {\bm A}+ {\bm B} ) \cdot {\bm b} }  
\prod_{l= \frac{{\rm w}_i \cdot ( {\bm A} + {\bm B})}{2}}^{\frac{{\rm w}_i \cdot (  {\bm B}- {\bm A} )}{2}-1} 
 \left( {\rm w}_i \cdot {\bm \varphi}-l \epsilon-\frac{a}{2} \epsilon \right)  \\
=&  \prod_{l= \frac{{\rm w}_i \cdot ( {\bm A} + {\bm B})}{2}}^{\frac{{\rm w}_i \cdot (  {\bm B}- {\bm A} )}{2}-1} 
 \left( {\rm w}_i \cdot \widehat{\bm \varphi}+ \left(  \frac{{\rm w}_i \cdot ( {\bm A} + {\bm B})}{2} -l -\frac{a}{2} \right) \epsilon  \right) e^{  ( {\bm A}+ {\bm B} ) \cdot \hat{\bm b} }  \\
=&\left[{\rm w}_i \cdot \widehat{\bm \varphi}-\frac{a}{2} \epsilon \right]^{-({\rm w}_i \cdot {\bm A})}
e^{  ( {\bm A}+ {\bm B} ) \cdot \hat{\bm b} } \,.
\label{eq:case11_1}
\end{aligned}
\end{equation}
We note that the special notation~$[x]^a$ is defined in~\eqref{eq:sqbracket}.

On the other hand, in case~\eqref{eq:cond2}, the left hand side of \eqref{eq:WWtrans1} can be written as
\begin{equation}
\begin{aligned}
& \text{Weyl transform of} \, \, e^{  ( {\bm A}+ {\bm B} ) \cdot {\bm b} }  
\prod_{l= \frac{{\rm w}_i \cdot (  {\bm B}-{\bm A} )}{2}}^{\frac{{\rm w}_i \cdot ( {\bm A} + {\bm B} )}{2}-1}  \left( {\rm w}_i \cdot {\bm \varphi}-l \epsilon-\frac{a}{2} \epsilon \right)  \\
&=  \prod_{l= \frac{{\rm w}_i \cdot ( {\bm B}- {\bm A} )}{2}}^{\frac{{\rm w}_i \cdot (  {\bm A}+ {\bm B} )}{2}-1} 
 \left( {\rm w}_i \cdot \widehat{\bm \varphi}+ \left(  \frac{{\rm w}_i \cdot ( {\bm A} + {\bm B})}{2} -l -\frac{a}{2} \right) \epsilon  \right) e^{  ( {\bm A}+ {\bm B} ) \cdot \hat{\bm b} }  \\
&=\left[{\rm w}_i \cdot \widehat{\bm \varphi}-\frac{a}{2} \epsilon \right]^{-({\rm w}_i \cdot {\bm A})} e^{  ( {\bm A}+ {\bm B} ) \cdot \hat{\bm b} }
\label{eq:case12_1}
\end{aligned}
\end{equation}
From \eqref{eq:case11_1} and \eqref{eq:case12_1}, we obtain the first line on the right hand side of~\eqref{eq:WWtrans1}.

Next, let us consider the second line on the right hand side of~\eqref{eq:WWtrans1}, {\it i.e.},  the case when the conditions~$ ( {\rm w}_i \cdot {\bm A} )( {\rm w}_i \cdot {\bm B} ) <0 $ and $ |{\rm w}_i \cdot {\bm A} | > |{\rm w}_i \cdot {\bm B} |$ are satisfied.
This case is divided into two subcases
\begin{align}
& 0 < -( {\rm w}_i \cdot {\bm B})<  {\rm w}_i \cdot {\bm A}
\label{eq:cond3}  \,,\\  
&\qquad\qquad 
\text{ or } \nonumber
\\
& 0 <  {\rm w}_i \cdot {\bm B}  < -({\rm w}_i \cdot {\bm A})  \,.
\label{eq:cond4}
\end{align}
Function~$ {\sf F}_a ( {\rm w}_i, {\bm m}; {\bm A}, {\bm B}) $ becomes
\begin{equation}
\begin{aligned}
  {\sf F}_a ( {\rm w}_i, {\bm m}; {\bm A}, {\bm B})  &=
\left\{
\begin{array}{cc}
{\displaystyle \prod_{l= \frac{{\rm w}_i \cdot ( {\bm B}- {\bm A} )}{2}}^{-\frac{{\rm w}_i \cdot (  {\bm A}+ {\bm B} )}{2}-1} }
 \left( {\rm w}_i \cdot {\bm \varphi}-l \epsilon-\frac{a}{2} \epsilon \right)
 & \text{for }  0 < - ({\rm w}_i \cdot {\bm B}) <  {\rm w}_i \cdot {\bm A} \,,\\
 &\\
{\displaystyle \prod_{l= -\frac{{\rm w}_i \cdot (  {\bm A}+{\bm B} )}{2}}^{\frac{{\rm w}_i \cdot (   {\bm B}-{\bm A} )}{2}-1} }
\left( {\rm w}_i \cdot {\bm \varphi}-l \epsilon-\frac{a}{2} \epsilon \right)
 & \text{for } 0 <  {\rm w}_i \cdot {\bm B}  < -({\rm w}_i \cdot {\bm A}) \,.
\end{array}
\right.
\end{aligned}
\end{equation}
With the condition \eqref{eq:cond3}, the left hand side of \eqref{eq:WWtrans1} can be written as
\begin{equation}
\begin{aligned}
& \text{Weyl transform of} \, \, e^{  ( {\bm A}+ {\bm B} ) \cdot {\bm b} }  
\prod_{l= \frac{{\rm w}_i \cdot ( {\bm B}- {\bm A} )}{2}}^{-\frac{{\rm w}_i \cdot (  {\bm A}+ {\bm B} )}{2}-1}  \left( {\rm w}_i \cdot {\bm \varphi}-l \epsilon-\frac{a}{2} \epsilon \right) \nonumber \\
&= 
e^{  ( {\bm A}+ {\bm B} ) \cdot \hat{\bm b} } \prod_{l= \frac{{\rm w}_i \cdot ( {\bm B}- {\bm A} )}{2}}^{-\frac{{\rm w}_i \cdot (  {\bm A}+ {\bm B} )}{2}-1}   \left( {\rm w}_i \cdot \widehat{\bm \varphi}+ \left(-  \frac{{\rm w}_i \cdot ( {\bm A} + {\bm B})}{2} -l -\frac{a}{2} \right) \epsilon  \right)  \nonumber \\
&=e^{  ( {\bm A}+ {\bm B} ) \cdot \hat{\bm b} }
\left[{\rm w}_i \cdot \widehat{\bm \varphi}-\frac{a}{2} \epsilon \right]^{{\rm w}_i \cdot {\bm B}} \,.
\label{eq:case21}
\end{aligned}
\end{equation}
With~\eqref{eq:cond3}, the left hand side of \eqref{eq:WWtrans1}  becomes
\begin{equation}
\begin{aligned}
& \text{Weyl transform of} \, \, e^{  ( {\bm A}+ {\bm B} ) \cdot {\bm b} }  
\prod_{l= -\frac{{\rm w}_i \cdot (  {\bm A}+{\bm B} )}{2}}^{\frac{{\rm w}_i \cdot (   {\bm B}-{\bm A} )}{2}-1}\left( {\rm w}_i \cdot {\bm \varphi}-l \epsilon-\frac{a}{2} \epsilon \right) \nonumber \\
&= 
e^{  ( {\bm A}+ {\bm B} ) \cdot \hat{\bm b} } 
\prod_{l= -\frac{{\rm w}_i \cdot (  {\bm A}+{\bm B} )}{2}}^{\frac{{\rm w}_i \cdot (   {\bm B}-{\bm A} )}{2}-1}\left( {\rm w}_i \cdot \widehat{\bm \varphi}+ \left(-  \frac{{\rm w}_i \cdot ( {\bm A} + {\bm B})}{2} -l -\frac{a}{2} \right) \epsilon  \right)  \nonumber \\
&=e^{  ( {\bm A}+ {\bm B} ) \cdot \hat{\bm b} }
\left[{\rm w}_i \cdot \widehat{\bm \varphi}-\frac{a}{2} \epsilon \right]^{{\rm w}_i \cdot {\bm B}} \,.
\label{eq:case22}
\end{aligned}
\end{equation}

Let us consider the third line on the right hand side of~\eqref{eq:WWtrans1}, {\it i.e.}, the case 
 $( {\rm w}_i \cdot {\bm A} )( {\rm w}_i \cdot {\bm B} ) \ge 0$. In this case ${\sf F}_a ( {\rm w}_i, {\bm m}; {\bm A}, {\bm B}) $ is simplified to   
\begin{equation}
\begin{aligned}
  {\sf F}_a ( {\rm w}_i, {\bm m}; {\bm A}, {\bm B})  &=1 \,.
\end{aligned}
\end{equation}

Putting everything together we obtain equation~(\ref{eq:WWtrans1}).

\bibliography{refs}

\end{document}